\documentclass[12pt]{report}
\usepackage[utf8]{inputenc}
\usepackage{graphicx}
\usepackage{comment}
\graphicspath{ {images/} }
\usepackage[]{amsmath}
\usepackage[]{amssymb}
\usepackage{amsthm}
\usepackage{mathtools}
\usepackage[a4paper, total={6in, 9in}]{geometry}
\usepackage[customcolors,norndcorners]{hf-tikz}
\usepackage{empheq}
\usepackage{physics}
\usepackage{multirow}
\usepackage{stackengine}
\usepackage{pdfpages}
\usepackage{tikz}

\usepackage{tocloft}

\usepackage{hyperref}

\usepackage{braket} 
\usepackage{changepage} 
\counterwithout{footnote}{chapter} 
\usepackage{enumitem} 

\usetikzlibrary{decorations.pathmorphing,arrows.meta}

\usetikzlibrary{arrows.meta,positioning,fit,backgrounds,shapes.geometric,shapes.misc}

\newsavebox{\tempbox}

\newcommand{\textbox}[1]
{\savebox{\tempbox}{#1}
 \ifdim\wd\tempbox<4cm\relax
   \makebox[4cm]{\usebox{\tempbox}}%
 \else
   \parbox{4cm}{\raggedright #1}%
 \fi}

\newtheorem{theorem}{Theorem}[section]
\newtheorem{corollary}{Corollary}[theorem]
\newtheorem{lemma}[theorem]{Lemma}
\newtheorem{proposition}[theorem]{Proposition}
\newtheorem{conjecture}{Conjecture}[section]
\newtheorem{corollaryconj}{Corollary}[conjecture]
\theoremstyle{definition}
\newtheorem{definition}{Definition}[section]
\theoremstyle{remark}

\usepackage{subcaption} 
\usepackage{bm} 

\usepackage{multicol,color,longtable}
\definecolor{darkred}{rgb}{0.65,0.15,0}
\hypersetup{pdfborder={0 0 0},colorlinks=true,urlcolor=blue,citecolor=blue,linkcolor=darkred,linktocpage=true}

\usepackage{cite}

\usepackage{tikz}
\usetikzlibrary{arrows}
\usepackage{pgfplots}
\pgfplotsset{compat=1.17}
\usepackage{mathrsfs}
\usepackage[Symbol]{upgreek}
\usepackage{dsfont}
\usepackage[vcentermath]{youngtab}
\usepackage{textcomp}
\usepackage{caption}

\usepackage{latexsym}

\newcommand{\be}{\begin{equation}}
\newcommand{\ee}{\end{equation}}
\newcommand{\dis}{\displaystyle}

\newcommand{\Eq}[1]{Equation~\eqref{#1}}
\newcommand{\Eqs}[1]{Equations~\eqref{#1}}

\newcommand{\Fig}[1]{Figure~\ref{#1}}
\newcommand{\Figs}[1]{Figures~\ref{#1}}
\DeclareMathOperator{\A}{Area}
\DeclareMathOperator{\hA}{\widehat\A}

\newcommand{\esps}{\phantom{\!\!\!\overset{|}{a}}}
\newcommand{\esp}{\phantom{\!\!\overset{\displaystyle |}{|}}}
\newcommand{\espD}{\phantom{\!\!\underset{\displaystyle |}{\cdot}}}

\newcommand{\R}{\mathbb{R}}

\begin{document}

\addtocounter{section}{0} \setcounter{part}{0}
\renewcommand\thepart{\arabic{part}}
\numberwithin{equation}{section}

\includepdf[pages=-]{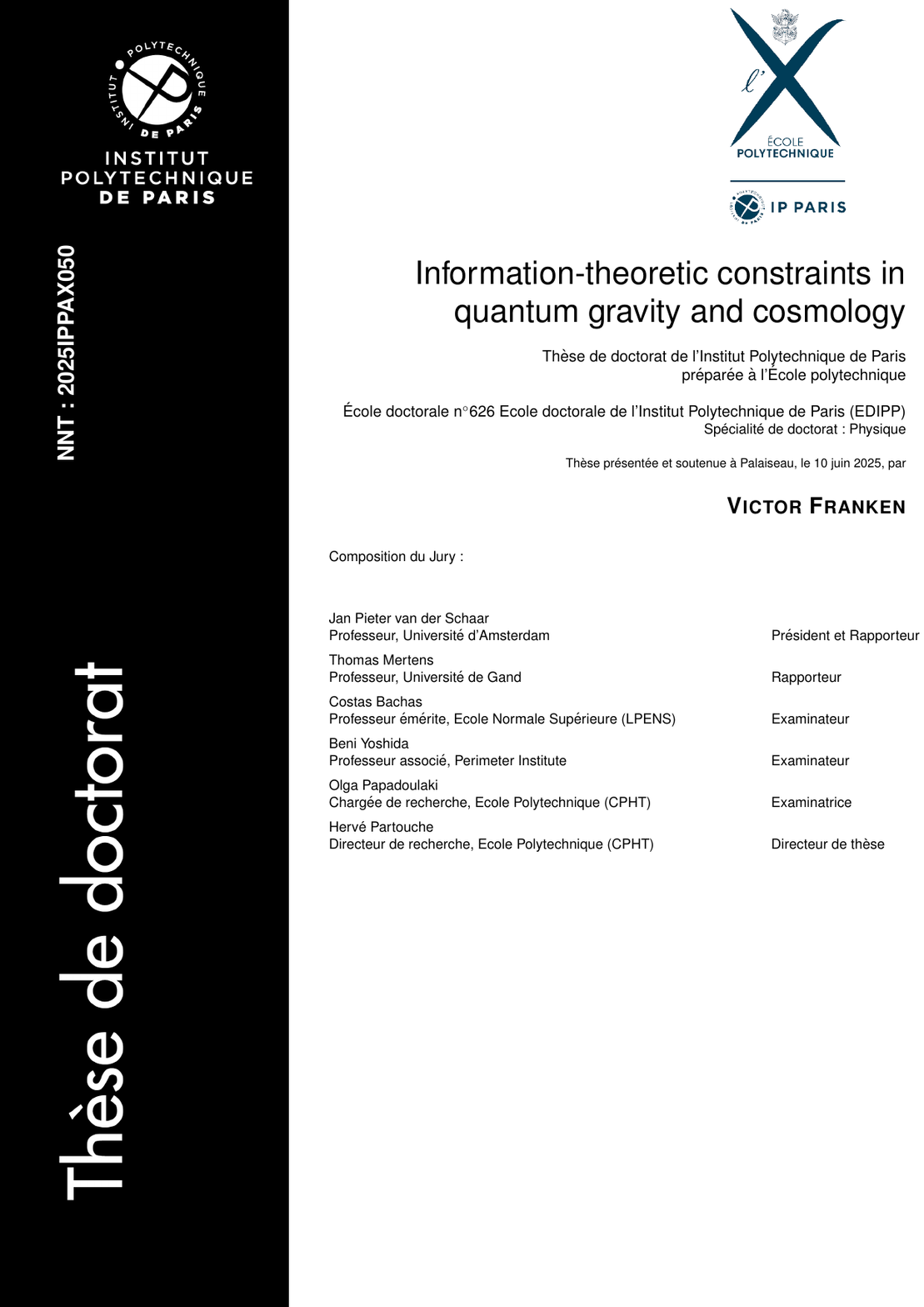}

\newpage

\thispagestyle{empty} 
\

\newpage

\begin{titlepage}
	\centering
	{\LARGE \textsc{Ecole Polytechnique}\par}
	\vspace{0.5cm}
	{\Large \textsc{PhD dissertation}\par}
	\vspace{1cm}
	{\huge\bfseries Information-theoretic constraints in quantum gravity and cosmology \par}
	\vspace{2cm}
	\par\vspace{1cm}
	{\Large Victor Franken\par}
	\vfill
	supervised by\par
	Hervé \textsc{Partouche}

	\vfill

	{\large June, 2025\par}
\end{titlepage}

\newpage
\thispagestyle{empty} 
\

\setcounter{page}{2}
\newpage

\chapter*{Acknowledgements}

Writing a thesis is, in many ways, an inward-looking process. So I am especially happy to begin this final milestone, the manuscript itself, by shining a light on all the people who made this journey possible.

First of all, I am grateful to Hervé Partouche for his thoughtful and dedicated guidance throughout my PhD journey. The first time we met, he told me that I could call him at any time of the day or night. He did not lie. He was always available to chat, whether it was to dig into some obscure intuition of mine or to advise me on my career choices. Thank you for teaching me the importance of patience, relentlessness, and humility in research, while giving me the confidence to explore my own research style and interests.

I had the immense good fortune of finding a second mentor in Nick Toumbas. Taking my first steps in research with Hervé and Nick has been a tremendous opportunity. I learned a lot from Nick's physical intuition and creative research style. Beyond our collaborations, a part of this thesis is based on works that were initiated in the US, thanks to Nick's invitation to join him. Thank you for your generosity and for the cherished memories of my visits to Cyprus.

I would like to express my gratitude to François Rondeau, who has been an amazing academic big brother, collaborator, and friend. I benefited from your experience as a young postdoc while learning together the process of independent research. We shared periods of deadlock and doubt, late nights on Skype discussing physics with the loudest background noise I have ever experienced, and exciting eureka moments.

I am grateful to my other collaborators Takato Mori, Arvin Shahbazi-Moghaddam, Sami Kaya, Patrick Tran, Lior Benizri, and Eftychios Kaimakkamis. I crossed paths with Takato at UC Berkeley by chance, and we laid down the main idea of our paper in just a few hours. This was quite an exciting first meeting which evolved into a great collaboration. I am inspired by his creativity, coupled with his remarkably broad knowledge of physics. During the same period, I was particularly inspired by Arvin's approach to research, always emphasizing the broad picture and fundamental lessons behind any specific problem. From the first day of my Bachelor studies, my academic destiny has been closely tied to Lior Benizri. Trying to keep up with him has made me a better physicist, and he inspired me to dare to be ambitious and excel as a physicist. 

I would like to acknowledge the "rapporteurs" of this PhD thesis, Thomas Mertens and Jan Pieter van der Schaar for their review of this manuscript. A special thanks to Thomas, who trusted me enough to soon welcome me in Ghent as a postdoc. I also thank Costas Bachas, Olga Papadoulaki, and Beni Yoshida for agreeing to evaluate this work.

I have been inspired by many professors, whose impact on me cannot be measured. In particular, I would like to recognize Emmanuel Thiran, for initiating and encouraging my desire to do research. His dedication to education and his contagious enthusiasm for physics sparked the interest of many future physicists, including me. Preparing for the physics olympiads and discussing dark matter with him has been a key part of my orientation towards physics. I am also happy to thank Christine Gaublomme for being such an amazing math teacher. The combination of her visible enjoyment of mathematics and her high expectations truly inspired me.

I had the privilege of meeting great researchers. For this, I would like to thank the Institute of Physics at the University of Amsterdam for their hospitality and in particular Jan Pieter van der Schaar and Cynthia Keeler for their interest in my research. Thanks to Marija Tomašević for inviting me and for her encouragement. I am grateful to Raphael Bousso and the Berkeley Center for Theoretical Physics for their hospitality, and in particular to Marco for making this visit possible. I would also like to thank François, Nick, and the department of physics at Cyprus University for making all of my visits very pleasant. Thanks to Dionysios Anninos for my short visit at King's College and for sharing his knowledge and vision of de Sitter space. I thank Glenn Barnich for kindly inviting me to Brussels. I am thankful for the great time I had with Thomas Mertens and his group in Ghent. Thanks to Sergio Aguilar-Gutierrez for our discussions and for inviting me to share my research with his group. My visit to Perimeter Institute was very enriching, thanks to Takato Mori, Rob Myers, Beni Yoshida, Céline Zwikel, Luca Ciambelli, and everyone I interacted with. I thank Lenny Susskind for inspiring me and sharing his thoughts about de Sitter space with Nick and me at Stanford.

My PhD experience would have been much less enjoyable without the supportive environment at CPHT. I thank Jean-René Chazotte and the amazing secretaries Florence Auger, Fadila Debbou, and Malika Lang for their constant support. I am grateful to Stéphane Munier for being on my doctoral committee and for supporting me in my master applications. A special thanks to Emilian Dudas for having been an amazing supervisor during my first year of master, for his kind advice and support, and for chairing my doctoral committee. I also thank the members of the string group:  Blaise, Emilian, Guillaume, Marios, and Olga for the friendly environment. I am thankful to Manuel Joffre for offering me the opportunity of tutoring his quantum mechanics class.

I want to thank Adrien L. and Matthieu for welcoming me in their office and accompanying me throughout almost all my PhD experience. The more than one-hour commuting to Palaiseau got the better of me many times, but their presence has been my main motivation for facing the RER and Lozère stairs during these three years. Thanks to the three Carrolians that have been amazing office mates during parts of my journey: David, Simon, and Adrien F. Thanks to Filippo, Gabriele, and Mikel for their kindness and for the nice moments in and out of the lab. Thanks to Magali for the laughs and Bueno expeditions. Thanks to Arthur, Clément, Fanny, Owen, Maddalena, Mathieu, Pierre, Thomas, and Victor for the great discussions and memories.

Thankfully, my life has not been all about physics during these past three years, thanks to some amazing friends. They have been present in the toughest and the happiest moments of my life, and believe in me more than I do. I list them here in the order they walked into my life. Thanks to Neil for being my oldest friend. We shared our first beer together. I hope we will share the last one, and many in between. Thanks to Charles for being as mad as me and for our precious walks in Paris at night. Thanks to Lior for our amazing complementarity, of brain and heart. Thanks to Simon B. for the great years in ULB and making me feel like I had never left Brussels. Thanks to Piet for our subtle mix of nerdiness and nights out. Thanks to Nico for remaining such a dear friend, wherever we live. Thanks to Chiara for our timeless tea times and making it so easy to be myself. Thanks to Noé for inspiring me in many ways and for creating together the most genuine memories. Thanks to Madeleine, Cedric, Abdel, Simon, and Rodin for making my first year at FBL one of the most unforgettable years of my life. Thanks to Oscar for being the secret ingredient to the greatest trips. I am grateful that fate brought us together, even a decade late. Thanks to Jeanne for her biting wit and our climbing sessions. Thanks to Juliette for our parallel universe and making me love theater. Thanks to Vi for our hopeless and fun duo. I would also like to thank Emile L. and Michelle for our climbing sessions and outings, Michael, Pablo, and Pan for the great physics and beers nights, Danaé, Emile E., and Neal for the good times and their support, and all those who have been a support and a source of happiness at one time or another in my life.     Finally, I want to thank Amelle for being an extraordinary partner. You have been my greatest support and always knew when I needed a little push to get through difficult moments. You gave me the confidence to pursue my goals. Thank you for your genuineness which makes every moment unique.

Des mots ne peuvent exprimer l'ampleur de ma reconnaissance pour ce qui constitue probablement les deux choses les plus précieuses que l'on puisse souhaiter: Un environnement sûr et heureux dans lequel grandir et un soutien inconditionnel dans la poursuite de ses rêves. Je dois ces deux choses et plus encore à ma famille. Merci à mes grands-parents pour leur soutien inconditionnel. En particulier, merci à ma grand-mère Nicole pour m'avoir transmis sa curiosité et sa combativité. Merci à ma grand-mère Monique pour m'avoir partagé son goût de l'esthétisme et des mots. Merci à mon grand-père Michel pour m'avoir communiqué sa passion de la voile et des sciences. Merci à mon grand-père Charles pour m'avoir légué son amour de la transmission et de la musique. Merci à toute ma famille proche pour leur présence et leur soutien, et en particulier à mon cousin Louis en qui je vois un frère. Merci à ma soeur Ariane pour avoir accompagné toutes les étapes de ma vie, et pour m'avoir servi d'example  par toutes ses qualités que je n'ai pas. Enfin, merci à mes parents Murielle et Laurent pour avoir fait de moi la personne que je suis aujourd'hui. Il est difficile de s'élever au-dessus de la confiance que nos parents nous transmettent. Merci d'avoir placé cette limite si haut que mon seul objectif ne puisse être que d'en être à la hauteur.

\vspace*{\fill}
\newpage

\thispagestyle{empty} 
\vspace*{\fill}
\begin{center}
\textit{\`{A} mes grands-parents,}
\end{center}
\vspace*{\fill}

\newpage

\

\newpage

\chapter*{Abstract}

Despite the success of the holographic principle in Anti-de Sitter (AdS) spacetime, an analogous holographic description of expanding spacetimes, such as our own, remains elusive. Resolving this problem would be a crucial step towards a deeper understanding of the early history of our universe.

In parallel, key progress has been made through the discovery of the connection between quantum information and the geometry of spacetime. Stemming from black hole thermodynamics, this idea has shed light on fundamental problems in the description of black holes and the AdS/CFT correspondence. In this thesis, we present new results on quantum information constraints in gravity that are relevant to cosmological models and demonstrate how this approach sheds light on cosmological holography.

A remarkable example of this framework is Bousso's conjecture that the number of degrees of freedom necessary for the description of a spatial region is bounded by the area of its boundary divided by $4G\hbar$, with $G$ being Newton's constant and $\hbar$ being Planck's constant. As the Bousso bound can be violated by quantum effects, it has been proposed that one should replace $\text{Area}/4G\hbar$ by the generalized entropy $S_{\rm gen}= \text{Area}/4G\hbar + S_{\rm out}$, where $S_{\rm out}$ is the entropy of quantum fields outside the spatial region, leading to a quantum Bousso bound. Another crucial conjecture is quantum focusing, which generalizes the classical property that light must focus in the presence of matter. We show that Jackiw-Teitelboim gravity—a two-dimensional toy model of quantum gravity—can be used to study the validity of these conjectures. In particular, we prove the quantum Bousso bound in this framework, as well as a weakened version of quantum focusing that suffices for all of its applications. Interestingly, we find violations of the unrestricted formulation of quantum focusing in this model.

A crucial implication of the Bousso bound is the static patch holography conjecture, where the causal region of an observer in de Sitter (dS) space is fully described by a dual theory living on the cosmological horizon. We propose a generalization of this conjecture to a broad class of closed Friedmann-Lemaître-Robertson-Walker spacetimes. We introduce a covariant prescription for computing holographic entanglement entropy in this holographic framework, leading to a formulation of subregion-subregion duality. In particular, we argue that entangling the holographic theories associated with complementary observers gives rise to the emergence of the region connecting their causal patches.

Finally, we discuss the connected wedge theorem, which establishes a relation between the causal structure of spacetime and information-theoretic constraints in the holographic dual. We argue that maintaining the consistency of static patch holography with this theorem leads to constraints on the causal structure of the dual theory. Our results suggest a novel relationship between static patch holography and the dS/CFT correspondence, an alternative approach to cosmological holography.

\chapter*{Résumé en français}

Malgré le succès du principe holographique en espace Anti-de Sitter (AdS), une description holographique d'un espace-temps en expansion comme le nôtre continue d'échapper à notre compréhension. Résoudre ce problème constituerait une étape cruciale vers une compréhension approfondie des premiers instants de notre univers.

En parallèle, des avancées majeures ont été réalisées grâce à la découverte du lien entre l’information quantique et la géométrie de l’espace-temps. Issue de la thermodynamique des trous noirs, cette idée a contribué à résoudre des problèmes fondamentaux dans la description des trous noirs ainsi que dans la correspondance AdS/CFT. Dans cette thèse, nous présentons des résultats concernant les contraintes imposées par l’information quantique en gravité et montrons comment celles-ci permettent de mieux comprendre l’holographie cosmologique.

Un exemple frappant de cette approche est la conjecture de Bousso, selon laquelle le nombre de degrés de liberté nécessaires à la description d’une région spatiale est limité par l’aire de sa frontière divisée par $4G\hbar$, où $G$ est la constante de Newton et $\hbar$ la constante de Planck. Cette borne de Bousso peut être violée par des effets quantiques, ce qui conduit à la proposition de remplacer $\mathrm{Aire}/4G\hbar$ par l’entropie généralisée $S_{\rm gen}=\mathrm{Aire}/4G\hbar+S_{\rm out}$, où $S_{\rm out}$ est l’entropie des champs quantiques situés à l’extérieur de la région spatiale considérée, menant ainsi à une borne de Bousso quantique. Une autre conjecture essentielle est le principe de focalisation quantique, qui généralise la propriété classique selon laquelle la lumière ne peut que se focaliser en présence de matière. Nous montrons que la gravité de Jackiw–Teitelboim (JT), un modèle simplifié de gravité quantique en deux dimensions, peut être utilisée pour tester la validité de ces conjectures. En particulier, nous démontrons la borne de Bousso quantique dans ce modèle, ainsi qu'une version restreinte du principe de focalisation quantique, suffisante pour toutes ses applications. Remarquablement, nous trouvons des violations de la formulation non restreinte du principe de focalisation quantique dans des modèles JT d’espaces Anti-de Sitter et de Sitter.

Une implication essentielle de la borne de Bousso est la conjecture holographique du patch statique, selon laquelle la région causale d’un observateur en espace de de Sitter (dS) est entièrement décrite par une théorie duale vivant sur l’horizon cosmologique. Nous proposons une généralisation de cette conjecture à une large classe d’espaces-temps fermés de Friedmann–Lemaître–Robertson–Walker. Nous introduisons une prescription covariante pour calculer l’entropie holographique d’intrication dans ce cadre, conduisant à une formulation de la dualité sous-région/sous-région. Celle-ci résout notamment des problèmes d’existence soulevés par les propositions existantes dans la littérature. Finalement, nous soutenons que l’intrication des théories holographiques associées à des observateurs complémentaires entraîne l’émergence de la région connectant leurs régions causales, de façon analogue à l’exemple bien connu du trou noir en espace Anti-de Sitter.

Enfin, nous discutons du théorème de la région connexe (" connected wedge theorem "), qui établit une relation entre la structure causale de l’espace-temps et des contraintes issues de la théorie de l’information dans le dual holographique. Ce résultat a été prouvé dans le contexte de la dualité AdS/CFT, mais peut aussi être démontré dans le cadre de l’information quantique selon des arguments qui ne dépendent pas de la géométrie de l’espace-temps. Nous soutenons que la cohérence de l’holographie du patch statique avec ce théorème impose des contraintes sur la structure causale de la théorie duale. Nos résultats suggèrent une nouvelle relation entre l’holographie du patch statique et la correspondance dS/CFT, qui constitue une approche alternative de l’holographie cosmologique.

\chapter*{List of publications}

\begin{itemize}
\item[\cite{Franken:2023pni}] V. Franken, H. Partouche, F. Rondeau, and N. Toumbas, \textit{Bridging the static patches: de Sitter holography and entanglement}, \textit{
JHEP} \textbf{08} (2023) 074.
\item[\cite{Franken:2023ugu}] V. Franken and F. Rondeau, \textit{On the quantum Bousso bound in JT gravity}, \textit{
JHEP} \textbf{03} (2024) 178.
\item[\cite{Franken:2023jas}] V. Franken, H. Partouche, F. Rondeau, and N. Toumbas, \textit{Closed FLRW holography: a time-dependent ER=EPR realization}, \textit{
JHEP} \textbf{05} (2024) 219.
\item[\cite{Franken:2024ruw}] V. Franken, \textit{de Sitter Connectivity from Holographic Entanglement}, Proceedings of Science \textbf{463} CORFU2023 (2024) 225. 
\item[\cite{Franken:2024wmh}] V. Franken and T. Mori, \textit{Horizon causality from holographic scattering in asymptotically dS$_3$}, \textit{
JHEP} \textbf{12} (2024) 119.
\end{itemize}
The following works were in preparation at the time of writing this manuscript.
\begin{itemize}
\item[\cite{QFC}] V. Franken, S. Kaya, F. Rondeau, A. Shahbazi-Moghaddam, and P. Tran,
\textit{Tests of restricted Quantum Focusing and a universal CFT bound}, arXiv:2510.13961.
\item V. Franken, E. Kaimakkamis, H. Partouche, and N. Toumbas, \textit{Universal Wheeler-DeWitt equation for flat minisuperspace models}, Work in progress.
\end{itemize}
This manuscript is based on \cite{Franken:2023pni,Franken:2023ugu,Franken:2023jas,Franken:2024ruw,Franken:2024wmh,QFC}.

\newpage
\thispagestyle{empty} 
\
\newpage

\setcounter{tocdepth}{1} 
\tableofcontents

\newpage

\chapter*{Preface}
\addcontentsline{toc}{part}{Preface} 

This PhD thesis summarizes four papers I contributed to over the past two and a half years, along with a forthcoming paper. By chance, these papers follow two different, albeit non-parallel, lines of research.

The first, rooted in black hole thermodynamics, seeks to establish fundamental laws in semiclassical gravity, drawing inspiration from quantum information theory. It aims to identify fundamental constraints on spacetime without relying on string theory or AdS/CFT.

The second line applies some of these tools to spacetimes where holography remains poorly understood, with the hope that these insights will advance our search for a holographic description of the universe.


Although connected, these areas of research differ in their motivations and methods. This diversity makes it challenging to unify these works into a single, self-contained document. A comprehensive review-style approach seems unsuitable, as it would favor one subject over the other. Nevertheless, I believe it is worthwhile to present a coherent approach that integrates all these projects meaningfully. After all, this is the approach I have followed.

I will try to present this coherent narrative, introducing these developments as a logical progression. Many important results from both lines of research will be omitted, so the present work should not be treated as a review of either of these two fields.

\chapter*{Introduction}\label{sec:intro}
\addcontentsline{toc}{part}{Introduction} 

Fundamental breakthroughs in physics often stem from a shift of language. When faced with a blind spot in our understanding or with incompatible existing theories, it sometimes becomes necessary to identify a new language that allows us to ask the right questions. While discovering the correct approach is rarely straightforward, unexplained patterns and guiding principles can illuminate these essential language shifts.

The development of differential geometry in the theory of gravitation provides a compelling example. Einstein first recognized fundamental principles such as the equivalence and relativity principles, leading him to abandon Euclidean geometry and the Newtonian concept of gravitational force. Instead, he and others found that the natural language to describe gravity is that of spacetime as a curved manifold. Similarly, the wave-like behavior of light led to the understanding that a fundamental description of matter required moving beyond the concept of a point-like particle. The language of waves and probability then enabled the formulation of postulates from which quantum mechanics emerges. Other examples include statistical mechanics for the description of gases, or group theory for particle physics.

In this thesis, we elaborate on this approach with quantum gravity and cosmology as our subject of study, and quantum information as our language. While gravity is the weakest interaction, it largely dominates at very large scales. Thus, spacetime geometry provides the best language for describing the universe at cosmic scales. Unfortunately, this description reaches its limits at very high energies—creating a critical issue for understanding the beginning of the universe. In its early moments, the prevailing energy scales must have led to strong quantum effects, both for matter and gravity. Moreover, insights from black hole physics suggest that quantum effects in gravity appear not only at very small scales, but also at macroscopic ones. Indeed, the radiation of black holes—whose wavelength can be very large—leads to fundamental paradoxes~\cite{PhysRevD.14.2460}. A theory of quantum gravity capable of solving them thus must have non-trivial effects at large scales.

As general relativity encounters its limits in regimes where spacetime is extremely curved, theoretical physicists are naturally led to study black holes. These objects serve as a window into quantum gravity. In particular, it has become clear that with the correct language, black holes in general relativity can reveal more about quantum gravity than expected from a low-energy effective theory. This language appears to be quantum information.

\section*{Quantum information and gravity}

The idea that quantum information plays a key role in gravitation dates back to Bekenstein and Hawking's realization that an entropy can be associated with black holes~\cite{Bekenstein1, Hawking:1975vcx}:
\begin{equation*}
    S_{\rm BH} = \frac{A_H}{4G},
\end{equation*}
where $A_H$ is the area of the horizon, and $G$ is Newton's gravitational constant. In this thesis, we use natural units where $c=\hbar=k_B=1$. General relativity does not include the usual framework needed to define entropy, such as partition functions or density matrices. However, Einstein equations lead to a set of thermodynamic laws~\cite{Bekenstein2, Bardeen:1973gs} for black holes, and Hawking showed that they evaporate by emitting a radiation which can be associated with $S_{\rm BH}$~\cite{Hawking:1975vcx}. Bekenstein and Hawking breakthrough has a surprising thermodynamic interpretation, called the central dogma: From the point of view of an external observer, a black hole can be described by a quantum system with $\exp(A_H/G)$ microstates. It was furthermore conjectured that the black hole entropy constitutes an upper bound for the information content of any spacetime region~\cite{PhysRevD.23.287}. Bekenstein suggested that the thermodynamic entropy of any spacetime region—that is, the number of degrees of freedom needed for describing it—is bounded from above by the area of its spatial boundary divided by $4G$. This statement is known as the Bekenstein bound.

Black hole thermodynamics has had immense consequences as it was the spark that ignited the holographic principle~\cite{tHooft:1993dmi, Susskind:1994vu}, which states that the quantum gravitational description of spacetime has an equivalent formulation in terms of a lower-dimensional quantum theory. The holography conjecture is realized in the context of string theory with the AdS/CFT correspondence~\cite{Maldacena:1997re, Gubser:1998bc, Witten:1998qj}, where anti-de Sitter (AdS) spacetime is dual to a conformal field theory (CFT) living on its asymptotic boundary. This has led to many theoretical successes concerning the description of quantum field theory (QFT) systems (and conformal field theories in particular) at strong coupling and quantum gravity. In particular, the AdS/CFT correspondence makes the connection between entropy and spacetime sharper, through a well defined prescription for the calculation of the Von Neumann entropy of gravitational systems. This formula relates the entropy of subsystems of the CFT to the area of extremal surfaces in the bulk of AdS~\cite{Ryu:2006bv,Hubeny:2007xt}. The computation of information-theoretic quantities in the bulk, such as entropy, also leads to an identification of the bulk region encoded in a given region of the CFT~\cite{Czech:2012bh,Hubeny:2012wa,Wall:2012uf}, which has been used to resolve the black hole information paradox~\cite{Penington:2019npb,Almheiri:2019psf}. These developments are part of a broader program which explores the fundamental structure of spacetime through the lens of quantum information~\cite{VanRaamsdonk:2009ar,VanRaamsdonk:2010pw,Maldacena:2013xja}, and which can be summarized with the slogan ``Entanglement builds bridges'' or ``ER=EPR''.\footnote{The ``ER'' stands for Einstein-Rosen bridge \cite{PhysRev.48.73} thus referring to a bridge in spacetime. The ``EPR'' stands for the Einstein-Podolsky-Rosen paradox \cite{PhysRev.47.777} that highlights the physical consequences of entanglement in quantum mechanics.}

While the Bekenstein-Hawking formula has been derived in string theory~\cite{Strominger:1996sh}, and holographic entanglement entropy was developed in AdS/CFT, there have been ongoing efforts in formulating information-theoretic constraints in semiclassical gravity. The importance of this approach extends beyond the peculiar AdS background, and it is worthwhile to formulate background-independent statements inspired by the Bekenstein-Hawking formula. The Bekenstein bound, which is connected to the holographic principle, is known to be violated in some contexts, due to the ambiguity of its formulation. In~\cite{Bousso:1999dw}, Bousso suggested the covariant entropy bound, or Bousso bound, which states that for a contracting null hypersurface $\mathcal{L}(\sigma)$ emanating orthogonally from a codimension-two spacelike surface $\sigma$, the following inequality is satisfied:
\begin{equation*}
    S(\mathcal{L}(\sigma)) \leq \frac{\text{Area}(\sigma)}{4G},
\end{equation*}
where $S(\mathcal{L}(\sigma))$ is the entropy on $\mathcal{L}(\sigma)$. This bound has been proven in the hydrodynamic regime, where $S(\mathcal{L}(\sigma))$ is well approximated by an integral over a local entropy density~\cite{Flanagan:1999jp,Bousso:2003kb}. Under appropriate conditions, the Bousso bound reproduces the Bekenstein bound and formulates a generalized version of the holographic principle~\cite{Bousso:2002ju}. Nevertheless, its validity depends on the Null Energy Condition (NEC) 
\begin{equation*}
    k^{\mu}k^{\nu}T_{\mu\nu}\geq 0,
\end{equation*}
where $k^{\mu}$ is a null vector and $T_{\mu\nu}$ is the stress energy tensor. The NEC can be violated by quantum effects~\cite{Epstein:1965zza,Lowe:1999xk}, which motivates the search for entropy and energy bounds robust to quantum fluctuations.

An effective strategy for producing entropy constraints in semiclassical gravity is to start with a classical statement such as Hawking's area theorem $dA_H\geq 0$, that is valid under the NEC, and replace the Bekenstein-Hawking entropy with the generalized entropy
\begin{equation*}
    S_{\rm gen} = \frac{\text{Area}}{4G} + S_{\rm out},
\end{equation*}
where $S_{\rm out}$ is the entropy of matter fields outside of the surface. Applying this to the area law leads to the Generalized Second Law (GSL) \cite{Bekenstein1,Bekenstein2,Bekenstein3}
\begin{equation*}
    dS_{\rm gen}\geq 0,
\end{equation*}
which was recently shown to hold to all orders in perturbative gravity~\cite{Faulkner:2024gst,Kirklin:2024gyl}. The Bekenstein-Hawking entropy and bulk field entropy are divergent quantities, due to the renormalization of Newton's constant and UV divergences, respectively. On the other hand, it is natural to define constraints on generalized entropy as it is divergence free~\cite{Susskind:1994sm,Witten:2021unn,Chandrasekaran:2022cip}. Motivated by these observations, Strominger and Thompson~\cite{Strominger:2003br} conjectured that applying the modification $\text{Area}/4G \rightarrow S_{\rm gen}$ to the Bousso bound leads to a quantum statement that holds at the semiclassical level. The validity of their
conjecture  was proven in two vacuum states of the Russo–Susskind–Thorlacius (RST) model~\cite{Russo:1992ax}, a two-dimensional model of an evaporating black hole in AdS. 

Similarly, it was conjectured in~\cite{Bousso:2015mna} that promoting the area of surfaces to their generalized entropy in the focusing theorem of general relativity—which states that the second derivative of the area along a null hypersurface must be negative—leads to a fundamental constraint of semiclassical gravity. The resulting Quantum Focusing Conjecture (QFC) defines the quantum expansion $\Theta$ as a functional derivative of $S_{\rm gen}$ along a null congruence, and states that $\Theta$ cannot increase. The QFC appears to be the strongest reasonable entropy bound in semiclassical gravity. No viable stronger bound is known, and it unifies a remarkable number of statements in QFT, semiclassical gravity, and holography. In particular, it entails the Bekenstein bound, the focusing theorem, the generalized second law, and a generalization of the singularity theorem~\cite{Bousso:2022tdb}. It also predicted the Quantum Null Energy Condition (QNEC) in flat space 
\begin{equation*}
    k^{\mu}k^{\nu}T_{\mu\nu} \geq \frac{1}{2\pi\mathcal{A}}S_{\rm out}'',
\end{equation*}
where $\mathcal{A}$ it the area element spanned by neighboring lightrays, which was later proven in quantum field theory~\cite{Bousso:2015wca,Balakrishnan:2017bjg,Koeller:2015qmn,Ceyhan:2018zfg,Kudler-Flam:2023hkl}. The QFC is also a fundamental assumption for the definition of subregion-subregion duality in AdS/CFT~\cite{Engelhardt:2014gca,Akers:2020pmf,Engelhardt:2021mue}, and it ensures its consistency with causality~\cite{Akers:2016ugt}. Moreover, it implies an alternative quantum generalization of the Bousso bound, that also relies on generalized entropy. Interestingly, Shahbazi-Moghaddam \cite{Shahbazi-Moghaddam:2022hbw} showed that a weaker statement is sufficient for all of these applications of the QFC, and proved it in braneworld holography~\cite{Shahbazi-Moghaddam:2022hbw}. This raises the question of whether the original QFC is a general statement, or if QFC violations that preserve the restricted QFC might be found.

\begin{figure}[ht]
    \centering

\begin{tikzpicture}[
    box/.style={draw, thick, minimum width=3.7cm, minimum height=1cm, align=center},
    colored box/.style={box, fill=orange!20},
    citation/.style={blue, font=\small},
    arrow/.style={->, >=latex, thick},
    scale=0.9
]

\node[box] (BH) at (0,0) {Black hole thermodynamics};

\node[box] (SBH) at (-6,-3) {$S_{\text{BH}} = \frac{A_{\text{H}}}{4G}$};
\node[box] (Entropy) at (0,-3) {Entropy bounds};
\node[box] (Sgen) at (6,-3) {$S_{\text{gen}} = \frac{A_{\text{H}}}{4G} + S_{\text{out}}$};

\node[box] (Holographic) at (-6,-6) {Holographic principle};
\node[colored box] (QuantumEntropy) at (3,-6) {Quantum entropy bounds};
\node[colored box] (QFC) at (6,-9) {QFC};
\node[box] (QNEC) at (6,-12) {QNEC};

\node[colored box] (QuantumCosmo) at (-6,-9) {Quantum cosmology};
\node[box] (AdSCFT) at (0,-9) {AdS/CFT};
\node[box] (QT) at (0,-12) {Quantum tasks};
\node[box] (QFT) at (6,-15) {QFT};

\node[box] (Algebraic) at (0,-15) {Algebraic approach};

\draw[arrow] (BH) -- node[citation, above, sloped] {\cite{Bekenstein1, Hawking:1975vcx}} (SBH);
\draw[arrow,dashed] (BH) -- node[citation, left] {\cite{PhysRevD.23.287}} (Entropy);
\draw[arrow,dashed] (BH) -- node[citation, above, sloped] {\cite{Bekenstein3}} (Sgen);

\draw[arrow,dashed] (SBH) -- node[citation, left] {\cite{tHooft:1993dmi, Susskind:1994vu}} (Holographic);
\draw[arrow] (Entropy) -- node[citation, above ,sloped, pos=0.5] {\cite{Bousso:2002ju}} (Holographic);
\draw[arrow,dashed] (Entropy) -- node[citation, above, sloped] {\cite{Strominger:2003br,
Bousso:2015mna}} (QuantumEntropy);

\draw[arrow,dashed] (Holographic) -- node[citation, left] {\cite{Susskind:2021omt}} (QuantumCosmo);
\draw[arrow,dashed] (Holographic) -- node[citation, above, sloped] {\cite{Maldacena:1997re}} (AdSCFT);

\draw[arrow] (QFC) -- node[citation, below, sloped , pos=0.5] {\cite{Bousso:2015mna}} (QuantumEntropy);
\draw[arrow] (QFC) -- node[citation, right] {\cite{Bousso:2015mna}} (QNEC);

\draw[arrow] (QFC) -- node[citation, above, pos=0.5] {\cite{Engelhardt:2014gca,Akers:2020pmf,Engelhardt:2021mue,Akers:2016ugt}} (AdSCFT);

\draw[arrow] (QT) -- node[citation, left, pos=0.5] {\cite{May:2019odp}} (AdSCFT);
\draw[arrow] (QFC) -- node[citation, above, sloped, pos=0.5] {\cite{May:2019yxi,May:2024epy}} (QT);
\draw[arrow] (Algebraic) -- node[citation,above, sloped, pos=0.5] {\cite{Balakrishnan:2017bjg}} (QNEC);

\draw[arrow] (QT) -- node[citation, above, sloped, pos=0.5] {\cite{Franken:2024wmh}} (QuantumCosmo);
\draw[arrow] (Algebraic) -- node[citation, above, sloped,  pos=0.5] {\cite{Chandrasekaran:2022cip}} (QuantumCosmo);

\draw[arrow] (Algebraic) -- (0,-16) -- (9,-16) --  (9,-3) -- (Sgen);
\draw[arrow,dashed] (Sgen) -- (9.3,-3) -- node[citation, right, pos=0.5] {\cite{Witten:2021unn,Chandrasekaran:2022eqq}}(9.3,-16.3) -- (0,-16.3) -- (Algebraic);

\draw[arrow,dashed] (Sgen) -- node[citation, right, pos=0.5] {\cite{Bousso:2015mna}} (QFC);

\draw[arrow] (QFT) -- node[citation, right, pos=0.5] {\cite{Bousso:2015wca}} (QNEC);
\draw[arrow] (QFT) -- node[citation, above, pos=0.5] {\cite{Haag:1992hx,Hollands:2009bke}} (Algebraic);

\end{tikzpicture}
    \caption{\footnotesize Diagrammatic picture of the (non-exhaustive) relations between different topics and statement in holography and semiclassical gravity mentioned in the introduction, with associated references. Full arrows denote a direct proof connecting the subjects or statements. Dashed arrows indicate that a statement or subject inspired or partially implies another. The shaded boxes correspond to the subjects to which this thesis contributes.}
    \label{fig:diagram}
\end{figure}
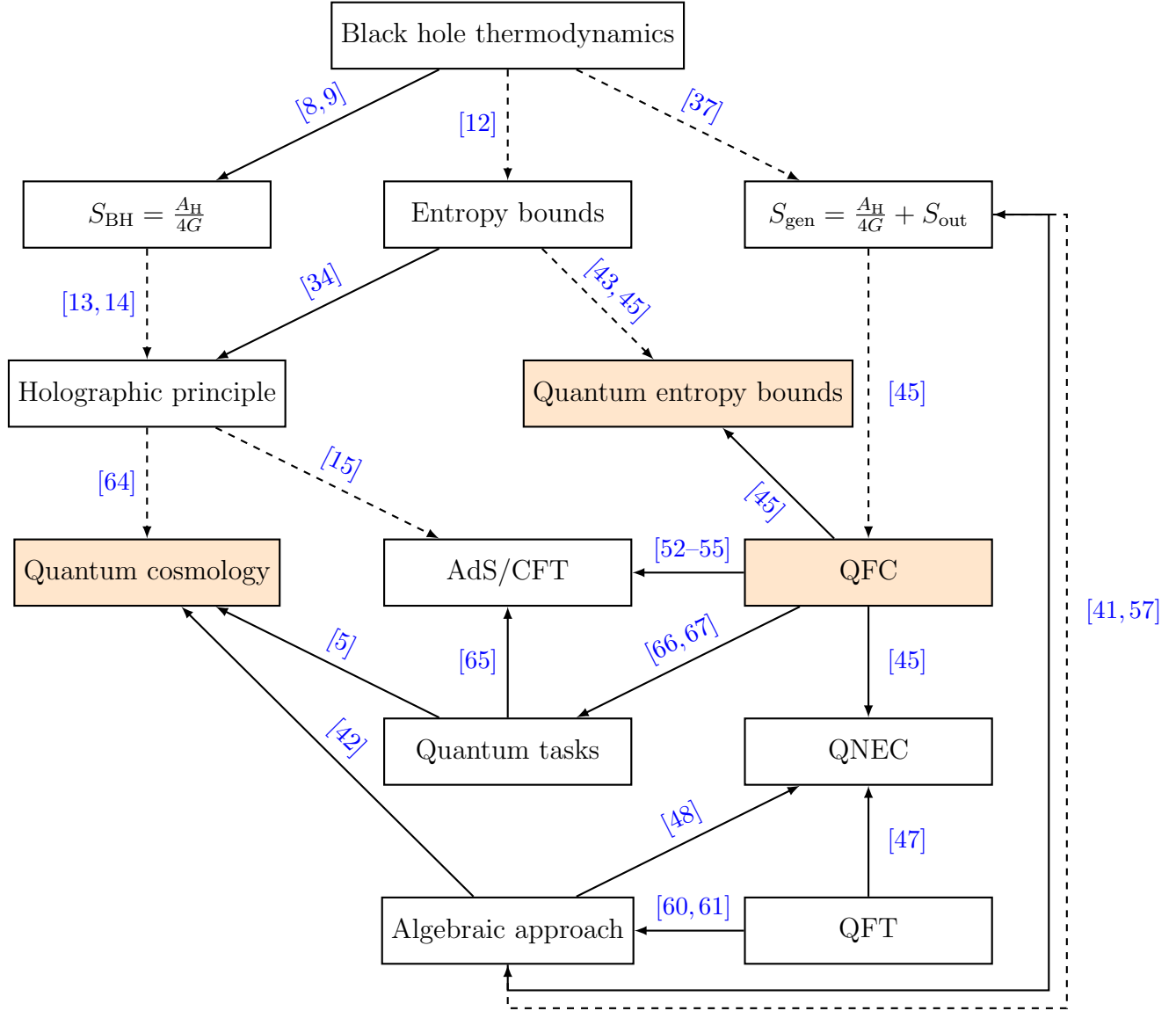

Historically, these past developments may be summarized as follows. Contradictions between the second law of thermodynamics and black hole physics led Bekenstein to postulate that the second law of thermodynamics should be a fundamental principle, leading to the formulation of the GSL which provides deep insights into quantum gravity. Subsequently, the natural appearance of generalized entropy in gravity without a fundamental explanation led to the postulate that generalized entropy should be a key concept in the quantum description of gravity. This postulate is now supported by many developments including those mentioned above. Although the connection between these results and AdS/CFT is well-established, they are background-independent. In particular, it is useful to explore possible generalizations of this connection in other backgrounds. Indeed, quantum gravity should not only apply to asymptotically AdS spacetimes, and one expects the usefulness of the QFC, entropy bounds, and other quantum information-inspired results to be relevant in non-AdS spacetimes. This idea is supported by the algebraic approach~\cite{Witten:2021unn,Chandrasekaran:2022cip,Chandrasekaran:2022eqq,Witten:2023qsv,Witten:2023xze}, inspired by the algebraic formulation of QFT~\cite{Haag:1992hx,Hollands:2009bke,Hollands:2014eia}, which suggests that generalized entropy is the natural notion of entropy in cosmological spacetimes~\cite{Chandrasekaran:2022cip}. Moreover, it can be used to prove the GSL \cite{Faulkner:2024gst,Kirklin:2024gyl} and energy constraints in semiclassical gravity \cite{Faulkner:2016mzt}. The connections between these many different subjects are depicted in the diagram of Figure~\ref{fig:diagram}.

\section*{Holography for cosmology}
In parallel to these developments, there has been compelling evidence that our universe has gone through two periods of accelerated expansion, namely the cosmic inflation in the early universe and the present period~\cite{SupernovaSearchTeam:1998fmf,SupernovaCosmologyProject:1998vns,Planck:2015fie}. Moreover, the current measured cosmological constant is $122$ orders of magnitude smaller than the QFT prediction \cite{Adler:1995vd}. A comprehensive understanding of both the early and present universe thus requires a quantum gravitational description, and extending the remarkable progress made in AdS/CFT to expanding universes remains one of the major challenges in modern high-energy physics. The simplest model of expanding spacetime in general relativity is the de Sitter (dS) universe, which is the maximally symmetric solution to Einstein's equations with a positive cosmological constant. The simplicity of this model and its large symmetry group make it a good starting point. Moreover, a cosmic no-hair theorem~\cite{PhysRevD.28.2118} states that homogeneous universes with positive cosmological constant evolve toward a de Sitter geometry. Unfortunately, it is famously hard to find de Sitter vacua in string theory~\cite{Danielsson:2018ztv}—see~\cite{Kachru:2003aw,McAllister:2024lnt} for developments in this direction—and no explicit holographic correspondence has been fully realized in de Sitter space.

Towards developing a quantum mechanical description of dS spacetime, dS holography emerges as a promising but enigmatic framework~\cite{Strominger:2001pn,Bousso:2001mw,Abdalla:2002hg,Alishahiha:2004md,Parikh:2004wh,Alishahiha:2005dj,McFadden:2009fg,Dong:2010pm,Anninos:2011ui,Anninos:2011af,Dong:2018cuv,Gorbenko:2018oov,Arias:2019pzy, Arias:2019zug, Lewkowycz:2019xse,Geng:2021wcq,Susskind:2021esx,Shaghoulian:2021cef,Susskind:2021omt,Coleman:2021nor,Susskind:2021dfc,Hikida:2022ltr,Svesko:2022txo,Levine:2022wos,Shaghoulian:2022fop,Lin:2022nss, Susskind:2022dfz, Susskind:2022bia,  Banihashemi:2022htw,Rahman:2022jsf,Goel:2023svz, Narovlansky:2023lfz,Susskind:2023rxm, Giveon:2023rsk, Franken:2023pni, Kawamoto:2023nki, Galante:2023uyf,Batra:2024kjl,Franken:2024wmh,Collier:2025lux}. One of its fundamental puzzles arises from the fact that dS spacetime is a closed universe, lacking a boundary in the traditional sense. Moreover, there is no global future-directed timelike Killing vector in this geometry. In other words, there is no global definition of time in de Sitter space \cite{Spradlin:2001pw}. Because of these problems, it is far from clear where the holographic degrees of freedom should be located, and whether such a holographic screen encodes the whole spacetime.

A natural extension of AdS/CFT to de Sitter spacetime is known as the dS/CFT correspondence~\cite{Strominger:2001pn, Bousso:2001mw, Balasubramanian:2001nb, Abdalla:2002hg,Anninos:2011ui}, which is achieved via analytical continuation. This approach situates the dual CFT at future or past null infinity, where conventional concepts of time and states are ill-defined. Consequently, standard concepts such as entanglement entropy, which are pivotal in understanding quantum gravity, become ambiguous. Instead of a meta-observer located at null infinity, one may consider a static observer following a worldline in the bulk. The timelike tube theorem~\cite{Borchers1961,osti_4665531,Strohmaier:2023opz,Witten:2023xze} ensures that such an observer is in principle capable of describing its entire causal region—its static patch, motivating an approach to de Sitter holography where the screen is a hypersurface very close to the observer's worldline \cite{Anninos:2011af,
Anninos:2017hhn,Anninos:2018svg,Coleman:2021nor,Blacker:2023oan,Loganayagam:2023pfb,Anninos:2024wpy}. See \cite{Alishahiha:2004md,Alishahiha:2005dj,Dong:2010pm,Dong:2018cuv,Gorbenko:2018oov,Geng:2021wcq} for another approach to de Sitter holography called the dS/dS correspondence.

A more recent approach builds on the Bousso bound, which supports the static patch holography conjecture~\cite{Susskind:2021omt,Susskind:2021dfc,Susskind:2021esx, Shaghoulian:2021cef,Shaghoulian:2022fop,Lin:2022nss,Susskind:2023hnj}. In this model, a holographic screen sits in the bulk near an observer's cosmological horizon and encodes the state of that observer's static patch. This approach provides a physical interpretation for the Gibbons-Hawking formula \cite{Gibbons:1977mu}, which associates an entropy to the cosmological horizon of de Sitter spacetime:
\begin{equation*}
    S_{\rm GH} = \frac{A_{\rm dS}}{4G},
\end{equation*}
where $A_{\rm dS}$ is the area of the de Sitter horizon. The thermodynamical interpretation of this formula is in general elusive, as the cosmological horizon is observer dependent. The Bekenstein-Hawking formula for black holes is naturally associated with an interpretation of the entropy as measuring our ignorance about the black hole interior. In the case of the cosmological horizon, it is not clear which side of the horizon is described by the Gibbons-Hawking entropy. In static patch holography however, this entropy is naturally identified with the entropy of the holographic screen located on the horizon. 

Besides the Bousso bound, static patch holography has been supported by the important role played by timelike boundaries in the definition of thermal ensembles in de Sitter space \cite{Banihashemi:2022htw,Banihashemi:2022jys}. Moreover, the apparently puzzling observer-dependence of the proposal may be supported by the recent realization that observers play a key role in the definition of observables in semiclassical gravity in de Sitter space \cite{Chandrasekaran:2022cip,Witten:2023xze}. Additionally, it has been understood that a CFT theory dual to AdS spacetime can be deformed to incorporate patches of dS into the AdS bulk, allowing for microstate counting derivations of the Gibbons-Hawking formula \cite{Shyam:2021ciy,Lewkowycz:2019xse,Coleman:2021nor,Batra:2024kjl,Aguilar-Gutierrez:2023zoi,Aguilar-Gutierrez:2024nst,Aguilar-Gutierrez:2024oea}. Nevertheless, the precise nature of a dual theory in pure de Sitter space remains uncertain.

While the double-scaled Sachdev-Ye-Kitaev (DSSYK) model—which has no spatial dimension—is one proposed candidate of holographic dual for two-dimensional de Sitter space~\cite{Susskind:2021esx,Lin:2022nss,Rahman:2022jsf,Goel:2023svz,Narovlansky:2023lfz,Verlinde:2024znh,Verlinde:2024zrh,Blommaert:2023opb, Blommaert:2023wad,Rahman:2024iiu,Milekhin:2023bjv,Xu:2024hoc,Milekhin:2024vbb}, it involves dimensional reduction, leaving the description of $(d+1)$-dimensional dS spacetimes for $d\ge 2$ unresolved. In particular, the dual theory in a higher-dimensional de Sitter spacetime is expected to be highly nonlocal, a feature that low-dimensional holography does not fully capture.

Locating holographic degrees of freedom on a hypersurface in the bulk has a cost: Fluctuations of the background are expected to backreact on the screen. It is far from clear how the static patch holography conjecture should be understood in strongly gravitating regimes. In this thesis, we focus on semiclassical considerations where the screen is located on a fixed hypersurface. 

\section*{Information-theoretic approach}

While the search for specific microscopic models like DSSYK continues, a parallel and complementary route is to leverage information-theoretic principles that should constrain any viable holographic theory. This approach takes inspiration from the AdS/CFT correspondence and attempts to find general principles that one would expect to hold in non-AdS holography. Our present discussion motivates an information-theoretic approach. In particular, extremal surfaces in AdS compute entanglement entropy and are at the core of subregion-subregion duality. Moreover, the QFC and other information-theoretic statements in semiclassical gravity dictate the dynamics of these surfaces, leading to non-trivial statements about de Sitter holographic entropy \cite{Franken:2024wmh}.

Even without prior knowledge of AdS/CFT, the consistency of extremal surface constructions with information-theoretic constraints could have suggested holographic entanglement entropy conjectures in AdS spacetimes. These constructions significantly constrain the possible entropy-like quantities in any holographic dual theory. In the specific case of AdS$_3$, one might have inferred the connection to conformal field theory by noting that extremal surface areas—which satisfy the Bekenstein-Hawking formula and all entropy inequalities—behave precisely like subsystem entropies in a CFT$_2$.

In static patch holography, the holographic screen's location in the bulk interior complicates the identification of the correct prescription for the holographic entanglement entropy, leading to various competing proposals~\cite{Susskind:2021esx,Shaghoulian:2021cef,Shaghoulian:2022fop, Franken:2023pni}. The two initial proposals are called the monolayer \cite{Susskind:2021esx} and bilayer \cite{Shaghoulian:2021cef} proposals. The setup consists of two holographic screens associated with two observers sitting at antipodal points of global slices of de Sitter space. In the bilayer prescription, the entanglement entropy of a dual subsystem is computed via an extremization problem in all three regions bounded by these screens. In the monolayer proposal on the other hand, the entropy receives contributions only from the exterior region between the two holographic screens. Discrepancies between the two approaches only appear for non-trivial subsystems and the problem of distinguishing them is tackled in \cite{Franken:2023pni}. A precise holographic entropy prescription opens the door to possible cross-checks with explicit models of dual theory, and to the study of spacetime emergence in static patch holography \cite{Franken:2023pni,Franken:2023jas}. At the same time, extensive work has been carried out on islands in de Sitter space, with the aim of calculating the entropy of spatial regions in de Sitter space, independently of the holographic screen~\cite{Chen:2020tes,Hartman:2020khs,Balasubramanian:2020xqf,Sybesma:2020fxg,Geng:2021wcq,Fallows:2021sge,Aalsma:2021bit,Langhoff:2021uct,Aguilar-Gutierrez:2021bns,Kames-King:2021etp,Aalsma:2022swk,Baek:2022ozg,Balasubramanian:2023xyd,Aguilar-Gutierrez:2023hls,Myers:2024zhb,Hao:2024nhd,Jiang:2024xnd,Jiang:2025hao}.

The emergence of spatial connectivity from entanglement has been extensively studied in AdS/CFT \cite{VanRaamsdonk:2009ar,VanRaamsdonk:2010pw,Pastawski:2015qua,Engelhardt:2023xer}. However, this approach neglects the non-trivial time evolution of spacetime. Specifically, one may wonder how the causal structure of spacetime emerges from the holographic quantum theory~\cite{May:2019yxi,May:2019odp,May:2021nrl, May:2022clu,Mori:2023swn,Caminiti:2024ctd,May:2024epy}. Let us consider a situation where we send information from a set of input points, process it, and share the outcome among output points.
This procedure, known as a relativistic quantum task, can have a local bulk realization but not necessarily on the boundary. Whether a background can locally perform the task is entirely determined by the causal structure and locations of input and output points. The connected wedge theorem~\cite{May:2019odp} in AdS$_3$/CFT$_2$ relates this causality statement to correlation, namely, that when a $2$-to-$2$ scattering from input and output points on the boundary is possible in the bulk, but not on the boundary, there must be $O(1/G)$ mutual information between certain boundary causal diamonds, which are fixed by the input and output locations and the boundary causality. The connected wedge theorem in AdS$_3$/CFT$_2$ has been proven both from the bulk, using general relativity~\cite{May:2019odp}, and from the boundary using quantum information without relying on the detailed nature of the bulk or boundary theory~\cite{May:2019yxi,May:2019odp}. This suggests that this theorem should be valid beyond the AdS/CFT correspondence. This opens the possibility for another application of quantum information as a constraining tool for holographic conjectures in a non-AdS holographic framework. For example, in \cite{Mori:2023swn}, the connected wedg theorem was generalized to AdS holography with a screen located on a cutoff surface inside the bulk.

\section*{Short summary and plan of the thesis}

This thesis is divided into three parts for clarity, although every Chapter builds on the previous ones. Chapter numbering is thus continued throughout the three parts.

In Part~\ref{part:preliminaries}, we review some background material that is used throughout this work. In Chapter~\ref{ch:entanglement}, we introduce the concept of entanglement entropy and thermodynamic entropy, which are presented as fine-grained and coarse-grained definitions, respectively. We highlight their basic properties. We briefly review the explicit formulas for computing entanglement entropy in a two-dimensional CFT on a curved background. In Chapter~\ref{ch:AdS/CFT}, we lay down the basic principles of geometric reconstruction in AdS/CFT. We present the Hubeny-Rangamani-Takayanagi (HRT) formula and introduce the idea that spatial connectivity in the bulk emerges from boundary entanglement. Chapter~\ref{ch:CWTAdS} reviews the connected wedge theorem mentioned earlier. In particular, we sketch the information-theoretic proof of this statement to argue that it must be valid in any holographic theory. Additionally, we briefly review the generalization of the theorem to braneworld holography.

In Part~\ref{part:QI}, we review and present new results related to information-theoretic constraints in semiclassical gravity. Chapter~\ref{ch:thermo} provides a pedagogical review of black hole entropy and the associated generalized second law, the focusing theorem of general relativity, the Bousso bound, and the holographic principle. We also review the hydrodynamic regime in which the Bousso bound has been proven, and interpret the assumptions made in this regime. In Chapter~\ref{ch:bounds}, we motivate the definition of generalized entropy and show how classical statements like focusing and the Bousso bound are modified when one replaces the area functional divided by $4G$ with $S_{\rm gen}$. In particular, we introduce the QFC and two conjectured generalizations of the Bousso bound due to Strominger and Thompson \cite{Strominger:2003br} and another due to Bousso, Fisher, Leichenauer, and Wall (BFLW) \cite{Bousso:2015mna}. We developed the significance of quantum focusing and introduce the restricted version of the QFC. We also present the QNEC as a corollary of the QFC.
Chapter~\ref{ch:JT} is mainly inspired by reference \cite{Franken:2023ugu}. We present a two-dimensional toy model of quantum gravity called Jackiw-Teitelboim (JT) gravity. In this model, the two-dimensional background is non-dynamical and the only backreacting quantity is a dilaton field, which plays the role of the size of a perpendicular dimension. The concept of generalized entropy can be defined in this model, and we motivate its physical relevance in JT gravity. In a toy model of de Sitter space, the background has two timelike boundaries. We present an entanglement entropy formula for finite two-dimensional backgrounds with two timelike boundaries that was derived in \cite{Franken:2023ugu}.

In Chapter~\ref{ch:proof2D}, we use JT gravity to probe the validity of conjectures in semiclassical gravity. Following \cite{Franken:2023ugu}, we prove the Strominger-Thompson and the BFLW quantum Bousso bounds in JT gravity coupled to conformal matter in any state. The proof of the Strominger-Thompson bound relies on the hydrodynamic regime and on the conformal symmetry of the matter sector. We derive a sufficient condition for the bound, which appears to be the QNEC. The QNEC can be proven using the conformal transformations of entropy and the stress tensor. The proof of the BFLW bound is analogous but does not rely on the hydrodynamic regime. Additionally, we discuss the connections between these two bounds, which were poorly understood in the previous literature. We argue that the Strominger-Thompson bound applies when matter can be treated in two independent sectors, one describing quantum matter and another that describes decoupled classical matter.

We then review one of the main results of \cite{QFC} and prove the restricted QFC in JT gravity. In particular, we show an upper bound for the first derivative of quantum expansion, which is non-zero in general but vanishes at any point where quantum expansion vanishes. Moreover, this upper bound is positive along null hypersurfaces that have classical and quantum expansions of opposite sign. This suggests that unrestricted quantum focusing may be violated in setups with strong quantum effects. We provide two explicit examples of violations in JT gravity coupled to a CFT. The first example is an extremal AdS black hole and the second is an evaporating de Sitter horizon. Our results strongly support the claim that unrestricted quantum focusing should not be seen as a fundamental constraint in quantum gravity. On the other hand, the restricted QFC gets additional support from our proof and its connection with boundary causality.

In Part~\ref{part:holography}, we apply some of the results introduced in Part~\ref{part:QI} to quantum cosmology. In Chapter~\ref{ch:SP}, using the Bousso bound and the island formula, we argue that a natural location for a holographic screen in de Sitter spacetime is the cosmological horizon. In this context, we introduce the monolayer \cite{Susskind:2021esx} and bilayer \cite{Shaghoulian:2021cef} proposals that have been conjectured to compute holographic entropy in static patch holography.
Chapter~\ref{ch:bridge} is mainly inspired by \cite{Franken:2023pni,Franken:2024ruw}. We formulate two covariant holographic entanglement entropy prescriptions inspired by the monolayer and bilayer proposals. We argue that a covariant version of the monolayer proposal leads to inconsistencies. On the other hand, the covariant formulation of the bilayer proposal is self-consistent and satisfies all the consistency checks we consider. Our covariant prescription includes quantum corrections due to bulk fields. We then argue that the bilayer proposal in its original form can lead to existence issues when considering non-trivial subsystems of the holographic screens. We suggest a modification of the definition of extremal surfaces, which takes into account bulk causality constraints. In particular, we explicitly impose that extremal surfaces must lie in the causal diamond of a spacelike slice bounded by the holographic screen, which leads to a constrained extremization problem. This can be defined by supplementing the area functional with Lagrange multipliers. We then apply our prescription to a single holographic screen and to the union of two complementary holographic screens. We argue that the bilayer proposal leads to a stronger holographic description, in which the full spacetime is encoded on two screens at the cosmological horizons. Analogously to the AdS double-sided black hole, global de Sitter space emerges from the entanglement between two disconnected holographic screens. Each of them encodes a static patch, and the inflationary region emerges from their entanglement. In other words, the connectivity of de Sitter space emerges from holographic entanglement.

Chapter~\ref{ch:FLRW} summarizes the results of \cite{Franken:2023jas}. We review the basics of closed Friedmann-Lemaître-Robertson-Walker (FLRW) cosmologies and then generalize the results of Chapter~\ref{ch:bridge} to a large class of bouncing and Big Bang/Big Crunch FLRW cosmologies. In the expanding phase, the two screens lie at the apparent horizons. In the contracting phase, there are an infinite number of possible trajectories for the holographic screens, which can be grouped into equivalence classes. In each class, the effective holographic theory can be derived from a pair of ``parent" screens on the apparent horizons. Some cosmologies escape our discussion,\footnote{These cosmologies correspond to geometries for which the Penrose diagram is wider than tall. This implies that one cannot find two worldlines such that the union of their causal patch contains at least one Cauchy slice.} and it is expected that two antipodal observers and their associated screens do not suffice to reconstruct these cosmologies. This picture entails a time-dependent realization of the ER=EPR conjecture, where an effective geometrical bridge connecting the screens via the minimal extremal surface emerges from entanglement. For the Big Crunch contracting cases, the screens disentangle and the geometrical bridge closes off when the minimal extremal trapped sphere hits the Big Crunch singularity at a finite time before the collapse of the Universe. Semiclassical, thermal corrections are incorporated in the cases of radiation-dominated cosmologies.

Chapter~\ref{ch:CWT} is based mainly on \cite{Franken:2024wmh}. We test the consistency of static patch holography and our entanglement entropy proposal with the connected wedge theorem. We examine scattering within the static patch of dS$_3$ spacetime. We show that associating bulk perturbations with a causal region constructed as a lightcone on the screen leads to violations of the theorem. To overcome this contradiction, we argue that the holographic theory defined on the screen should be highly nonlocal, such that bulk perturbations emanating from the screen are not associated with local operators. In particular, we show that maintaining consistency with the connected wedge theorem leads to a concept of screen causality which is induced from null infinities. Specifically, signals propagating in the static patch, encoded nonlocally on the screen, are associated with localized points at null infinity. This leads to apparently superluminal lightcones on the screen. We argue that this construction is minimal in at least a large number of non-trivial cases. Finally, we prove the connected wedge theorem in static patch holography using this notion of causality for asymptotically dS$_3$ spacetime, with a timelike holographic screen in the causal patch of an observer. Our results suggest a novel connection between static patch holography and the dS/CFT correspondence.

\part{Preliminaries\label{part:preliminaries}}
\addtocontents{toc}{\protect\begin{adjustwidth}{1cm}{0cm}}

\chapter{Entanglement entropy}
\label{ch:entanglement}

One of the main focuses of this PhD thesis is the entanglement structure of quantum and gravitational theories. A crucial aspect of this approach will be the study of entanglement entropy, a measure of correlations in quantum systems. In this chapter, we introduce some basic concepts and definitions that will be used throughout this work.

\section{von Neumann entropy}
\label{sec:EE}

In quantum mechanics, the state of a quantum system is described by a vector $\ket{\Psi}$ in a Hilbert space $\mathscr{H}$. Consider two complementary subsystems $A$ and $A^c$ such that
\begin{equation}
\mathscr{H} = \mathscr{H}_A \otimes \mathscr{H}_{A^c}.
\end{equation}
It is not true in general that the state $\ket{\Psi}$ can be written as a tensor product of two states, \emph{ie} $\ket{\Psi} = \ket{\Psi}_A \otimes \ket{\Psi}_{A^c}$ where $\ket{\Psi}_{A^{(c)}}\in\mathscr{H}_{A^{(c)}}$. When this is the case, systems $A$ and $A^c$ are said to be in a pure state, which physically means that the description of the full system can be reduced to the description of the union of its subsystems. This is always true in classical physics. At the quantum level, however, $A$ and $A^c$ might be entangled, meaning that the state $\ket{\Psi}$ fails to factorize. In such a case, $\ket{\Psi}$ takes the form $\ket{\Psi}=\sum_i c_i \ket{\psi_i}_A\otimes\ket{\psi_i}_{A^c}$ where $\{\ket{\psi_i}_{A^{(c)}}\}_i$ is a basis of states for $\mathscr{H}_{A^{(c)}}$, and each subsystem is described by a density matrix
\begin{equation}
\rho_A = \tr_{A^c} \ket{\Psi}\bra{\Psi},
\end{equation}
and similarly for $A^c$. Expectation values of observables on $A$ are computed by
\begin{equation}
\langle \hat{O}_A \rangle =\tr(\rho_A \hat{O}_A).
\end{equation}
The time evolution of an isolated subsystem is described by a unitary evolution
\begin{equation}
\rho_A \rightarrow U_A\rho_A U_A^{\dagger},
\end{equation}
which cannot transform a pure state into a mixed state, and vice versa. A very useful measure of entanglement is called \textit{entanglement entropy}, or von Neumann entropy, which is defined as
\begin{equation}
S(A) = -\tr_A\left( \rho_A\ln\rho_A \right).
\end{equation}
One should view this quantity as a measure of the number of degrees of freedom that are shared by $A$ and $A^c$. When $S(A)=0$, the two systems are completely independent, and their states are pure. Otherwise, $A$ and $A^c$ are in a mixed state, and their correlation is measured by $S(A)$, which is equal to $S(A^c)$ and bounded from above by $\min\left(\ln\left[\dim\mathscr{H}_A\right],\ln\left[\dim\mathscr{H}_{A^c}\right]\right)$. Note that unitary evolution leaves entanglement entropy unchanged, as one can see from the cyclic property of the trace:
\begin{equation}
\frac{dS}{dt}=0.
\end{equation}
In quantum mechanics with a finite-dimensional Hilbert space, entanglement entropy is well-defined. In continuum field theory, however, the situation becomes more complex. In general, the Hilbert space itself is not factorizable,\footnote{Physically, a complete basis of modes cannot be decomposed into two bases of modes on both sides of the boundary. Doing so would ignore modes spreading over the boundary.} and the notion of trace might be ill-defined, so one needs to introduce a regularization scheme.

One way to do so is to assume that degrees of freedom are located on a lattice, with $\alpha$ indexing the sites. Here, a spatial subregion $A$ is defined as the set of sites that belong to $A$, and its Hilbert space is $\mathscr{H}_A = \bigotimes_{\alpha\in A}\mathscr{H}_{\alpha}$. Another approach is to introduce UV cutoffs. While entanglement entropy might present UV divergences due to short-wavelength fluctuations across the boundary between $A$ and $A^c$, entropies can still be defined relationally. In other words, entropy differences are free of divergences. In practice, one smooths out the boundary between regions $A$ and $A^c$ over a length $\delta$, such that the contributions within a short distance $\delta$ across the boundary are excluded.
\begin{equation}
S(A) = S_{\rm reg}(A) + f(\delta),
\end{equation}
where $S_{\rm reg}(A)$ is the regulated entropy from which we have subtracted the UV divergent part. $f(\delta)$ is a function that diverges logarithmically in the limit $\delta\rightarrow 0$, and should be independent of $A$.

In the context of continuum quantum field theory (QFT) on an arbitrary background, spatial subsystems are studied by considering some foliation of spacetime into spatial slices $\Sigma$. A slice is a codimension-one hypersurface, which is said to be spatial if every pair of points $p,p'$ in $\Sigma$ is spacelike separated.\footnote{A codimension-$n$ object has $n$ dimensions less than the manifold in which it is embedded in.} When $\Sigma$ covers the full space, it is called a Cauchy slice, and a subsystem of a Cauchy slice $\Sigma$ may be denoted $\Sigma_A$. The domain of dependence $D(\Sigma_A)$, or causal completion, is defined as the set of points $p$ such that any causal curve (timelike or lightlike) passing through $p$ must pass through $\Sigma_A$. The description of any region $\Sigma'_A$ whose domain of dependence is $D(\Sigma'_A)=D(\Sigma_A)\equiv D_A$ can be obtained from $\Sigma_A$, \emph{ie} their density matrices are related by a unitary transformation. In particular,
\begin{equation}
S(\Sigma_A) = S(\Sigma'_{A}).
\end{equation}
Now, let us consider subsystems $A$ and $B$ whose union is not necessarily in a pure state. If $A$ and $B$ are entangled, one may ask how much information is obtained about $A$ if one performs a measurement on $B$. This is measured by \textit{mutual information}:
\begin{equation}
I(A:B) = S(A) + S(B) - S(A\cup B).
\end{equation}
A crucial property of entanglement entropy is that mutual information is always positive, which implies that the sum of the entropy of two subsystems is always greater than the entropy of their union. This property is called weak subadditivity,
\begin{equation}
S(A) + S(B) \geq  S(A\cup B).
\end{equation}
Another useful inequality is the Araki-Lieb inequality:
\begin{equation}
S(A\cup B) \geq \left|S(A)-S(B)\right|.
\end{equation}
Finally, let us keep in mind that entropy must satisfy strong subadditivity, stating that given three subsystems $A,B,C$,
\begin{equation}
\label{eq:SSA}
S(A\cup B\cup C) + S(A) \leq S(A\cup B) + S(A\cup C).
\end{equation}

\section{Thermodynamic entropy}

Until now, we have focused on entropy as a measure of entanglement. However, the notion of entropy was historically introduced as a way to measure the number of microstates $N$ of a system:
\begin{equation} \label{eq:thermo} S_{\rm th} = \ln N. \end{equation}
From our previous definitions, it is clear that entanglement entropy does not count degrees of freedom. For example, a pure state with an infinite number of degrees of freedom would still have vanishing entropy. However, it is possible to relate these two approaches in some thermodynamical ensembles.

The microcanonical ensemble of thermodynamics is characterized by the ensemble of states of an isolated system whose total energy is exactly specified. Denoting the number of eigenstates $\ket{E_i}$ of the Hamiltonian in the interval $[E,E+dE]$ $N$, the density matrix is given by
\begin{equation} \rho_{\rm micro} = \frac{1}{N}\sum_{i=1}^N\ket{E_i}\bra{E_i}. \end{equation}
In this case, the entanglement entropy exactly corresponds to the number of microstates of the system,
\begin{equation} S_{\rm micro} = \ln N. \end{equation}

In the canonical ensemble, or thermal state, the system is coupled to a heat bath that fixes its temperature $T$. In such a system, the density matrix is given by
\begin{equation} \label{eq:thermalrho} \rho_{\rm thermal} = \sum_{i=1}^N \frac{e^{-\beta E_i}}{Z}\ket{E_i}\bra{E_i},\quad Z = \sum_{i=1}^Ne^{-\beta E_i}, \end{equation}
where $\beta=1/T$. The entropy is given by
\begin{equation} \label{eq:entropyth} S_{\rm thermal} = \beta\langle{E}\rangle + \ln Z, \end{equation}
which is consistent with the thermodynamic definition of entropy. This state maximizes the entropy for a given expectation value of the energy.

It is sometimes useful to consider the following pure state, called the \textit{thermofield double state}:
\begin{equation} \label{eq:TFD} \ket{\Psi(\beta)} = \frac{1}{\sqrt{Z}}\sum_{i=1}^Ne^{-\frac{\beta E_i}{2}}\ket{E_i}_A\otimes\ket{E_i}_B, \end{equation}
where $\ket{E_i}_A$ and $\ket{E_i}_B$ are the eigenstates of two copies of the system. Tracing out one of the two systems leads to the thermal density matrix \eqref{eq:thermalrho}.

More generally, one may interpret thermodynamic entropy as a coarse-grained entropy. Entanglement entropy, sometimes called fine-grained entropy, requires knowledge of the microscopic details of the theory. A natural definition of coarse-grained entropy, consistent with thermodynamics, is due to Jaynes~\cite{Jaynes:1957zza,Jaynes:1957zz}.

Consider a setup where we do not have access to the microscopic details of a theory, but where we can measure some coarse-grained observables such as energy, temperature, and pressure. Given a set of such observables $\hat{A}_i$, we define ${\Tilde{\rho}}$ as the set of density matrices $\Tilde{\rho}$ consistent with the observables we consider:
\begin{equation} \tr(\Tilde{\rho} \hat{A}_i) = \langle \hat{A}_i \rangle. \end{equation}
Then, \textit{thermodynamic entropy} is defined as
\begin{equation} S_{\rm th} = \max_{\Tilde{\rho}} S(\Tilde{\rho}). \end{equation}
A direct consequence of this definition is that for any subsystem $A$,
\begin{equation}  S(A)\leq S_{\rm th}(A) . \end{equation}
Additionally, while entanglement entropy is a conserved quantity under time evolution, the second law of thermodynamics states that coarse-grained entropy cannot decrease,
\begin{equation} \label{eq:2ndlaw} dS_{\rm th} \geq 0. \end{equation}

\section{Entanglement entropy in \texorpdfstring{\boldmath CFT$_2$}{CFT2}}
\label{sec:SCFT}

Our introduction to entanglement entropy has been quite abstract, particularly in the context of QFT. Deriving explicit expressions for the entropy of subregions in QFT appears to be extremely difficult. However, there is a class of QFTs where one can leverage the large (infinite, in fact) symmetry to write explicit formulas: two-dimensional conformal field theories (CFT$_2$)~\cite{DiFrancesco:1997nk}.

A CFT is a QFT which is invariant under conformal transformations $g_{\mu\nu} \rightarrow \Lambda(x) g_{\mu\nu}$. In two dimensions, conformal symmetry implies symmetry under the Poincaré group as well as scale invariance, and its group is infinite-dimensional. An important quantity associated with any CFT is the central charge $c$, which may be interpreted as an effective number of degrees of freedom.

 We start with the simplest case of a static state on a flat background $\mathbb{R}^{1,1}$ with coordinates $(t,x)$. The entanglement entropy of an interval $A=\{(t,x), t=0, x \in [x_1, x_2]\}$ reads 
\begin{equation} S(A) = \frac{c}{6}\ln \frac{(x_2-x_1)^2}{\delta^2}, \end{equation}
where $\delta$ is the UV regulator mentioned in Section~\ref{sec:EE}. See \emph{eg}~\cite{Rangamani:2016dms} for a derivation. An expression for a CFT defined on $\mathbb{R}_t \times \mathbb{S}^1$ with spatial sections of length $L$ can also be found:
\begin{equation} \label{eq:CFTS1} S(A) = \frac{c}{3}\ln\left(\frac{L}{\pi\delta}\sin\frac{x_2-x_1}{L}\right). \end{equation}

If the CFT is defined instead on a curved background, some subtleties need to be taken care of. Consider the metric: \begin{equation}
ds^2 = e^{2\omega_x} dx^+ dx^-, \end{equation}
where $\omega_x$ is the conformal factor for coordinates $x$. Let $\Sigma$ be a spacelike slice with endpoints $x_1 = (x_1^+, x_1^-)$ and $x_2 = (x_2^+, x_2^-)$. We consider the limit where right- and left-moving modes are decoupled and can be treated separately, which is the case for free massless scalars, for example~\cite{Fiola:1994ir}. All left-moving modes along the interval $[x_1^+, x_2^+]$ pass through $\Sigma$, and all right-moving modes along $[x_1^-, x_2^-]$ also pass through $\Sigma$. We consider the vacuum state built on the $x^{\pm}$ coordinates, where a complete set of left- (right-) moving modes is constructed both inside and outside the intervals $[x_1^+, x_2^+]$ and $[x_1^-, x_2^-]$. See Figure~\ref{fig:CFTent}.

\begin{figure}[ht] \centering \includegraphics[width=0.6\linewidth]{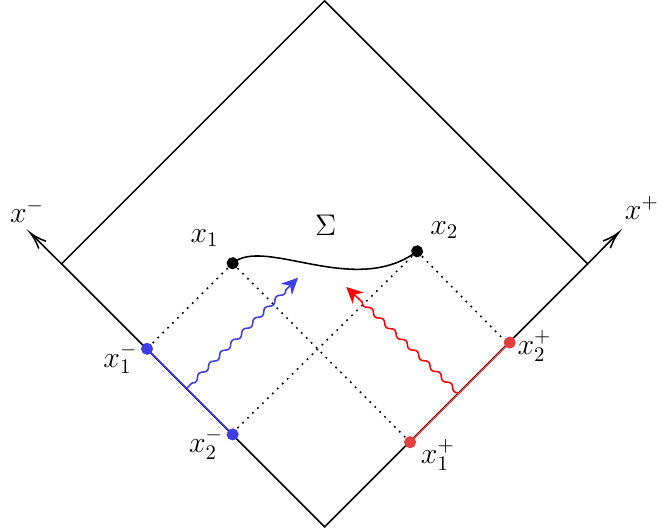} \caption{\footnotesize A spacelike slice $\Sigma$ with endpoints $x_1=(x_1^+, x_1^-)$ and $x_2=(x_2^+, x_2^-)$ in a patch of spacetime. The right-moving modes are depicted in blue and the left-moving modes are depicted in red.} \label{fig:CFTent}
\end{figure}

The entropy receives independent contributions from left- and right-moving modes. The entropy contribution from the left-moving modes is given by 
\begin{equation} \label{eq:entropy_left} S^{\rm(left)}_x(\Sigma) = \frac{c}{12}\ln\left[\frac{(x_2^+-x_1^+)^2}{\epsilon(x_1)\epsilon(x_2)}\right], \end{equation} 
where the subscript $x$ denotes the coordinate system in which the vacuum is constructed and $\epsilon(x)$ is the UV cutoff at $x$. We assume that the UV cutoff at $x$ is the same for left- and right-moving modes. One may replace $\epsilon^2$ with $\epsilon_{\rm L}\epsilon_{\rm R}$ to maintain generality. Summing the contributions from the left- and right-moving modes, the total entropy is given by: 
\begin{equation} S_x(\Sigma) = \frac{c}{12}\ln\left[\frac{(x_2^+-x_1^+)^2(x_2^--x_1^-)^2}{\epsilon^2(x_1)\epsilon^2(x_2)}\right], \end{equation} 
The UV cutoffs are related to the cutoffs in inertial coordinates $\delta$ by the following relation: \begin{equation} \epsilon(x)=e^{-\omega_x(x)}\delta(x), \end{equation} which enables us to express $S_x$ in the boost-invariant form~\cite{Fiola:1994ir}: \begin{equation} \label{eq:CFTcurved} S_x(\Sigma) = \frac{c}{6}\left(\omega_x(x_1)+\omega_x(x_2)\right)+\frac{c}{12}\ln\left[\frac{(x_2^+-x_1^+)^2(x_2^--x_1^-)^2}{\delta_1^2\delta_2^2}\right], \end{equation} where $\delta_i = \delta(x_i)$. In curved backgrounds, vacuum states defined by expanding modes in different coordinates are inequivalent, which is reflected in the expression for entanglement entropy. Consider the vacuum state built in the coordinates $y^{\pm}$ with the associated metric $ds^2 = e^{2\omega_y} dy^+ dy^-$. $y^{\pm}$ coordinates are related to $x^{\pm}$ coordinates by the functions $y^{\pm}(x^{\pm})$. The entropy of $\Sigma$ in the $y$-vacuum is~\cite{Fiola:1994ir} \begin{equation} \label{eq:transf} S_y(\Sigma) = \frac{c}{12}\ln\left[\frac{(y^+(x_2^+)-y^+(x_1^+))^2(y^-(x_2^-)-y^-(x_1^-))^2}{\epsilon^2(y(x_1))\epsilon^2(y(x_2))}\right], 
\end{equation} 
where \begin{equation}
\epsilon(y) = e^{-\omega_y(y)}\delta(y). \end{equation} See Appendix \ref{app:entanglement_entropy} for a discussion of entanglement entropy in the presence of reflective boundaries.

\chapter{Geometry reconstruction in AdS/CFT}
\label{ch:AdS/CFT}

Building on the holographic conjecture~\cite{tHooft:1993dmi,Susskind:1994vu}, Maldacena established the foundations of the duality between quantum gravity in Anti-de Sitter (AdS) spacetime and Conformal Field Theories (CFT) in~\cite{Maldacena:1997re}. This duality, known as the AdS/CFT correspondence~\cite{Maldacena:1997re,Witten:1998qj,Gubser:1998bc}, posits an exact mapping between states and observables in both theories, even though their interpretations differ significantly.

In this work, we focus on a specific aspect of this duality, summarized by the following question: In the semiclassical limit $G\rightarrow 0$, where gravity is well described by general relativity, how does the geometric picture emerge from the conformal field theory? Another related question is what types of CFTs correspond to a classical spacetime, as opposed to, for example, a quantum superposition of spacetimes?

One of the most exciting recent developments in theoretical physics has been the realization that quantum information provides the correct framework for addressing these questions. In this chapter, we review some key elements of this approach. For excellent reviews on related topics, see \textit{eg}~\cite{VanRaamsdonk:2016exw,Rangamani:2016dms,Harlow:2018fse}.

\section{AdS geometry}

AdS spacetime is the maximally symmetric solution to Einstein equations with a negative cosmological constant. It can be constructed as a hypersurface embedded in a spacetime with two timelike dimensions:
\begin{equation}
    ds^2 = -dX_0^2 - dX_1^2 + \sum_{i=2}^{d}dX_i^2.
\end{equation}
In this $(d+2)$-dimensional manifold, AdS is defined as the hypersurface
\begin{equation}
    -X_0^2 - X_1^2 + \sum_{i=2}^{d}X_i^2 = -l_{\rm AdS}^2.\label{eq:AdS}
\end{equation}
This definition makes the SO$(d-1,2)$ isometry group of AdS explicit. The AdS radius $l_{\rm AdS}$ is related to the cosmological constant as $\Lambda=-d(d-1)/(2l_{\rm AdS})$. We set $l_{\rm AdS}=1$. The metric of dS is the induced metric on the hypersurface \eqref{eq:AdS}. Global coordinates cover the full geometry:
\begin{equation}
    ds^2=-(1+r^2)dt^2+\frac{dr^2}{1+r^2}+r^2d\Omega^2_{d-1},
\end{equation}
where $d\Omega^2_{d-1}$ is the measure of a ($d-1$)-dimensional sphere of radius $1$. The asymptotic boundary of AdS, at $r\rightarrow \infty$, has the geometry of a cylinder $\mathbb{R}_t\times\mathbb{S}^{d-1}$. In AdS/CFT, we often use the Poincaré coordinates
\begin{equation}
\label{eq:Poincaré}
    ds^2=\frac{1}{z^2}(dz^2-dt^2+\sum_{i=1}^{d-1}dx^2_i).
\end{equation}
The asymptotic boundary of AdS is $z\rightarrow 0$.

\section{Entanglement builds bridges}
\label{sec:ER=EPR}

A typical realization of the correspondence between geometry and entanglement was introduced by Van Raamsdonk~\cite{VanRaamsdonk:2009ar, VanRaamsdonk:2010pw}, based on the earlier work~\cite{Maldacena:2001kr}.

Take two non-interacting copies of a CFT defined on a cylinder $\mathbb{R}_t \times \mathbb{S}^{d-1}$~\cite{Witten:1998zw}, and denote them $A$ and $B$. It was shown in~\cite{Maldacena:2001kr} that when the CFT is in a thermal state, the holographic dual depends on the temperature $\beta^{-1}$. At low temperatures, the holographic dual is a gas of particles in AdS${}_{d+1}$. At high temperature, the holographic dual is a Schwarzschild-AdS black hole in AdS${}_{d+1}$.

Consider a thermofield double state, as defined in equation~\eqref{eq:TFD}, constructed from the two copies $A$ and $B$. At low temperatures, the ground state dominates, and the entanglement between $A$ and $B$, as given by \eqref{eq:entropyth}, is low. At high temperature, the state is highly mixed, and the entanglement is large. The holographic duals of the low- and high-temperature limits are~\cite{Maldacena:2001kr, Balasubramanian:2014hda}: \begin{itemize} \item $\beta \rightarrow \infty$: Two disconnected AdS${}_{d+1}$ spacetimes. See Figure~\ref{fig:prodAdS}.\footnote{The blue triangular Penrose diagram in Figure \ref{fig:ER=EPR} depicts only a Poincaré patch of AdS, covered by the Poincaré coordinates \eqref{eq:Poincaré}.} \item $\beta \rightarrow 0$: A maximally extended Schwarzschild-AdS${}_{d+1}$ black hole. See Figure~\ref{fig:AdSBH}. \end{itemize}

\begin{figure}[ht] \begin{subfigure}[t] {0.48\linewidth}\centering \includegraphics[width=0.7\linewidth]{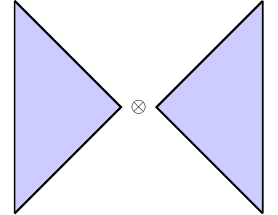} \caption{} \label{fig:prodAdS} \end{subfigure} \begin{subfigure}[t] {0.48\linewidth}\centering \includegraphics[width=0.6\linewidth]{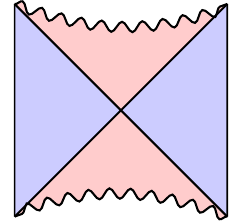} \caption{} \label{fig:AdSBH} \end{subfigure} 
\caption{\footnotesize Penrose diagrams of the gravity duals to the thermofield-double state in the low- and high-temperature limits. The two asymptotically AdS spacetimes are depicted in blue, with the two vertical lines being the two conformal boundaries, and the Einstein-Rosen bridge connecting them is depicted in red.} \label{fig:ER=EPR} 
\end{figure}

In the high-temperature limit, there are two asymptotically AdS regions, bounded by an $\mathbb{R}_t \times \mathbb{S}^{d-1}$, connected by the interior of the black hole. In other words, entangling two copies of a CFT, each individually dual to an AdS spacetime, creates a spatial bridge between the two asymptotic regions. Stated more generally, connectivity between disconnected boundary regions emerges from the entanglement between them. This idea was then generalized in~\cite{Maldacena:2013xja}, with the slogan \textit{ER$=$EPR}: Any quantum state with Bell-type bipartite entanglement is associated with a spatial connection between entangled parties.

\section{Holographic entanglement entropy}
\label{sec:const}

In an attempt to generalize the ideas of Van Raamsdonk, one may try to elaborate on the AdS/CFT dictionary to find the geometric dual to entanglement entropy. In particular, the Bekenstein-Hawking formula for black hole entropy was derived in the framework of string theory~\cite{Strominger:1996sh}, encouraging the idea that geometric quantities in the bulk should be related to precise quantum features on the boundary dual theory. This was addressed by Ryu and Takayanagi~\cite{Ryu:2006bv,Ryu:2006ef}, who proposed a formula to compute the entanglement entropy of subsystems of the CFT in time-independent settings.

What Ryu and Takayanagi showed is that equation~\eqref{eq:CFTS1} for the entropy of a CFT subsystem $A$ exactly matches the length of the geodesic in AdS$_3$ that connects the endpoints of the subsystem $A$ on the boundary, up to a multiplicative factor. This factor, using the AdS$_3$/CFT$_2$ identification $c=3l_{\rm AdS}/2G$~\cite{Brown:1986nw}, appears to be $1/4G$. This remarkable identification establishes the connection between the AdS/CFT correspondence and previous works on black hole entropy and entropy bounds in semiclassical gravity. It was later generalized to general states (including the time-dependent ones) in~\cite{Hubeny:2007xt}. In this section, we review the HRT formula~\cite{Hubeny:2007xt} as well as its generalizations~\cite{Wall:2012uf,Faulkner:2013ana,Engelhardt:2014gca,Akers:2019lzs}. Throughout this work, we will refer to homologous surfaces, which we define as follows. 

\begin{definition}[Homology condition] 
\label{def:hom}
Let $D$ be a bulk subregion. A codimension-two surface $\gamma\in D$ is homologous to the codimension-two subsystem $A\in\partial D$ if there exists an achronal slice $\mathcal{C}(A)$ such that $\partial\mathcal{C}(A) = A\cup \gamma$.\footnote{We denote $\partial R$ the boundary of a region $R$.} $\mathcal{C}(A)$ is called the \textit{homology region}
\end{definition} 

Let $\mathcal{S}$ be the conformal boundary of AdS$_{d+1}$, \textit{ie} the holographic screen. We denote $\Sigma\vert_{\mathcal{S}}$ a spacelike slice of $\mathcal{S}$. The proposal of Hubeny, Rangamani, and Takayanagi~\cite{Rangamani:2016dms}, later proven using a path integral computation~\cite{Lewkowycz:2013nqa,Dong:2016hjy}, goes as follows. 

\begin{definition}[Extremal surface] \label{def:HRT} Let $A$ be a codimension-two spacelike subsystem of $\Sigma\vert_{\mathcal{S}}$. The entanglement entropy of $A$ as a spatial subsystem of the quantum theory defined on the boundary $\mathcal{S}$ is given by \begin{equation} 
\label{eq:HRT}
S(A) = \min\mathrm{ext}\left[\frac{\text{Area}(\gamma)}{4G}\right] + O(G^0),
\end{equation} 
where $\gamma$ must be homologous to $A$. The area is extremized, and if multiple surfaces that extremize it exist, we consider the one with the smallest area. The resulting surface, denoted $\gamma_{\rm e}(A)$, is called the HRT surface or \textit{extremal surface} associated with $A$. See Figure~\ref{fig:HRT}. \end{definition}

\begin{figure}[ht]
    \centering
    \includegraphics[width=0.4\linewidth]{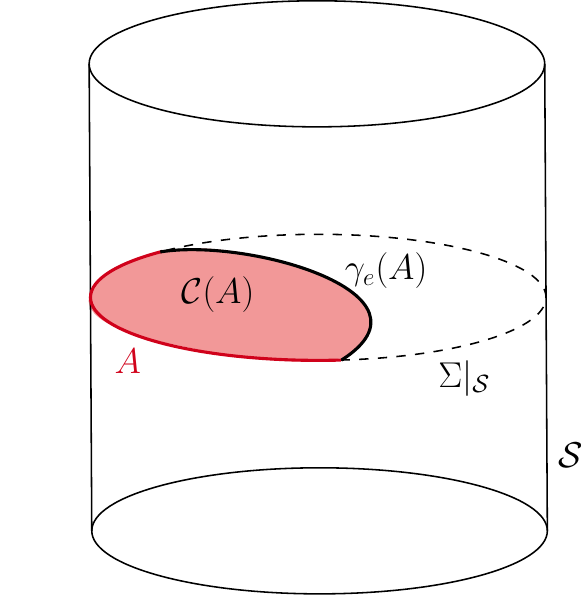}
    \caption{\footnotesize Depiction of AdS$_3$, which has the topology of a cylinder. The dual CFT$_2$ is located on the boundary of the cylinder which plays the role of a holographic screen $\mathcal{S}$. A subsystem $A$ of spatial slice $\Sigma\vert_{\mathcal{S}}$ (dashed line) is depicted by the red curve. The extremal surface $\gamma_{\rm e}(A)$ is the black curve while its homology region $\mathcal{C}(A)$ is the red region.}
    \label{fig:HRT}
\end{figure}

Assuming that $\Sigma\vert_{\mathcal{S}}$ is in a pure state, we have $S(A)=S(A^c)$. This implies that $\gamma_{\rm e}(A)=\gamma_{\rm e}(A^c)$, and that the homology region $\mathcal{C}(A)$ must always lie on a Cauchy slice $\Sigma$ whose boundary is $\Sigma\vert_{\mathcal{S}}$~\cite{Headrick:2014cta}. In the definitions above, we explicitly implemented this condition by considering a restricted version of the usual HRT formula~\cite{Marolf:2019bgj, Grado-White:2020wlb}, which is analogous to the original definition, except that it considers bulk surfaces defined on achronal slices whose boundaries are identified with the screen $\Sigma\vert_{\mathcal{S}}$, instead of slices with the looser condition that the boundary contains the subsystem $A$. In AdS, the two prescriptions are equivalent when the surface lies in a smooth region of spacetime~\cite{Marolf:2019bgj}. The restricted formulation is necessary in some backgrounds to ensure strong subadditivity and entanglement wedge nesting~\cite{Grado-White:2020wlb}. From another perspective, restricted extremization is important when dealing with systems that are not isolated, such as evaporating black holes, where time evolution on the boundary alters the entropy~\cite{Penington:2019npb, Almheiri:2019psf} .

It can be shown that the holographic entanglement entropy \eqref{eq:HRT} satisfies all usual properties of entanglement entropy presented in Section~\ref{sec:EE}~\cite{Headrick:2007km,Hayden:2011ag,Wall:2012uf,Headrick:2013zda}. However, the formula above appears to be suboptimal for checking these constraints. Wall introduced in~\cite{Wall:2012uf} a reformulation of the HRT formula, which greatly simplifies a number of proofs while providing a nice complementary approach to the construction of extremal surfaces.

\begin{definition}[Maximin surfaces]
    Consider $A\in\Sigma\vert_{\mathcal{S}}$ and let $\gamma_{\rm min}(A,\Sigma)$ be the codimension-two surface homologous to $A$ that has minimal area on $\Sigma$ with $\partial\Sigma=\Sigma\vert_{\mathcal{S}}$. The \textit{maximin surface} $\gamma_{\rm m}(A)$ is the surface $\gamma_{\rm min}(A,\Sigma_m)$ which is maximal when varying $\gamma_{\rm min}(A,\Sigma')$ over all slices $\Sigma'$ such that $\partial\Sigma'=\Sigma\vert_{\mathcal{S}}$. The maximin surface must be stable in the sense that, considering a small variation of the slice $\Sigma_m$, the new slice must have a minimal surface which is in the neighborhood of $\gamma_{\rm m}$ with no greater area.  
\end{definition}
\begin{theorem}[Maximin=Extremal~\cite{Wall:2012uf}]
\label{thm:HRT=m}
    A maximin surface $\gamma_{\rm m}(A)$ is an extremal surface $\gamma_{\rm e}(A)$, and conversely.
\end{theorem}
 In other words, equation~\eqref{eq:HRT} is equivalently rewritten as
\begin{equation}
\label{eq:maximin}
    S(A) = \max_{\Sigma'\in\{\Sigma,\partial\Sigma=\Sigma\vert_{\mathcal{S}}\}}\min_{\gamma\in\Sigma'}\left[\frac{\text{Area}(\gamma)}{4G}\right] + O(G^0),
\end{equation}
with the additional stability assumption.

The ER=EPR idea is realized explicitly by the HRT formula, for example, through the following thought experiment~\cite{VanRaamsdonk:2009ar,VanRaamsdonk:2010pw}. Consider the vacuum state of a CFT defined on a sphere. Divide the sphere into two hemispheres $A$ and $B$. One can define a family of states where the entanglement of $A$ goes from $0$ to its initial value for the vacuum state. The HRT formula \eqref{eq:HRT} implies that decreasing the entropy of $A$ must result in a shrinkage of the minimal extremal surface anchored to $\partial A = \partial B$. Hence, the spatial region connecting $A$ and $B$ is pinched out as the entropy decreases, continuously disconnecting the two asymptotic regions. In the limit where $S(A)\rightarrow 0$, $A$ and $B$ become further and further apart, until spacetime becomes disconnected.

\section{Entanglement wedge reconstruction}
\label{sec:EW}

In the last two Sections, we have argued that entanglement in the CFT is closely related to the geometric structure of the dual spacetime. More precisely, the connectivity of the bulk emerges from entanglement. Moreover, the HRT formula tells us that boundary entanglement does more than ensure connectivity. For a given CFT state, one can compute the entanglement entropy of any subsystem and determine the associated dual spacetime using knowledge of the set of associated extremal surfaces. There is, however, a limitation to this procedure, which is that some regions of spacetime may not be penetrable by extremal surfaces. Clearly, this region, the \textit{entanglement shadow}, cannot be reconstructed using this procedure~\cite{Engelhardt:2013tra}.

The question of geometry reconstruction can be expressed more generally~\cite{Bousso:2012sj,Bousso:2012mh,Czech:2012bh,Hubeny:2012wa}. Given a state of the total CFT and some subsystem $A$, which regions can be reconstructed from knowledge of the density matrix $\rho_A$?

A natural answer would be the causal wedge, which is defined as
\begin{equation} D(A) = J^+(\mathcal{D}(A))\cap J^-(\mathcal{D}(A)), \end{equation}
where $\mathcal{D}(A)$ is defined as the causal diamond of $A$ on the boundary geometry, and $J^{\pm}$ are the bulk causal past and causal future. An example of causal wedge is depicted in Figure~\ref{fig:EW}. We mentioned in Section~\ref{sec:EE} that the state on $\mathcal{D}(A)$ can be obtained uniquely from $A$. On the CFT side, one may compute response functions in $\mathcal{D}(A)$, which corresponds to creating some perturbation and measuring the response at a later point. This type of correlator is necessarily sensitive to the content of $D(A)$, which can send and receive signals from $\mathcal{D}(A)$. However, one may argue that the bulk region encoded in $A$ must be larger than $D(A)$. Indeed, in addition to weak and strong subadditivity, a very important property of the HRT surface of $A$ is that it always lies outside the causal wedge of $A$~\cite{Wall:2012uf}. Since computing $S(A)$ provides some information about a surface lying outside $\mathcal{D}(A)$, $D(A)$ cannot be the bulk region dual to $A$. In particular, $A$ not only encodes information about the area of the extremal surface $\gamma_{\rm e}(A)$, but also about the area of the extremal surface of any subsystem of $A$. It has been argued that the bulk region encoded in $A$ identifies with the entanglement wedge, which is defined as follows~\cite{Wall:2012uf,Czech:2012bh,Headrick:2014cta}.

\begin{definition}[Entanglement wedge] 
\label{def:EW}
The \textit{entanglement wedge} of a subsystem $A\in\Sigma\vert_{\mathcal{S}}$ is defined as \begin{equation}
W(A) = D(\mathcal{C}(A)). 
\end{equation}
\end{definition}
The conjecture that all the information about the entanglement wedge $W(A)$ is contained in $A$ has received strong support from tensor network models of holography~\cite{Almheiri:2014lwa,Pastawski:2015qua,Dong:2016eik,Harlow:2016vwg,Hayden:2016cfa,Cotler:2017erl}, see~\cite{Harlow:2018fse} for a review.

Analogs of entropy inequalities can be derived for the entanglement wedge. One of them follows from the fact that $\gamma_{\rm e}$ lies outside of the causal wedge, see Figure~\ref{fig:EW}:

\begin{equation} \label{eq:CinW} D(A)\subset W(A). \end{equation}

\begin{figure}[ht] \centering \includegraphics[width=0.4\linewidth]{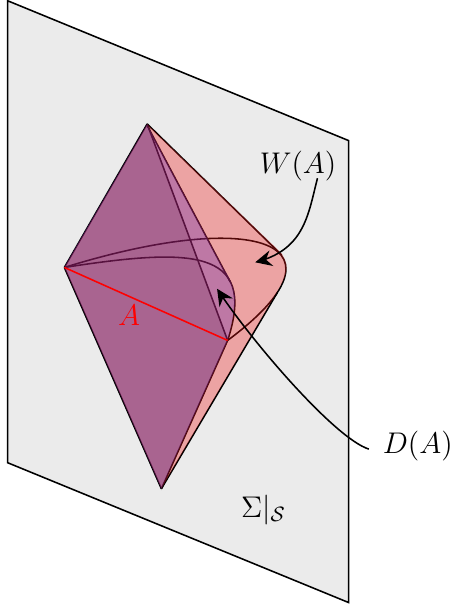} \caption{\footnotesize Schematic representation of the causal wedge $D(A)$ (in blue) and entanglement wedge $W(A)$ (in red) of a spacelike subsystem of the holographic screen. The boundary of $W(A)$, $\gamma_{\rm e}(A)$, always lies outside of $D(A)$ such that $D(A)\subset W(A)$.} \label{fig:EW} \end{figure}

This condition was proven geometrically using the maximin formula~\cite{Wall:2012uf}, but can also be justified using CFT causality~\cite{Headrick:2014cta}. Consider a CFT perturbation whose support lies in $\mathcal{D}(A)\cup\mathcal{D}(A^c)$ with $A^c\cup A = \Sigma\vert_{\mathcal{S}}$. Let $\Sigma\vert_{\mathcal{S}}'\in\mathcal{D}(A)\cup\mathcal{D}(A^c)$ be in the past of the support of the perturbation. If $A'\in \Sigma\vert_{\mathcal{S}}'$ has $\mathcal{D}(A')=\mathcal{D}(A)$, then $S(A)=S(A')$ while $S(A')$ cannot depend on the perturbation by definition. Hence, $\gamma_{\rm e}(A)$ cannot be causally related to $\mathcal{D}(A)$, which implies \eqref{eq:CinW}. Moreover, the same holds for $\mathcal{D}(A^c)$, implying the stronger constraint

\begin{equation} \gamma_{\rm e}(A) \subset Q(A), \end{equation}

where $Q(A)$ is called the \textit{causal shadow} of $A$ and is defined as the region that is spacelike separated from $D(A)\cup D(A^c)$.

Another important property is analogous to weak subadditivity for the entropy and states that the entanglement wedge of a boundary region $A$ must contain the entanglement wedge of any subsystem of $A$. This is called \textit{entanglement wedge nesting}:

\begin{equation} \label{eq:EWnesting} A\subset B \Rightarrow W(A)\subset W(B). \end{equation}

Finally, entanglement wedges must not allow for superluminal communication on the boundary. This is called the \textit{Boundary Causality Condition} (BCC)~\cite{Engelhardt:2016aoo}. In fact, it can be shown that both the BCC and $D\subset W$ can be derived from entanglement wedge nesting~\cite{Akers:2016ugt}. At the classical level, these can be proven using the focusing theorem and the null energy condition in the framework of general relativity, see Sections \ref{sec:GSL} and \ref{sec:focusing}.

At the semiclassical level, when accounting for quantum fluctuations of matter, the HRT formula \eqref{eq:HRT} should include higher-order terms, and finding general constraints on this entropy to ensure entanglement wedge nesting will be the subject of Section~\ref{sec:QFC}. In the next section, we generalize the HRT formula to take into account these higher-order terms.

\section{Beyond the classical regime}
\label{sec:HRTQ}

Up to this point, we have worked in the semiclassical regime $G(\hbar)\rightarrow 0$, operating at leading order in $G(\hbar)$.\footnote{We work in natural units. Where necessary, we include $\hbar$ factors in brackets.} The resulting quantities are purely classical, and on the gravity side, they derive from geometrical terms. However, both bulk geometry and fields can fluctuate, leading to contributions of order $(G(\hbar))^0$ and higher. 

In the following Chapters, we will often focus on the \textit{semiclassical regime}, where fields can be expanded perturbatively in $G(\hbar) \sim 1/c_{eff}$. When studying the entanglement entropy of boundary subregions, it is useful to take inspiration from the generalized second law of thermodynamics and define the so-called generalized entropy~\cite{Faulkner:2013ana,Barrella:2013wja,Engelhardt:2014gca}. See Section~\ref{sec:GSL} for a discussion of the generalized second law.

Let $\gamma$ be a codimension-two spacelike surface that splits a Cauchy slice $\Sigma$ into two parts, $\Sigma_{\rm in}$ and $\Sigma_{\rm out}$. The generalized entropy is defined as
\begin{equation} \label{eq:Sgen}
S_{\rm gen}(\gamma) = \frac{\text{Area}(\gamma)}{4G} + S(\Sigma_{\rm out}), \end{equation}
where $S(\Sigma_{\rm out})$ is the fine-grained entropy of fields in $\Sigma_{\rm out}$ (see Section~\ref{sec:EE}). Note that $S(\Sigma_{\rm out})=S(\Sigma_{\rm in})$, so the concepts of inside and outside of $\gamma$ can be interchanged. We sometimes use the simpler notation $S_{\rm out} \equiv S(\Sigma_{\rm out})$.

It was conjectured in~\cite{Engelhardt:2014gca} that, in an expansion in $G(\hbar)$, the entanglement entropy of $A$ is given by
\begin{equation} 
\label{eq:QES}
S(A)=\min\mathrm{ext}\left[\frac{\text{Area}(\gamma)}{4G}+S(\mathcal{C}(A))\right],
\end{equation}
where $\mathcal{C}(A)$ is the homology region such that $\partial\mathcal{C}(A)=A\cup \gamma$. The resulting surface is called the \textit{quantum extremal surface} associated with $A$.

Hence, beyond the classical limit, the area of $\gamma$ divided by $4G$ should be replaced with the generalized entropy \eqref{eq:Sgen} in the definition of extremal surfaces (see Definition \ref{def:HRT}). Note that the quantum extremal surface may not be close to the extremal surface. For an evaporating black hole, the extremal surface associated with the black hole is the empty surface, while the quantum extremal surface lies close to the horizon after Page time~\cite{Penington:2019npb,Almheiri:2019psf}.

The above formula was shown to satisfy a number of consistency constraints, such as the entropy inequalities of Section~\ref{sec:EE} and the entanglement wedge constraints of Section~\ref{sec:EW}~\cite{Engelhardt:2014gca,Akers:2016ugt}. Moreover, quantum extremal surfaces have been shown to recover the Page curve for evaporating black holes in AdS/CFT~\cite{Almheiri:2014lwa,Penington:2019npb}. There is also strong evidence for the validity of \eqref{eq:QES} from path integral computations~\cite{Penington:2019kki,Dong:2017xht}.

In~\cite{Akers:2019lzs}, it was shown that theorem \ref{thm:HRT=m}, which states that extremal surfaces are maximin surfaces, remains valid at all orders in $G(\hbar)$ when replacing the area term in the definitions of extremal and maximin surfaces with the generalized entropy. While quantum extremal surfaces lead to a Page curve for the black hole entropy, they do not truly resolve the information paradox that concerns the entropy of Hawking radiation. Unitarity should ensure that the entropy of Hawking radiation matches the entropy of the black hole, but a complete answer should also be able to derive the former. This problem was solved using the island formula~\cite{Almheiri:2019hni}.
\begin{definition}[Island formula]
    Let $R$ be a non-gravitating region coupled to a gravitational system. The entanglement entropy of $R$ is computed by
    \begin{equation}
    \label{eq:island}
        S(R) = \min \mathrm{ext} \left[\frac{\text{Area}(\partial I)}{4G}+S(R \cup I)\right],
    \end{equation}
    where $I$ is a spacelike slice in the gravitational theory called the \textit{island}.
\end{definition}
This formula is supported by path integral computations in semiclassical gravity~\cite{Penington:2019kki,Almheiri:2019qdq}, and can be formally derived in the framework of quantum error correcting codes~\cite{Almheiri:2014lwa,Harlow:2016vwg,Dong:2016eik}. The main point of the island formula is that the generalized entropy \eqref{eq:Sgen} can be applied to a disconnected region $\Sigma_{\rm out}=R \cup I$. A disconnected region may be dominant (minimal) when distant regions $R$ and $I$ contain entangled matter. Then, considering $R\cup I$ instead of $R$ alone decreases the semiclassical contribution $S(\Sigma_{\rm out})$ and increases the area term. In the example of evaporating black holes: At late times, the semiclassical entropy of the collected radiation $R$ is greater than the generalized entropy associated with the union of $R$ and an island that spans the interior of the black hole.

\chapter{Quantum tasks in AdS/CFT}
\label{ch:CWTAdS}

In the previous chapter, our understanding of the connection between quantum information and gravity in AdS/CFT has been almost exclusively tied to the notion of entanglement entropy. In this chapter, we discuss another approach to AdS/CFT which relies on quantum information even more explicitly.\footnote{Another important aspect of this connection, which we do not discuss in this thesis, is holographic complexity~\cite{Brown:2019rox,Susskind:2014rva,Stanford:2014jda,Couch:2016exn,Brown:2015bva,Brown:2015lvg,Belin:2021bga,Belin:2022xmt}.}

The traditional formulation of AdS/CFT is an equality between the partition functions of the two theories. Another approach, introduced in \cite{May:2019yxi}, is operational: There must be an equivalence between the operations that can be performed on the boundary and in the bulk. For example, if a quantum circuit mapping a state $\rho_{\rm in}$ to another $\rho_{\rm out}$ exists in the bulk, an alternative realization must exist on the boundary.

The bulk has one more spatial dimension that the boundary, so it is expected that the causal structure of the bulk is richer than the boundary causal structure. In particular, one might expect that some operations are realizable in the bulk but not on the boundary, or that the bulk could be used as a shortcut to perform complicated tasks on the boundary. A simple example would be the task of sending a signal from one boundary point $c$ to a future boundary point $r$. Can the bulk be used to send the signal faster? The answer appears to be no, due to a theorem by Gao and Wald \cite{Gao:2000ga} which states that the fastest null geodesic connecting boundary points must lie within the boundary. In the following Chapter, we review how more complicated information processing involving direct signal exchanges can be realized in the bulk, but not on the boundary. The operational formulation of AdS/CFT implies that an analog of the bulk operation must exist on the boundary. The fact that it cannot be realized through direct signals implies that it must be realized nonlocally. This puts strong constraints on the boundary state.

\section{Holographic quantum tasks}\label{sec:asympt-task}
\begin{figure}
    \centering
    \includegraphics[width=0.8\linewidth]{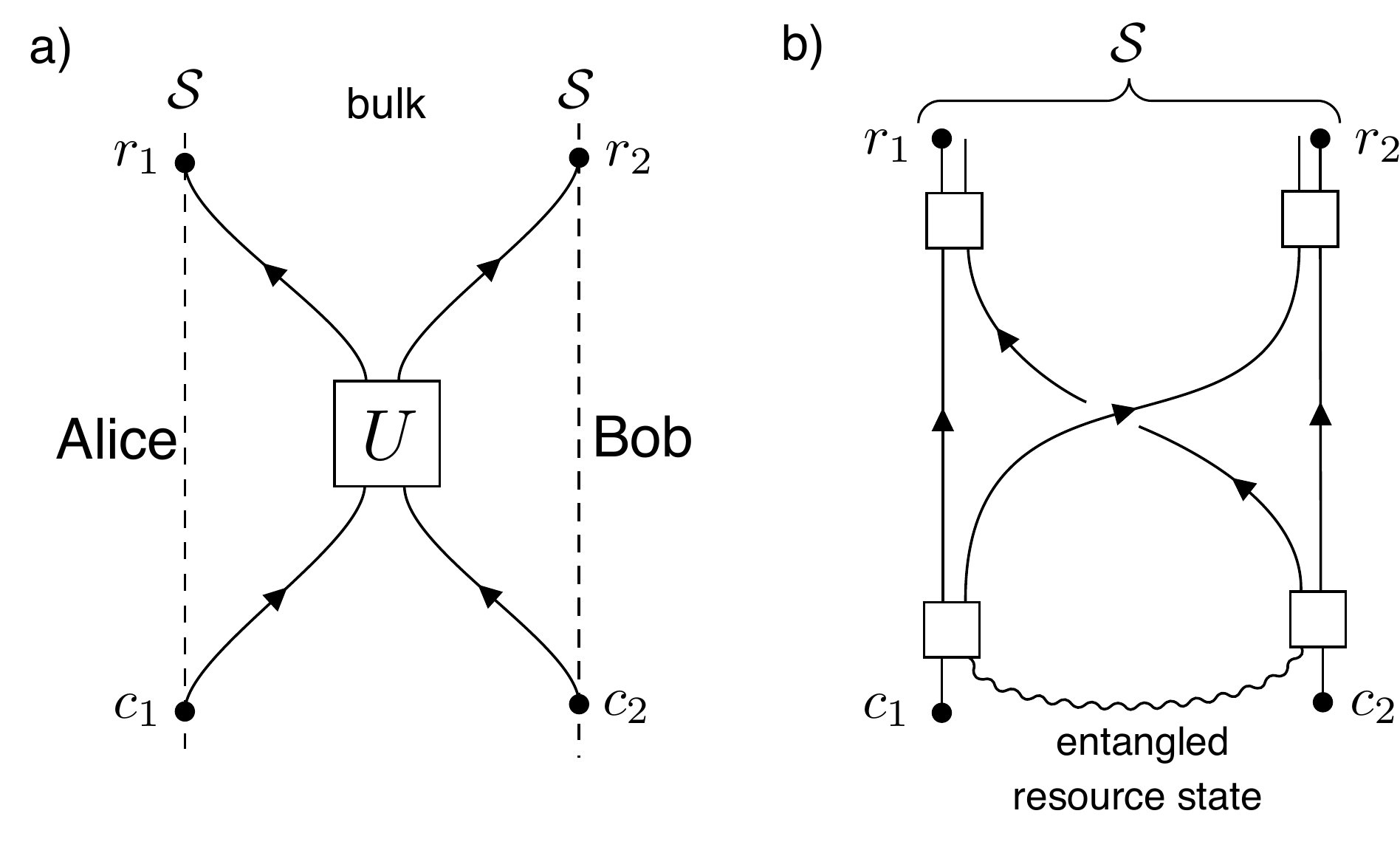}
    \caption{\footnotesize Two realizations of a 
    quantum task. a) A local implementation: 
    Applying a unitary gate $U$ between Alice and Bob's qubits via {a} direct scattering in the bulk. b) An equivalent, nonlocal realization: the task can be achieved on the boundary/screen without direct contact using an entangled resource state between Alice and Bob. White boxes represent some local operations. This is known as a nonlocal quantum computation.}
    \label{fig:NLQC}
\end{figure}

Let $\ket{\Psi}$ be the state of all fields defined on an initial slice of some manifold $\mathcal{M}$. Let $A=\{A_1,A_2,...,A_p\}$ and $B=\{B_1,B_2,...,B_q\}$ be a set of input and output systems in $\mathcal{M}$, respectively. The systems $\{A_1,A_2,...,A_p\}$ and $\{B_1,B_2,...,B_q\}$ are assumed to be located at input and output points $\{c_1,c_2,...,c_p\}$ and $\{r_1,r_2,...,r_q\}$. A \textit{quantum task} is a quantum channel that maps $A$ to $B$.\footnote{A quantum channel is the generalization of time evolution to open systems in quantum mechanics. In particular, it describes the evolution of reduced density matrices under the unitary evolution of the global state $\ket{\Psi}$.}

An \textit{holographic quantum task} corresponds to a quantum task defined on a manifold endowed with quantum fields that have a holographic dual theory defined on its boundary, and which takes its input and output points to be localized on the boundary \cite{May:2019yxi} .\footnote{A quantum task for which input and output points can be identified with boundary points is called an asymptotic quantum task.} A dual boundary quantum task can be defined on the boundary manifold $\partial\mathcal{M}$ with the same input and output systems $A$ and $B$.\footnote{We assume that systems $A$ and $B$ can be localized on the set of points $c_i$ and $r_i$. This is not possible in practice due to the covariant entropy bound discussed in Chapter~\ref{sec:Bousso}. Holographic quantum tasks can be generalized to quantum tasks between input and output regions \cite{May:2021nrl}.} The AdS/CFT correspondence implies that an asymptotic quantum task in $\mathcal{M}$ is possible if and only if the associated boundary quantum task on $\partial\mathcal{M}$ is possible \cite{May:2019yxi}.

Let Alice and Bob each receive some information. They input their information (\textit{eg} a qubit or a bit) at their respective input locations $c_1,c_2$, then process them through a holographic evolution, to send the shared information to their respective output locations $r_1,r_2$. We first consider performing a certain quantum task in the bulk through direct scattering. The input and output points are placed on the boundary so that the causal future of the input points and the causal past of the output points have an intersection in the bulk, which is the scattering region:
\begin{equation}
    J_{12\rightarrow 12} = J^+(c_1)\cap J^+(c_2)\cap J^-(r_1) \cap J^-(r_2). 
\end{equation}
We configure the holographic evolution such that a unitary gate acts locally on the (qu)bits in the scattering region to accomplish the task. This protocol accomplishes the quantum task through {a} direct scattering as shown in Figure~\ref{fig:NLQC}.\footnote{{Note that in principle, any unitary process can be implemented in a holographic setting.}} 
Now, let us turn to the boundary perspective. Since we work in holography, the task must also be achievable on the boundary, however, there are cases where no direct scattering is possible on the boundary.
Thus, we need to find a way to perform the same task nonlocally. Such a task is known as a nonlocal quantum computation~\cite{PhysRevLett.90.010402, Buhrman_2014}. In general, entanglement between Alice and Bob is required to accomplish the task nonlocally (Figure~\ref{fig:NLQC}).

\section{The connected wedge theorem}

By considering a particular task called the \textbf{B}${}_{\bm{84}}$, it can be proven that accomplishing the task nonlocally requires a finite correlation between Alice and Bob. In the \textbf{B}${}_{\bm{84}}$ task, a classical bit $q\in\{0,1\}$ is received at $c_2$, and a qubit in the state $H^q\ket{b}$ is received at $c_1$. $H $ is the Hadamard gate
\begin{equation}
\label{eq:Hadamard}
    H = \frac{1}{\sqrt{2}}\begin{pmatrix}
        1 & 1 \\
        1 & -1
    \end{pmatrix},
\end{equation}
and $b\in\{0,1\}$. We take $b$ and $q$ to be independent random variables. The \textbf{B}${}_{\bm{84}}$ consists of sending $b$ and $q$ to $r_1$ and $r_2$. When $J_{12\rightarrow 12}$ is not empty, this task can be performed locally as follows:
\begin{enumerate}
    \item Send $q$ and $H^q\ket{b}$ from $c_2$ and $c_1$ to $J_{12\rightarrow 12}$.
    \item Inside $J_{12\rightarrow 12}$, knowing $q$, apply $H^q$ to $H^q\ket{b}$. The outcome is $b$.
    \item Send $b$ from $J_{12\rightarrow 12}$ to $r_1$ and $r_2$.
\end{enumerate}
This is what we call a task through direct scattering, or $2$-to-$2$ scattering. 

As mentioned above, the task may be performed even when $J_{12\rightarrow 12}=\varnothing$. The quantum circuit of Figure \ref{fig:B84} shows a nonlocal realization of the task.\footnote{Recall that $X=\ket{0}\bra{1}+\ket{1}\bra{0}$, $Z=\ket{0}\bra{0}-\ket{1}\bra{1}$, and the CNOT gate denoted by a dot and a cross connecting two black lines is defined as $\mathrm{CNOT} = \ket{00}\bra{00} + \ket{01}\bra{01} + \ket{10}\bra{11} + \ket{11}\bra{10}$.} By nonlocal, we mean that the information has been processed without applying any unitary gate in $J_{12\rightarrow 12}$. There is a strong prerequisite for this nonlocal task, which is that a Bell pair must be accessible to $c_1$ and $c_2$. 
\begin{figure}[ht]
    \centering
    \includegraphics[width=0.5\linewidth]{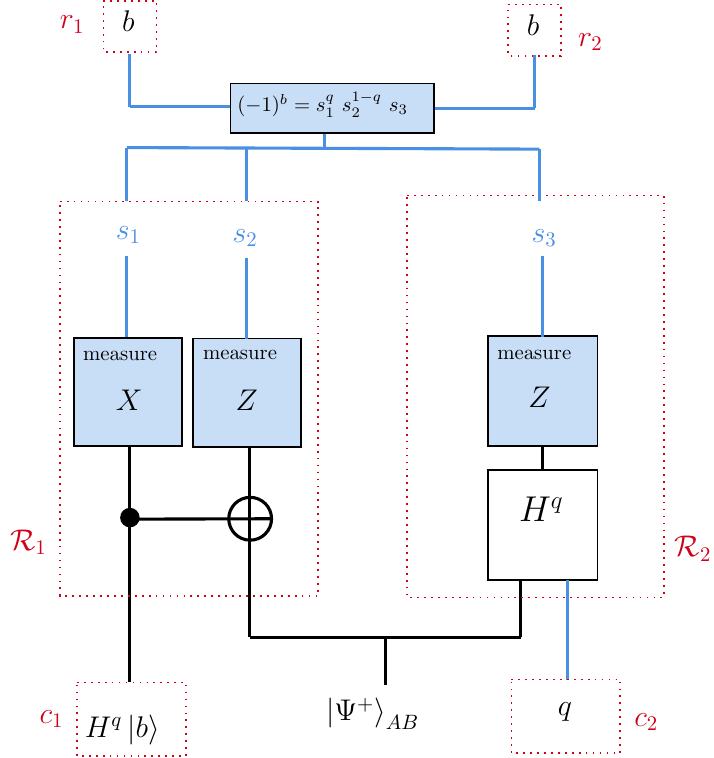}
    \caption{\footnotesize Circuit diagram of the \textbf{B}${}_{\bm{84}}$ task. Black and blue line indicate quantum and classical inputs/outputs, respectively. Blue boxes correspond to measurements or classical operations. Operations taking place in specific regions are indicated by a dotted red box. The resource state is a Bell pair $\ket{\Psi}^+_{AB}=\frac{1}{\sqrt{2}}(\ket{00}+\ket{11})$.  }
    \label{fig:B84}
\end{figure}
The relevant correlation must be available after the input is received and in the past of both outputs. In particular, the unitary operations performed in the \textbf{B}${}_{\bm{84}}$ task are local, and must take place in regions $\mathcal{R}_1$ and $\mathcal{R}_2$ which are in the causal past of $r_1$ and $r_2$ and in the future of $c_1$ and $c_2$, respectively:
\begin{equation}
    \mathcal{R}_i =J^+(c_{i})\vert_\mathcal{S}\cap J^-(r_1)\vert_\mathcal{S} \cap J^-(r_2)\vert_\mathcal{S},
\end{equation}
with $i\in\{1,2\}$. We call $\mathcal{R}_1$ and $\mathcal{R}_2$ the decision regions, and depict them schematically in Figure \ref{fig:B84}.

In our holographic setup, information is transferred between bulk or boundary points by encoding it in localized perturbations that propagate through the bulk or the boundary. Assuming that the bulk theory allows the completion the \textbf{B}${}_{\bm{84}}$ task locally in the bulk, this implies that $J_{12\rightarrow 12}\neq \varnothing$. As mentioned above, there are choices of input and output points such that $J_{12\rightarrow 12}\neq \varnothing$ while the boundary scattering region is empty. In such a case, the task cannot be completed locally on the boundary. The operational equivalence between the bulk and boundary requires the task to be performed nonlocally as in Figure \ref{fig:B84}. For this, there must be a finite mutual information between the decision regions $\mathcal{R}_{1}$ and $\mathcal{R}_{1}$. This leads to the \textit{conncted wedge theorem}:
\begin{theorem}[Connected wedge theorem~\cite{May:2019yxi}]\label{th:CWT}
    Consider two input points $c_1,c_2$ and output points $r_1,r_2$ on the boundary of an asymptotically AdS$_3$ spacetime. If a 2-to-2 scattering $c_1,c_2\rightarrow r_1,r_2$ is possible in the bulk, \emph{\textit{ie}}
    \begin{equation}
        J_{12\rightarrow 12} \neq \varnothing,
    \end{equation}
    but not on the asymptotic boundary $\mathcal{S}$, \emph{\textit{ie}}
    \begin{equation}
        J^+(c_1)\vert_\mathcal{S}\cap J^+(c_2)\vert_\mathcal{S}\cap J^-(r_1)\vert_\mathcal{S} \cap J^-(r_2)\vert_\mathcal{S} = \varnothing,
    \end{equation}
    then the decision regions $\mathcal{R}_{1,2}$ on the boundary must have a connected entanglement wedge. Equivalently,
    \begin{equation}
        I(\mathcal{R}_1:\mathcal{R}_2)=O(1/G).
    \end{equation}
\end{theorem}
The equivalent statement in terms of the entanglement wedge $W(\mathcal{R}_1\cup \mathcal{R}_2)$ can be shown using concepts from the previous chapter. Assuming that the scattering is not possible on the boundary, the decision regions cannot overlap. Let $\tilde{\mathcal{R}}_1$ and $\tilde{\mathcal{R}}_2$ be the two disconnected regions such that $\tilde{\mathcal{R}}_1\cup \tilde{\mathcal{R}}_2 = (\mathcal{R}_1\cup \mathcal{R}_2)^c$ on some spacelike slice of the boundary. Mutual information of order $O(1/G)$ implies that $\gamma_{\rm e}(\mathcal{R}_1\cup \mathcal{R}_2) < \gamma_{\rm e}(\mathcal{R}_1) \cup \gamma_{\rm e}(\mathcal{R}_2)$. Then, $\mathcal{R}_1\cup \mathcal{R}_2 = \gamma_{\rm e}(\tilde{\mathcal{R}}_1)\cup \gamma_{\rm e}(\tilde{\mathcal{R}}_2)$. See Figure \ref{fig:connected}. Conversely, when $\gamma_{\rm e}(\mathcal{R}_1\cup \mathcal{R}_2) = \gamma_{\rm e}(\mathcal{R}_1) \cup \gamma_{\rm e}(\mathcal{R}_2)$, $I(\mathcal{R}_1:\mathcal{R}_2)$ can only come from higher order corrections and must be $O(1)$.
\begin{figure}[ht]
\begin{subfigure}[t]{0.48\linewidth}
\centering
\includegraphics[width=4cm]{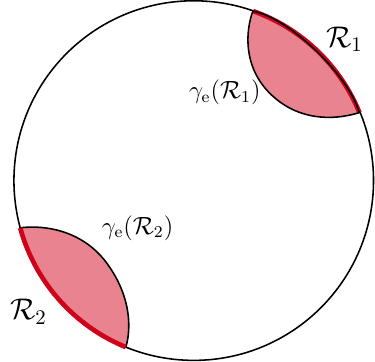}
\caption{$I(\mathcal{R}_1:\mathcal{R}_2)=O(1)$}
\end{subfigure}
\begin{subfigure}[t]{0.48\linewidth}
\centering
\includegraphics[width=4cm]{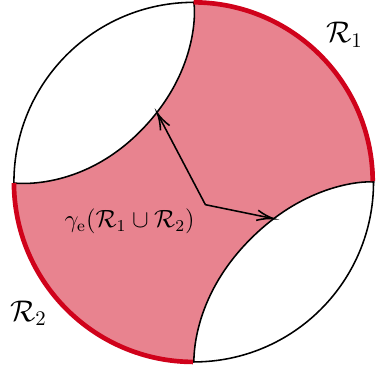}
\caption{$I(\mathcal{R}_1:\mathcal{R}_2)=O(1/G)$}
\end{subfigure}
    \caption{\footnotesize The two figures show a Cauchy slice of AdS$_3$. The entanglement wedge of $\mathcal{R}_1\cup \mathcal{R}_2$ is depicted in red for two different scalings of their mutual information.}
    \label{fig:connected}
\end{figure}

A proof of Theorem \ref{th:CWT} is given in \cite{May:2019odp,May:2022clu}. We sketch it briefly. The argument relies on the task \textbf{B}${}^{\times n}_{\bm{84}}$, defined as $n$ \textbf{B}${}_{\bm{84}}$ tasks performed in parallel. The parameters $b$ and $q$ of each task are independent. Without any entangled resource state accessible, it can be shown that the success probability $p_{\rm sucess}$ of \textbf{B}${}^{\times n}_{\bm{84}}$ without using $J_{12\rightarrow 12}$ is bounded from above by $\cos^{2n}(\pi/8)$ \cite{Tomamichel_2013}. Conversely, assume $p_{\rm sucess}\geq 1-\epsilon$. Let $\rho_{AB}$ be a two-partite state with $A$ accessible to $c_1$ and $B$ accessible to $c_2$. Then \cite{May:2019odp}, 
\begin{equation}
\label{eq:const}
    I(A:B) \geq -2\ln\left[2(\epsilon+\cos^{2n}(\pi/8))\right].
\end{equation}
In the holographic context, considering the task on the boundary, $A$ and $B$ are identified with $\mathcal{R}_1$ and $\mathcal{R}_2$, respectively.\footnote{One caveat is that we ignore the complementary regions of the decision regions. As pointed out in~\cite{May:2019odp} and the erratum of~\cite{May:2019yxi}, GHZ-type (tripartite) resource states among the complementary regions and $\mathcal{R}_1\cup\mathcal{R}_2$ can offer a protocol that does not need a finite mutual information between $\mathcal{R}_1$ and $\mathcal{R}_2$. This is however unlikely to be the case for holographic scattering as we do not expect a large amount of the GHZ-type entanglement in holography~\cite{Susskind:2014yaa,Nezami:2016zni,Mori:2024gwe}.} Since the scattering region in the bulk is assumed to be non-empty, the success rate for the bulk task is high: $p_{\rm sucess}\geq 1-\epsilon$. In the semiclassical limit $G\rightarrow 0$, we take $\epsilon\rightarrow 0$ and $n\rightarrow \infty$. The bulk being completely fixed and classical, the task can be performed perfectly as described below \eqref{eq:Hadamard}. The only constraint on $n$ comes from the condition that sending too many signals into the bulk might lead to a backreaction closing off the scattering region. By taking $n$ as large as $o(1/G)$, backreaction can be ignored.\footnote{The notation $f(x)=o(g(x))$ as $x\rightarrow \infty$ means that for any constant $\epsilon>0$ there exists $x_0$ such that $|f(x)|\leq \epsilon g(x)$ for $x\geq x_0$.}

From the boundary point of view, where no scattering region is available, the constraint \eqref{eq:const} reads
\begin{equation}
    I(\mathcal{R}_1:\mathcal{R}_2)\gtrsim n  \Rightarrow I(\mathcal{R}_1:\mathcal{R}_2) = O(1/G).
\end{equation}
Note that this does not rely on the large-$N$ separation between the area term and the quantum correction of the mutual information.\footnote{This point is important, since later the dual quantum theory remains unknown so holographic entanglement entropy may not have a nice split into the area term of order $1/G$ (or possibly higher order) and quantum corrections of order unity.
There is possibly a term growing with a rate that is sub-linear but larger than any $1/G^a$ with $a<1$. Even if this is the case, the argument in the main body says the mutual information must be {at least as large as $1/G$,} implying a geometric correlation from the area terms. We thank Alex May for pointing out a loophole in the first version of \cite{Franken:2024wmh}, and explaining this to us.}

This discussion is based on quantum information and does not rely on the details of the theory. Thus, it should work in any holographic spacetime. In this paper, we use the connected wedge theorem as a guiding principle to constrain or check various proposals related to the holographic duality.

We note that for the asymptotic quantum task in AdS$_3$/CFT$_2$, there is a gravitational proof based on the focusing theorem (see Section~\ref{sec:focusing}) and the quantum extremal surface formula of the holographic entanglement entropy~\cite{May:2019odp}. We will follow the same strategy for the dS case in Chapter~\ref{ch:CWT}. We also note that the connected wedge theorem was originally claimed to be valid in any dimension~\cite{May:2019odp,May:2019yxi,May:2021nrl}. It was pointed out in~\cite{May:2022clu} that the geometric and quantum information arguments are not valid above three bulk dimensions. For the same reasons, the results of Chapter~\ref{ch:CWT} are also strictly restricted to dS$_3$.

\section{Holographic scattering from a non-asymptotic boundary}

While the original proposal of the connected wedge theorem considers an asymptotic quantum task, where the nonlocal quantum computation takes place on the asymptotic boundary, there is no reason not to consider more general holographic spacetimes such as non-AdS and/or a holographic screen not located at the asymptotic boundary.

The authors of~\cite{Mori:2023swn} extended the connected wedge theorem to a braneworld or cutoff surface in an asymptotically AdS$_3$ spacetime. In these setups, the holographic screen $\mathcal{S}$ is located somewhere other than the asymptotic boundary. It turns out that the causality based on the induced metric on $\mathcal{S}$ leads to an apparent violation of the connected wedge theorem. The resolution presented in \cite{Mori:2023swn} is to fill behind the hypersurface with a fictitious asymptotically AdS space,\footnote{In the work~\cite{Mori:2023swn}, the focus was on a braneworld/cutoff AdS so just extending the original spacetime beyond the hypersurface was sufficient to resolve the puzzle. In general, one can glue an arbitrary fictitious spacetime with a fictitious asymptotic boundary as long as it satisfies the Israel junction condition~\cite{Israel:1966rt} to be a smooth spacetime satisfying the Einstein equation.} and extend the scattering trajectories to the fictitious asymptotic boundary to define fictitious input and output points $\tilde{c}_{1,2}\in J^-(c_{1,2})$, $\tilde{r}_{1,2}\in J^+(r_{1,2})$. See Figure~\ref{fig:brane-ind} for its illustration.
The authors have shown that the boundary domains of dependence defined from the induced causality, denoted by $\hat{J}^\pm$, align with the connected wedge theorem.
Ultimately,~\cite{Mori:2023swn} proposes the following refined connected wedge theorem:
\begin{theorem}[Refined connected wedge theorem]\label{thm:refined-CWT}
    Two input points $c_1,c_2$ and output points $r_1,r_2$ are on a holographic screen $\mathcal{S}$, which is not necessarily the conformal boundary of asymptotically AdS$_3$ spacetime. Suppose a 2-to-2 scattering $c_1,c_2\rightarrow r_1,r_2$ is possible in the bulk, \emph{\textit{ie}}
    \begin{equation}
        J_{12\rightarrow 12} = J^+(c_1)\cap J^+(c_2)\cap J^-(r_1) \cap J^-(r_2) \neq \varnothing,
    \end{equation}
    but not on the holographic screen $\mathcal{S}$, \emph{\textit{ie}}
    \begin{equation}
        \qty[J^+(\tilde{c}_1)\cap J^+(\tilde{c}_2)\cap J^-(\tilde{r}_1)\cap J^-(\tilde{r}_2)] \cap \mathcal{S} = \varnothing,
    \end{equation}
    where the boundary causality determining $\hat{J}^\pm (p)$ is given by the induced lightcones $\hat{J}^{\pm}(p)=J^\pm (\tilde{p})\cap\mathcal{S}$ from a fictitious point $\tilde{p}$ on the fictitious asymptotic boundary.
    Then, the decision regions $\mathcal{R}_{1,2}=\hat{J}^+(c_{1,2})\cap \hat{J}^-(r_1) \cap \hat{J}^-(r_2)$ on the screen $\mathcal{S}$ must have a connected entanglement wedge, implied from
    \begin{equation}
        I(\mathcal{R}_1:\mathcal{R}_2)=O(1/G).
    \end{equation}
\end{theorem}

While the fictitious spacetime and boundary behind the holographic screen are not necessarily unique, the induced causality from a local point on the fictitious boundary is anticipated from the apparent nonlocality/superluminality of the boundary theory based on holographic renormalization group flow~\cite{Freedman:1999gp,Girardello:1998pd,Distler:1998gb,McGough:2016lol}. The fictitious boundary behind the holographic screen serves as the `true' UV boundary, and a fictitious local excitation on the UV boundary induces an effective, apparently nonlocal excitation dual to a localized signal in the bulk. This concept of the induced lightcone identifies causal and entanglement structures consistent with the holographic description.
The connected wedge theorem serves as a nontrivial check for the induced lightcone proposal.

\begin{figure}
    \centering
    \includegraphics[width=0.3\linewidth]{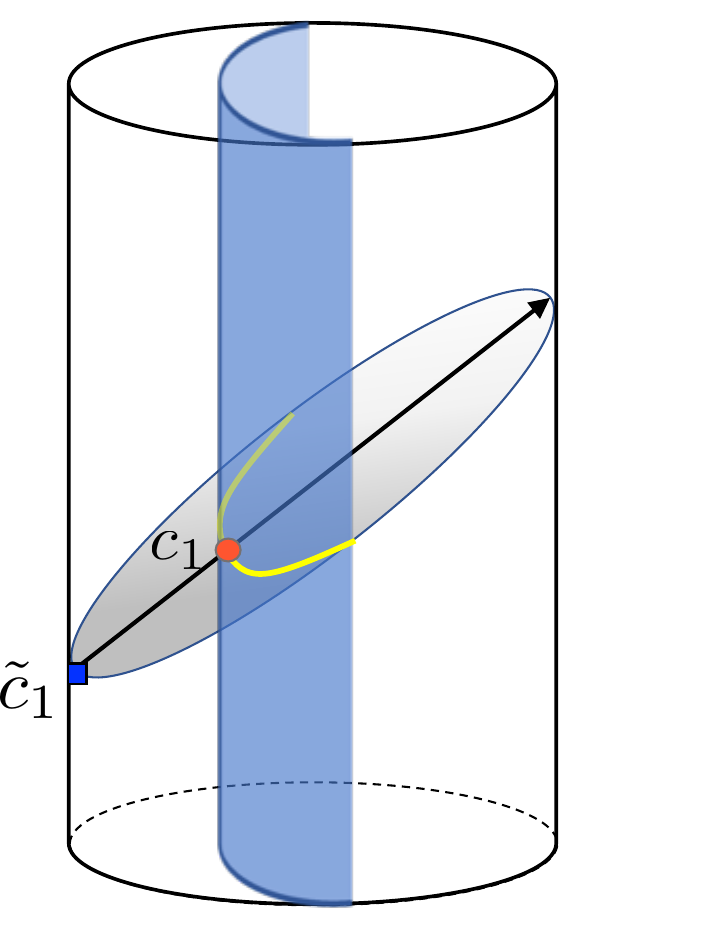}
    \caption{\footnotesize Given an input point $c_1$ on the holographic screen $\mathcal{S}$ (blue surface), a fictitious input point $\tilde{c}_1$ is defined on the fictitious asymptotic boundary. The intersection of the bulk lightcone emanating from $\tilde{c}_1$ and the screen $\mathcal{S}$ defines the induced lightcone on $\mathcal{S}$. Figure taken from \cite{Mori:2023swn}.}
    \label{fig:brane-ind}
\end{figure}

We emphasize that this induced lightcone approach in the light of the connected wedge theorem amounts to identifying the UV boundary that describes a nonlocal boundary theory locally. This determines the causality associated with holographic entanglement on the holographic screen. 

\addtocontents{toc}{\protect\end{adjustwidth}}
\part{Gravitational constraints from quantum information \label{part:QI}}
\addtocontents{toc}{\protect\begin{adjustwidth}{1cm}{0cm}}

\chapter{Thermodynamical properties of spacetime}
\label{ch:thermo}

In this chapter, we review the second law of black hole thermodynamics and the associated energy condition. We then introduce the focusing theorem of general relativity. Finally, we present the covariant entropy bound and holographic principle.

\section{Generalized second law of thermodynamics}
\label{sec:GSL}

In 1972, Bekenstein~\cite{Bekenstein1} made a groundbreaking leap in our understanding of gravity by making a very simple point. In general relativity, the no-hair theorem~\cite{nohair1,nohair2,nohair3} states that a black hole in equilibrium has only three degrees of freedom: its mass, electric charge, and angular momentum. From the point of view of an observer, there is no way to determine the number of degrees of freedom of objects previously dropped into the black hole. In other words, the existence of black holes in general relativity seems to contradict the second law of thermodynamics \eqref{eq:2ndlaw}.

A few months earlier, Hawking~\cite{Hawking:1971tu,Hawking:1971vc} had shown that the area of a black hole horizon, denoted $A_H$, cannot decrease:
\begin{equation}
\label{eq:areathm}
    {dA_H} \geq 0.
\end{equation}
This can be shown under the \textit{null energy condition} (NEC) 
\begin{equation}
\label{eq:NEC}
    k^{\mu}k^{\nu}\braket{T_{\mu\nu}}\geq 0,
\end{equation}
where $k^{\mu}$ is a null vector and $T_{\mu\nu}$ is the stress-energy tensor. The NEC is broadly satisfied in classical field theories. Moreover, it serves as a crucial assumption in many other proofs of constraints for classical spacetimes \cite{PhysRevLett.14.57,PhysRevLett.61.1446,Friedman:1993ty,Farhi:1986ty,Gao:2000ga,PhysRevLett.37.879,PhysRevD.46.603,Olum:1998mu,Visser:1998ua,Penrose:1993ud}. 

This led Bekenstein to conjecture that one should associate an entropy with black holes, proportional to the area of their event horizons. This intuition was further developed~\cite{Bekenstein2, Bardeen:1973gs} by defining a complete set of black hole thermodynamic laws. Hawking derived explicitly the multiplicative factor relating $A_H$ to the entropy by showing that black holes evaporate~\cite{Hawking:1974rv,Hawking:1975vcx}, producing a thermal radiation with associated thermodynamic entropy
\begin{equation}
\label{eq:BekensteinHawking}
    S_{\rm BH} = \frac{A_H}{4G}.
\end{equation}
A more general notion of gravitational entropy, which holds for diffeomorphisms invariant gravity theories beyond general relativity, can be defined as~\cite{Wald:1993nt,Iyer:1994ys}
\begin{equation}
    \label{eq:WaldS}
    S_{\rm Wald} = -2\pi \int_Hdx^{d-1}\sqrt{-h}\frac{\partial L}{\partial R^{\mu\nu}_{~~\rho\sigma}}\varepsilon_{\mu\nu}\varepsilon^{\rho\sigma},
\end{equation}
where we integrate over the horizon surface, $h$ is the trace of the induced metric on the horizon, $L$ is the gravitational Lagrangian, and $\varepsilon_{\mu\nu}$ is the binormal to the horizon normalized such that $\varepsilon_{\mu\nu}\varepsilon^{\mu\nu}=-2$.\footnote{The binormal is the cross product of a tangent and a normal vector of the surface.} This expression, called \textit{Wald entropy}, is the Noether charge associated with the Killing vector generating a Killing bifurcate horizon.\footnote{A Killing horizon is the null hypersurface generated by a null Killing vector. In other words a spacetime has a Killing horizon when there is a Killing vector that maps a null hypersurface onto itself. A bifurcate Killing horizon is the intersection of two Killing horizons.} Wald entropy and Killing horizons are central tools in many proofs of theorem in black hole thermodynamics~\cite{Wall:2018ydq}, and $S_{\rm Wald}$ reduces to the Bekenstein-Hawking formula in general relativity.

Returning to the initial problem posed by Bekenstein, it was conjectured in \cite{Bekenstein3} that a \textit{generalized second law} (GSL) of thermodynamics should combine contributions from both the black hole and the exterior system:
\begin{equation}
    dS_{\rm gen} \geq 0,
\end{equation}
where
\begin{equation}
\label{eq:Sgenth}
    S_{\rm gen} = \frac{A_H}{4G} + S_{\rm out}.
\end{equation}
Here $S_{\rm out}$ is the entropy of matter fields outside of the black hole. While the NEC is almost always satisfied in classical field theory, it can be violated in quantum field theories~\cite{Epstein:1965zza}. In particular, Hawking radiation crucially relies on the violation of the NEC close to the horizon~\cite{PhysRevD.13.2720}. Nevertheless, strong evidence supports the generalized second law, as it has been demonstrated in various contexts~\cite{Wall:2009wm, Wall:2011hj}, and has recently been shown to hold to all orders in the semiclassical regime~\cite{Faulkner:2024gst,Kirklin:2024gyl}.

\section{Focusing theorem}
\label{sec:focusing}

Another example of a spacetime property demonstrated under the NEC is the \textit{focusing theorem}. This theorem will be used frequently in this work, as it provides a tool to study the dynamics of horizons and null hypersurfaces.

Consider a congruence of lightrays $\mathcal{N}(\gamma)$ emanating orthogonally from a codimension-two spacelike hypersurface $\gamma$. Let $\lambda$ be the affine parameter of the generating lightrays, such that $d/d\lambda$ is null and orthogonal to the congruence. In particular, we define the normal vector
\begin{equation}
    k^{\mu} = \pm\left(\frac{d}{d\lambda}\right)^{\mu},
\end{equation}
with sign $+$ for future-directed lightrays and $-$ for past-directed lightrays. For any choice of affine parameter, the normal vector must satisfy the geodesic equation
\begin{equation}\label{eq:k_geo_eq}
    k^{\mu}\nabla_{\mu}k^{\nu} = 0,
\end{equation}
with $\nabla_{\mu}$ the covariant derivative compatible with the metric $g_{\mu\nu}$. The induced metric $h_{ab}$ and the extrinsic curvature $K_{ab}$ of the $d$-dimensional hypersurface $\mathcal{N}(\gamma)$ with normal vector $k^{\mu}$ are:
\begin{eqnarray}
    h_{ab} &=&\frac{\partial x^{\mu}}{\partial y^{a}}\frac{\partial x^{\nu}}{\partial y^{b}}{g}_{\mu\nu},\\
    K_{ab} &=& \frac{\partial x^{\mu}}{\partial y^{a}} \frac{\partial x^{\nu}}{\partial y^{b}} \nabla_{\mu} k_{\nu},
\end{eqnarray}
where $\{y^{a},~a=0,...,d-1\}$ is a set of coordinates on $\gamma$. The \textit{expansion parameter} of the congruence is defined as the trace of the extrinsic curvature tensor\cite{Wald:1984rg}:
\begin{equation}
\label{eq:expansion}
    \theta = h^{ab}K_{ab} = \nabla_{\mu}k^{\mu}.
\end{equation}
An equivalent definition is
\begin{equation}
\label{eq:A}
    \theta = \lim_{\mathcal{A}\rightarrow 0}\frac{1}{\mathcal{A}}\frac{d\mathcal{A}}{d\lambda},
\end{equation}
where $\mathcal{A}$ is the area element spanned by neighboring lightrays. A \textit{lightsheet} $\mathcal{L}$ is a null congruence with non-positive expansion $\theta \leq 0$. In classical general relativity, the Raychaudhuri equation describes the evolution of the expansion parameter along a congruence:
\begin{equation}
\label{eq:Raychaudhuri}
\frac{d\theta}{d\lambda}=-\frac{1}{d-1}\theta^2-\sigma_{\mu\nu}\sigma^{\mu\nu} - k^{\mu}k^{\nu}R_{\mu\nu} ,
\end{equation}
where $\sigma^{\mu\nu}$ is the shear tensor~\cite{Wald:1984rg}. Assuming the NEC, $k^{\mu}k^{\nu}R_{\mu\nu}$ must be positive due to Einstein's equations. $\sigma_{\mu\nu}\sigma^{\mu\nu}$ is non-negative. This leads to the following theorem.
\begin{theorem}[Focusing theorem]
\label{thm:focus}
    In a spacetime satisfying the NEC, the expansion is non increasing at all regular points of a congruence of lightrays emanating orthogonally from a codimension-two hypersurface:\footnote{Regular points refer to non-caustic points, \emph{ie} points at which $\theta$ is finite.}
\begin{equation}
    \frac{d\theta}{d\lambda}\leq 0.
\end{equation}
\end{theorem}
The focusing theorem implies that, under the NEC, a congruence which has $\theta\leq 0$ at $\lambda=0$ must be a lightsheet. We emphasize that, as the NEC can be violated in quantum field theories, so too can the focusing theorem be violated in spacetimes sourced by quantum fields.

\section{Bousso bound and hydrodynamic regime}
\label{sec:Bousso}

As we will discuss shortly, the works of Bekenstein and Hawking raise the question of whether we may associate a counting of degrees of freedom to the area of a black hole. Interestingly, the black hole entropy is not substantially very different from the expected value of the entropy for any more common system of similar mass and size. In particular, Bekenstein~\cite{PhysRevD.23.287} conjectured an upper bound for the thermodynamic entropy of an arbitrary system:
\begin{equation}
\label{eq:Bekenstein}
    S_{\rm th}\leq \frac{A}{4G},
\end{equation}
where $A$ is the area of the smallest sphere containing the whole system. In other words, the \textit{Bekenstein bound} conjectures that the Bekenstein-Hawking entropy \eqref{eq:BekensteinHawking} saturates the bound on the entropy content of any system whose spherical boundary has an area $A_H$. 

Although rich in insights, the Bekenstein bound is known to be violated by gravitationally collapsing objects and very large regions of the universe. Moreover, it presents limitations due to the ambiguity of its statement. In particular, $A$ refers to the boundary of some spatial region, but which one, exactly? General relativity requires covariance and thus there should not be any preferred spatial region. The fact that equation~\eqref{eq:Bekenstein} refers to any spatial region leads to violations in cosmology~\cite{Fischler:1998st}.\footnote{Roughly, the idea is that over cosmological scales, the entropy density should be constant. However, volume grows faster than area. So the bound must be violated at large enough scales~\cite{Bousso:2003kb}.} Moreover, there is no specification of what constitutes the interior or exterior of a surface. In a closed universe, this becomes particularly ambiguous. Again, one would require a covariant prescription to decide which side of the boundary is considered.

Bousso proposed a resolution of these conceptual issues in~\cite{Bousso:1999xy}, where he realized that
\begin{itemize}
    \item From the contradiction presented in~\cite{Fischler:1998st}, it is clear that one cannot consider any spatial region for a given boundary. Covariance forbids us from selecting any preferred spacelike slice. The only sensible solution would then be to focus on lightlike hypersurfaces.
    \item Considering lightlike hypersurfaces leads to a natural notion of interior and exterior. From a given codimension-two boundary $\gamma$, there are four distinct null directions in which one could construct a lightlike hypersurface orthogonal to $\gamma$. See Figure~\ref{fig:lightsheet}. Two of them are of positive expansion and two are of negative expansion. We consider the direction in which the area grows as the exterior.
\end{itemize}
\begin{figure}[ht]
    \centering
    \includegraphics[width=0.5\linewidth]{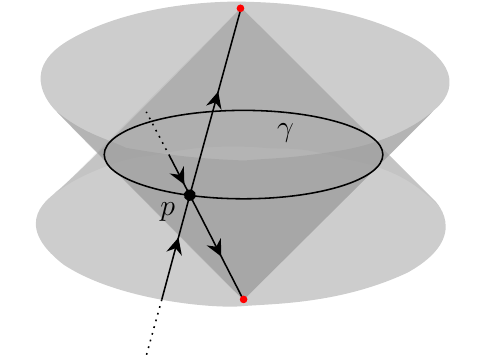}
    \caption{\footnotesize From a codimension-two surface, one can construct four orthogonal null hypersurfaces. Two of them are contracting, and the two others expanding. In this example, we depict by red points the two caustics at which the two contracting hypersurfaces end. Every point \( p \in \gamma \) generates four lightrays that emanate orthogonally from \( \gamma \).}
    \label{fig:lightsheet}
\end{figure}
From these two insights emerges the covariant entropy bound, or \textit{Bousso bound}, stated as follows~\cite{Flanagan:1999jp}. 

\begin{theorem}[Covariant entropy bound] \label{th:Bousso}
   In a spacetime satisfying the NEC, let $\mathcal{L}(\gamma_1,\gamma_2)$ be a lightsheet that emanates and terminates orthogonally from two codimension-two spacelike hypersurfaces $\gamma_1$ and $\gamma_2$. Then,
   \begin{equation}
   \label{eq:BB}
       S_{\rm th}(\mathcal{L}(\gamma_1,\gamma_2)) \leq \frac{\text{Area}(\gamma_1)-\text{Area}(\gamma_2)}{4G},
   \end{equation}
   where $S_{\rm th}(\mathcal{L}(\gamma_1,\gamma_2))$ is the thermodynamic entropy of the lightsheet.
\end{theorem}
One can always choose to take $\gamma_2$ to be a caustic of the lightsheet where $\theta=-\infty$, such that the bound takes the form $S_{\rm th}(\mathcal{L}(\gamma))\leq\text{Area}(\gamma)/4G$ where $\gamma$ is the generating surface. 

Although we referred to the Bousso bound as a theorem, it has only been proven in the \textit{hydrodynamic regime}, where the entropy in a given region can be computed by integrating an entropy density~\cite{Flanagan:1999jp,Bousso:2003kb}. Let $\vec{y}=(y^1,\cdots, y^{d-1})$ be a set of coordinates on $\gamma_1$. Each geodesic is labeled by $\vec{y}$. Hence, $(\vec{y},\lambda)$ is a coordinate system on $\mathcal{L}(\gamma_1,\gamma_2)$. The main assumption of the hydrodynamic regime is that there exists a local entropy current $s^{\mu}$, such that the entropy density on the lightsheet is given by
\begin{equation}
\label{eq:hydros}
    s = -k^{\mu}s_{\mu}.
\end{equation}
It can be shown that the entropy passing through the lightsheet is~\cite{Flanagan:1999jp}
\begin{align}\label{eq:4D_entropy_flux}
    S_{\rm th}(\mathcal{L}(\gamma_1,\gamma_2)) &=\int_{\gamma_1} d^{d-1}y \sqrt{h(y,0)}\int_0^1 d\lambda~\frac{\mathcal{A}(\vec{y},\lambda)}{\mathcal{A}(y,0)}~s(y,\lambda),
\end{align}
where we normalized $\lambda$ such that $\lambda=0$ on $\gamma_1$ and $\lambda=1$ on $\gamma_2$, and $h(\vec{y},\lambda)$ is the determinant of the induced metric on the fixed $\lambda$ slice of the lightsheet, and $s(\vec{y},\lambda)$ is the entropy density on the lightsheet. 

Clearly, this definition of entropy is coarse-grained and cannot hold when the size of the lightsheet becomes smaller than some microscopic length scale. To ensure the validity of \eqref{eq:4D_entropy_flux}, two conditions are imposed~\cite{Flanagan:1999jp,Bousso:2003kb, Strominger:2003br}:
\begin{enumerate} 
\item The ``\emph{gradient of entropy density}'' condition: 
\begin{equation}
\label{eq:cond1_hydro}
\left| s(\lambda)' \right| \leq 2\pi k^{\mu} k^{\nu}  T_{\mu\nu}, \end{equation} 
where $' \equiv \frac{d}{d\lambda} = k^{\mu} \nabla_{\mu}$. 
\item 
The ``\emph{isolated system}'' condition: \begin{equation} 
\label{eq:cond2} 
s(0) \leq -\frac{\theta(0)}{4G}. \end{equation}
\end{enumerate} 
The gradient of entropy density condition requires that the gradient of the entropy density be less than the energy density. This ensures that the entropy density only varies significantly over scales greater than the wavelength of any field mode contributing to the entropy. importantly, this condition also implies the null energy condition (NEC).

The second condition states that the Bousso bound holds for surfaces that are infinitesimally close to $\gamma_1$. As argued in~\cite{Bousso:1999xy}, it does not impose any significant restriction, as the entropy current can always be adjusted locally around $\gamma$, without violating the condition on the gradient of the entropy density, and without significantly altering the total entropy $S(\mathcal{L}(\gamma_1, \gamma_2))$. Physically, this second condition requires the lightsheet to be isolated, in the sense that modes that significantly contribute to the entropy should be contained within $\mathcal{L}(\gamma_1, \gamma_2)$ and not extend beyond $\gamma_1$. Note that this condition couples to the focusing theorem implies that the congruence is a lightsheet.

Under assumptions \eqref{eq:cond1_hydro} and \eqref{eq:cond2}, the Bousso bound can be proven in any dimension~\cite{Flanagan:1999jp,Bousso:2003kb}. Although we presented arguments for the physical relevance of these two conditions, the proofs~\cite{Flanagan:1999jp,Bousso:2003kb} do not constitute a fundamental derivation of Bousso's conjecture. In particular, they do not account for the nonlocal aspects of entropy. They do, however, rule out a very large class of potential counterexamples, and provide good evidence for the bound in cosmological contexts, where the hydrodynamic regime is a very good approximation~\cite{Bousso:2003kb}.

An important check is that the Bousso bound implies the generalized second law. Consider a foliation of the horizon using spacelike surfaces $\gamma(\lambda)$, and take $k^{\mu}$ to be past-directed such that $\gamma_2 = \gamma(\lambda_2)$ is in the past of $\gamma_1 = \gamma(\lambda_1)$ where $\lambda_1 \leq \lambda_2$. Writing $S_{\rm th}(\lambda)$ as the entropy that has passed through the horizon in the past of $\lambda$, the generalized second law, $S_{\rm gen}(\lambda_1) \geq S_{\rm gen}(\lambda_2)$, is rewritten as \begin{align} \frac{\text{Area}(\gamma_1)-\text{Area}(\gamma_2)}{4G} &\geq S_{\rm th}(\lambda_1) - S_{\rm th}(\lambda_2), \end{align} which is guaranteed by the Bousso bound \ref{th:Bousso}.\footnote{Since we have taken $k^{\mu}$ to be past-directed, $\theta \leq 0$ follows when the NEC is satisfied, according to the Hawking area law.}

The success of the Bousso bound, which is well-defined for arbitrary null hypersurfaces, suggests that the connection between spacetime and quantum information extends well beyond black hole thermodynamics.

\section{Holographic principle}
\label{sec:holography}

Another crucial check that the Bousso bound should satisfy is whether it implies the Bekenstein bound in the appropriate limit. Although we previously mentioned that light-like hypersurfaces are necessary for producing a well-defined bound, statements can also be formulated for spacelike surfaces in some cases. The Bousso bound helps us identifying them.

As noted in Chapter~\ref{ch:entanglement}, thermodynamical entropy grows under unitary evolution. We also noted that the density matrices of spatial slices sharing the same causal diamond are related by unitary transformations. In particular, these slices share the same boundary $\gamma$. Let $\Sigma(t)$ be a complete foliation in achronal slices of $D(\Sigma)$ parametrized by $t \in (-\infty, \infty)$, where $\partial \Sigma = \partial \Sigma(t) = \gamma$, and $\Sigma = \Sigma(t_0)$ is a specific slice chosen arbitrarily from the foliation,\footnote{An achronal slice is a codimension-one hypersurface that is lightlike or spacelike.} such that $\Sigma(t_2)$ lies entirely in the future of $\Sigma(t_1)$ if and only if $t_1 < t_2$, and $D(\Sigma) = \left\{\Sigma(t)\right\}$. Then, the second law of thermodynamics \eqref{eq:2ndlaw} implies 
\begin{equation} S_{\rm th}(\Sigma(t_1)) \leq S_{\rm th}(\Sigma(t_2)),\quad t_1 \leq t_2. \end{equation} 
We assume that there exists a future-directed lightsheet $\mathcal{L}(\gamma)$ emanating orthogonally from $\gamma$ and that has no other boundary than $\gamma$, \emph{ie} that ends at a caustic. Then, a foliation $\Sigma(t)$ as defined above exists such that 
\begin{equation} 
\mathcal{L}(\gamma) = \lim_{t \rightarrow \infty} \Sigma(t).
\end{equation} 
By applying the Bousso bound and the second law of thermodynamics to the lightsheet $\mathcal{L}(\gamma)$ we find
\begin{equation} 
S_{\rm th}(\Sigma) \leq S_{\rm th}(\mathcal{L}(\gamma)) \leq \frac{\text{Area}(\gamma)}{4G}. \end{equation} 
This concludes the proof of the \textit{spacelike projection theorem}~\cite{Bousso:1999xy}:

\begin{theorem}[Spacelike projection theorem] 
\label{th:spacelike}
Let $\gamma$ be a codimension-two spacelike surface and $\Sigma$ an achronal slice with $\partial \Sigma = \gamma$. If $\gamma$ has a future-directed lightsheet $\mathcal{L}(\gamma)$ that has no other boundary than $\gamma$, \begin{equation} 
S_{\rm th}(\Sigma) \leq \frac{\text{Area}(\gamma)}{4G}. 
\end{equation}
\end{theorem} 
Considering the statistical definition of thermodynamic entropy \eqref{eq:thermo}, this statement implies that the number of microstates in a spacelike region is upper-bounded by the area of the boundary of the system divided by $4G$~\cite{Bousso:2002ju}. Although there is no fundamental derivation of the Bousso bound, its broad validity suggests that any candidate for an underlying theory that would describe matter and gravity at the fundamental level should be consistent with the Bousso bound in the low-energy limit.

The spacelike projection theorem can also be formulated in terms of the number of bits required to describe the system, given by $N_{\rm bits} = S_{\rm th}/\ln 2$. We thus formulate the so-called \textit{holographic principle}~\cite{Bousso:2002ju} as follows:

\textit{Let $\Sigma$ be an achronal region of spacetime whose boundary $\gamma$ has a future-directed lightsheet that belongs to $D(\Sigma)$. The number of bits necessary to describe $\Sigma$ satisfies} 
\begin{equation} 
\label{eq:holo}
N_{\rm bits}(\Sigma) \leq \frac{\text{Area}(\gamma)}{4G \ln 2}. 
\end{equation} 
This bound should serve as a constraint on quantum gravity theory candidates, where the Bousso bound should appear explicitly. Physically, the condition states that a spatial region (that can evolve to a lightsheet under unitary evolution) has a fundamental description whose number of degrees of freedom is restricted by the area of its boundary. In quantum field theory, one would expect this quantity to be related to the volume of the region. The holographic principle thus implies a rather high redundancy in the effective description. The most natural way for this constraint to be satisfied is to conjecture that the fundamental description of a given region of spacetime lies on its boundary, with one degree of freedom per Planck unit area. Our discussion therefore connects with the earlier works of Susskind and 't Hooft that led to the same idea~\cite{tHooft:1993dmi, Susskind:1994vu}.

The holographic principle is now very well understood in AdS, thanks to the AdS/CFT correspondence~\cite{Maldacena:1997re, Witten:1998qj, Gubser:1998bc}. The point of our discussion is to emphasize that the insights raised by Bousso and others are expected to hold in any spacetime, paving the way for a holographic description of non-AdS backgrounds.

\chapter{Entropy and energy bounds in semiclassical gravity}
\label{ch:bounds}

In the previous chapter, we have laid down several statements in classical gravity that are proven or conjectured to hold under the null energy condition. In this chapter, we present a strategy that has proven to be be highly effective in producing gravitational constraints that hold at the quantum level. We apply it to the focusing theorem and Bousso bound.

\section{Generalized entropy as a discovery tool for quantum gravity}
\label{sec:Sgen}

In the previous section, we discussed how the area law for black holes \eqref{eq:areathm} fails to account for Hawking radiation. In particular, near a black hole horizon, the NEC is violated, leading to a decrease in the horizon area. This issue can be resolved by defining the generalized entropy \eqref{eq:Sgenth}, which is expected to satisfy the generalized second law even in the presence of strong quantum effects.

The notion of generalized entropy has also been employed for holographic entanglement entropy, where loop corrections are incorporated by extremizing $S_{\rm gen}$ instead of the usual area term. See Section~\ref{sec:HRTQ}.

Concurrently, entropy bounds were formulated to constrain the information content of spacetime subregions. In particular, the Bekenstein bound was improved by the Bousso bound, which is based on null hypersurfaces. This bound successfully generalizes the Bekenstein bound while remaining valid in broader regimes. However, similar to the black hole area law, it is expected to be violated in the presence of evaporating black holes~\cite{Lowe:1999xk} and is not well-defined in regimes where the entropy cannot be approximated by a fluid description. Moreover, the set of sufficient conditions for the Bousso bound \eqref{eq:cond1_hydro},\eqref{eq:cond2} explicitly assumes the NEC, which is violated by Hawking radiation.

Following~\cite{Wall:2010jtc,Myers:2013lva,Engelhardt:2014gca}, we consider generalized entropy for arbitrary codimension-two surfaces \eqref{eq:Sgen}. The past developments mentioned above suggest a strategy to develop gravitational constraints that remain valid in the semiclassical regime: one takes a classical theorem or conjecture concerning the area of surfaces and their evolution in time, that holds in the classical regime when the NEC is valid, and replaces the area of the surface divided by $4G$ with the generalized entropy:
\begin{equation}
\label{eq:strategy}
\frac{\text{Area}(\gamma)}{4G} \rightarrow S_{\rm gen}(\gamma).
\end{equation}
Alternatively, one may replace the classical area term with a quantum area
\begin{equation}
\label{eq:quarea}
\text{Area}_{\rm qu}(\gamma) = \text{Area}(\gamma) + 4GS(\Sigma_{\rm out}).
\end{equation}
This strategy has yielded statements robust to quantum fluctuations, such as quantum extremal surfaces computing holographic entanglement~\cite{Engelhardt:2016aoo}, the generalized second law of thermodynamics~\cite{Wall:2009wm, Wall:2011hj}, or a quantum version of Penrose's singularity theorem~\cite{Wall:2010jtc}. It is therefore compelling to apply it to the two classical statements we made in the previous section: the focusing theorem and the Bousso bound.

Beyond the past successes of generalized entropy, there has been strong evidence that generalized entropy is a well-defined quantity in quantum gravity and that it should be seen as a measure of entanglement. As mentioned in Section~\ref{sec:EE}, entanglement entropy of quantum fields is divergent due to short-range correlations between degrees of freedom on both sides of the boundary. Moreover, the area term is only well-defined in the classical limit $G\rightarrow 0$. Hence, both terms in the definition of generalized entropy exhibit divergences.

In~\cite{Susskind:1994sm}, this issue was studied in the context of four-dimensional black holes. It was noted that the divergence in $S(\Sigma_{\rm out})$ goes as $\text{Area}(\partial \Sigma_{\rm out})/\epsilon^{2} = A_H/\epsilon^2$, where $\epsilon$ is the UV cutoff. Moreover, they argued that the physical Newton constant, which takes into account quantum fluctuations and is renormalized with a UV cutoff, is related to the bare Newton constant $G$ by
\begin{equation}
\label{eq:Gren}
\frac{1}{G_{\rm R}} - \frac{1}{G} \propto \frac{1}{\epsilon^2} + \dots
\end{equation}
In particular, they showed that in a regularized approach where fluctuations in the metric are taken into account, one finds an area term $\text{Area}/4G_{\rm R}$, where $G_{\rm R}$ is related to the bare Newton constant $G$ appearing in the Einstein-Hilbert action as in equation~\eqref{eq:Gren}. Thus, the divergences in the area term $\text{Area}/4G$ and the field entropy term $S(\Sigma_{\rm out})$ cancel each other out exactly.

The modern understanding of the finiteness of generalized entropy stems from developments in algebraic quantum field theory~\cite{Leutheusser:2021frk,Witten:2021unn,Chandrasekaran:2022eqq}. Roughly speaking, algebraic quantum field theory studies the mathematical structure of observables in a given spacetime region. In quantum field theory, observables form a type III algebra in which the notion of a renormalized trace does not exist.\footnote{By a renormalized trace, we mean any map from the algebra of observables to $\mathbb{R}^+$ that is linear, invariant under unitary evolution of observables, and cyclic.} See~\cite{Sorce:2024pte} for a pedagogical review. In other words, entanglement entropy cannot be formally defined in quantum field theory. However,~\cite{Leutheusser:2021frk,Witten:2021unn} showed that coupling quantum field theory outside of a black hole with quantum fluctuations of the black hole mass leads to a type II algebra in which traces can be defined. In particular, the resulting trace has been shown to be unique up to an additive constant, and consistent with Bekenstein's generalized entropy \ref{eq:Sgenth}~\cite{Chandrasekaran:2022cip}. These statements have been generalized to arbitrary spacetime regions (\emph{ie}, to the generalized entropy associated with arbitrary surfaces)~\cite{Jensen:2023yxy}, and the interpretation of generalized entropy as a count of states was provided in~\cite{Akers:2024bel}. In summary, generalized entropy is better defined than its two constituent components and is now assumed to be a well-defined object in quantum gravity, associated with counting the number of degrees of freedom.

\section{Quantum Focusing}
\label{sec:QFC}

Let us define a quantum analog to the classical expansion parameter $\theta$ \eqref{eq:expansion}. As in equation~\eqref{eq:4D_entropy_flux}, we consider $(\vec{y},v)$, a coordinate system on some null hypersurface $\mathcal{N}$ emanating orthogonally from a codimension-two surface $\gamma$. Surface $\gamma$ has a coordinate $\vec{y}$ and generating geodesics have an affine parameter $v$.\footnote{We temporarily don't use $\lambda$ as the affine parameter for reasons that will become clear below.} See Figure~\ref{fig:generators}. The affine parameter is taken such that $v=0$ at $\gamma$ and increases away from $\gamma$. A function $V(\vec{y})\geq 0$ defines a slice $\gamma_V$ of $\mathcal{N}$ as 
\begin{equation}
    \gamma_V = \{(\vec{y},v)\in\mathcal{N}, v=V(\vec{y})\}.
\end{equation}
\begin{figure}
    \centering
    \includegraphics[width=0.5\linewidth]{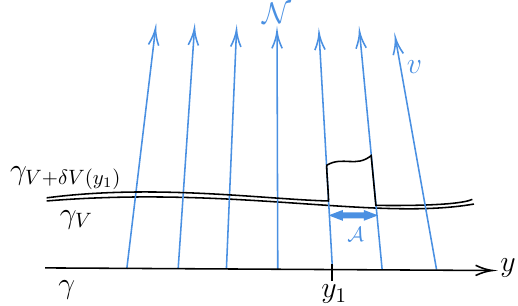}
    \caption{\footnotesize Two-dimensional picture of a surface $\gamma$ generating a null congruence $\mathcal{N}$. $y$ is a coordinate system of $\gamma$, indexing lightrays generating the congruence—generators—which are associated with an affine parameter $v$. A slice $\gamma_V$ of $\mathcal{N}$ is defined by a function $V(y)$ such that $v=V(y)$ on each generator $y$. A perturbation of the slice $\gamma_V$ is built by deforming the function $V$ in a neighborhood of the generator $y_1$ with area element $\mathcal{A}$. The resulting function is $V+\delta V(y_1)$ and defines the slice $\gamma_{V+\delta V(y_1)}$.}
    \label{fig:generators}
\end{figure}Each surface $\gamma_V$ splits a Cauchy slice into two parts that we call $\Sigma_{V,\rm in}$ and $\Sigma_{V,\rm out}$. We define the generalized entropy associated with $\gamma_V$:
\begin{equation}
    S_{\rm gen}[V] = \frac{\text{Area}(\gamma_V)}{4G} + S(\Sigma_{V,\rm out}).
\end{equation}
Instead of infinitesimal variations along one lightray, we consider variations with respect to $V$ and define the \textit{quantum expansion} as~\cite{Bousso:2015mna}
\begin{equation}
    \Theta[V;\vec{y}] = \frac{4G}{\sqrt{h_V}}\frac{\delta S_{\rm gen}[V]}{\delta V(\vec{y})},
\end{equation}
where $h_V$ is the determinant of the induced metric on $\gamma_V$. This factor ensures that $\Theta$ is defined per unit of geometric area, such that
\begin{equation}
    \lim_{G\rightarrow 0} \Theta[V,\vec{y}] = \theta(v=V(\vec{y})),
\end{equation}
where $\theta(v=V(\vec{y}))$ is the classical expansion of the generating lightray emanating from $(\vec{y},0)$, at $v=V(\vec{y})$.

Let $V_{\lambda}$ be a one-parameter family of positive functions satisfying $V_0(\vec{y})=0$ and
\begin{equation}
    \frac{\partial}{\partial \lambda}V_{\lambda}(\vec{y}) \geq 0,
\end{equation}
thus providing a foliation of the null hypersurface $\mathcal{N}$~\cite{Shahbazi-Moghaddam:2022hbw}. These definitions, as well as the analogy with the classical focusing theorem, lead to the following statement:

\begin{conjecture}[Quantum Focusing]
\label{conj:QFC}
In the semiclassical regime, the quantum expansion cannot increase along null hypersurfaces:
    \begin{equation}
    \frac{\partial\Theta[V_{\lambda},\vec{y}]}{\partial \lambda} \leq 0 \quad \forall \vec{y}.
\end{equation}
\end{conjecture}
The \textit{Quantum Focusing Conjecture} (QFC)~\cite{Bousso:2015mna} is a nonlocal statement: perturbations in $\Sigma_{V,\rm out}$ far away from $\mathcal{N}$ can modify the quantum expansion. Note also that it is cutoff-independent as it is entirely defined from the generalized entropy.

Although the QFC has not been proven yet, it is the strongest plausible semiclassical constraint we currently know. In particular, it implies a number of corollaries for which we have very good evidence. In this chapter, we focus on two of them: the Bousso bound and the quantum null energy condition. In addition to those, the QFC is a central assumption in many proofs of properties of quantum extremal surfaces~\cite{Akers:2016ugt,Brown:2019rox,Akers:2019lzs}. For example, the properties that the entanglement wedge always includes the causal wedge of a boundary subregion \eqref{eq:CinW}, strong subadditivity \eqref{eq:SSA}, entanglement wedge nesting \eqref{eq:EWnesting}, as well as the equivalence between quantum maximin surfaces and quantum extremal surfaces \ref{thm:HRT=m}, are valid in the semiclassical regime for quantum extremal surfaces under the assumption of the QFC. Crucially, all these properties are necessary conditions to preserve causality on the boundary theory. See~\cite{Bousso:2015eda,Engelhardt:2021mue,Bousso:2022hlz,Bousso:2022tdb,Bousso:2023sya} for other applications of the QFC. See also~\cite{Akers:2020pmf, Akers:2023fqr, Bousso:2024iry} for recent alternative definitions of the QFC based on other definitions of entropy. In the second part of this thesis, the usefulness of these properties of extremal surfaces appears to extend beyond AdS/CFT and in particular in the development of de Sitter holography. Exploring these semiclassical constraints is therefore of strong interest not only in semiclassical gravity and AdS/CFT but also in the search for a holographic description of cosmological spacetimes.

Shahbazi-Moghaddam~\cite{Shahbazi-Moghaddam:2022hbw} noted that a weakened version of the QFC is sufficient to prove all of its known applications discussed above. In particular, the following corollary to the QFC is all that is required.

\begin{corollaryconj} \label{cor:QFC} Let $\alpha$ be a subsystem of the generating codimension-two spacelike hypersurface $\gamma$. The QFC implies: \begin{equation} \Theta[0,\vec{y}] \leq 0 \quad \forall \vec{y}\in\alpha \Rightarrow \Theta[V,\vec{y}]\leq 0,\quad V\vert_{\gamma\backslash\alpha} = 0, \label{eq:corQFC} \end{equation} where $V\vert_{\gamma\backslash\alpha}=0$ imposes that $V=0$ on all generators of $\gamma$ that are not in $\alpha$. \end{corollaryconj}

Thus, if the quantum expansion is negative on some subsystem $\alpha$ of $\gamma$, then it will remain negative on all later points of the lightrays generated from $\alpha\subset\gamma$. This constraint is ensured by the following condition~\cite{Shahbazi-Moghaddam:2022hbw}:

\begin{conjecture}[Restricted Quantum Focusing] \label{conj:rQFC} At the semiclassical level, if the quantum expansion vanishes on some spacelike Section of a null hypersurface $\mathcal{N}$, it cannot increase away from this section: \begin{equation} \Theta[V_{\lambda},\vec{y}]=0 \quad \Rightarrow \quad \frac{\partial\Theta[V_{\lambda},\vec{y}]}{\partial\lambda}\leq 0. \end{equation} \end{conjecture}
Restricted quantum focusing is weaker than the QFC because the quantum expansion may increase. However, it cannot become positive. This is enough to ensure the validity of Corollary \ref{cor:QFC}:

Let $V_{\lambda}$ be a family of functions for which $\Theta[V_{\lambda},\vec{y}]$ is differentiable in $\lambda$. A violation of Corollary \ref{cor:QFC} would imply that there exists a null hypersurface and generator $\vec{y}$ such that $\Theta[V_0,\vec{y}]\leq 0$ and $\Theta[V_{\lambda_1},\vec{y}] > 0$ for some value $\lambda_1>0$. Since $V_{\lambda}$ is a differentiable function of $\lambda$, there must be a value $0<\lambda_0\leq \lambda_1$ at which $\Theta[V_{\lambda_0},\vec{y}]=0$ and $\partial\Theta[V_{\lambda_0},\vec{y}]/\partial\lambda > 0$, thus violating the restricted QFC. Therefore, restricted quantum focusing implies \eqref{eq:corQFC}.

In~\cite{Shahbazi-Moghaddam:2022hbw}, the restricted QFC was proven in braneworld holography. Moreover, proving the QFC in the same setup does not appear to be possible. So far, no counterexample to the QFC has been found, so we might still hope for the unrestricted statement of the QFC to be a universal statement in semiclassical gravity. However, the existence of a weaker bound—the restricted QFC—which is sufficient to prove all known applications of the QFC strongly motivates the idea that only the restricted QFC generally holds.

In Chapter~\ref{ch:proof2D}, we prove restricted focusing in two-dimensional quantum gravity and argue for violations of unrestricted focusing.

\section{Quantum Null Energy Condition}

Perhaps the strongest argument for the validity of the QFC or its restricted version is that it leads to a nontrivial and provable prediction in quantum field theory. In this section, we briefly review this statement and discuss its physical significance.

It is instructive to consider specific limits of the QFC. We already argued that the restricted QFC is valid, at least in braneworld setups. In general, the off-diagonal term of the QFC is ensured by strong subadditivity~\cite{Bousso:2015mna}. In other words,
\begin{equation}
\label{eq:QFCoff-diag}
    \frac{\delta}{\delta V(\vec{y}_1)}\left(\frac{4G}{\sqrt{h_{V(\vec{y}_2)}}}\frac{\delta S_{\rm gen}[V]}{\delta V(\vec{y}_2)}\right) \leq 0 ,\qquad \vec{y}_1\neq \vec{y}_2.
\end{equation}
The remaining nontrivial statement is therefore the diagonal part of the QFC. In such a case, all variations are studied along the generator $\vec{y}_1=\vec{y}_2$ of a surface $\gamma$, and it is useful to use the affine parameter $\lambda$ of this lightray. Then, we write
\begin{align}
\label{eq:QFCdiag}
\Theta'\leq 0,\quad  \mathrm{where} \quad  \Theta = \theta + 4G S'_{\rm out},
\end{align}
and $' = d/d\lambda=k^{\mu}\nabla_{\mu}$ is the derivative with respect to $\lambda$ on the generator $\vec{y}_1$, and all quantities are implicitly associated with a point $p=(\lambda,\vec{y}_1)$. Using the Raychaudhuri equation~\eqref{eq:Raychaudhuri} and Einstein equations, we expand \eqref{eq:QFCdiag} as
\begin{equation}
\label{eq:quRach}
 \Theta' =-\frac{1}{d-1}\theta^2-\sigma_{\mu\nu}\sigma^{\mu\nu} - 8\pi G k^{\mu}k^{\nu}\braket{T_{\mu\nu}} + \frac{4G(\hbar)}{\mathcal{A}}\left(S_{\rm out}'' - \theta S'_{\rm out}\right)\leq 0,
\end{equation}
where we added the factor of $\hbar$ to make the different limits more explicit. In the limit $\hbar \rightarrow 0$, one recovers the NEC \eqref{eq:NEC}.

Taking the limit $\theta=\sigma^2=0$ while keeping $\hbar$ finite yields an interesting result:
\begin{theorem}[Quantum null energy condition]
    Let $p$ be a point in spacetime, $k^{\mu}$ a null vector at $p$, and $\Sigma$ a spacelike slice whose boundary contains $p$. At $p$, if the shear tensor and expansion scalar vanish,
    \begin{equation}
    \label{eq:QNEC}
    k^{\mu}k^{\nu}\braket{T_{\mu\nu}} \geq \frac{1}{2\pi\mathcal{A}}S''(\Sigma),
    \end{equation}
    where $'=d/d\lambda$ denotes differentiation with respect to null perturbations of a region of area element $\mathcal{A}$ around $p$.
\end{theorem}
The \textit{Quantum Null Energy Condition} (QNEC) is particularly useful in maximally symmetric spacetimes where every point is contained in a Rindler, de Sitter, or Poincaré horizon, such that the condition $\theta=0$ is satisfied at every point of spacetime for a suitable choice of $k^{\mu}$~\cite{Bousso:2015wca}.

Interestingly, this limit of the QFC does not depend on the gravitational constant $G$ and is unaffected by higher curvature terms in the gravitational action~\cite{Bousso:2015mna}. In other words, it is not a statement about quantum gravity but about quantum field theory. In fact, it can be proven within this framework~\cite{Bousso:2015wca,Balakrishnan:2017bjg,Koeller:2015qmn,Ceyhan:2018zfg,Kudler-Flam:2023hkl}. Thus, the QNEC provides a generalization of the NEC that is valid in any state of quantum field theory. Starting from a theorem in general relativity, we have conjectured a statement in semiclassical gravity, which implies a new result in quantum field theory. This is not only a strong argument in favor of the QFC, but it also illustrates the power of the rather simple strategy presented in equation~\eqref{eq:strategy} and above.

\section{Quantum Bousso Bounds}
\label{sec:QBB}

As noted in Section~\ref{sec:Sgen}, the Bousso bound is not expected to hold in spacetimes that violate the NEC. It is therefore very tempting to apply the strategy we have followed in this chapter, which is to start from a classical statement and replace the area of codimension-two surfaces by their quantum area \eqref{eq:quarea}. Two different approaches to this idea have been followed in~\cite{Strominger:2003br,Bousso:2015mna}. 

The first approach~\cite{Strominger:2003br} was to only modify the right-hand side of \eqref{eq:BB}, replacing the area difference with a quantum area difference. Doing so suggests the following generalization of the notion of a lightsheet. We define a \textit{quantum lightsheet} $\mathcal{L}_{\rm qu}$ as a null congruence with non-positive quantum expansion, $\Theta\leq 0$. Note that, analogous to the focusing theorem, assuming the QFC implies that the definition of a quantum lightsheet reduces to the constraint $\Theta(0)\leq 0$ on the generating surface. We do not use this assumption here.

\begin{conjecture}[Strominger-Thompson quantum Bousso bound]
\label{conj:STQBB}
     Let $\mathcal{L}_{\rm qu}(\gamma_1,\gamma_2)$ be a quantum lightsheet that emanates and terminates orthogonally from two codimension-two spacelike hypersurfaces $\gamma_1$ and $\gamma_2$. Then,
   \begin{equation}
   \label{eq:STBB}
       S_{\rm th}(\mathcal{L}_{\rm qu}(\gamma_1,\gamma_2)) \leq \frac{\text{Area}_{\rm qu}(\gamma_1)-\text{Area}_{\rm qu}(\gamma_2)}{4G},
   \end{equation}
   where $S_{\rm th}(\mathcal{L}_{\rm qu}(\gamma_1,\gamma_2))$ is the thermodynamic entropy of the lightsheet.
\end{conjecture}
The right-hand side of the \textit{Strominger-Thompson quantum Bousso bound} is cutoff-independent and well-defined in semiclassical gravity. The existence of a well-defined $S_{\rm th}(\mathcal{L}_{\rm qu}(\gamma_1,\gamma_2))$ remains an issue, and one needs to provide a definition of this quantity in order to verify the validity of \eqref{eq:STBB}. In their original paper~\cite{Strominger:2003br}, Strominger and Thompson proved the validity of their conjecture in two vacuum states of the Russo–Susskind–Thorlacius (RST) model~\cite{Russo:1992ax}, a two-dimensional model of an evaporating black hole in AdS. In~\cite{Franken:2023ugu}, we showed that in the hydrodynamic regime, where $S_{\rm th}(\mathcal{L}_{\rm qu}(\gamma_1,\gamma_2))$ is expressed in terms of an entropy density, the Strominger-Thompson bound can be proven in two-dimensional quantum gravity, see Chapter~\ref{ch:proof2D}. 

As $S_{\rm th}(\mathcal{L}_{\rm qu}(\gamma_1,\gamma_2))$ does not have a semiclassical definition in terms of fine-grained quantities, the Strominger-Thompson bound cannot be treated as a fundamental constraint. However, it has the advantage of maintaining the initial objective of the Bousso bound, which is to impose an upper bound on the number of degrees of freedom in a spacetime region in terms of gravitational quantities.

More recently,~\cite{Bousso:2015mna} noted that the QFC implies a quantum version of the Bousso bound which is completely well-defined in semiclassical gravity and cutoff-independent. Consider a null hypersurface emanating from the surface $\gamma_1$, a subsystem $\alpha\subset\gamma_1$, and a surface $\gamma_2$ that differs from $\gamma_1$ only on generators from $\alpha$. Corollary \ref{cor:QFC} of the QFC implies
\begin{equation}
    \Theta[0,\vec{y}]\leq 0 \quad \forall\vec{y}\in\alpha\Rightarrow S_{\rm gen}(\gamma_1) \leq S_{\rm gen}(\gamma_2).
\end{equation}
In other words, if the quantum expansion is initially decreasing on a subset $\alpha$ of $\gamma$, then the generalized entropy of $\gamma$ is upper-bounded by the generalized entropy of a surface $\gamma'$ containing $\gamma\backslash\alpha$. Taking $\alpha = \gamma_1$ leads to the \textit{Bousso-Fisher-Leichenauer-Wall quantum Bousso bound} (BFLW).
\begin{conjecture}[BFLW quantum Bousso bound]
\label{conj:BFLW}
    Let $\mathcal{L}_{\rm qu}(\gamma_1,\gamma_2)$ be a quantum lightsheet that emanates and terminates orthogonally from two codimension-two spacelike hypersurfaces $\gamma_1$ and $\gamma_2$. Then,
   \begin{equation}
   \label{eq:BFLWQBB}
       S(\mathcal{L}_{\rm qu}(\gamma_1,\gamma_2)) \leq \frac{\text{Area}(\gamma_1)-\text{Area}(\gamma_2)}{4G}.
   \end{equation}
   We defined
   \begin{equation}
       S(\mathcal{L}_{\rm qu}(\gamma_1,\gamma_2)) = S(\Sigma_2) - S(\Sigma_1),
   \end{equation}
   where $S(\Sigma_i)$ is the entropy of fields in $\Sigma_i$, a spacelike slice bounded by $\gamma_i$.
\end{conjecture}
There is no general proof of this statement, except as a byproduct of the restricted QFC in braneworld holography~\cite{Shahbazi-Moghaddam:2022hbw}. This proposal is very similar to the original Bousso bound. However, the entropy associated with the lightsheet is fine-grained and the inequality is cutoff-independent. Instead of replacing the right-hand side of the bound while keeping the coarse-grained entropy on the left-hand side, as Strominger and Thompson did, the BFLW quantum Bousso bound clarifies the definition of the entropy associated with a null surface in a way that leads to a cutoff-independent statement. 

The inequality \eqref{eq:BFLWQBB} may be rewritten as 
\begin{equation}
        \frac{\text{Area}_{\rm qu}(\gamma_1)-\text{Area}_{\rm qu}(\gamma_2)}{4G} \geq 0.
\end{equation}
Thus, the BFLW bound is weaker than the Strominger-Thompson bound. However, the latter is more difficult to interpret as it mixes generalized entropy with coarse-grained entropy. While the BFLW bound is motivated by our confidence in the validity of the QFC, it is far from clear how the Strominger-Thompson bound could emerge from first principles.

We conclude this section by mentioning the alternative approach of Bousso, Casini, Fisher, and Maldacena (BCFM)~\cite{Bousso:2014sda} which also proposes a modification of the left-hand side of the classical Bousso bound, but without resorting to generalized entropies. In their work, the entropy of the lightsheet $S(\mathcal{L}(\gamma_1,\gamma_2))$ is defined as the difference $\Delta S$ between the fine-grained entropies of the state under consideration and the vacuum, restricted to $\mathcal{L}(\gamma_1,\gamma_2)$. A proper definition of $\Delta S$ may be found in~\cite{Casini:2008cr,Bousso:2014sda}. 

Note that the BCFM bound applies to lightsheets in their classical sense, contrary to the Strominger-Thompson and BFLW bounds that apply to quantum lightsheets. The quantity $\Delta S$ is divergence-free and the bound is proven for any portion of the lightsheet. In particular, $\Delta S$ is well defined for lightsheets of arbitrary sizes, whereas the hydrodynamic limit, which is used in the Strominger-Thompson bound, breaks down at small scales. However, this definition of entropy on the lightsheet is only valid in the weak gravity limit, \emph{ie} when the spacetime geometry in the presence of matter is well approximated by the vacuum geometry. Beyond this limit, the backreaction of matter on the spacetime geometry becomes non-negligible, and $\Delta S$ is no longer well defined. Indeed, the meaning of ``same'' lightsheets in two different geometries is unclear. 

\chapter{A two-dimensional model of quantum gravity}
\label{ch:JT}

A successful strategy towards quantum gravity has been to consider lower dimensions, where simplified models of gravity can be quantized exactly~\cite{Teitelboim:1983ux,Jackiw:1984je,Henneaux:1985nw, Banks:1990mk,Ikeda:1993fh,Louis-Martinez:1993bge,Grumiller:2002nm,Saad:2019lba,Mertens:2022irh}. Within this framework, significant progress has been made in understanding holography~\cite{Sachdev:2010um,Almheiri:2014cka,Maldacena:2016hyu,Maldacena:2016upp,Jensen:2016pah,Engelsoy:2016xyb}, the evaporation of black holes~\cite{Russo:1992ax,Callan:1992rs,Fiola:1994ir,Spradlin:1999bn,Mertens:2019bvy}, the reconstruction of entanglement wedges~\cite{Almheiri:2019psf,Almheiri:2019qdq,Almheiri:2019yqk,Penington:2019kki}, quantum cosmology~\cite{Cotler:2019nbi,Maldacena:2019cbz,Chen:2020tes}, and information-theoretic constraints in gravity~\cite{Strominger:2003br,Wall:2009wm,Wall:2011kb,Franken:2023ugu}.

In this chapter, we introduce a two-dimensional model of quantum gravity where the generalized entropy can be computed exactly in the semiclassical regime. We employ this framework in the next chapter to study entropy constraints in $(1+1)$-dimensional semiclassical gravity.

\section{Jackiw–Teitelboim gravity}

The two-dimensional Einstein-Hilbert action is topological, thus making two-dimensional gravity trivial. The best known solution to this problem is to couple the Ricci tensor to a scalar field $\phi$, known as the dilaton. The most general two-derivative dilaton gravity action can always be reduced to
\begin{equation}
    I = -\frac{1}{16\pi G}\int_{\mathcal{M}} dx^2 \sqrt{-g}(\phi R + U(\phi)),
\end{equation}
where $U(\phi)$ is the dilaton potential~\cite{Mertens:2022irh} and $R$ is the Ricci scalar of the two-dimensional background $\mathcal{M}$. Jackiw–Teitelboim (JT) gravity is a simplified quantum gravity model obtained by assuming a linear potential $U(\phi)= - \Lambda \phi$~\cite{Jackiw:1984je, Teitelboim:1983ux}, providing a solvable model for quantum gravity~\cite{Henneaux:1985nw,Saad:2019lba}. When focusing on Anti-de Sitter (AdS) and de Sitter (dS) space, we set $\Lambda=-1/l_{\rm (A)dS}^2$ and $\Lambda=+1/l_{\rm (A)dS}^2$, respectively. The complete JT gravity action is~\cite{Jackiw:1984je, Teitelboim:1983ux, Cotler:2019nbi}
\begin{equation}
\label{eq:JT}
    I_{\rm JT} = - \frac{1}{16\pi G}\int_{\mathcal{M}}d^2x \sqrt{-g}((\phi_0+\phi) R -2\Lambda \phi)- \frac{1}{8\pi G}\int_{\partial\mathcal{M}} dy \sqrt{-h}(\phi_0+\phi) K,
\end{equation}
where $y$ is the coordinate on the boundary $\partial\mathcal{M}$, $h$ is the trace of the induced metric on $\partial\mathcal{M}$, and $K$ is the trace of the extrinsic curvature on the boundary. We have added a topological component $\phi_0$ and the Gibbons-Hawking-York boundary term~\cite{York:1972sj,Gibbons:1976ue,Mertens:2022irh}.

Interestingly, the action \eqref{eq:JT} can be derived from the dimensional reduction of a higher-dimensional Einstein-Hilbert action with the metric ansatz\footnote{See, for example, Appendix $A$ of~\cite{Svesko:2022txo} for a derivation.}
\begin{equation}
\label{eq:sphred}
    d s^2 = \hat{g}_{MN}d X^Md X^N = g_{\mu\nu}(x) d x^{\mu}d x^{\nu} + l_d^2\Phi^{2/(d-1)}(x)d\Omega_{d-1}^2,
\end{equation}
where $\{X^{M},~M=0,...,n\}$ are coordinates on the original $(d+1)$-dimensional manifold $\hat{\mathcal{M}}$, $\hat{g}_{MN}$ is the metric tensor on $\hat{\mathcal{M}}$, and $\Phi = \phi_0+\phi$ is the complete dilaton field. We introduce the radius of curvature $l_d$ of (A)dS$_{d+1}$, which is related to the higher-dimensional cosmological constant $\hat{\Lambda}$ by $\hat{\Lambda}=\pm d(d-1)/(2l_d^2)$. Similarly, the two-dimensional Newton constant $G$ is given by 
\begin{equation}\label{eq:2d_Newton_cste}
\frac{1}{G} = \frac{S_{d-1}(l_d)}{\hat{G}},
\end{equation}
where $\hat{G}$ is the gravitational constant in $\hat{\mathcal{M}}$, and $S_{d-1}(l_d)=2\pi^{d/2}~l_d^{d-1}/\Gamma(d/2)$ is the surface area of the $(d-1)$-sphere of radius $l_n$.

In the case $d=2$, the JT gravity action corresponds to the dimensional reduction of an empty (A)dS background. In such a case, $\phi_0=0$. When spherically reducing a higher-dimensional metric with $d>2$, the dilaton $\Phi$ can be seen as an expansion around an extremal black hole with associated dilaton value $\phi_0$. From the point of view of the two-dimensional background, $\phi_0$ is a topological term in the action and does not change the equations of motion. However, the dimensional reduction confers an interpretation to the dilaton as the area associated with points in $\mathcal{M}$. In particular, let $x$ be a point in $\mathcal{M}$, associated with a sphere $\gamma(x)$ in $\hat{\mathcal{M}}$. Then,
\begin{equation} 
\label{eq:JTarea}
\frac{\text{area}(\gamma(x))}{4\hat{G}} = \frac{\Phi(x)}{4G}. 
\end{equation}
Thus, we impose
\begin{equation} 
\Phi\geq 0, 
\end{equation}
so that taking $\phi_0\neq 0$ allows for negative values of $\phi$, changing the topology of the two-dimensional geometry.

The gravitational JT action can be coupled to a matter action $I_{\rm matter}$, associated with a stress tensor $T_{\mu\nu}$. In the semiclassical limit, where the stress tensor $T_{\mu\nu}$ is replaced by its expectation value $\braket{T_{\mu\nu}}$ in some unspecified state, the equation of motion obtained by varying $I_{\rm JT}+I_{\rm matter}$ with respect to $g^{\mu\nu}$ is
\begin{equation} 
\label{eq:dilaton_eom}
(g_{\mu\nu}\nabla^2-\nabla_{\mu}\nabla_{\nu}{-}g_{\mu\nu}\Lambda)\phi=8\pi G\Braket{T_{\mu\nu}}, 
\end{equation}
which does not depend on the value of the topological term $\phi_0$. The only difference between AdS and dS JT gravity is the sign of the third term on the left-hand side. By contracting with any null vector $k^{\mu}$, we obtain the same result in both AdS and dS JT gravity:
\begin{equation}\label{eq:contracted_eom} \phi''\equiv k^{\mu}k^{\nu}\nabla_{\mu}\nabla_{\nu}\phi=-8\pi G k^{\mu}k^{\nu}\Braket{T_{\mu\nu}}. \end{equation}
Varying the action with respect to the dilaton yields the equation of motion for the metric:
\begin{equation} \label{eq:eomLambda} R=2\Lambda \end{equation}
thus fixing the background geometry as (Anti-)de Sitter spacetime.

In the absence of matter, $T_{\mu\nu}=0$, the solutions are
\begin{align} 
\label{eq:solJT} 
ds^2 &= -e^{2\omega_x}dx^+dx^-\\
\Phi &= \Phi_* \equiv \phi_0 + \phi_r\frac{1+\Lambda x^+x^-}{1-\Lambda x^+x^-}, \end{align}
where $\phi_r>0$ and the Kruskal conformal factor is defined as
\begin{equation} \label{eq:Kruskal} e^{2\omega_x} = \frac{4}{(1-\Lambda x^+x^-)^2}. \end{equation}
This metric and dilaton are solutions to the equations of motion \eqref{eq:eomLambda} and \eqref{eq:contracted_eom} for both signs of the cosmological constant.\footnote{In general, the AdS solution has a dimensionless mass parameter proportional to the ADM mass, that we take to be $1$ here for simplification~\cite{Moitra:2019xoj}.}

The geometry of de Sitter JT gravity depends on the choice of $\phi_0$. For $\phi_0=0$, the resulting geometry is known as the \textit{half reduction} of de Sitter spacetime. In particular, the condition $\Phi\geq 0$ implies
\begin{equation}
    -\frac{1}{\Lambda} \leq x^+x^- \leq \frac{1}{\Lambda},
\end{equation}
where $x^+x^-=1/\Lambda$ corresponds to the past and future null infinity $\mathcal{I}^{\pm}$ and $x^+x^-=-1/\Lambda$ corresponds to the spatial boundaries. From a higher-dimensional perspective, the timelike boundaries correspond to the poles of dS$_3$. We will discuss the de Sitter geometry in more depth in Chapter~\ref{sec:dS}. Spatial slices are segments bounded by the poles, with comoving observers located there. Each observer has an associated cosmological horizon at $x^+ x^-=0$, which bounds their causal patch. The Penrose diagram for two-dimensional de Sitter space in the half reduction model is depicted in Figure~\ref{fig:PenroseJT}, which can be interpreted as a $\mathbb{Z}_2$-orbifold of dS$_2$.

\begin{figure}[ht]
    \centering
    \includegraphics[width=0.4\linewidth]{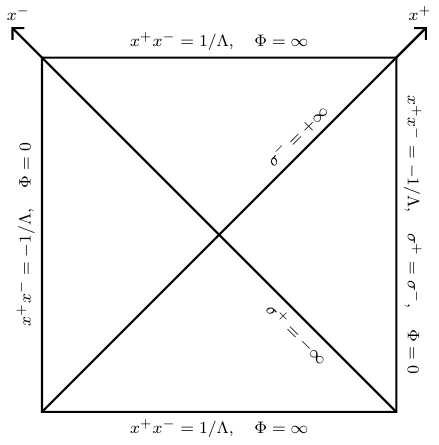}
    \caption{\footnotesize Penrose diagram for two-dimensional de Sitter space in the half reduction model. It features two timelike boundaries, represented by the two vertical lines. The diagonal lines represent the past and future cosmological horizons for observers at the poles. Key coordinate values and dilaton information are provided.}
    \label{fig:PenroseJT}
\end{figure}

For $\phi_0>0$, the geometry is known as the \textit{full reduction} of de Sitter space, corresponding to a Schwarzschild-de Sitter spacetime with both a cosmological and black hole horizon. In the full reduction model, $x^+ x^-$ remains bounded from above by $1/\Lambda$, corresponding to $\mathcal{I}^{\pm}$. However, it is no longer bounded below, as $x^+x^- \rightarrow -\infty$ corresponds to the black hole horizons. In fact, the full reduction model can be maximally extended such that it is periodic in space and includes the black hole region of the Schwarzschild-de Sitter space~\cite{Svesko:2022txo}, see Figure~\ref{fig:full}.

\begin{figure}[t]
    \centering
    \includegraphics[width=0.7\linewidth]{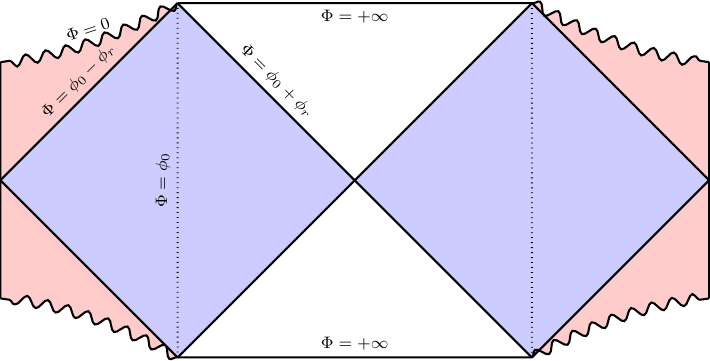}
    \caption{\footnotesize  Penrose diagram for two-dimensional de Sitter space in the full reduction model. The left and right edges are identified, making any complete spacelike slice topologically a circle. The two blue-shaded regions are causally accessible to observers at $x^+x^-=-1/\Lambda$ and are bounded by both a cosmological and black hole horizon. The red shaded regions are the black hole interiors, with past and future singularities depicted by the wavy lines.  }
    \label{fig:full}
\end{figure}
The solution \eqref{eq:Kruskal} with negative $\Lambda$ represents a maximally extended eternal AdS$_2$ black hole. The past and future horizons lie at $x^+x^-=0$, and the mass of the black hole is given by
\begin{equation}
    M=\frac{\phi_{r}}{16\pi G}.
\end{equation}
The conformal boundary is located at $x^+x^-=1/\Lambda$. The relevant portion of the Penrose diagram is depicted in Figure~\ref{fig:JTAdS}.
\begin{figure}[ht]
    \centering
    \includegraphics[width=0.4\linewidth]{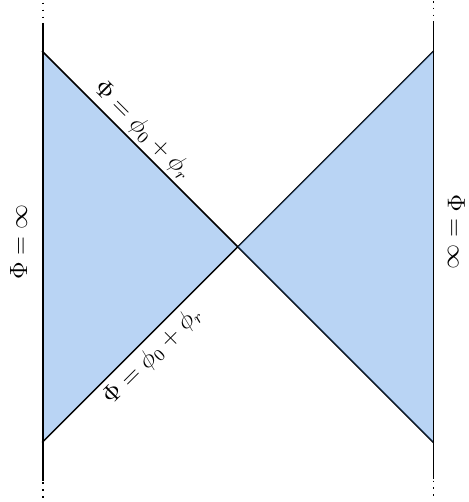}
    \caption{\footnotesize Relevant portion of an eternal black hole in AdS JT gravity. The dotted lines indicate that the Penrose diagram is incomplete. The vertical lines correspond to conformal boundaries while the diagonal lines are the past and future horizons.}
    \label{fig:JTAdS}
\end{figure}

\section{Vacuum states and backreaction}
\label{sec:vacuum}

Let us take the matter content to be a CFT$_2$ with central charge $c$, described by an action $I_{\rm CFT}$. We write $T_{\mu\nu}=\tau_{\mu\nu}+ T^{\rm(quantum)}_{\mu\nu}$ where $\tau_{\mu\nu}$ is the classical stress tensor and $T^{\rm(quantum)}_{\mu\nu}$ corresponds to the quantum matter contributions. Working in conformal coordinates, 
\begin{equation}
    ds^2 =-e^{2\omega}dx^+dx^-,
\end{equation}
and imposing general covariance of the quantum stress tensor and the conservation law $\nabla^{\mu}T_{\mu\nu}=0$ results in the general expression~\cite{PhysRevD.13.2720,Fabbri:2005mw}
\begin{eqnarray}
\label{eq:stress}
\braket{T_{\pm\pm}(x^{\pm})}&=&\braket{\tau_{\pm\pm}}+\frac{c}{12\pi}\left(\partial_{\pm}^2\omega-(\partial_{\pm}\omega)^2\right) -\frac{c}{24\pi}t_{\pm}(x^{\pm}),\\
\braket{T_{+-}(x^+,x^-)}&=&-\frac{c}{12\pi}\partial_+\partial_-\omega,
\end{eqnarray}
where $\tau_{\mu\nu}$ is the state-independent contribution to the stress tensor that is blind to the geometry, and $t_{\pm}$ is a state-dependent contribution~\cite{Fabbri:2005mw}. 
Moreover, equation~\eqref{eq:eomLambda} implies that $\partial_{\pm}^2\omega-(\partial_{\pm}\omega)^2=0$, which simplifies our expressions: $\braket{T_{\pm\pm}(x^{\pm})}=\braket{\tau_{\pm\pm}} -\frac{c}{24\pi}t_{\pm}(x^{\pm})$

An effective action can be constructed such that the quantum mechanical part of the stress tensor arises from its variation. This is achieved through the nonlocal Polyakov action $I_{\rm Polyakov}$~\cite{Polyakov:1981rd,PhysRevD.13.2720,PhysRevD.15.2088,Fabbri:2005mw}:\footnote{The nonlocality of \eqref{eq:Polyakov} reflects the fact that the stress tensor \eqref{eq:stress} is state-dependent.}
\begin{equation} 
\label{eq:Polyakov}
    I_{\rm Polyakov} = - \frac{1}{16\pi G}\int d^2x_1d^2x_2\sqrt{g(x_1)g(x_2)}\left[\frac{c}{48}R(x_1)G(x_1,x_2)R(x_2)\right],
\end{equation}
where $G(x_1,x_2)$ is the Green function satisfying
\begin{equation}
    \nabla^2_{x_1}G(x_1,x_2) = \frac{1}{\sqrt{-g(x_1)}}\delta(x_1-x_2)
\end{equation}
We consider the semiclassical limit $G\rightarrow 0$ where the background satisfies the classical Einstein equations with $T_{\mu\nu}$ replaced by its expectation value $\braket{T_{\mu\nu}}$. To ensure that quantum effects on the dilaton are subdominant, we take the limit $c\rightarrow \infty$ while keeping $cG$ fixed.

Defining the normal ordered stress tensor~\cite{Fabbri:2005mw}
\begin{equation}
    :T_{\pm\pm}(x^{\pm}): ~\equiv~ T_{\pm\pm}(x^{\pm}) - \bra{0_x}T_{\pm\pm}(x^{\pm})\ket{0_x},
\end{equation}
where $\ket{0_x}$ is the vacuum state with respect to $x^{\pm}$ coordinates, we find $:T_{\pm\pm}(x^{\pm}): = T_{\pm\pm}(x^{\pm}) - \tau_{\pm\pm}(x^{\pm})$, and 
\begin{equation}
\label{eq:normalorder}
    \braket{:T_{\pm\pm}(x^{\pm}):}=-\frac{c}{24\pi}t_{\pm}(x^{\pm}).
\end{equation}
These expressions hold only in $x^{\pm}$ coordinates, since the stress-energy tensor transforms non-trivially under the conformal transformation $x^{\pm} \rightarrow y^{\pm}(x^{\pm})$. In particular, $\tau_{\pm\pm}$ transforms as a rank-$2$ tensor, 
\begin{equation}\label{eq:transtau}
    \tau_{\pm\pm}(y^{\pm}) = \left(\frac{dx^{\pm}}{dy^{\pm}}\right)^2 \tau_{\pm\pm}(x^{\pm}),
\end{equation}
while the functions $t_{\pm}$ satisfy the anomalous transformation law:
\begin{equation}
\label{eq:transft}
t_{\pm}(y^{\pm})=\left(\frac{dx^{\pm}}{dy^{\pm}}\right)^2 t_{\pm}(x^{\pm})+\{x^{\pm},y^{\pm}\},
\end{equation}
where $\{x^{\pm},y^{\pm}\}$ is the Schwarzian derivative defined by
\begin{equation}
\{x^{\pm},y^{\pm}\}=\frac{\dddot x^{\pm}}{\dot x^{\pm}}-\frac{3}{2}\left(\frac{\ddot x^{\pm}}{\dot x^{\pm}}\right)^2,
\end{equation}
with $\dot x^{\pm}=dx^{\pm}/dy^{\pm}$. This results in the following transformation law for the stress tensor:
\begin{equation}
\label{eq:transfT}
\braket{T_{\pm\pm}(y^{\pm})}=\left(\frac{dx^{\pm}}{dy^{\pm}}\right)^2 \braket{T_{\pm\pm}(x^{\pm})}-\frac{c}{24\pi}\{x^{\pm},y^{\pm}\}.
\end{equation}
From equation~\eqref{eq:normalorder}, a \textit{vacuum state} is defined as a state where $\braket{:T_{\pm\pm}(y^{\pm}):}=0$ in some coordinate system $y^{\pm}$. Thus, $\tau_{\pm\pm}$ corresponds to the vacuum expectation value of $T_{\pm\pm}(y^{\pm})$ in the vacuum state defined with respect to the coordinates $y^{\pm}$, where $t_{\pm}(y^{\pm})=0$. All vacuum states are related by conformal transformations of the lightcone coordinates in which they are defined. We also refer to them as \emph{conformal vacua}.

While the classical JT action arises from a dimensional reduction of the Einstein-Hilbert action, the semiclassical theory—when described in terms of the CFT$_2$ action—does not directly follow from such a reduction. Therefore, backreacted geometries in JT gravity should not be seen as a canonical dimensional reduction of semiclassical gravity in higher dimensions.

Two typical states of interest can be defined in both dS and AdS JT gravity~\cite{Spradlin:1999bn}.\\

\noindent \textbf{Hartle-Hawking/Bunch-Davies vacuum:}\\

The best known state was defined by \textit{Hartle-Hawking} (HH)~\cite{PhysRevD.13.2188}:
\begin{equation}
\label{eq:HH}
    \braket{T^{(x)}_{\pm\pm}(x^{\pm})} = 0,
\end{equation}
which is the vacuum state with respect to Kruskal coordinates. This corresponds to the state in which the interior of the horizon is in thermal equilibrium with its surroundings, namely, when there is equal ingoing and outgoing radiation at the horizon. This state is called the \textit{Bunch-Davies} (BD) vacuum in de Sitter space. Another set of useful coordinates is the \textit{static coordinates} defined by the metric
\begin{align}
    ds^2 &= -e^{2\omega_{\sigma}}d\sigma^+d\sigma^-,
\end{align}
where the static conformal factor is
\begin{equation}
\label{eq:static}
    e^{2\omega_{\sigma}} = \frac{1}{ \mathrm{sh}_{\Lambda}^2\left(\frac{\Lambda}{2}(\sigma^+-\sigma^-)\right)},
\end{equation}
and $ \mathrm{sh}_+=\cosh$ for de Sitter spacetime ($\Lambda=+1$) and $ \mathrm{sh}_-=\sinh$ for Anti-de Sitter spacetime ($\Lambda=-1$). The two coordinate systems are related by:
\begin{equation}\label{eq:relation_x_sigma}
x^{\pm}=\pm \frac{1}{\sqrt{|\Lambda|}}e^{\pm\sqrt{|\Lambda|}\sigma^{\pm}}.
\end{equation}
Static coordinates define the natural time flow for a free-falling observer outside of an AdS black hole, or an observer in de Sitter spacetime. The transformation law \ref{eq:transfT} yields
\begin{equation}
    \braket{T^{(x)}_{\pm\pm}(\sigma^{\pm})} = \frac{c|\Lambda|}{48\pi}.
\end{equation}
Thus, this vacuum state is perceived as thermal by a static observer, at temperature $\beta^{-1}=\sqrt{|\Lambda|}/2\pi$~\cite{Pedraza:2021cvx}. This is the Unruh effect. The HH and BD vacua preserve the symmetries of AdS and dS spacetimes, ensuring that the stress-tensor does not lead to any dynamical backreaction~\cite{Pedraza:2021cvx}. In particular, the backreaction leads to a shift
\begin{equation}
    \Phi \rightarrow \Phi + \frac{cG}{3}.
\end{equation}

\noindent \textbf{Boulware/Static vacuum:}\\

Another vacuum state of interest~\cite{Boulware} is constructed in static coordinates. The resulting vacuum
\begin{equation}
\label{eq:Boulware}
    \braket{T_{\pm\pm}^{(\sigma)}(\sigma^{\pm})} = 0,
\end{equation}
is called \textit{Boulware} in AdS and \textit{static} in dS, and it is the analog of the Rindler vacuum in Minkowski space. In Kruskal coordinates,
\begin{equation}
    \braket{T^{(\sigma)}_{\pm\pm}(x^{\pm})}=-\frac{c}{48\pi (x^{\pm})^2}.
\end{equation}
The stress tensor diverges at the horizon and leads to a negative energy density, which is interpreted as a Casimir energy when placing a boundary at the horizon~\cite{Spradlin:1999bn}. This allows for a non-trivial flux of energy at $\mathcal{I}^+$. Note that the NEC is strongly violated in this state.\\

\noindent \textbf{Unruh vacuum:}\\

Inspired by black hole evaporation, it is possible to break the symmetry between the right- and left-moving sectors of matter fields~\cite{Spradlin:1999bn,Aalsma:2019rpt}. Such a state describes the evaporation of the horizon and is built as a hybrid of the HH/BD and Boulware/static vacua, for the incoming and outgoing modes, respectively. Considering the right region bounded by a horizon in Figures \ref{fig:PenroseJT},\ref{fig:full}, \ref{fig:JTAdS}, such a state is defined by:
\begin{align}\label{eq:t_unruh_vac_1}
T_{++}^{(\sigma)}(x^{+})&=-\frac{c}{48\pi(x^{+})^2}, & T_{++}^{(\sigma)}(\sigma^{+})&=0,
\\
\label{eq:t_unruh_vac_2}
T_{--}^{(x)}(x^{-})&=0, & T_{--}^{(x)}(\sigma^{-})&=\frac{c|\Lambda|}{48\pi}.
\end{align}
This is called the \textit{Unruh} state~\cite{PhysRevD.14.870}. It explicitly violates the NEC. We will see in the next chapter that it can be used to probe the validity of the QFC and other semiclassical conjectures.

The Unruh vacuum yields a divergent stress tensor at the past horizon. It is considered physical because it is regular at the future horizon, while the past horizon should be discarded in a physically sensible model, and replaced by collapsing matter. It has also been argued to be a reasonable state in de Sitter spacetime~\cite{Greene:2005wk,Aalsma:2019rpt}. The modeling of black hole evaporation in JT gravity has been a very active subject of research, which we will not develop further here. See 
\emph{eg}~\cite{Spradlin:1999bn,Engelsoy:2016xyb,Mertens:2019bvy,Mertens:2022irh} for further reading.

\section{Generalized entropy}

Considering the semiclassical model introduced in the previous section, one can compute exactly the fine-grained entropy of spacelike regions as introduced in Section~\ref{sec:SCFT}. Moreover, using equation~\eqref{eq:JTarea}, one can define an analog of area functionals in JT gravity. This leads to the definition of generalized entropy of a point $p$ in JT gravity:
\begin{equation} \label{eq:quant_area_point} 
S_{\rm gen}(p) = \frac{\Phi(p)}{4G} + S(\Sigma_{p}) = 4G~\text{Area}_{\rm qu}(p), 
\end{equation}
where $\Sigma_{p}$ is a spacelike surface defined such that $\partial \Sigma_{p} = p$. We can also make sense of the quantities describing null hypersurfaces introduced in the previous chapters. A lightray emanating from $p$ in JT gravity is naturally associated with a null hypersurface in the higher-dimensional geometry as a congruence emanating orthogonally from $\gamma(p)$. Applying the definition \eqref{eq:expansion} for $\gamma(p)$ and the ansatz \ref{eq:sphred}, one finds~\cite{Franken:2023ugu} 
\begin{equation}
\theta = \frac{1}{\Phi}k^{\mu}\nabla_{\mu}\Phi = \frac{\Phi'}{\Phi}, 
\end{equation} 
where we set $\lambda=0$ at $\gamma$. Comparing this equation with the definition of an area element \eqref{eq:A} yields
\begin{equation} 
\mathcal{A}(\lambda) = \mathcal{A}(0)\frac{\Phi(\lambda)}{\Phi(0)}. 
\end{equation}
$\mathcal{A}(0)$ is sometimes rescaled to $1$ such that $\mathcal{A}$ is interpreted as an area decrease factor instead of an area element. Furthermore, the two-dimensional QNEC \eqref{eq:QNEC} is given by \begin{equation} \label{eq:QNEC2D} 2\pi k^{\mu}k^{\nu}\braket{T_{\mu\nu}} \geq S'',
\end{equation}
where the normal vector $k^{\mu}$ and stress tensor $T_{\mu\nu}$ are implicitly associated with a point $p$, and $S''$ is computed on a slice $\Sigma_p$. This framework provides a powerful tool for semiclassical gravity: Concepts of generalized entropy, expansion, quantum focusing, etc., can be treated as physical quantities in JT gravity, even though there are no dimensionful codimension-two surfaces in this framework. This simplifies computations that would otherwise be highly non-trivial in higher-dimensional geometries.

However, the structure of JT gravity as a quantum theory of gravity is quite different from conventional quantum gravity, and it is not clear whether quantities such as generalized entropy and quantum expansion in JT gravity remain sensible objects that carry crucial information about the fundamental structure of our model. In the previous chapter, we argued at length about the fundamental role of generalized entropy and quantum focusing in quantum gravity. Before studying the behavior of these quantities in JT gravity, we should ensure that generalized entropy is a meaningful quantity in our framework. This question will be discussed in more depth in~\cite{QFC}, but we briefly address it here.

In~\cite{Almheiri:2014cka}, a first law of black hole thermodynamics was derived in JT gravity, where generalized entropy naturally appears. This quantity was shown to satisfy a generalized second law~\cite{Moitra:2019xoj}. In~\cite{Pedraza:2021cvx}, the analysis was extended to general quantum extremal surfaces, which were shown to satisfy a generalized second law of thermodynamics that implies properties such as the inclusion of the causal wedge in the entanglement wedge \eqref{eq:CinW}. Moreover, the Wald entropy \eqref{eq:WaldS} in semiclassical JT gravity reproduces exactly the generalized entropy,\footnote{In JT gravity, any point is a bifurcated Killing horizon due to the symmetries of (A)dS, allowing the Wald entropy to be computed at any point~\cite{Pedraza:2021cvx}.}
\begin{equation}
S_{\rm Wald} = S_{\rm gen}.
\end{equation}
Generalized entropy is the natural notion of entropy that arises in algebraic quantum field theory when coupling quantum gravity with quantum field theory. In this context, the entropy is uniquely defined up to a state-independent additive constant. A similar analysis was performed in AdS JT gravity~\cite{Penington:2023dql,Kolchmeyer:2023gwa,Penington:2024sum}. The Hilbert space of JT gravity coupled to matter is acted on by two algebras associated with the left and right conformal boundaries, which can be seen in Figure~\ref{fig:JTAdS}. These algebras are of type II, analogous to the higher-dimensional black hole coupled to quantum matter. These two algebras are commutants: any operator that commutes with all operators in the left algebra must be an operator of the right algebra, and vice versa. This statement is closely related to the causal structure of the two-dimensional spacetime. In particular, it is the algebraic analog of the boundary causality condition (see below equation~\eqref{eq:EWnesting}) that ensures that no causal curve can connect the two entanglement wedges associated with two boundary regions that are causally disconnected. It was also shown in~\cite{Penington:2023dql} that semiclassical states have boundary entanglement entropies that are given by the quantum area of quantum extremal surfaces. To summarize, generalized entropy has been shown to be the uniquely defined notion of entropy $S_{\rm vN}$ associated with the algebra of observables in JT gravity,
\begin{equation} 
S_{\rm vN}=S_{\rm gen},
\end{equation} 
and quantum extremal surfaces compute boundary entanglement entropy. Moreover, the algebraic approach leads to fundamental constraints on boundary causality, which are ensured by quantum focusing. This will be our main motivation for studying the generalized entropy and proving the restricted QFC for JT gravity \ref{conj:rQFC} in Chapter~\ref{ch:proof2D}~\cite{QFC}.

\section{Entropy in the half reduction model}

In the half reduction model of de Sitter JT gravity, the conformal field theory is defined on a curved background , as shown in \eqref{eq:solJT}, with spatial boundaries at the poles where $x^+x^-=-1/\Lambda$. Although de Sitter spacetime does not have boundaries, the half-reduction model can be viewed as a spherical reduction of dS$_3$ with reflective boundary conditions for matter fields $\psi_i$ at the poles:
\begin{equation}
    \left.\psi_i\right|_{x^+x^-=-1/\Lambda}=0.
\end{equation}
When the background has infinite spatial extent, equation~\eqref{eq:CFTcurved} can be used to compute the entanglement entropy of spatial slices in an arbitrary two-dimensional curved background. This is the case of the double-sided AdS black hole. Analogously, equation~\eqref{eq:CFTS1} can be used when the background is spatially periodic, which is the case of the full-reduction model in the absence of matter.\footnote{In the presence of matter this is not always the case, as we will see in Chapter~\ref{ch:proof2D}.} 

These formulas do not apply in the half reduction model with reflective boundary conditions~\cite{Franken:2023ugu}. First, we expect the von Neumann entropy of a Cauchy slice to vanish, as it is associated with a pure state. Second, the entropy of a slice bounded by a point $p$ and one of the boundaries should not depend on the endpoint at the boundary. Indeed, the reflection of right- and left-moving modes at the boundaries ensures that all spacelike slices joining the point $p$ and the boundary are crossed by the same modes, and thus have the same entropy. In particular, these slices are related by unitary evolution and share the same causal diamond, as illustrated by the blue-shaded triangle in Figure~\ref{fig:BD}. The entropy formula \eqref{eq:CFTcurved} does not satisfy either of these two conditions, since the half reduction model is not infinitely extended. 

Considering reflecting boundary conditions at two spatial boundaries parameterized by an arbitrary function $x^-=f_x(x^+)$, the entanglement entropy $S(\Sigma_p)$ associated with a point $p$ located at $(x^+,x^-)$ in the vacuum constructed with respect to coordinates $x^{\pm}$ is derived in Appendix \ref{app:reflec_bound}:
\begin{equation}
\label{eq:entropybdy}
    S_x(\Sigma_p) = \frac{c}{6}\omega(x^+,x^-)+\frac{c}{12}\ln\left[f_x'(f_x^{-1}(x^-))\frac{(x^+ -f_x^{-1}(x^-))^2}{\delta^2}\right],
\end{equation}
where $f_x'\equiv df_x/dx^+$ and $\delta$ is the UV cutoff in inertial coordinates.\footnote{ The function $f_x$ is restricted to be a $C^1$ involution on $\mathbb{R}_*$, with $0< f'_x <\infty$ and $x^+f_x(x^+)\leq 0$. See Appendix \ref{app:reflec_bound}} The proof relies on a careful mapping between a complete set of modes, which contains left- and right-moving modes, to a complete set of modes along one of the two axes $x^+=0$ or $x^-=0$, containing only right-moving modes or left-moving modes, respectively~\cite{Franken:2023ugu}. This takes into account the correlation between left- and right-moving modes that span the spacetime, due to reflections on the boundary, as illustrated in Figure~\ref{fig:BD}. In particular, the term $f'_x(f_x^{-1}(x^-))$ in the entropy formula above arises from the correlation between the UV cutoff for modes before and after their reflection on the boundary.

Applying this formula to the Bunch-Davies state, we find a relation between the entropy and the dilaton solution in the absence of matter $\Phi_*$, as given in \eqref{eq:solJT}:
\begin{align}\label{eq:entang_entropy}
    S_{x}(\Sigma_p) = \frac{c}{12}\ln\left[\frac{4}{\Lambda\delta^2}\frac{(1+\Lambda x^+x^-)^2}{(1-\Lambda x^+x^-)^2}\right] = \frac{c}{6}\ln\Phi_* + \text{constant},
\end{align}
where we used that $\phi_0=0$ in the half reduction model. The reflective boundaries ensure that the state is symmetric with respect to the $x^+$ and $x^-$ coordinates, and that the entropy of $\Sigma_p$ does not depend on the position of its endpoint on the boundary, see Figure~\ref{fig:BD}. This formula can be generalized to any vacuum state, as shown in Appendix \ref{sec:changevac}. In particular, the entropy in the vacuum constructed in coordinates $y^{\pm}(x^{\pm})$ is given by
\begin{align}
\begin{split}
\label{eq:transf_bdy}
    S_{y}(\Sigma_p) &= \frac{c}{6}\omega_x(x^+,x^-) + \frac{c}{12}\ln\left[f'_x(f_x^{-1}(x^-))\frac{\left(y^+(x^+) -y^+(f_x^{-1}(x^-))\right)^2}{\delta^2}\right]\\
    &-\frac{c}{12}\ln\left[\frac{dy^+}{dx^+}(x^+)\frac{dy^+}{dx^+}\left(f_x^{-1}(x^-)\right)\right].
\end{split}
\end{align}
\begin{figure}
\centering
\includegraphics[width=0.4\linewidth]{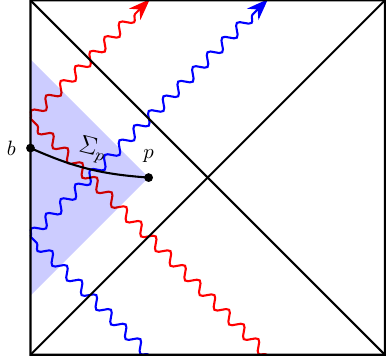}
\caption{\footnotesize
    Penrose diagrams of the half reduction model, with reflective boundary conditions. They imply that the causal diamond of a spacelike slice $\Sigma_p$ ending on one boundary is a triangle, depicted by the blue shaded region, and that $S(\Sigma_p)$ is independent of the location of $b$ along the left edge of the triangle. Examples of radiation emanating as left-moving modes from $\cal{I}^-$ are shown. The red one crosses $\Sigma_p$ as a left-moving mode, while the blue one crosses $\Sigma_p$ as a right-moving mode.
   }
   \label{fig:BD}
\end{figure}

\chapter{Quantum Bousso bounds and restricted quantum focusing in two dimensions}
\label{ch:proof2D}

We use the computational control over $S_{\rm gen}$ provided by JT gravity to prove the Strominger-Thompson and BFLW quantum Bousso bounds in JT gravity coupled to a CFT$_2$ in the large $c$ limit, as well as the restricted QFC. Our proofs extend to any matter content, if one assumes the QNEC. this chapter is mainly inspired by~\cite{Franken:2023ugu} and the forthcoming work~\cite{QFC}. We focus on semiclassical JT gravity for its simplicity, but our derivations are expected to hold for any two-dimensional dilaton theory.

\section{Proof of two quantum Bousso bounds}
\label{sec:proofQBB}

In semiclassical JT gravity, the classical Bousso bound \ref{th:Bousso} is not universally valid. Indeed, the gradient of entropy density assumption \eqref{eq:cond1_hydro} relies on the validity of the NEC, which is explicitly violated by quantum states of matter such as the Boulware/static vacuum (equation~\eqref{eq:Boulware}) or the Unruh vacuum (equation~\eqref{eq:t_unruh_vac_1}). In particular, the classical part of the stress tensor, $\braket{\tau_{\mu\nu}}$, is assumed to satisfy the NEC but the normal ordered stress tensor proportional to $t_{\mu\nu}$ is not lower bounded by any value.

To remedy this, Strominger and Thompson conjectured the bound \ref{conj:STQBB} which in two dimensions reads
\begin{equation}
       S_{\rm th}(\mathcal{L}_{\rm qu}(x_1,x_2)) \leq \frac{\text{Area}_{\rm qu}(x_1)-\text{Area}_{\rm qu}(x_2)}{4G},
   \end{equation}
for any quantum lightsheet $\mathcal{L}_{\rm qu}(x_1,x_2)$, without assuming the NEC. Initial and final surfaces $\gamma_1$ and $\gamma_2$ are replaced by initial and final points $x_1=(x_1^+,x_1^-)$ and $x_2=(x_2^+,x_1^-)$. In two dimensions, null geodesics always propagate either along $x^+$ or $x^-$, such that $k^{\mu}=(k^+,0)$ or $k^{\mu}=(0,k^-)$ respectively. The geodesic equation~\eqref{eq:k_geo_eq} yields the differential equation
\begin{align}
    \partial_{\pm}k^{\pm} + 2\partial_{\pm}\omega k^{\pm} &=0,
\end{align}
which is solved by
\begin{align}
    \label{eq:kexplicit}
    k^{\pm} &= C e^{-2\omega},
\end{align}
where the constant $C$ can be normalized to $1$ by an appropriate rescaling of $\lambda$. Following~\cite{Strominger:2003br}, we consider the hydrodynamic regime \eqref{eq:hydros} where $S_{\rm th}(\mathcal{L}_{\rm qu}(x_1,x_2))$ can be expressed as an integral over a local entropy current $s^{\mu}$. The two-dimensional entropy density $s$ is related to the $d$-dimensional entropy density $s^{(d)}$ by
\begin{equation}\label{eq:2d_entropy}
s(\lambda)=S_{d-1}(l_d)\Phi(\lambda)~s^{(d)}(\lambda).
\end{equation}
Equation~\eqref{eq:4D_entropy_flux} applied to an ansatz metric \eqref{eq:sphred} is then given by 
\begin{equation}
    S_{\rm th}(\mathcal{L}_{\rm qu}(x_1,x_2)) = \int_0^1 d\lambda ~s(\lambda),
\end{equation}
where we have normalized $\lambda$ such that it is $0$ at $x_1$ and $1$ at $x_2$. Following~\cite{Strominger:2003br}, we adapt the two conditions for the hydrodynamic regime \eqref{eq:cond1_hydro} and \eqref{eq:cond2} for semiclassical spacetimes:
\begin{itemize}
    \item The first classical condition \eqref{eq:cond1_hydro} is unchanged, without introducing the quantum mechanical part of the stress tensor:\footnote{A spherical reduction of the higher-dimensional equivalent of this first condition leads to the constraint $\Phi(\lambda)\left|(s(\lambda)\Phi^{-1}(\lambda))^{'}\right|\leq 2\pi k^{\mu}k^{\nu}\braket{\tau_{\mu\nu}}(\lambda)$. This inequality is stronger than \eqref{eq:cond1_QNEC}.}
    \begin{equation}
    \label{eq:cond1_QNEC}
    \left|s^{'}(\lambda)\right|\leq 2\pi k^{\mu}k^{\nu}\braket{\tau_{\mu\nu}}(\lambda),
    \end{equation}
    which corresponds to imposing the classical condition in the non-backreacted geometry. It is well defined since $\tau_{\mu\nu}$ always satisfies the NEC.
    \item The modification $\text{Area}\rightarrow \text{Area}_{\rm qu}$ is applied to the second classical condition \eqref{eq:cond2}, implying that the QBB is initially satisfied:
    \begin{equation}
    \label{eq:cond2_QNEC}
        s(0) \leq -\frac{\text{Area}_{\rm qu}'(0)}{4G}.
    \end{equation}
    Since $s(\lambda)$ is positive, this condition implies that the quantum area must be initially non-increasing. This is satisfied by a quantum lightsheet.
\end{itemize}
From these conditions, we derive a sufficient condition for Strominger and Thompson's bound \cite{Franken:2023ugu}.
\begin{proposition}[Sufficient condition]
\label{prop:sufc}
    In semiclassical JT gravity, assuming the hydrodynamic regime with associated conditions \eqref{eq:cond1_QNEC} and \eqref{eq:cond2_QNEC}, the inequality
    \begin{equation}
    \label{eq:sufc}
    2\pi k^{\pm}k^{\pm}\braket{:T_{\pm\pm}:}\geq S''
\end{equation}
is a sufficient condition for Conjecture \ref{conj:STQBB}.
\end{proposition}
We make the dependence on $p$ and $\Sigma_p$ implicit to simplify the notations.
\begin{proof}
    This follows directly from integrating equation~\eqref{eq:cond1_QNEC} and using equation~\eqref{eq:cond2_QNEC} as an initial condition as well as the sufficient condition \eqref{eq:sufc}. See Appendix \ref{app:proofsufc} for a detailed computation.
\end{proof}
Interestingly, equation~\eqref{eq:sufc} closely resembles the QNEC \eqref{eq:QNEC2D}. In particular, it is equivalent to it under the assumption $\braket{\tau_{\mu\nu}}\geq 0$. The route taken in~\cite{Franken:2023pni} consists of studying the transformation rules of the normal ordered stress tensor and second derivative of the entropy under a change of conformal vacuum, \emph{ie} for vacua constructed in coordinates $x^{\pm}$ and $y^{\pm}(x^{\pm})$. Under a change of coordinates $x^{\pm}\rightarrow y^{\pm}(x^{\pm})$, the metric becomes
\begin{align}
d s^2 &= -e^{2\omega_x(x^+,x^-)}dx^+dx^- = -e^{2\omega_x(x^+,x^-)}\frac{dx^+}{dy^+}\frac{dx^-}{dy^-}dy^+dy^-\\
&= -e^{2\omega_y(y^+,y^-)}dy^+dy^-,
\end{align}
from which we get the transformation of the conformal factor:
\begin{equation}
\label{eq:transfo_conf_factor}
    \omega_{y}(y^+(x^+),y^-(x^-)) = \omega_{x}(x^+,x^-) -\frac{1}{2}\ln\left[\frac{dy^+}{dx^+}\frac{dy^-}{dx^-}\right].
\end{equation}
The proof of the Strominger-Thompson bound then relies on the following property.

\begin{lemma}
    In a two-dimensional CFT defined on a curved background, the quantity
    \begin{equation}
    \mathcal{Q} = 2\pi k^{\pm}k^{\pm}\braket{:T_{\pm\pm}:} - S'' - \frac{6}{c}(S')^2
\end{equation}
    is a scalar under the change of vacuum state and vanishes in these states.
\end{lemma}
    \begin{proof}
See~\cite{Franken:2023ugu} for a detailed proof. The argument relies on studying the transformation law of the entropy $S(\Sigma)$ under a change of conformal vacuum, given in \eqref{eq:transf} and \eqref{eq:transf_bdy}. From these relations, one computes the transformation laws for the first and second derivatives of the entropy. We consider the case where the derivation is taken along a lightsheet going in the $x^+$ direction, \emph{ie} $'\equiv d/d\lambda = k^+\partial/\partial x^+$. Combining the transformation laws for the first and second derivatives, we find the transformation law for the quantity $S''+\frac{6}{c}(S')^2$ introduced by Wall in~\cite{Wall:2011kb}:
\begin{equation}\label{eq:trans_law_S''}
S_x''+\frac{6}{c}(S_x')^2\rightarrow S_y''+\frac{6}{c}(S_y')^2 = S_x''+\frac{6}{c}(S_x')^2-\frac{c}{12}\{y^+,x^+\}\left(\frac{\partial x^+}{\partial \lambda}\right)^2.
\end{equation}
Similarly, we compute the $t_{\pm}(x^{\pm})$ function in the vacuum defined by $t_{\pm}(y^{\pm})=0$ using equation~\eqref{eq:transft}. As for the entropy, we make the vacuum state explicit by writing the $t_{\pm}$ function in the vacuum defined in coordinates $y^{\pm}$ as  $t^{(y)}_{\pm}$, \emph{ie} $t^{(y)}_{\pm}(y^{\pm})=0$. The transformation under a change of vacuum is
\begin{equation}
   t^{(x)}_{\pm}(x^{\pm}) = 0 \rightarrow t^{(y)}_{\pm}(x^{\pm}) = \left\{y^{\pm},x^{\pm}\right\}.
\end{equation}
From the transformation laws of $S$, its derivatives, and $t_{+}$, we find that the quantity 
\begin{equation}
    \mathcal{Q} = - \frac{c}{12} k^{+}k^{+}t^{(x)}_{+} + S_x'' + \frac{6}{c}(S_x')^2
\end{equation}
is a scalar under conformal transformation.

In the HH/BD vacuum, $t_{\pm}(x^{\pm})=0$ and
\begin{equation}
\label{eq:BDeq}
    S'' + \frac{6}{c}(S')^2=0,
\end{equation}
as follows from equation~\eqref{eq:entang_entropy}. Thus $\mathcal{Q}=0$. Because $\mathcal{Q}$ is a scalar under conformal transformations, this must be true in any vacuum state.
 \end{proof}
 
 To conclude, we have shown that
\begin{equation}
    S''+\frac{c}{12} k^{\pm}k^{\pm}t_{\pm}  = - \frac{6}{c}(S')^2 \leq 0
\end{equation}
in any vacuum state, completing the proof of the Strominger-Thompson quantum Bousso bound.

At the time of the publication of~\cite{Franken:2023ugu}, we were not aware of a proof of the QNEC in JT gravity coupled to a CFT$_2$ by Almheiri, Mahajan, and Maldacena, in~\cite{Almheiri:2019yqk}. Their proof relies on the existing proof that the QNEC is valid in flat space~\cite{Bousso:2015wca,Balakrishnan:2017bjg,Koeller:2015qmn,Ceyhan:2018zfg,Kudler-Flam:2023hkl}. Since all two-dimensional metrics are related by a Weyl rescaling, one can use the conformal transformation of the stress tensor and entropy to derive the QNEC in an arbitrary background. In particular, they show that
\begin{equation}
\label{eq:WallQNEC}
    \mathcal{Q}\geq 0
\end{equation}
for any state in JT gravity minimally coupled to a CFT$_2$. Hence, our proof of the Strominger-Thompson conjecture is extended to all states in JT gravity coupled to a CFT$_2$ in the large $c$ limit, while saturation of the bound may be obtained in a conformal vacuum.

Without matter, the full reduction model is either spatially periodic or infinitely extended, and one would need at least two points to define the boundary of a non-trivial slice. Hence, we always consider one of these two points to be fixed and spacelike separated from the second one moving along the lightsheet. This produces the same transformation laws ensuring that $\mathcal{Q}$ is invariant. This situation is depicted for the dS full reduction model in Figure~\ref{fig:Penrose_diag_full_LS}, in the case where the fixed point $a$ lies at $x_a^{\pm}=\pm\infty$. This is a preferred point since it is the only one spacelike separated from any lightsheet contained in the white and/or blue regions of Figure~\ref{fig:full}.\footnote{Another possibility would be to consider that both endpoints of the slice belong to two lightrays forming a disconnected lightsheet.}

The backreaction of matter on the dilaton modifies the effective geometry, potentially leading to the creation of effective boundaries. This is the case in the Unruh-de Sitter vacuum~\cite{Aalsma:2019rpt, Kames-King:2021etp, Aalsma:2021bit},  see Figure~\ref{fig:Penrose_diag_full_LS} and Section~\ref{sec:dSCE} for more details. In this example, there is a conformal boundary at $x^+\rightarrow +\infty$, which acts as a weakly gravitating region where the radiation emanating from the past cosmological horizon can be collected. We then places the fixed point $a$ at the effective boundary, as in Figure~\ref{fig:Penrose_diag_full_LS}.
\begin{figure}[ht]
    \centering
\includegraphics[width=0.65\linewidth]{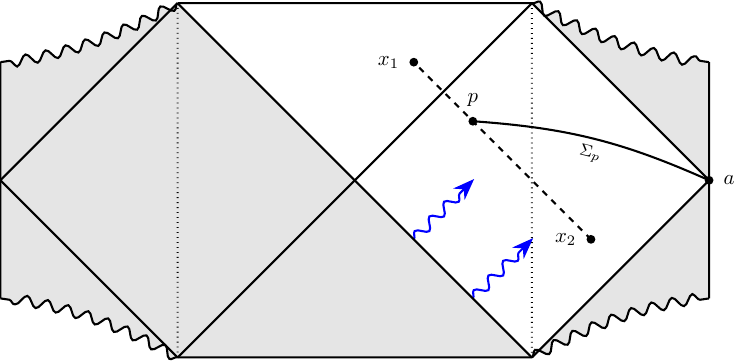}
    \caption{\footnotesize Penrose diagram for two-dimensional de Sitter space in the full reduction model. A portion of a lightray bounded by two points $x_1$ and $x_2$ is depicted by the dashed line, as well as a Cauchy slice $\Sigma_p$ bounded by a point $p$ of the lightray and the point $a$ at spatial infinity. In the Unruh-de Sitter vacuum, the dilaton diverges on the past cosmological horizon and the black hole horizon of the static patch of the antipode, eliminating the gray shaded regions from the backreacted solution~\cite{Aalsma:2019rpt,Aalsma:2021bit}. The past horizon emits radiation (in blue) crossing $\Sigma_p$ as right-moving modes, and ending up in the weakly gravitating region at $x^+\rightarrow\infty$.}
    \label{fig:Penrose_diag_full_LS}
\end{figure}

In Section~\ref{sec:QBB}, we introduced the BFLW quantum Bousso bound~\cite{Bousso:2015mna}, which takes the classical statement of the Bousso bound \ref{th:Bousso} and replaces the thermodynamic entropy on the left-hand side by a fine-grained definition of the entropy, as in equation~\eqref{eq:BFLWQBB}. The resulting bound can be obtained by considering Strominger-Thompson's bound and imposing $s(\lambda)=0$. Formally removing the coarse-grained entropy density from the conditions \eqref{eq:cond1_QNEC} and \eqref{eq:cond2_QNEC} gives
\begin{align}
    k^{\mu}k^{\nu}\braket{\tau_{\mu\nu}}&\geq 0,\\
    \text{Area}_{\rm qu}'(0)&\leq 0,
\end{align}
from which one can follow through with our proof and recover the BFLW quantum Bousso bound, without relying on the hydrodynamic limit or the QFC. Thus, our sufficient condition and proof of its validity extend to the BFLW quantum Bousso bound, with the only assumptions being that the classical part of the stress tensor satisfies the NEC, and initial quantum non-expansion.

\section{Proof of the restricted QFC}

An expanded formulation of $\Theta'$ is found using the equations of motion for the dilaton \eqref{eq:contracted_eom}:
\begin{equation}
     \Theta' = -8\pi G k^{\mu}k^{\nu}\frac{\Braket{T_{\mu\nu}}}{\Phi}-\theta^2+4G\left(\frac{S''}{\Phi}-\frac{\theta S'}{\Phi}\right).
\end{equation}
Assuming $\Phi\geq 0$ and applying the two-dimensional QNEC \eqref{eq:sufc}, which was proven in JT gravity coupled to conformal matter~\cite{Almheiri:2019yqk}, and in holographic quantum field theories \cite{Koeller:2015qmn}, we obtain
\begin{equation}
\label{eq:rQFCbis}
    \Theta' \leq -\theta\Theta.
\end{equation}
This inequality demonstrates that vanishing quantum expansion implies non-positive quantum expansion, which is the statement of the restricted quantum focusing conjecture. Notably, the bound \eqref{eq:rQFCbis} also implies quantum non-expansion at classical extremal surfaces.

While equation~\eqref{eq:rQFCbis} implies restricted focusing, it explicitly allows for violations of the QFC. Conformal invariance serves as a strong constraint that allows us to prove various results such as the QNEC~\cite{Almheiri:2019yqk}, quantum Bousso bounds, the generalized second law~\cite{Wall:2011kb}, and even restricted quantum focusing. However, the QFC appears not to be provable, even in this context. These results provide good evidence that, at least in JT gravity, restricted quantum focusing is a fundamental constraint that follows from the fundamental description of the theory, while the QFC has no reason to be true.

\section{Counter-examples to the QFC}

The lower bound \eqref{eq:rQFCbis} provides a necessary condition for violations of the QFC:
\begin{equation}
    \label{eq:necessarycond}
    \Theta' > 0 \Rightarrow \theta\Theta < 0.
\end{equation}
Although this condition is not sufficient, it is restrictive enough to hint at where to look if one wants to find a violation of the QFC. Condition \eqref{eq:necessarycond} implies that a surface violating the QFC should lie on a lightsheet that is not a quantum lightsheet, or the opposite. This can happen in backgrounds with strong quantum effects, such as an evaporating horizon. In this section, we present two examples of matter states in JT gravity that violate the QFC. We work in units where $l_{\rm (A)dS}=1$ and the sign of $\Lambda$ will be fixed to $-$ in Section~\ref{sec:AdSCE} and to $+$ in Section~\ref{sec:dSCE}.

\subsection{AdS black hole in equilibrium with a bath}
\label{sec:AdSCE}

Black holes in AdS differ significantly from the usual picture of Schwarzschild black holes in asymptotically flat space. Indeed, a lightlike signal can cross the whole spacetime in a finite amount of time. Thus, one needs to equip AdS with boundary conditions, which are usually taken to be reflective. This results in black hole radiation returning into the black hole in a finite time. To study black hole evaporation, one replaces these conditions with absorbing boundary conditions, such that all radiation emitted by the black hole can be collected in some non-gravitating reservoir, see \emph{eg}~\cite{Rocha:2008fe,VanRaamsdonk:2013sza,Almheiri:2018xdw,Penington:2019npb, Almheiri:2019psf,Almheiri:2019yqk}. The CFT state describing the black hole evolution is non-unitary and has irreversible dynamics. However, the product of the CFT and the reservoir Hilbert spaces has a unitary evolution. This construction is central to the resolution of the information paradox through entanglement wedge reconstruction in AdS/CFT~\cite{Penington:2019npb,Almheiri:2019psf}. 

Following~\cite{Almheiri:2019yqk,Almheiri:2019psf}, we consider an extremal black hole in AdS, in equilibrium with a bath modeling the asymptotically flat region far away from the black hole where gravity can be neglected. Near-extremal AdS black holes in JT gravity emerge from the dimensional reduction of the near-horizon geometry of a near-extremal Reissner-Nordstrom black hole, leading to the geometry depicted in Figure~\ref{fig:JTAdS}. On the other hand, extremal black holes emerge from the dimensional reduction of an extremal Reissner-Nordstrom black hole~\cite{Spradlin:1999bn,Mertens:2022irh}, for which the two horizons merge. An extremal black hole does not radiate, \emph{ie} it has zero temperature. 

This toy model is constructed in two dimensions, with a CFT$_2$ defined on half of two-dimensional Minkowski space glued along its conformal boundary to the Poincaré patch of an extremal AdS$_2$ black hole in JT gravity, see Figure~\ref{fig:AdS}. The half of flat space is described in null coordinates $(U_{\rm flat},V_{\rm flat})$ with $V_{\rm flat}>U_{\rm flat}$, while the black hole side is described by Poincaré coordinates $(U,V)$ where $U>V$, with metric
\begin{equation}
    ds^2 = -\frac{4dUdV}{(U-V)^2}.
\end{equation}
This geometry is found in the near-horizon limit of a higher-dimensional black hole, having only one boundary at $U=V$. The horizons are located at $U-V\rightarrow\infty$, at an infinite proper distance from the boundary. From the higher-dimensional point of view, this is related to the length of the black hole throat being infinitely long in the extremal limit. This geometry is depicted in the left part of the diagram in Figure~\ref{fig:AdS}.
\begin{figure}[ht]
    \centering
    \includegraphics[width=0.5\linewidth]{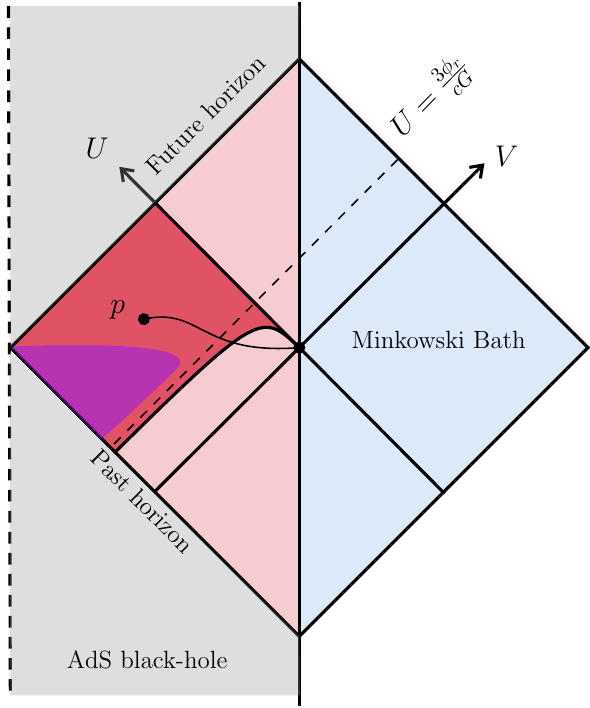}
    \caption{\footnotesize Penrose diagram of a two-dimensional extremal black hole in equilibrium with a bath at zero temperature. The blue region is half of Minkowski space and the red region is the Poincaré patch of the black hole. An achronal slice with free endpoint $p$ and fixed endpoint at $(0,0)$ is depicted. The region where $\theta\Theta <0$ associated with $p$ along the $V$ direction, bounded by curves \eqref{eq:bdy} and \eqref{eq:bdy2}, is depicted in dark red. The QFC violation region is shaded in purple.}
    \label{fig:AdS}.
\end{figure}

We consider the black hole to be in Poincaré vacuum $T_{UV}=T_{UU}=T_{VV}=0$.\footnote{Note that the vanishing of $T_{UV}$ is unusual, even for vacuum states. From \eqref{eq:stress}, the off diagonal term of the stress tensor is finite in general. $T_{UV}=0$ is specific to the Poincaré coordinates.} The bath is taken to be at zero temperature as it is in equilibrium with the extremal black hole, such that $T_{U_{\rm flat}V_{\rm flat}}=T_{U_{\rm flat}U_{\rm flat}}=T_{V_{\rm flat}V_{\rm flat}}=0$. Transparent boundary conditions imply that $(U_{\rm flat},V_{\rm flat})=(U,V)$ with a unique vanishing stress tensor~\cite{Almheiri:2019yqk}. The solution to the dilaton equation of motions \eqref{eq:contracted_eom} in the gravitating region $U>V$ is
\begin{equation}
\label{eq:dilAdS}
    \Phi = \phi_0 + \frac{2\phi_{\rm r}}{U-V}.
\end{equation}
In \eqref{eq:Sgen}, we approximated the gravitational entropy to be given by the area term. To ensure its validity, we impose $U-V\ll \phi_r$~\cite{Maldacena:2019cbz}. The entanglement entropy of a slice $\Sigma_p$ bounded by a point $p=(U,V)$ in AdS and a point $(U_a,V_a)$ in the bath is given by
\begin{equation}
\label{eq:SAdS}
    S(\Sigma_p) = \frac{c}{12}\log\left[\frac{(U-U_a)^2(V-V_a)^2}{(U-V)^2}\right] + {\rm constant}.
\end{equation}
Let us consider the generalized entropy associated with a slice bounded by a point fixed at $(0,0)$ and a free point $p$. The classical expansion $\theta$ of $p$ in the $V$ direction vanishes at $V=U$. Analogously, the quantum expansion $\Theta$ vanishes along the curves
\begin{align}
    V&=U \label{eq:bdy}\\
    V&=-\frac{U^2}{\frac{3\phi_{\rm r}}{cG}-U}. \label{eq:bdy2}
\end{align}
Equation~\eqref{eq:bdy2} asymptotes to $U=3\phi_r/cG$, which goes to $0$ when $\phi_r\ll cG$ and to $\infty$ when $\phi_r\gg cG$. We find QFC violations in the region $\theta\Theta < 0$, for a large range of parameters. Violations are only found in the regime $cG\gtrsim\max( \phi_0, \phi_r)$, without restriction on the ratio $\phi_0/\phi_r$. We provide a schematic picture of the violation region in Figure~\ref{fig:AdS}, and plot the function $\Theta'$ along the null direction $-V$ at fixed $U=1$ in Figure~\ref{fig:AdSplot}.
\begin{figure}[ht]
    \centering
    \includegraphics[width=0.8\linewidth]{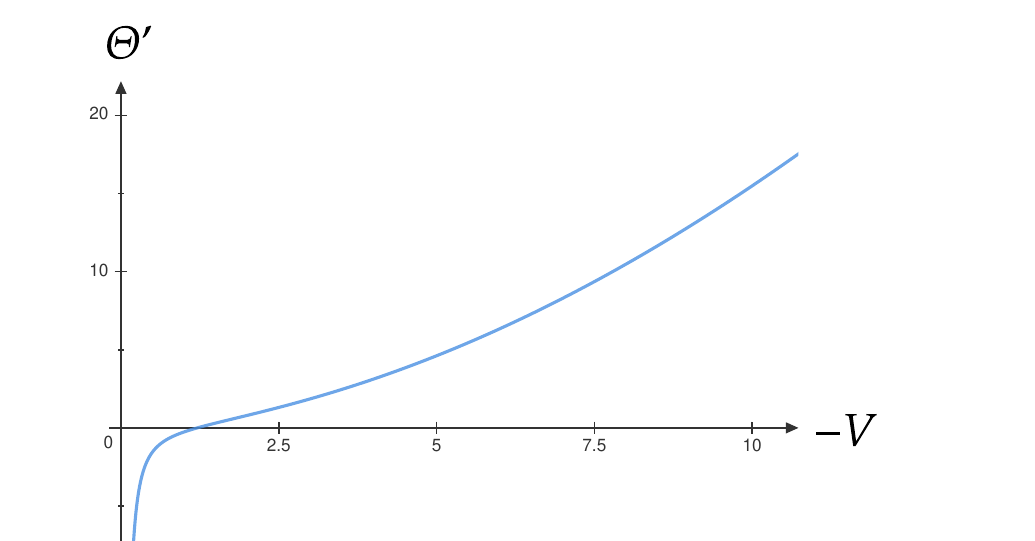}
    \caption{\footnotesize Plot of $\Theta'$ at fixed $U=1$, along the $-V$ direction. We have taken the constants $\phi_0=1,\phi_r=100,cG=1000$.}
    \label{fig:AdSplot}
\end{figure}
The analysis for the classical and quantum expansion is analogous for lightrays along $U$, as expressions \eqref{eq:dilAdS} and \eqref{eq:SAdS} are symmetric around $U=-V$. 

Considering a black hole at finite temperature in the Hartle-Hawking vacuum is also possible, although we find that QFC violations require a specific choice of Polyakov cosmological constant in the JT gravity action. In general, the Polyakov action \eqref{eq:Polyakov} always allows the addition of a cosmological constant term proportional to $\lambda\Lambda$ depending on some $\lambda$ parameter~\cite{Fabbri:2005mw}. Standard conventions, which we followed in this work, ignore this cosmological term by taking $\lambda=0$. We find QFC violations for a specific gauge choice $\lambda=1$, which eliminates semiclassical corrections to the dilaton.

\subsection{Evaporating de Sitter horizon}
\label{sec:dSCE}

Our second example violating the QFC is obtained by considering an evaporating de Sitter cosmological horizon, modeled in the full reduction model, characterized by a non-vanishing topological term $\phi_0>0$ as written in the action \eqref{eq:JT}. The geometry contains both a black hole and a cosmological horizon, surrounding the static patch of a free-falling observer.

The Kruskal coordinates $(x^+,x^-)$ cover the static patches of two antipodal observers together with the expanding region of the de Sitter geometry behind their cosmological horizons. The metric in Kruskal coordinates \eqref{eq:Kruskal} with $\Lambda=1$ is given by:
\begin{equation}
d s^2=-\frac{4}{(1-x^+x^-)^2}d x^+ d x^-.
\end{equation}
The metric in null static coordinates \eqref{eq:static} with $\Lambda=1$ is
\begin{equation}
    d s^2 = -\frac{1}{\cosh^2\left(\frac{\sigma^+-\sigma^-}{2}\right)} d\sigma^+ d\sigma^-.
\end{equation}
We consider the Unruh vacuum state \eqref{eq:t_unruh_vac_1}, \eqref{eq:t_unruh_vac_2} introduced in Section~\ref{sec:vacuum}. The equations of motion of the dilaton \eqref{eq:dilaton_eom} are solved by
\begin{equation}\label{eq:sol_Unruh_vac}
    \Phi(x^+,x^-) = \phi_0+ \phi_r\frac{1+x^+x^-}{1-x^+x^-} + \frac{cG}{6}\left(1 - \frac{1+x^+x^-}{1-x^+x^-}\log x^+\right),
\end{equation}
with $x^+>0$. The dilaton diverges on the past cosmological horizon $x^+=0$ of a static observer, and on its future black hole horizon $x^+=+\infty$. It diverges to $+\infty$ and $-\infty$ at future infinity for $x<x_0^+$ or $x>x_0^+$ respectively, where $x_0^+\equiv e^{6 \phi_r/(cG)}$. The Penrose diagram of the backreacted geometry is depicted in Figure~\ref{fig:Penrose_diag_UdS}.
\begin{figure}[ht]
\centering
\includegraphics[width=0.75\linewidth]{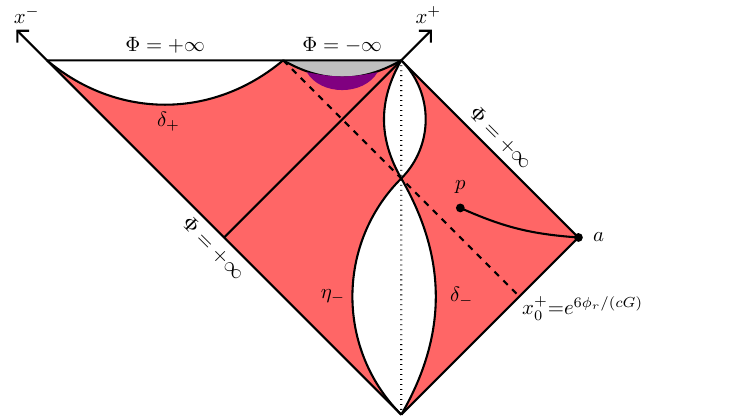}
    \caption{\footnotesize Penrose diagram for the backreacted geometry in the Unruh-de Sitter vacuum. The trajectory of a free-falling observer corresponds to the vertical dotted line. The region where $\Phi<0$ is shaded in gray. Considering a lightray along $x^+$, the regions where $\theta \Theta <0$ are depicted in red. The QFC violation region is shaded in purple. A spacelike slice between a point $p$ in the bulk and the point $a$ at spatial infinity is also depicted.}
    \label{fig:Penrose_diag_UdS}
\end{figure}

The sign of the backreacted classical expansion $\theta$ of a given lightray in the $x^+$ direction can be easily obtained by computing $\partial_+ \Phi$. One finds that $\partial_+ \Phi \leq 0$ for $\delta_-(x^+)<x^-<\delta_+(x^+)$, where $\delta_{\pm}(x^+)$ are the curves parameterized by: 
\begin{equation}
\delta_{\pm}(x^+)=\frac{1}{x^+}\left(\log\frac{x^+}{x_0^+}\pm\sqrt{1+\left(\log\frac{x^+}{x_0^+}\right)^2}\right).
\end{equation}

The entanglement entropy of a spacelike slice $\Sigma_p$ bounded by two points $p=(x^+,x^-)$ and $a=(x_a^+,x_a^-)$ is given by~\cite{Aalsma:2021bit}:
\begin{equation}
S(\Sigma)=\frac{c}{12}\log\left[
\frac{x^+ x_a^+ ~(x^- - x_a^-)^2~[\log(x_a^+/x^+)]^2}{(1-x^+ x^-)^2(1-x_a^+ x_a^-)^2}\right] + {\rm constant}.
\end{equation}
$a$ is arbitrary and will be sent to spatial infinity at the end of the computation, see Figure~\ref{fig:Penrose_diag_UdS}. One obtains the sign of the quantum expansion $\Theta$ of a lightray in the $x^+$ direction by computing $\partial_+ S_{\rm gen}$. After sending $x_a^{\pm}\rightarrow\pm\infty$, one finds that $\partial_+ S_{\rm gen}\geq 0$ for $\eta_-(x^+)<x^-<\eta_+(x^+)$, where $\eta_{\pm}(x^+)$ are the curves parameterized by:
\begin{equation}
\eta_{\pm}(x^+)=\frac{1}{x^+}\left(-\log\frac{x^+}{x_0^+}\pm\sqrt{1+\left(\log\frac{x^+}{x_0^+}\right)^2}\right).
\end{equation}
The curves $\delta_{\pm}$ and $\eta_-$ are drawn on the Penrose diagram of Figure~\ref{fig:Penrose_diag_UdS}, while the curve $\eta_+$ lies in the region where $\Phi<0$, depicted in gray.

The regions where $\theta \Theta <0$ along $x^+$ are shaded in red. They correspond to the regions where violations of the QFC along $x^+$ can occur. 

In the regime $cG\gtrsim\max( \phi_r,\phi_0)$, violations of the QFC along $x^+$ are found. We schematically depict the violation region in this regime in Figure~\ref{fig:Penrose_diag_UdS}, and plot the functions $\Theta'$ and $\Phi$ along the null direction $x^+$ at fixed $x^-=0.5$ in Figure~\ref{fig:UdSplot}.
\begin{figure}[ht]
    \centering
    \includegraphics[width=0.6\linewidth]{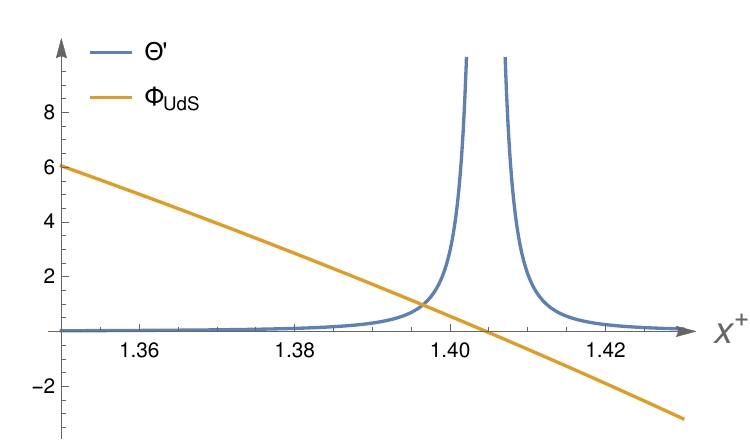}
    \caption{\footnotesize Plot of $\Theta'$ and $\Phi$ at fixed $x^-=0.5$, along the $x^+$ direction. We have taken the constants $\phi_0=10,\phi_r=1,cG=100$.}
    \label{fig:UdSplot}
\end{figure}
The region where $\Phi<0$ is not considered physical and should be removed from the spacetime. Violations of the QFC occur in the physical region where $\Phi>0$.

\subsection{Discussion}

We considered JT gravity coupled to a quantum field theory and showed that restricted quantum focusing is ensured by the QNEC. We then provided two examples of matter states that yield QFC violations. Interestingly, these violations appear in regimes where $cG\gtrsim\max(\phi_0,\phi_r)$. This contradicts the interpretation of JT gravity coupled to matter as a dimensional reduction of semiclassical gravity in higher dimensions. Indeed, requiring matter quantum contributions to be small compared to classical terms, and that the dynamic dilaton field corresponds to a perturbation around an extremal black hole, implies
\begin{equation}
\phi_0 \gg \phi_r \gg cG.
\end{equation}
Thus, QFC violations appear only when treating JT gravity coupled to matter as a genuine two-dimensional theory (still imposing $G\rightarrow 0$, $c\rightarrow\infty$ with $cG$ fixed to ensure the semiclassical limit), allowing quantum matter effects to be large.

This remark is not specific to the examples displayed here. Indeed, applying the strengthened QNEC \eqref{eq:WallQNEC} to $\Theta'$ leads to
\begin{equation} 
\label{eq:QFCexp} \Theta' \leq -\theta^2 - \frac{4G}{\Phi}\left(\frac{6}{c}\frac{(S')^2}{\Phi}+\theta S'\right),
\end{equation}
which can be formulated as another necessary condition for QFC violations:
\begin{align} 
\label{eq:thetaprim} 
\Theta'\Phi^2 &\leq -(\Phi'+2GS')^2 - \frac{24G}{c}(S')^2\left(\Phi-\frac{cG}{6}\right). 
\end{align} 
QFC violations require $\Phi\geq \frac{cG}{6}$. Consider a conformal vacuum state in the large $c$ limit. In this case, $t_{\pm}$ is independent of $c$, and
\begin{equation} 
\label{eq:phicG} 
\Phi(x) = \phi_0 + \phi_r \mathcal{F}(x) + cG \mathcal{G}(x), 
\end{equation}
where $\phi_r\mathcal{F}(x)$ is the solution to the classical equations of motion without matter, and $\mathcal{G}(x)$ is the $cG$-independent solution to the equations
\begin{equation} -\frac{\partial^2 \mathcal{G}(x) }{(\partial x^{\pm})^2}+ 2\frac{\partial\omega}{\partial x^{\pm}}\frac{\partial\mathcal{G}(x)}{\partial x^{\pm}} = -\frac{1}{3}t_{\pm}. 
\end{equation} 
Inserting \eqref{eq:phicG} into \eqref{eq:thetaprim}, we find
\begin{equation} 
\Theta'\vert_{\frac{cG}{\phi_0},\frac{cG}{\phi_{\rm r}}\ll 1} \leq 0. 
\end{equation}
It is thus expected to find violations of the QFC only in the $cG\gtrsim\max(\phi_0,\phi_r)$ regime.

\addtocontents{toc}{\protect\end{adjustwidth}}
\part{Probing de Sitter holography with quantum information \label{part:holography}}
\addtocontents{toc}{\protect\begin{adjustwidth}{1cm}{0cm}}

\chapter{Static patch holography}
\label{ch:SP}

We start with a brief review of the de Sitter geometry and relevant coordinate systems. See~\cite{Galante:2023uyf,Spradlin:2001pw,Anninos:2012qw} for excellent reviews of de Sitter spacetime. Sections \ref{sec:holodS} and \ref{monobil} lay out the basic ideas behind static patch holography and two proposals that have been put forward to compute holographic entanglement entropy.

\section{de Sitter spacetime}
\label{sec:dS}

The metric of $(d+1)$-dimensional \textit{de Sitter} space dS$_{d+1}$ in global coordinates is given by
\begin{equation}
 ds^2=- d\tau^2+\cosh^2\tau\,  d\Omega^2_{d},
\end{equation}
where $\tau\!\in\!(-\infty,+\infty)$ is the global time and 
\begin{equation}
    d\Omega_d^2\!=\! d\theta_1^2+\sin^2\theta_1  d\theta_2^2+\cdots+\sin^2\theta_1\cdots\sin^2\theta_{d-1} d\theta^2_{d}
\end{equation}
is the metric on the unit $d$-dimensional sphere $\mathbb{S}^d$. dS$_{d+1}$ is the maximally symmetric solution to the Einstein equations with a positive cosmological constant $\Lambda=(d-1)d/2l_{\rm dS}^2$. The radius of curvature $l_{\rm dS}$ is set to unity for simplicity. A slice of constant global time corresponds to a $d$-dimensional sphere with scale factor $\cosh\tau$, which takes infinite values at $\tau=\pm\infty$ and reaches its minimal value at $\tau=0$. 
By means of a time reparametrization
\begin{equation}
\cosh\tau=\frac{1}{\cos\sigma},
\label{tau-sigma}
\end{equation}
the metric admits the following conformal form
\begin{align}
 ds^2&=\frac{1}{\cos^2\sigma}\left(- d\sigma^2+ d\Omega^2_{d}\right)\!,\nonumber \\
&=\frac{1}{\cos^2\sigma}\left(- d\sigma^2+ d\theta^2+\sin^2\theta_1\, d\Omega^2_{d-1}\right)\!,
\label{eq:conf}
\end{align}
where $\sigma\in(-\pi/2,\pi/2)$ is the conformal time and $ \mathrm{sign} ~\tau= \mathrm{sign} ~\sigma$. We define $\theta=\theta_1$ to simplify notation. Using the conformal coordinates $(\sigma,\theta)$, where $\theta\in[0,\pi]$ when $d\ge 2$, we can construct the Penrose diagram for de Sitter space, which is depicted as a square depicted in Figure~\ref{fig:Penrose_dS}. 
\begin{figure}[ht]
    \centering
\includegraphics[width=0.4\linewidth]{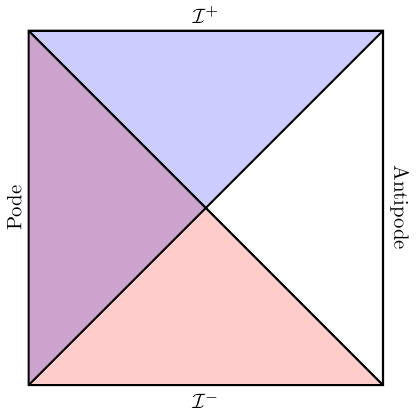}
    \caption{\footnotesize Penrose diagram for de Sitter space. The diagonal lines are the past and future horizons for an observer on the north and south poles (pode and antipode). The regions $O^+$ (purple or blue) and $O^-$ (purple or red) intersect in the purple region, the northern causal patch. The southern causal patch is the blank region.}
    \label{fig:Penrose_dS}
\end{figure}
Each horizontal slice of constant $\sigma$ represents an $\mathbb{S}^d$. Every point of the diagram represents an $\mathbb{S}^{d-1}$, except for points on the left and right vertical sides, which are actual points, corresponding to the north and south poles of $\mathbb{S}^{d}$. We will refer to these points as the pode and the antipode, respectively. The spacelike surfaces $\mathcal{I}^-$ and $\mathcal{I}^+$ at $\sigma=\mp \pi/2$ are the past and future null infinities. They correspond to the surfaces where null geodesics originate and terminate, respectively. 

Throughout this section, we consider a comoving observer at the pode. The purple and blue shaded regions of Figure~\ref{fig:Penrose_dS}, called $O^+$, are the parts of spacetime to which the observer at the pode will ever be able to send a signal. The purple and red shaded regions, called~$O^-$, are the parts of spacetime from which the observer at the pode will ever be able to receive a signal. The intersection $O^+\cap O^-$ of these two regions, which is purple, is called the northern causal diamond, or \textit{static patch} associated with the observer at the pode. We denote it $P_{\rm L}$. Likewise, we can construct the southern causal diamond associated with an observer at the antipode, $P_R$, which is represented by the blank region in the Penrose diagram. These two patches are causally disconnected from each other, and they are bounded by \textit{cosmological horizons} depicted by the diagonal lines in Figure~\ref{fig:Penrose_dS}.\footnote{Any perturbation or addition of energy in de Sitter space will lead the causal patches to intersect \cite{Gao:2000ga}.}

The global and conformal coordinates cover the entire de Sitter spacetime. However, in these coordinate systems the cosmological horizons associated with the comoving observers at the pode and antipode are not manifest. For this purpose, one introduces the static coordinates, in which the metric of dS$_{d+1}$ takes the form
\begin{equation}
     ds^2 = -(1-r^2)  dt^2 + \frac{ dr^2}{1-r^2}+ r^2  d\Omega_{d-1}^2,
    \label{staticds}
\end{equation}
with $t\in(-\infty,+\infty)$ being the static time and $r\in[0,1]$. These static coordinates cover only the causal patch associated with the pode. The three remaining patches of the Penrose diagram can also be covered by independent sets of $(r,t)$ coordinates. Note that in the blue and red triangular regions of Figure~\ref{fig:Penrose_dS}, $r$ takes values from $1$ to infinity and becomes timelike, while $t$ is spacelike. As depicted in Figure~\ref{fig:Penrose_diag_glob_coord},
\begin{figure}[ht]
    \centering
\includegraphics[width=0.4\linewidth]{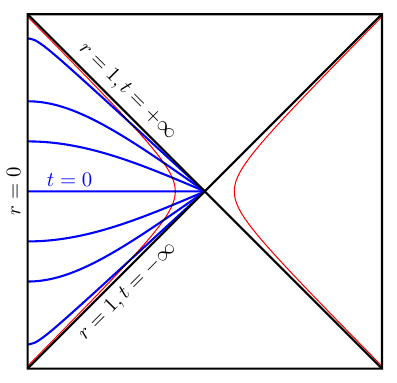}
    \caption{\footnotesize Constant static time $t$ slices (blue curves) covering the causal patch associated with the pode. The stretched horizons are represented by the red hyperbolas.}
    \label{fig:Penrose_diag_glob_coord}
\end{figure}
the cosmological horizon associated with the pode is located at $r=1$, while the pode is at $r=0$. The metric is manifestly static, independent of $t$, with $\partial_t$ being a Killing vector associated with the isometry $t\rightarrow t + \text{constant}$. In static coordinates, $\partial_t$ can only be used to define a consistent unitary Hamiltonian evolution in the northern causal patch. In particular, this patch can be foliated by constant $t$ slices, with the state of the system evolving unitarily from slice to slice. Notice that these surfaces end on the bifurcate horizon, which lies at the intersection of the diagonals of the Penrose diagram, and approach the past and future horizons as $t\to \mp \infty$ (see Figure~\ref{fig:Penrose_diag_glob_coord}).

Requiring the regularity of the near horizon limit of the static patch geometry $\eqref{staticds}$ in Euclidean signature leads to the conclusion that the static patch is associated with a temperature
\begin{equation}
\label{eq:TdS}
    T_{\rm dS} = \frac{1}{2\pi}.
\end{equation}
This is the temperature associated with the horizon, which is in equilibrium with the static patch, see discussion of the Hartle-Hawking/Bunch Davies vacuum below equation~\eqref{eq:HH}. The temperature of the static patch can also be derived by showing that a detector in the static patch will measure radiation at temperature $T_{\rm dS}$ coming from the horizon~\cite{Spradlin:2001pw}. 

An alternative definition of dS$_{d+1}$ space, which will be very useful in the following, is given as a hypersurface embedded in the $(d+2)$-dimensional Minkowski spacetime
\begin{align}
    d s^2 &= \eta_{MN}X^MX^N,
\end{align}
where $\eta_{MN}=\text{diag}(-1,1,1,\dots)$ and $M,N=0,1,\dots,d+1$. dS spacetime is then defined by considering the induced metric on the hypersurface
\begin{equation}
\label{eq:dS_hyperb}
   1 = -X_0^2 +X_1^2 + \dots X_{d+1}^2.
\end{equation}
This makes the SO$(d+1,1)$ symmetry group of dS$_{d+1}$ explicit. One can show that the metrics \eqref{eq:conf} and \eqref{staticds} satisfy this constraint by parametrizing the embedding coordinates as
\begin{alignat}{2}
        X_0 & =  \tan \sigma &&= \sqrt{1-r^2} \sinh t, \\
        X_3 & =  \frac{\cos\theta}{\cos\sigma} &&= - \sqrt{1-r^2}\cosh t, \label{eq:X3}\\
        X_1 & =  \frac{\sin\theta}{\cos\sigma} \cos\varphi && = r \cos\varphi, \\
        X_2 & =  \frac{\sin\theta}{\cos\sigma} \sin\varphi && = r \sin\varphi.
\end{alignat}
Note that we matched the time direction between the conformal patch and the right static patch. Static coordinates of the pole patch are found by taking a $+$ sign in \eqref{eq:X3}. The static coordinates above only cover the right static patch as $X_3\ge 0$ in the static coordinates above, which corresponds to $\theta\in[\pi/2,\pi]$.

Finally, one can also define a \textit{stretched horizon}, which is a hyperbola $r=1-\varepsilon$ in static coordinates, where $\varepsilon$ is a small positive constant. The latter is located at an infinitesimal proper distance from the horizon and serves as a regularized, timelike boundary for the static patch. This stretched horizon plays the role of a cutoff surface regulating the temperature, which from the point of view of a static accelerating observer (following the line $r=\rm constant$) becomes arbitrarily large as the horizon is approached. As seen in Figure~\ref{fig:Penrose_diag_glob_coord}, the constant $t$ slices intersect the stretched horizon at distinct points, spanning the whole hyperbola as $t$ evolves from $-\infty$ to $+\infty$. 

\section{A holographic description of de Sitter spacetime?}
\label{sec:holodS}

As discussed in Section~\ref{sec:holography}, the spacelike projection theorem \ref{th:spacelike} implies the holographic principle \eqref{eq:holo}, which suggests that any spacetime region $R$ (whose boundaries have negative expansion in the future direction) has a fundamental description in terms of a non-gravitating quantum theory defined on the boundary. Taking $R$ to be the entire spacetime, this implies that the dynamical degrees of freedom associated with the universe are located on its timelike boundary. For Anti-de Sitter spacetime, this is explicitly realized in string theory~\cite{Maldacena:1997re,Witten:1998qj,Gubser:1998bc}, where the bulk description is dual to a CFT defined on the asymptotic boundary of AdS. Here, we are interested in the quantum description of an expanding universe, whose simplest model is de Sitter space. 

de Sitter spacetime is famously endowed with a closed spatial topology: there is no timelike boundary. So, is there really a holographic description of de Sitter space? There are three ways to approach this question:
\begin{enumerate}
    \item One could conclude that there is no holographic description of de Sitter space. This is quite an unsatisfactory resolution and leaves many questions unanswered. In particular, the covariant entropy bound and its generalizations are background independent, so we expect the holographic principle to apply at least in subregions of dS. A breakdown of this statement at very large scales would require some canonical explanation. 
    \item A moderate approach is to seek a comparison with AdS/CFT and postulate that dynamical degrees of freedom describing dS must be located on the only boundary of this spacetime, at null infinities. This was the approach of Strominger~\cite{Strominger:2001pn}, who postulated that the holographic dual is located at null infinity $\mathcal{I}^+$, the conformal spacelike boundary~\cite{Strominger:2001pn, Bousso:2001mw,Anninos:2011ui}. This can be seen as a sort of a Euclidean continuation of AdS/CFT, known as the \emph{dS/CFT correspondence}. However, one of the largest differences between dS/CFT and AdS/CFT is that the notion of time is lacking in the former. Thus, the dual CFT is considered to be exotic, like non-unitary and/or having an imaginary central charge~\cite{Hikida:2022ltr}. Even though the dS spacetime itself is Lorentzian, the dual CFT defined on $\mathbb{S}^d$ at future null infinity $\mathcal{I}^+$ (which is a conformal boundary) is Euclidean. We will not develop the dS/CFT correspondence in this thesis, although we will briefly come back to it in Chapter~\ref{ch:CWT}.
    \item While counter-intuitive, the simplest approach might be that the holographic principle, unaltered, does hold in de Sitter spacetime. Then, one is led to conclude that there are no dynamical degrees of freedom in the quantum description of global de Sitter space~\cite{Susskindconf,Shaghoulian:2023odo,Engelhardt:2025vsp,Abdalla:2025gzn,Harlow:2025pvj,Marolf:2020xie,McNamara:2020uza,Dong:2024tjx,Usatyuk:2024mzs,Usatyuk:2024isz}. Let us make this statement sharper. The island formula \eqref{eq:island} has been extensively applied in cosmological settings~\cite{Chen:2020tes,Hartman:2020khs,Balasubramanian:2020xqf,Sybesma:2020fxg,Geng:2021wcq,Fallows:2021sge,Aalsma:2021bit,Langhoff:2021uct,Aguilar-Gutierrez:2021bns,Kames-King:2021etp,Aalsma:2022swk,Baek:2022ozg,Balasubramanian:2023xyd,Aguilar-Gutierrez:2023hls,Myers:2024zhb,Hao:2024nhd,Jiang:2024xnd,Jiang:2025hao}. A simple setup consists of entangling the matter content of a dS universe to some non-gravitating and disconnected reservoir $R$. Then, taking the island $I$ to be a Cauchy slice of dS minimizes the generalized entropy since Cauchy slices of dS do not have a boundary, and $R\cup I$ is in a pure state. That $S(R)=0$ for any state of dS implies that it is impossible to entangle oneself with a de Sitter universe. The only possible explanation is that de Sitter space only has one state~\cite{Penington:2019npb,Almheiri:2019psf,Shaghoulian:2023odo}, \textit{ie} it cannot have any dynamical degree of freedom.\footnote{This is not necessarily true when the system coupled to dS is gravitating. In particular, when $R$ plays the role of a gravitating observer, the Hilbert space becomes non-trivial~\cite{Balasubramanian:2023xyd}.} The Bousso bound leads to the same conclusion. See \cite{McNamara:2020uza,Dong:2024tjx,Usatyuk:2024isz,Usatyuk:2024mzs} for other evidence.
\end{enumerate}
In this work, we focus our attention on the third interpretation. Even if we accept the idea that the Hilbert space of de Sitter space is trivial, a satisfying quantum theory of the universe should be able to explain the dynamics we observe in the real world. In particular, such a theory should describe observables, which implicitly assumes the choice of an observer.\footnote{It assumes at least the choice of one or more worldlines along which measurements are made.} See~\cite{Anninos:2011af,
Anninos:2017hhn,Anninos:2018svg,Coleman:2021nor,Blacker:2023oan} for an observer-based approach to de Sitter holography. As observers in dS do not have access to the full universe, we are naturally led to study subregions of de Sitter space—static patches—rather than global de Sitter. Both the island formula and the Bousso bound predict non-trivial Hilbert spaces for these regions. A fascinating idea that has emerged in the past few years is that the observer plays a crucial role in de Sitter quantum gravity. In particular, explicitly including an observer in the theory is necessary to get a non-trivial algebra of observables~\cite{Chandrasekaran:2022cip,Witten:2023qsv,Witten:2023xze,Kudler-Flam:2023qfl,Chen:2024rpx,Jensen:2024dnl,DeVuyst:2024uvd,Akers:2025ahe,Abdalla:2025gzn}. See \textit{eg}~\cite{Balasubramanian:2023xyd, Maldacena:2024spf,Mirbabayi:2023vgl,Anninos:2024wpy,Abdalla:2025gzn,Harlow:2025pvj,Marolf:2020xie} for other approaches to this idea.

The discussion above motivates the view that it is natural to consider the static patch of an observer in order to probe non-trivial behaviors of gravity in de Sitter space. In particular, the static patch is self-contained in the sense that an observer may neglect everything happening outside the cosmological horizon. The region of the universe they observe is effectively bounded by their cosmological horizon. Thus, from their point of view, the observable universe has a boundary on which one should place holographic degrees of freedom, according to the holographic principle. Moreover, it was shown in~\cite{Gibbons:1977mu,PhysRevD.15.2752} that the horizon of the observer, which is the boundary of the static patch, is associated with a gravitational entropy proportional to its area $A_H$,
\begin{equation}
\label{eq:GH}
	S_{\rm GH}=\frac{A_H}{4G\hbar}.
\end{equation}
See~\cite{Anninos:2020hfj,Maldacena:2024spf} for generalizations of this formula. The fact that thermodynamic quantities can be associated with the horizon of the static patch is indicative of holographic behavior. In particular, it suggests that degrees of freedom located on the cosmological horizon should be able to encode everything happening in some region of de Sitter space, but which one? The discussion above indicates that an intuitive answer would be the static patch, where the degrees of freedom describing it emerge from the choice of worldline associated with the static patch observer.\footnote{A more conservative approach consists of associating this entropy with a Brown-York boundary~\cite{Brown:1986nw}, which is a fictitious boundary located inside the static patch that allows for a covariant definition of thermodynamic quantities. See~\cite{Banihashemi:2022htw} for a discussion of this approach and its connection with holography, and~\cite{Anninos:2024wpy} for an observer-based approach.} See~\cite{Bousso:1999dw,Bousso:2000nf,Banks:2006rx} for early related arguments.

This intuition is supported by the spacelike projection theorem \ref{th:spacelike}~\cite{Susskind:2021omt}. See~\cite{Franken:2023pni} for a detailed argument. Let us consider an observer whose worldline follows a timelike geodesic in the bulk. All geodesics are equivalent under the SO$(d+1,1)$ symmetry group of de Sitter space, so we can choose a coordinate system in which the observer worldline follows the right edge of the diagram. This corresponds to a static observer at the south pole (or antipode) of the $d$-sphere. We denote its static patch $P_R$. There is a unique worldline such that $P_R$ does not overlap with its static patch, corresponding to the left edge of the Penrose diagram. We denote this complementary static patch $P_{\rm L}$. For every closed codimension-two spacelike surface $\sigma$ inside one of the static patches, the two null directions of non-positive expansion are directed towards the interior of the static patch. In particular, any spacelike slice $\Sigma$ that does not intersect the cosmological horizon satisfies all assumptions of the spacelike projection theorem \ref{th:spacelike} and holographic principle \eqref{eq:holo}. Moreover, surfaces that are not entirely contained in a static patch fail to satisfy the conditions of the holographic principle, as all future directed null hypersurfaces in the exterior region are of positive expansion. Thus, a timelike codimension-one hypersurface in a static patch that spans $\sigma\in(-\pi/2,\pi/2)$ defines a holographic surface $\sigma$ on every Cauchy slice $\Sigma$. We introduce the following definition of holographic screen.
\begin{definition}[Holographic screen in the static patch]
\label{def:ap-sp}
    The holographic screen $\mathcal{S}$ associated with an observer is a codimension-one convex timelike hypersurface in the static patch, anchored to the observer's worldline endpoints.
\end{definition}
The convexity condition is not necessary in this chapter, but will be explained in Chapter~\ref{ch:CWT}. When referring to a holographic screen without mentioning an observer, $\mathcal{S}$ may be the union of holographic screens associated with different observers. In particular, we may denote $\mathcal{S}_{\rm L}$ a holographic screen in $P_{\rm L}$ and associated with an observer at the pode, and $\mathcal{S}_{\rm R}$ a holographic screen in $P_R$ associated with an observer at the antipode. Then, $\mathcal{S}=\mathcal{S}_L\cup \mathcal{S}_{\rm R}$ is a holographic screen encoding two disconnected regions of spacetime. See Figure \ref{fig:screen}.
\begin{figure}
    \centering
    \includegraphics[width=0.35\linewidth]{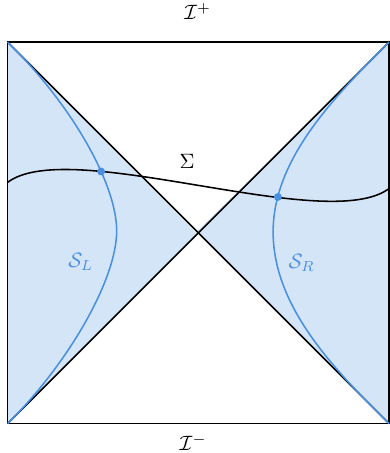}
    \caption{\footnotesize Two holographic screens $\mathcal{S}_L$ and $\mathcal{S}_R$ in the complementary static patches $P_L$ and $P_R$ (sheded in blue). In the two static patches, any closed codimension-two surface has a future directed lightsheet directed towards the pole. An example of Cauchy slice $\Sigma$ is depicted by a black curve.}
    \label{fig:screen}
\end{figure}
\begin{conjecture}[Static patch holography~\cite{Susskind:2021omt}]
\label{conj:SPH}
The semiclassical gravity description of the region inside a holographic screen $\mathcal{S}$ is dual to a quantum theory defined on $\mathcal{S}$.
\end{conjecture}
We insist on the semiclassical limit in the above statement, since the holographic screen does not lie in an asymptotic region where backreaction can be ignored. In particular, defining subsystems of a holographic screen in a covariant way, and computing their entanglement entropy to all orders in the semiclassical regime goes beyond the scope of this thesis.

Inside a static patch $P$, the location of $\mathcal{S}\in P$ as a holographic screen is arbitrary. However, the area of closed surfaces on a given Cauchy slice $\Sigma$, as well as the size of the slice subsystem that lies inside $\mathcal{S}$, increase as we get closer to the cosmological horizon, such that pushing $\mathcal{S}$ towards the horizon maximizes the number of degrees of freedom that are holographically encoded. Hence, on any spacelike slice $\Sigma$, the surface with the greatest number of holographic degrees of freedom is located at the cosmological horizon of the static patch, and describes $\Sigma\cap P_R$ or $\Sigma\cap P_{\rm L}$. This leads to the original formulation of static patch holography~\cite{Susskind:2021omt}, which locates the screen $\mathcal{S}$ of an observer on its cosmological horizon, call it $\mathcal{H}$.\footnote{To be more precise, the screen is located on the stretched horizon to put a UV cutoff on the theory.} This conjecture was developed in~\cite{Susskind:2021omt,Susskind:2021dfc, Susskind:2021esx, Lin:2022nss, Susskind:2023hnj}. See~\cite{Shyam:2021ciy, Lewkowycz:2019xse, Coleman:2021nor, Banihashemi:2022htw, Banihashemi:2022jys,Batra:2024kjl} for additional supportive evidence and~\cite{Aguilar-Gutierrez:2023zqm,
Aguilar-Gutierrez:2024rka,
Aguilar-Gutierrez:2024nau,
Aguilar-Gutierrez:2024yzu,
Aguilar-Gutierrez:2024nst,
Aguilar-Gutierrez:2024oea,
Jorstad:2022mls,
Auzzi:2023qbm,
Anegawa:2023wrk,
Aguilar-Gutierrez:2023tic,
Anegawa:2023dad,
Baiguera:2023tpt} for extensive literature on holographic complexity in static patch holography. Interestingly, locating the holographic screen on the cosmological horizon provides an intuitive explanation for the Gibbons-Hawking formula \eqref{eq:GH}. If a quantum theory dual to the static patch is located on the cosmological horizon, then $S_{\rm GH}$ has a direct thermodynamical interpretation as the size of the Hilbert space of the holographic theory.

A precise realization of this duality is still lacking although promising results have been found in lower-dimensional cosmological models~\cite{Shaghoulian:2021cef, Shaghoulian:2022fop, Susskind:2021esx, Susskind:2022dfz, Susskind:2022bia, Rahman:2022jsf, Goel:2023svz, Narovlansky:2023lfz, Rahman:2023pgt,Verlinde:2024znh,Verlinde:2024zrh,Blommaert:2023opb,Blommaert:2023wad,Rahman:2024iiu,Milekhin:2023bjv,Milekhin:2024vbb,Xu:2024hoc,Heller:2024ldz,Blommaert:2024ydx,Blommaert:2024whf,Bossi:2024ffa,Blommaert:2025avl,Mahapatra:2025fpx}. Following a dimensional reduction, the dual theory is conjectured to be the double-scaled Sachdev-Ye-Kitaev (DSSYK) model, with a number of nontrivial matches, including correlation functions, the partition function, quasinormal modes, hyperfast scrambling and the algebra of observables. In this work, we do not consider any explicit proposal of holographic dual and keep our discussion general.

The main goal of the Chapter~\ref{ch:bridge} is to study the geometric structure associated with the union of holographic screens. In particular, we study the geometric dual to the union of two-screen system
\begin{equation}
    \mathcal{S}= \mathcal{S}_L\cup \mathcal{S}_R
\end{equation}
associated with complementary static patches. 

\section{Monolayer and bilayer proposals}
\label{monobil}

Following the static patch holography conjecture, two proposals have been advanced by  Susskind and Shaghoulian~\cite{Susskind:2021esx, Shaghoulian:2021cef, Shaghoulian:2022fop} to compute entanglement entropies in the holographic dual description on $\mathcal{S}$, thus generalizing the Ryu-Takayanagi formula~\cite{Ryu:2006bv}, to de Sitter space.  For these proposals to be meaningful, it is necessary to assume that considering subsystems of $\mathcal{S}$ is sensible. In particular, we assume that a spacelike slice $\Sigma\vert_{\mathcal{S}}$ of the screen $\mathcal{S}$ is associated with a density matrix $\rho_{\Sigma\vert_{\mathcal{S}}}$ which acts on the Hilbert space of the dual theory. We further assume that a subsystem $A\in \Sigma\vert_{\mathcal{S}}$ corresponds to a subsystem of the Hilbert space on which $\rho_{\Sigma\vert_{\mathcal{S}}}$ acts~\cite{Sanches:2016sxy}. We seek a prescription for computing $S(A) = S(\rho_A)$ where $\rho_A = \tr_{A^c}\left(\rho_{\Sigma\vert_{\mathcal{S}}}\right)$.

The standard HRT prescription must be suitably adapted to encode the entanglement structure in the de Sitter case. A key challenge is that there is bulk spacetime on both sides of the screen $\mathcal{S}$. Perturbatively in $G(\hbar)$, the prescriptions in~\cite{Susskind:2021esx, Shaghoulian:2021cef, Shaghoulian:2022fop} incorporate the leading contributions, which are of order $(G(\hbar))^{-1}$. Let us examine the two proposals:

\noindent $\bullet$ {\bf Monolayer proposal \cite{Susskind:2021esx}:} \textit{The entanglement entropy of a subregion $A$ of the two screens and its complement is $1/(4G)$ times the area of an extremal surface homologous to $A$,\footnote{See definition \ref{def:hom}. If there are multiple such extremal surfaces, we choose the one with the minimal area.} and lying between the two sets of degrees
of freedom; in this case between the two cosmological horizons.}

\noindent $\bullet$ {\bf Bilayer proposal \cite{Shaghoulian:2021cef}:} \textit{The entanglement entropy of a subregion $A$ of the two screens and its complement is $1/(4G)$ times the sum of the areas of the extremal surfaces anchored to $A$ in each of the three subregions of the bulk, that is, in the exterior and the two interior regions.}

\noindent Both proposals represent natural modifications of the HRT prescription, where the bulk manifold admits a holographic dual on its boundary and the fine-grained entropy of subregions of the boundary theory is given in terms of an extremal area in the bulk. In particular, extremal surfaces should be anchored on the screen and extend in bulk regions bounded by the screen. Considering the whole de Sitter space would be inconsistent since it is not bounded by the horizons.\footnote{A Cauchy slice is topologically a sphere with no boundary.} Indeed, the interior regions and the exterior region are the only regions bounded by the horizons.
In~\cite{Susskind:2021esx, Shaghoulian:2021cef, Shaghoulian:2022fop}, it was observed that applying an HRT-like prescription to the left (right)  interior region leads to a vanishing entropy for the single-screen system $\mathcal{S}{\rm L}$ (or $\mathcal{S}{\rm R}$). Conversely, applying an HRT-like prescription to the exterior region gives the expected result to leading order, namely the area of the screen divided by $4G$. This was the main motivation for the monolayer proposal. Alternatively, one could argue that since both the interiors and the exterior region are bounded by horizons, one should look for extremal surfaces in all three regions and add the contributions. This leads to the bilayer proposal of~\cite{Shaghoulian:2021cef}, which is reminiscent of AdS wedge holography~\cite{Akal:2020wfl,Geng:2020fxl}. The bilayer prescription also yields the expected result for the entropy of a single-screen system at the horizon of de Sitter space~\cite{Shaghoulian:2021cef, Shaghoulian:2022fop}. In the following Chapter, we will argue that the bilayer proposal is the only consistent prescription.

\chapter{Bridging the static patches: Connectivity of de Sitter space}
\label{ch:bridge}

Motivated by the insights provided by quantum extremal surfaces in AdS/CFT, and their crucial role in the unraveling of novel gravitational laws, we can specify the monolayer and bilayer proposals. To accomplish this, we formulate them in a covariant way and define the associated entanglement wedges for subsystem of the screens~\cite{Franken:2023pni}. We consider a pair of screens $\mathcal{S}_L\cup\mathcal{S}_R=\mathcal{S}$ as defined in \ref{def:ap-sp}, without fixing $\mathcal{S}$ to lie at a fixed location~\cite{Franken:2024wmh}. Our main goal here is to test the validity of the monolayer and bilayer proposals by establishing a formal covariant definition of an HRT-like geometric quantity (\textit{ie} consistent with all properties presented in Sections \ref{sec:const} and \ref{sec:EW}) that is consistent with the Gibbons-Hawking formula \eqref{eq:GH} and the purity of the full holographic system. If such a definition can be found, it is expected to have a dual interpretation as an entropy in the dual quantum theory. In particular, any candidate for an explicit holographic theory defined on the horizon should be able to define a quantity dual to this geometric construction. In Chapter~\ref{ch:CWT}, we present an additional information-theoretic constraint that applies to holographic entanglement entropy, imposing additional non-trivial constraints on the causal structure of the dual quantum theory.

Moreover, leveraging entanglement wedge reconstruction, we address the following question: How can a closed and connected de Sitter spacetime emerge from a holographic theory that only describes open and finite regions of spacetime? We argue in this chapter that, as in AdS/CFT, the connectivity of de Sitter space emerges from the entanglement between disconnected holographic screens.

\section{Covariant monolayer and bilayer proposals}

Consider an arbitrary Cauchy slice $\Sigma$ of dS$_{d+1}$. We define the screens $\mathcal{S}_{\rm L}$ and $\mathcal{S}_{\rm R}$ associated with two antipodal observers, each lying in their respective patches $P_{\rm L}$ and $P_R$. Let $\mathcal{S}$ denote the union of the two screens, \textit{ie} $\mathcal{S}=\mathcal{S}_{\rm L}\cup \mathcal{S}_{\rm R}$. An arbitrary spacelike slice of $\mathcal{S}_L\cup \mathcal{S}_{\rm R}$ is denoted $\Sigma\vert_{\mathcal{S}_{\rm L}}\cup \Sigma\vert_{\mathcal{S}_{\rm R}}$, associated with a Cauchy slice $\Sigma$ of the bulk. Importantly, $\Sigma\vert_{\mathcal{S}_{\rm L}}$ and $\Sigma\vert_{\mathcal{S}_{\rm R}}$ are spacelike regions and they are spacelike separated. See Figure~\ref{fig:qucorr}. The boundary of the middle region (exterior) is denoted $\mathcal{S}_{\rm E}= \mathcal{S}=\mathcal{S}_L\cup\mathcal{S}_{\rm R}$. 
\begin{figure}
    \centering
    \includegraphics[width=0.35\linewidth]{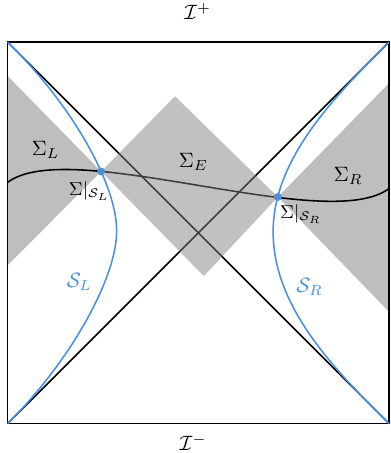}
    \caption{\footnotesize Penrose diagram of dS. The two blue lines represent the two screens, and the three parts of a complete slice $\Sigma$ are depicted by black lines. The worldlines of two antipodal observers follow the two vertical lines. The bulk regions $D_I$ are shown as shaded regions.}
    \label{fig:qucorr}
\end{figure}

We denote $\Sigma_I$, with $I=\rm {L,R,E}$, as any achronal slice such that $\partial \Sigma_I = \Sigma\vert_{\mathcal{S}_I}$ for $I\in\{\rm L,E,R\}$. We denote as $D_I=D(\Sigma_I)$ the bulk region spanned by slices $\Sigma_I$. See Figure~\ref{fig:qucorr}. We now state the covariant version of the monolayer and bilayer proposals, generalized to timelike screens:
\begin{definition}[$D_I$-Homology in dS]
\label{def:homol}
    Following definition \ref{def:hom}, we say that $\gamma$ is $D_I$-homologous to $A\in\Sigma\vert_{\mathcal{S}_{\rm L}}\cup \Sigma\vert_{\mathcal{S}_{\rm L}}$ if $\gamma\in D_I$ is homologous to $A_I = A\cap\mathcal{S}_I$. The homology region is then denoted $\mathcal{C}_I(A)$ and we impose that $\mathcal{C}_I(A)\in\Sigma'_I$ where $\Sigma_I'$ is an achronal slice such that $\partial\Sigma_I' = \Sigma\vert_{\mathcal{S}_I}$.
\end{definition}
\begin{conjecture}[Covariant monolayer and bilayer proposals]
\label{con:ent_pr}
    Let $A$ be a subsystem of $\Sigma\vert_{\mathcal{S}_{\rm L}}\cup\Sigma\vert_{\mathcal{S}_{\rm R}}$. According to the monolayer proposal, the entanglement entropy of $A$ is given by
    \begin{equation}
    	\label{eq:mono}
            S_{\rm mono}(A)= \min \mathrm{ext}\left[\frac{{\rm Area}(\gamma_E)}{4G}\right] +O(G^0).
    \end{equation}
    According to the bilayer proposal,
    \begin{equation}
    	\label{eq:bil}
            S_{\rm bil}(A)= \min \mathrm{ext}\left[\sum_{I=L,E,R}\frac{{\rm Area}(\gamma_I)}{4G}\right] +O(G^0),
    \end{equation}
    where $\gamma_I$ must be $D_I$-homologous to $A$. The resulting surfaces are called the extremal surfaces $\gamma_{\rm e}(A;D_I)$.\footnote{When there is no such surface in some region $D_I$, one has to extend the notion of extremal surface, see Section~\ref{sec:ext} and~\cite{Franken:2023pni,Franken:2024wmh}.} 
\end{conjecture}
    Quantum corrections due to the entropy of bulk fields can be taken into account by replacing the sum over the area terms with the generalized entropy~\cite{Faulkner:2013ana, Engelhardt:2014gca, Franken:2023pni}
    \begin{align}
    S_{\rm mono,gen}(\gamma_E)&= \frac{{\rm Area}(\gamma_E)}{4G} + S\left({\cal {C}}_{E}(A) \right),\\
    S_{\rm bil, gen}(\gamma_L,\gamma_E,\gamma_R)&= \sum_{I=L,E,R}\frac{{\rm Area}(\gamma_I)}{4G} + S\left(\bigcup_{I}{\cal {C}}_{I}(A) \right),
    \label{eq:SgenDS}
\end{align}
for the monolayer and bilayer, respectively. The resulting surfaces are called quantum extremal surfaces. The necessity of joint extremization was already pointed out at the classical level in~\cite{Shaghoulian:2021cef, Shaghoulian:2022fop}. Moreover, this choice of semiclassical correction is always minimal compared to the sum of the entropy of bulk fields on the three slices, by weak subadditivity. Definition \eqref{eq:SgenDS} is reminiscent of the island formula \eqref{eq:island}. Since the homologous surface $\gamma_I$ lies on a slice $\Sigma'_I$, it is spacelike and has a real area. Timelike extremal homologous surfaces may exist but should not be considered, as they lead to imaginary entropy.\footnote{See however~\cite{Doi:2023zaf} for recent discussions on timelike entanglement entropy.} Moreover, homologous extremal surfaces intersecting $\mathcal{S}$ may exist, but should not be considered either. 

As mentioned below the static patch holography conjecture \ref{conj:SPH}, the definition of the screen and associated entropies is restricted to the semiclassical limit. Thus, while in AdS/CFT quantum extremal surfaces compute entanglement entropy non-perturbatively, we do not have any control beyond the semiclassical limit here. In particular, we only consider first order corrections, which are computed by~\cite{Faulkner:2013ana,Akers:2020pmf}
\begin{align}
    S_{\rm mono}(A) &= \frac{{\rm Area}(\gamma_{\rm e}(A;D_E))}{4G} + S\left({\cal {C}}_{E}(A) \right) + O(G^1),\\
    S_{\rm bil}(A) &= \sum_{I=L,E,R} \frac{{\rm Area}(\gamma_{\rm e}(A;D_I))}{4G} + S\left(\bigcup_{I}{\cal {C}}_{I}(A) \right) + O(G^1),\label{eq:qubil}
\end{align}
where $\gamma_{\rm e}(A;D_I)$ are the $D_I$-homologous surfaces extremizing the classical area term, and $\mathcal{C}_I(A)$ are the associated homology regions.

Building on our attempt to define holographic entanglement entropy and entanglement wedges in static patch holography, we assume that the bulk region dual to the reduced density matrix $\rho_{A}$ is defined by the following entanglement wedge. Our motivation is identical to the one presented above definition \ref{def:EW}.
\begin{definition}[Entanglement wedge in dS]
    The entanglement wedge $W(A)$ of $A$ is given by the union of the causal diamonds of all homology regions $\mathcal{C}_I(A)$. Each of them is denoted $W_I(A)=D(\mathcal{C}_I(A))\subset D_I$. In the monolayer proposal, there is only one homology region and
    \begin{equation}
    \label{eq:EWm}
        W_{\rm mono}(A) = D(\mathcal{C}_{\rm E}(A)) = W_{\rm E}(A).
    \end{equation}
    In the bilayer proposal, three homology regions are found, and we define
    \begin{equation}
    \label{eq:EWb}
        W_{\rm bil}(A) = \bigcup_{I} D(\mathcal{C}_I(A)) = \bigcup_I W_I(A).
    \end{equation}
\end{definition}
By analogy with AdS/CFT, we have defined the entanglement wedge as the causal diamond of the bulk region connecting the subsystem $A$ to its extremal surface. The physical motivation for this is that $A$ has information about the entropy of any subsystem of $A$. By constructing all the associated extremal surfaces, one expects to be able to reconstruct the homology region. In our setup, this construction excludes any point that is timelike separated from $A$. However, the causal diamond of the union of the three homology regions is larger than their independent causal diamonds. In particular, the entanglement wedge we defined above is causally equivalent to $D(\bigcup_{I}\mathcal{C}_I(A))$. Therefore, we may have alternatively defined the entanglement wedge as $W(A)=D(\bigcup_{I}\mathcal{C}_I(A))$. 

Note that we adopt the minimal choice concerning the entanglement wedge structure in the monolayer and bilayer cases. This choice implies in particular that entanglement wedges in the monolayer case do not extend into the interior regions. One may argue that the absence of extremal surfaces in the static patch does not necessarily imply that entanglement wedges cannot cover them. However, without any extremal surface in the static patches, there is no preferred location for the entanglement wedge in these regions, other than the entire interior. This would amount to artificially including the entanglement shadow (See Section~\ref{sec:EW}) of $\Sigma\vert_\mathcal{S}$ into the entanglement wedge. This would be contradictory as it would lead to the conclusion that the entanglement wedge of a spatial subsystem of the screen whose size approaches zero would still covers the full static patch. It would be interesting to explore the possibility that the monolayer proposal of~\cite{Susskind:2021esx} could be supplemented by a consistent set of rules concerning the structure of the entanglement wedges, allowing them to extend into the interior regions. However, it is difficult to see how such a consistent set of rules could be formulated, since in all other well-understood examples in the literature, the entanglement wedge structure depends on the location of the extremal surfaces, and the monolayer proposal does not involve any extremal surface in the interior regions. 

\section{Preference for the bilayer proposal}

In~\cite{Franken:2023pni}, the analysis of the holographic entanglement entropy of different subsystems was carried out with the monolayer and bilayer proposals. In this section, we briefly review the main arguments favoring the bilayer proposal, leaving detailed computations to Section~\ref{sec:EWdS}.

\textbf{Entanglement wedge nesting}
\vspace{1em}\\
Assuming the two alternative definitions of entanglement wedge \eqref{eq:EWm} and \eqref{eq:EWb}, the monolayer and bilayer proposals lead to fundamentally different entanglement wedge structures. 

The monolayer prescription states that the entanglement entropy is related to the area of extremal surfaces lying entirely in the exterior region, and not in the interior ones. As a result, 
\begin{equation}
\label{eq:contEW}
    W_{\rm mono}(A)\subseteq D_E,
\end{equation}
for any $A\in\Sigma\vert_{\mathcal{S}}$. However, the starting point of this whole construction is the static patch holography conjecture, which may be stated in terms of entanglement wedges as
\begin{equation}
    W_{\rm mono}(\Sigma\vert_{\mathcal{S}_{\rm L}}) \supseteq D_L,
\end{equation}
for any spacelike slice $\Sigma\vert_{\mathcal{S}_{\rm L}}$ of the screen $\mathcal{S}_{\rm L}$, and analogously for $D_R$. The regions $D_{\rm L}\cup D_R$ and $D_E$ have no intersection, therefore the property \eqref{eq:contEW} of the monolayer proposal is in contradiction with $W_{\rm mono}(\Sigma\vert_{\mathcal{S}_{\rm L}}) \subseteq W_{\rm mono}(\Sigma\vert_{\mathcal{S}})$. This latter constraint comes from entanglement wedge nesting, which is a trivial constraint from the point of view of causality on $\mathcal{S}$. Hence, it is expected to hold for any holographic entanglement entropy proposal.

To escape the above contradiction, the monolayer proposal should be endowed with a non-trivial subregion-subregion duality. However, as explained in the previous section, it is unclear how one could covariantly define a dual bulk region, different from the entanglement wedge. On the other hand, we show in Section~\ref{sec:EWdS} that the bilayer proposal is consistent with the assumption $W_{\rm bil}(\Sigma\vert_{\mathcal{S}_{\rm L}})\subset W_{\rm bil}(\Sigma\vert_{\mathcal{S}})$. In particular, we discuss the fact that this inclusion is strictly not saturated.

\textbf{Generalized entropy}
\vspace{1em}\\
Consider two holographic screens $\mathcal{S}_{\rm L}$ and $\mathcal{S}_{\rm R}$ located at the cosmological horizons $\mathcal{H}_{\rm L}$ and $\mathcal{H}_{\rm R}$ of de Sitter space. Take $A=\Sigma\vert_{\rm L}$ to be the bifurcate horizon $E$. We will show in Section~\ref{sec:EWdS} that the geometric contribution to $S(A)$ is exactly $\text{Area}(A)/4G$. Taking into account quantum corrections, the monolayer proposal includes a contribution $S(\mathcal{C}_{\rm E}(E))$. We will show that $\mathcal{C}_{\rm E}(E)=\varnothing$, such that $S_{\rm mono}(E) = \text{Area}(E)/4G + O(G^1)$. In other words, with the monolayer prescription, the entropy at one loop of an horizon is equal to its area divided by $4G$. This is in contradiction with the formula
\begin{equation}
\label{eq:SgendS}
    S_{\rm gen} = \frac{A_{\rm dS}}{4G} + S_{\rm out} + O(G),
\end{equation}
which was argued for in~\cite{Maeda:1997fh,Chandrasekaran:2022cip,Banihashemi:2022htw,Balasubramanian:2023xyd}. In our notations, $A_{\rm dS}=\mathrm{Area}(E)$ and $S_{\rm out} = S(\Sigma_{\rm L})$ where $\Sigma_{\rm L}$ is a slice bounded by $E$ in the left static patch. In particular, the entropy of the static patch computed from the algebraic approach of~\cite{Chandrasekaran:2022cip} is exactly the generalized entropy \eqref{eq:SgendS}. This is also supported by the thermodynamic approach of~\cite{Banihashemi:2022htw}. Thus, the monolayer proposal fails to account for the quantum correction due to fields in the static patch. On the other hand, we show in Section~\ref{sec:EWdS} that the bilayer proposal exactly reproduces \eqref{eq:SgendS}.

\textbf{Purity of the full screen system}
\vspace{1em}\\
Since the slices ${\mathcal {C}}_{\rm E}$ that are relevant for the computation of the semiclassical entropy in the monolayer proposal are not complete bulk Cauchy slices, the semiclassical entropy on them cannot vanish, except for special cases where ${\cal {C}}_{\rm E}$ collapses to a subregion of the screens. In the latter cases, however, the geometrical area terms associated with non-trivial subsystems are not zero. As a result, the monolayer entropies associated with dual subsystems cannot vanish at the quantum level. This applies in particular to the union of the two screens, implying that
\begin{equation}
    S_{\rm mono}(\Sigma\vert_{\mathcal{S}}) > 0
\end{equation}
at the quantum level. In other words, the monolayer proposal implies that the state of the two-screen system cannot be pure. 

On the other hand, the union of the three slices ${\cal {C}}_{I}$ in the interior and the exterior regions can amount to a Cauchy slice, leading to the vanishing of the semiclassical entropy contribution (since the bulk state on a full Cauchy slice is taken to be pure). If the quantum extremal surfaces are of zero area, we would get a vanishing bilayer entropy. As we will see in Section~\ref{sec:EWdS}, this turns out to be the case for the two-screen system, implying that  
\begin{equation}
    S_{\rm bil}(\Sigma\vert_{\mathcal{S}})=0,
\end{equation} 
including quantum corrections. The bilayer proposal thus predicts that the two-screen system is in a pure state. One can see that computing the semiclassical entropies independently in the three regions and adding them as separate contributions to the generalized entropy would yield a larger entropy, justifying our choice. The bilayer proposal is therefore consistent with the statement that the Hilbert space of de Sitter space is trivial. In particular, the island formula is non-perturbative and we expect that $S_{\rm bil}(\Sigma\vert_{\mathcal{S}})=0$ beyond the perturbative regime.

\textbf{Entropy of two aligned arcs in dS$_3$}
\vspace{1em}\\
In dS$_3$, the holographic screen $\mathcal{S}$ is foliated by pairs of circles.
It was shown in~\cite{Shaghoulian:2021cef,Franken:2023pni} that a subsystem $A=A_{\rm L}\cup A_{\rm R}$ defined as the union of two arcs $A_{\rm L}\in\Sigma\vert_{\mathcal{S}_{\rm L}}$ and $A_{\rm R}\in\Sigma\vert_{\mathcal{S}_{\rm R}}$ can have arbitrarily small entropy in the monolayer proposal, independently of their size. This requires very special correlations for the degrees of freedom on the two arcs and their complements on each screen, $\bar A_{\rm L}$ and $\bar A_{\rm R}$. For example, to get an almost vanishing  entropy for the two arc system $A_{\rm L}\cup A_{\rm R}$, the total state on the screens needs to assume a form close to a factorized one, $\Psi=\Psi_1(A_{\rm L},A_{\rm R})\Psi_2(\bar{A}_{\rm L},\bar {A}_{\rm R})$, so that when we take a partial trace over the degrees of freedom of the complementary arc subsystem $\bar{A}_{\rm L}\cup \bar{A}_{\rm R}$ we get an almost pure density matrix. This quasi-factorization should hold irrespective of the location of $A_{\rm L}$ on the left screen and of its size. On the other hand, the bilayer entropy receives $O(1/G)$ contributions from the interior regions $D_{\rm L}$ and $D_{\rm R}$, so it seems that the bilayer proposal is more consistent and natural from this point of view.

Let us mention that there is further supporting evidence for the bilayer proposal in the literature. As discussed in~\cite{Shaghoulian:2022fop}, this proposal seems to reproduce features of entanglement entropies in SYK. It would be interesting to compute the entropies of subsystems in this double-scaling limit of the SYK model and compare the results with those obtained by applying the bilayer proposal. 

For all the reasons presented above, we only consider the bilayer proposal in the remaining of this work. See~\cite{Franken:2023pni} for further details and discussions on the monolayer proposal.

\section{Covariant holographic entanglement entropy}
\label{sec:ext}

A crucial motivation for formulating the bilayer proposals in a covariant way is that causality and other entropy conditions leads to constraints on the location of extremal surfaces in the bulk. In particular, purity of the screen state implies that $\gamma_{\rm e}(A,D_I)=\gamma_{\rm e}(A^c,D_I)$ for any subsystem $A$. This implies that all extremal surfaces must lie on a Cauchy slice that contains $\Sigma\vert_{\mathcal{S}}$. In other words, we only consider homologous surfaces with
\begin{equation}
\label{eq:}
    \gamma_I \in D_I.
\end{equation}
The way the monolayer and bilayer proposals were applied in their original formulation~\cite{Susskind:2021esx,Shaghoulian:2022fop,Shaghoulian:2021cef} consisted in assuming a preferred Cauchy slice and finding the minimal surface on this slice, in the spirit of the Ryu-Takayanagi formula. This approach is only valid on time-symmetric slices of the screen, \emph{ie} when $\Sigma\vert_{\mathcal{S}}$ is located on the bifurcate horizon. In~\cite{Franken:2023pni}, it was shown that the extremization problem is not well-defined in general. In particular, for some choices of subsystem $A$, there is no extremal surface $\gamma_{\rm e}(A,D_E)$, $D_E$-homologous to $A$ such that $\gamma_{\rm e}(A,D_E)\in D_E$~\cite{Franken:2023pni}. We review this result in Section~\ref{subsec:arc}. Moreover, when extremal surfaces exist, they are not always equivalent to the maximin surface~\cite{Franken:2024wmh}. We propose a natural modification of the definition of extremal surface, called C-extremal surface, which is always well-defined and leads to consistent results~\cite{Franken:2023pni}, see Appendix \ref{app:proofs}. 

We do not to prove which is the correct prescription. To answer this question, one should derive it from a path integral computation. We do not have any physical reason to expect that the maximin method should be equivalent to any holographic entropy prescription. In particular, we argue in Section~\ref{sec:discussion} that it leads to contradictory results in FLRW holography. On the other hand, we expect holographic entropy prescriptions in non-AdS spacetimes to involve an extremization process. Indeed, a large number of results in holography, such as entanglement wedge nesting, the boundary causality condition, and the inclusion of the causal wedge in the entanglement wedge~\cite{Akers:2016ugt} heavily rely on the fact that surface computing entanglement entropy is of vanishing (quantum) expansion in all directions.\footnote{In particular, all these properties rely on the classical focusing theorem or restricted quantum focusing conjecture~\cite{Bousso:2015mna,Shahbazi-Moghaddam:2022hbw}, which constrains the evolution of the expansion or quantum expansion along null hypersurfaces.} These conditions emerge from fundamental constraints such as bulk and boundary causality, and we expect them to be satisfied beyond AdS/CFT.

Since the extremization problem of equation~\eqref{eq:bil} may not have a solution, we
must relax our definition of extremality while maintaining consistency with weak and strong subadditivity, entanglement wedge nesting, etc.  In particular, we would like to define a notion of surface that is extremal ``with respect to'' $D_I$. Surely, extremal surfaces should be included in this definition. Additionally, we would like to consider surfaces for which the area takes an extreme value with respect to all neighboring surfaces, without having a vanishing expansion. For example, the function $f:[1,2]\rightarrow[1,4]:x\rightarrow x^2$ has no extrema, but it has a non-extremal minimum and a non-extremal maximum at $x=1$ and $x=2$, respectively. These boundary extremal surfaces can be found using the method of Lagrange multipliers. 

As an example, consider the problem of \textit{constrained extremization} of the area functional of a SO$(d)$-symmetric surface restricted to some bounded bulk region $D$.\footnote{We impose this symmetry to simplify our discussion, but the computation generalizes to arbitrary surfaces.} Any such surface is parameterized by $(\sigma,\theta)$ in conformal coordinates \eqref{eq:conf}, and its area is denoted $\text{Area}(\sigma,\theta)$. Let $f_i(\sigma,\theta)$ be the set of functions that define $D$ with $f_i\geq 0$. We supplement the area functional with terms proportional to Lagrange multipliers:
\begin{equation}
	\label{eq:Lagrange}
 \widehat{\text{Area}}(\sigma,\theta,\nu_i,a_i) = \text{Area}(\sigma,\theta) + \sum_i  \nu_i \left(f_i(\sigma,\theta)-a_i^2\right).
\end{equation}
$\widehat{\text{Area}}$ is defined in a domain without boundary as $\sigma,\theta,\nu_i,a_i$ all span $\mathbb{R}$. Solving the problem of extremizing $\widehat{\text{Area}}$ over all these variables give the sphere with extremal area with respect to $D$, that is either a true extrema inside $D$ (associated with $\nu_i=0$ for all $i$), or a boundary extremal surface on the boundary $\partial D$ (associated with $a_i=0$ for at least one $i$). Indeed, the equations of motions for $(\sigma,\theta)$ read
\begin{align}
\begin{split}
\label{eq:eomL}
    \frac{\partial \text{Area}}{\partial\sigma} + \sum_i \nu_i \frac{\partial f_i}{\partial\sigma} &= 0,\\
    \frac{\partial \text{Area}}{\partial\theta} + \sum_i \nu_i \frac{\partial f_i}{\partial\theta}  &= 0.
\end{split}
\end{align}
The equations of motion for $\nu_i$ lead to the constraint $f_i-a_i^2=0$ for all $i$, and the equations of motion for $a_i$ read $\nu_i a_i=0$ for all $i$.
\begin{itemize}
    \item If the surface lies in the interior of $D$, denoted by $D^0$, $a_i\neq 0$ and $\nu_i=0$ for all $i$. Thus, the constrained extremization problem reduces to an unconstrained extremization whose solutions are true extremal surfaces.
    \item If the surface lies on $\partial D$, $a_{i_0}=0$ for some $i_0$ and $\nu_i=0$ for $i\neq i_0$.\footnote{In general, $a_i$ might vanish for more than one $i$. This would be the case at a corner of the boundary, where two boundary regions $f_{i_0}=0$ and $f_{i_1}=0$ meet. This does not modify the present discussion.} Equations~\eqref{eq:eomL} become
    \begin{align}
        \begin{split}
            \frac{\partial \text{Area}}{\partial\sigma} + \nu_{i_0} \frac{\partial f_{i_0}}{\partial\sigma} &= 0,\\
            \frac{\partial \text{Area}}{\partial\theta} + \nu_{i_0} \frac{\partial f_{i_0}}{\partial\theta} &= 0,
        \end{split}
    \end{align}
    which exactly correspond the constrained extremization of the area functional along the boundary $f_{i_0}=0$. The solution to this system is therefore a boundary extremal surface that is not extremal in an enlarged region containing $D$, but which is extremal along $\partial D$.
\end{itemize}
In Appendix \ref{app:Lagrange}, this method is applied to the explicit examples of an extremization constrained to a causal diamond in dS$_3$.

As in the original HRT formula, we select the surface of minimal area among the constrained extremal surfaces. We consider compact and bounded regions $D$ of spacetime, such that $\partial D$ has a closed topology. By the extreme value theorem, there exists surfaces where the area functional takes its maximal and minimal value. If this boundary minimal surface has a smaller area than any extremal surface in the interior of $D$, then it is the minimal constrained-extremal surface. We therefore summarize our definition of constrained extremization as follows.
\begin{definition}
\label{def:Rext}
\textbf{[Constrained extremization/C-extremization]}
    The C-extremal surface of $A\in\partial D$ in region $D$, denoted as $\gamma_{c}(A;D)$, is a spacelike codimension-two surface $\gamma$, $D$-homologous to $A$ that satisfies one of the two following conditions.
    \begin{enumerate}
        \item It is the extremal surface of $A$, 
        $\gamma_{\rm e}(A,D)$.
        \item A positive measure subset\footnote{By a positive measure subset of $\gamma$, we mean that a subset of $\gamma$ of non-vanishing area lies on $\partial D$. This is the negation of the statement that $\gamma\in D \backslash \partial D$ except for a countable number of points $p\in\gamma$.} of $\gamma$ lies on $\partial D$ and $\gamma$ is the spacelike surface with minimal area in $D$. See Figure~\ref{fig:const_ext} for a schematic example.
    \end{enumerate}
    When there are multiple C-extremal surfaces, we chose the one with the smallest area.
\end{definition}
\begin{figure}[ht]
    \centering
    \includegraphics[width=0.6\linewidth]{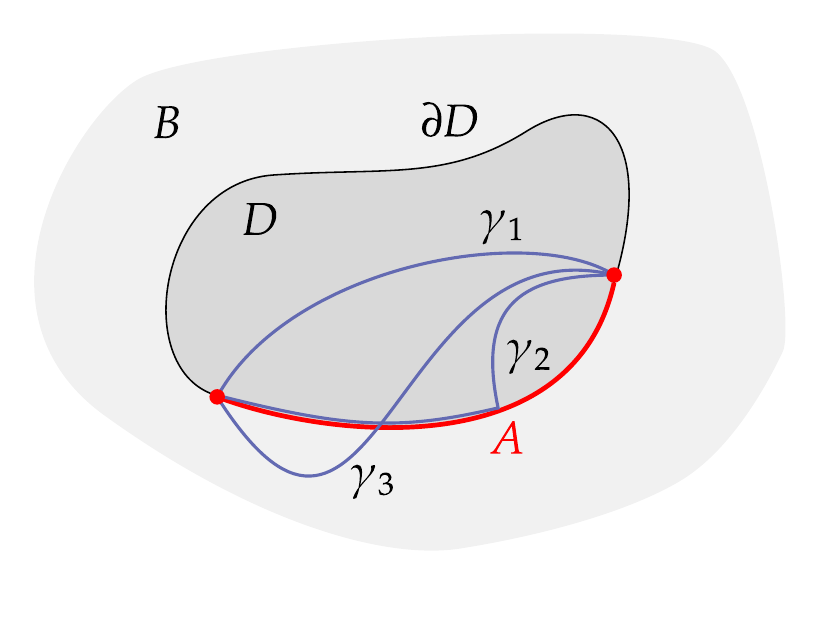}
    \caption{\footnotesize Schematic example of constrained extremization. Consider a region $D$ of the bulk $B$. We look for constrained extremal surfaces (curves here) in $D$ and homologous to a subsystem $A$ of $\partial D$. $\gamma_1$ is extremal and has no positive measure subset on $\partial D$, so it is a constrained extremal curve. $\gamma_2$ is not extremal but has a positive measure subset on $\partial D$ while being the curve of minimal length, so it is a constrained extremal curve. Finally, $\gamma_3$ is extremal in $B$ but is not fully contained in $D$ so it is not a constrained extremal curve in $D$. }
    \label{fig:const_ext}
\end{figure}
In~\cite{Hao:2024nhd}, a very similar formula was derived in the context of islands computations in de Sitter space. By embedding a dS$_2$ braneworld in an AdS$_3$ spacetime, it was shown that the correct island for a non-gravitating region is not extremal but instead globally minimizes the area functional in the gravitating region, and its boundary sits on the boundary of the gravitating region. This result is explicitly derived using AdS/CFT. This result provides additional motivation for constrained extremization in the context of holographic entanglement entropy. 

From now on, we promote the extremization process used in Conjecture \ref{con:ent_pr} to a constrained extremization process. We thus formulate our holographic entanglement entropy proposal:
\begin{conjecture}[Holographic entanglement entropy in static patch holography]
    Let $A$ be a subsystem of $\Sigma\vert_{\mathcal{S}_{\rm L}}\cup\Sigma\vert_{\mathcal{S}_{\rm R}}$. The entanglement entropy of $A$ is given by
    \begin{equation}
    	\label{eq:FPRT}
            S(A)= \min \mathrm{ext}_{D_I}\left[\sum_{I=L,E,R}\frac{{\rm Area}(\gamma_I)}{4G}\right] +O(G^0),
    \end{equation}
    where $\gamma_I$ must be $D_I$-homologous to $A$, and $\mathrm{ext}_{D_I}$ corresponds to a constrained extremization in $D_I$. The resulting surfaces are called the C-extremal surfaces $\gamma_{\rm c}(A;D_I)$. Their homology region are denoted by $\mathcal{C}_I(A)$. The entanglement wedge of $A$ is defined by
    \begin{equation}
        \label{eq:dSEW}
        W(A) = \bigcup_I D(\mathcal{C}_I(A))
    \end{equation}
\end{conjecture}
This prescription can be generalized to take into account quantum corrections by replacing the area functional with the generalized entropy \eqref{eq:SgenDS}.

In Appendix \ref{app:proofs}, we define maximin surfaces in $D_I$, and we prove two statements:
\begin{itemize}
    \item Theorem \ref{ineq}: \textit{Maximin, extremization, and C-extremization are inequivalent. Moreover, none of them implies another.}
    \item Theorem \ref{th:patch}: \textit{Maximin, extremization, and C-extremization are equivalent in the interior of the holographic screen of an observer.}
\end{itemize}
In other words, issues with the non-existence of extremal surfaces only appear in $D_E$. We will extensively use theorem \ref{th:patch} in Chapter~\ref{ch:CWT}.

\section{Bulk reconstruction in static patch holography}
\label{sec:EWdS}

We now apply our covariant prescription to simple subsystems: The one-screen system and the two-screen system where we locate $\mathcal{S}$ on the two cosmological horizons. In all instances, we consider SO$(d)$-symmetric slices $\Sigma$ such that $\Sigma\vert_{\mathcal{S}_{\rm L}}$ and $\Sigma\vert_{\mathcal{S}_{\rm R}}$ are spherical screens. Let us denote $\sigma_{\rm L}$ and $\sigma_{\rm R}$ as the conformal times of $\Sigma\vert_{\mathcal{S}_{\rm L}}$ and $\Sigma\vert_{\mathcal{S}_{\rm R}}$, respectively. Since the screens lie on a Cauchy slice, we always have $\sigma_{\rm L}\sigma_{\rm R}\geq 0$. While we do expect each screen to encode its own static patch, we explicitly check that the entanglement wedge of a screen covers its interior. Moreover, it is not clear a priori if and how the exterior region is encoded in some spatial subsystem of $\mathcal{S}$. We argue that the union of the two screens encodes the exterior region. Analogously to the double-sided black hole in AdS, the inflationary region emerges from the entanglement between two disconnected holographic screens.

\subsection{Single-Screen System: The de Sitter central dogma}

Consider $A=\Sigma\vert_{\mathcal{S}_{\rm L}}$. In this section, we show that a constrained extremization procedure reproduces the Gibbons-Hawking formula \eqref{eq:GH}. However, we show that there is an ambiguity in the entanglement wedge. We then briefly discuss quantum corrections.

The subsystems with respect to each region are:
\begin{equation}
     A_{\rm L} =\Sigma\vert_{\mathcal{S}_{\rm L}},\quad A_{\rm E}= \Sigma\vert_{\mathcal{S}_{\rm L}}, \quad  A_{\rm R} = \varnothing.
\end{equation}
The extremal surfaces in the two interiors both are the empty surface $\varnothing$. Indeed, the empty surface is extremal by definition, its area is vanishing, and $A_{\rm L} \cup \varnothing = \partial \Sigma_{\rm L}'$ for any slice $\Sigma'_{\rm L}$. Thus, the associated homology region is $\Sigma_{\rm L}$ and $W_{\rm L}=D_{\rm L}$. Similarly, $A_{\rm R} \cup \varnothing = \varnothing$ such that $W_{\rm R}=\varnothing$.

Since there are no geometrical contributions to the entropy of $A$ from the interior regions, the entropy is fully captured by $\gamma_{\rm c}(A;D_{\rm E})$. Consider all possible choices of slices $\Sigma'_{\rm E}$ in the exterior region, which by definition have boundary $\Sigma\vert_{\mathcal{S}}$. They span a common causal diamond, which is the magenta rectangle in Figure~\ref{fig:Penrose_dS_ext}.

\begin{figure}[ht]
    \centering
    \includegraphics[width=0.3\linewidth]{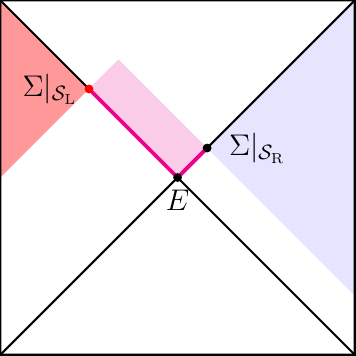}
    \caption{\footnotesize Entanglement wedge (in red) of the system $\Sigma\vert_{\mathcal{S}_{\rm L}}$. The blue triangle is $D_{\rm R}$, in which the entanglement wedge does not extend. The thick magenta line corresponds to the union of all C-extremal surfaces that have the same area as $\Sigma\vert_{\mathcal{S}_{\rm L}}$. Depending on the choice of C-extremal surface, the entanglement wedge can have a non-trivial extent in the region $D_{\rm E}$ shaded in magenta.}
    \label{fig:Penrose_dS_ext}
\end{figure}

Since $A_{\rm E}=\Sigma\vert_{\mathcal{S}_{\rm L}}$ has no boundary, the homologous surface $\gamma_{\rm E}$ lying on some slice $\Sigma'_{\rm E}$ must be closed. We look for minimal extremal surfaces that are ${\rm SO}(d)$-symmetric.\footnote{ In other words, we assume that the ${\rm SO}(n)$ symmetry is not spontaneously broken when extremizing the area functional.} All ${\rm SO}(n)$-symmetric surfaces are spheres represented by points on the Penrose diagram, and their areas depend on their positions on the diagram. In Appendix~\ref{A1}, we derive that in the exterior causal diamond $D_{\rm E}$, the only $\mathbb{S}^{d-1}$ of extremal area is the bifurcate horizon~\cite{Shaghoulian:2021cef}, denoted $E$ in Figure~\ref{fig:Penrose_dS_ext}. Notice that it is a sphere of minimal area in the diamond.\footnote{On the other hand, with respect to global de Sitter space, $E$ is a saddle point of the area functional. Indeed, it is a minimax surface. That is, one first finds the maximal area sphere for each Cauchy slice connecting the pode and the antipode and then takes the minimal area one within this set. The fact that $E$ is a minimax rather than a maximin of the area functional was one of the motivations for the bilayer and monolayer proposals \cite{Shaghoulian:2021cef}.} It also lies on the particular slice $\Sigma'_{\rm E}$ represented on the Penrose diagram as the lightlike line segments from $\Sigma\vert_{\mathcal{S}_{\rm L}}$ to $E$ and from $E$ to $\Sigma\vert_{\mathcal{S}_{\rm L}}$. This slice is the part of the boundary of $D_{\rm E}$ that lies on the horizons. 
However, all spheres along this slice have the same area as $E$. Therefore, they are also minimal-area surfaces in $D_{\rm E}$, and thus C-extremal, even though they do not extremize the area functional (except $E$). In Appendix~\ref{A2}, we show that these extra spheres of minimal area that lie on the boundary of $D_{\rm E}$ at the horizons are solutions of the constrained extremization problem \ref{def:Rext}. Since they have degenerate areas equal to that of the screen $\Sigma\vert_{\mathcal{S}_{\rm L}}$, they all yield the same geometrical contribution to the entropy of $S(A)$, equal to the area of $A$ divided by $4G$. We conclude that 
\begin{equation}
    S(\Sigma\vert_{\mathcal{S}_{\rm L}}) =\frac{\mathrm{Area}(\Sigma\vert_{\mathcal{S}_{\rm L}})}{4G} +{O}(G^{0}).
\end{equation}
To leading order, the entropy of $\Sigma\vert_{\mathcal{S}_{\rm L}}$ is independent of $\sigma_{\rm L}$, $\sigma_{\rm R}$ and is equal to the Gibbons-Hawking entropy of the horizon \cite{Gibbons:1977mu}, irrespective of the choice among the minimal area spheres. A similar argument holds for $A=\Sigma\vert_{\mathcal{S}_{\rm R}}$.

However, choosing one or another of the minimal-area spheres leads to different entanglement wedges in the exterior region. Indeed, letting $\gamma_{\rm c}(A,D_{\rm E})$ be a minimal sphere on the left lightlike segment between $\Sigma\vert_{\mathcal{S}_{\rm L}}$ and $E$, the only possible choice for $\mathcal{C}_{\rm E}(A)$ is the smaller segment between $\Sigma\vert_{\mathcal{S}_{\rm L}}$ and $\gamma_{\rm c}(A,D_{\rm E})$ on the Penrose diagram (because the boundary of $\mathcal{C}_{\rm E}(A)$ must be $\Sigma\vert_{\mathcal{S}_{\rm L}} \cup \gamma_{\rm c}(A,D_{\rm E})$ and $\Sigma\vert_{\mathcal{S}_{\rm L}}$ and $\gamma_{\rm c}(A,D_{\rm E})$ are lightlike separated). As a result, the corresponding entanglement wedge in the exterior is the lightlike segment between $\Sigma\vert_{\mathcal{S}_{\rm L}}$ and this minimal sphere. The entanglement wedge is confined to the left horizon. In particular, if $\gamma_{\rm c}(A,D_{\rm E})=\Sigma\vert_{\mathcal{S}_{\rm L}}$, the entanglement wedge shrinks to $\Sigma\vert_{\mathcal{S}_{\rm L}}$ and does not extend into the exterior region. On the other hand, if $\gamma_{\rm c}(A,D_{\rm E})$ is a minimal area sphere on the right horizon between $E$ and $\Sigma\vert_{\mathcal{S}_{\rm R}}$, $\mathcal{C}_{\rm E}(A)$ can be chosen to be any slice bounded by $\Sigma\vert_{\mathcal{S}_{\rm L}}\cup \gamma_{\rm c}(A,D_{\rm E})$. All these slices have the same causal diamond. The corresponding entanglement wedge is the common causal diamond of the various $\mathcal{C}_{\rm E}(A)$, which is a smaller rectangle inside the magenta rectangle of Figure~\ref{fig:Penrose_dS_ext}. In the special case $\gamma_{\rm c}(A,D_{\rm E})=\Sigma\vert_{\mathcal{S}_{\rm R}}$, the entanglement wedge assumes its largest possible extent, covering the whole colored rectangular region. This would imply that the full causal diamond in the exterior region can be reconstructed holographically in terms of degrees of freedom on $\Sigma\vert_{\mathcal{S}_{\rm L}}$.    

Hence, it is important to figure out which minimal area sphere is actually the correct one. The classical degeneracy can be lifted at the quantum level. We expect the actual quantum extremal surface to be close to the classical extremal surface for which the first-order corrected entropy is the smallest.\footnote{By doing so, we neglect the backreaction of the semiclassical corrections on the classical surface.} 

Let $\gamma_{\rm E}(\bar{\sigma})$ and $\mathcal{C}_{\rm E}(\bar{\sigma})$ be one of the minimal area spheres on any of the two horizons and the associated homology region, and denote $\bar{\sigma}$ as its conformal time. We apply Equation \eqref{eq:SgenDS} and add to the geometrical contribution the semiclassical entropy of $\mathcal{C}_{\rm L}(A)\cup \mathcal{C}_{\rm E}(\bar{\sigma})\cup\mathcal{C}_{\rm R}(A)=\Sigma_{\rm L}\cup \mathcal{C}_{\rm E}(\bar{\sigma})$. The slice $\Sigma_{\rm L}\cup \mathcal{C}_{\rm E}(\bar{\sigma})$ is unitarily related to any slice bounded by $\gamma_{\rm E}(\bar{\sigma})$, denoted $\Sigma_{\rm L}(\bar{\sigma})$. Hence,
\begin{equation}
    S\big(\Sigma_{\rm L} \cup \mathcal{C}_{\rm E}\big)=S\big(\Sigma_{\rm out}(\bar{\sigma})\big).
\end{equation}
The generalized entropy of $A$, defined in Equation \eqref{eq:SgenDS}, for this choice of minimal surface, is 
\begin{align}
    S_{\rm gen}(\varnothing,\gamma_{\rm E}(\bar{\sigma}),\varnothing)
    &=\frac{{\rm Area}(\gamma_{\rm E}(\bar{\sigma}))}{4G}+S(\Sigma_{\rm out}(\bar{\sigma})) = \frac{\mathrm{Area}_{\rm qu}(\gamma_{\rm E}(\bar{\sigma}))}{4G}.
\end{align} 
The dominant minimal area sphere $\gamma_{\rm c}(A,D_{\rm E})$ is the one for which the total quantum area is the smallest. The GSL ensures that this surface is the one with the lowest $|\bar{\sigma}|$. Hence, in any sufficiently asymmetric state where the GSL is not saturated, the GSL tends to favor the bifurcate horizon $E$ as the quantum extremal surface. However, beyond the semiclassical limit, the screen may fluctuate away from the horizon, which prevents us from using the GSL. It would be interesting to formulate a general statement about the behavior of $S_{\rm gen}$ along a holographic screen.

In states satisfying the de Sitter symmetries, such as the BD state \eqref{eq:HH}, we expect the quantum extremal surface to suffer from the same degeneracies as the classical extremal surface. For example, in the Bunch-Davies state of the half reduction model presented in Chapter~\ref{ch:JT}, the generalized entropy of a point is
\begin{equation}
\label{eq:BDSgen}
    S_{\rm gen} = \frac{\Phi_*}{4G}+\frac{c}{6}\log\Phi_* + \mathrm{constant},
\end{equation}
where 
\begin{equation}
    \Phi_* = \phi_r\frac{1+\Lambda x^+x^-}{1-\Lambda x^+x^-}
\end{equation}
is the dilaton solution without matter. We used Equation \eqref{eq:entang_entropy} for the entanglement entropy of a slice bounded by a point $(x^+,x^-)$ in the BD state. The quantum extremal surface is computed by solving $\partial S_{\rm gen} / \partial x^{\pm}=0$. From Equation \eqref{eq:BDSgen}, it is clear that studying the variation of $S_{\rm gen}$ is equivalent to studying the variation of the dilaton itself. Thus, our classical analysis directly maps to an analysis of quantum extremal surfaces in the Bunch-Davies state.

Note that the degeneracy discussed here disappears when considering a time-symmetric slice of the screens, where $\Sigma\vert_{\mathcal{S}_{\rm L}}$ is located on the bifurcate horizon $E$. For $A=E$, the exterior region $D_{\rm E}$ shrinks to $E$ itself. Then, $S(E)=\frac{\mathrm{Area}(E)}{4G}+S(\Sigma_{\rm L})$ as described in Equation \eqref{eq:SgendS}.

\subsection{The Pair of Holographic Screens is Pure and Encodes Everything}
\label{sec:2screens}

Consider $A=\Sigma\vert_{\mathcal{S}}$. From definition \ref{def:homol},
\begin{equation}
 A_{\rm L} =\Sigma\vert_{\mathcal{S}_{\rm L}},\quad A_{\rm E}= \Sigma\vert_{\mathcal{S}_{\rm L}}\cup\Sigma\vert_{\mathcal{S}_{\rm R}}, \quad  A_{\rm R} = \Sigma\vert_{\mathcal{S}_{\rm R}}.
\end{equation}
All these surfaces are closed. Hence, in each region $I=$ L, E, R we are looking for an extremal surface without boundary lying on a slice $\Sigma'_I$ satisfying $\partial \Sigma'_I = \Sigma\vert_{\mathcal{S}_{I}}$. The empty surface is a subset of any slice $\Sigma'_I$ and has minimal extremal area by definition. It satisfies the homology constraint in every region:
\begin{equation}
	A_I \cup \varnothing = \partial \mathcal{C}_I,
\end{equation}
where $\mathcal{C}_I$ can be any of the slices $\Sigma'_I$. Hence, the entropy of the two-screen system, consisting of some spatial slice of the full dual theory, is 
 \begin{equation}
	S(\Sigma\vert_{\mathcal{S}})= 0+  O\left(G\right),
\end{equation}
where we used $\text{Area}(\varnothing)=0$, $\Sigma'=\Sigma'_{\rm L}\cup\Sigma'_{\rm E}\cup\Sigma'_{\rm R}$, and $S\left(\Sigma' \right)=0$ since $\Sigma'$ is a Cauchy slice. In other words, the full holographic system has zero entropy at next-to-leading order. Following the covariant prescription, the part of the entanglement wedge in each region $I=$ L, E, R is given by the causal diamond of $\mathcal{C}_I=\Sigma'_I$, such that
\begin{equation}
	W(\Sigma\vert_{\mathcal{S}}) = D_{\rm L} \cup D_{\rm E} \cup D_{\rm R}.
\end{equation} 
This region is spanned by equivalent Cauchy slices $\Sigma'$. Thus, $\Sigma\vert_{\mathcal{S}}$ encodes all Cauchy slices passing through $\Sigma\vert_{\mathcal{S}}$. The fact that the entanglement wedge of the two holographic screens covers the exterior region in addition to their static patches leads to an extension of the statement of static patch holography: 
\begin{conjecture}[Extended static patch holography]
\label{conj:double}
    Global de Sitter spacetime can be encoded
	holographically in terms of a quantum theory defined on the two cosmological horizons. The states of the interior regions $D_{\rm L}$ and $D_{\rm R}$ are encoded on $\Sigma\vert_{\mathcal{S}}$. The exterior region $D_{\rm E}$ emerges from the entanglement between $\Sigma\vert_{\mathcal{S}_{\rm L}}$ and $\Sigma\vert_{\mathcal{S}_{\rm R}}$.    
\end{conjecture}
This story is reminiscent of the double-sided black hole in AdS presented in Section~\ref{sec:ER=EPR}. Disconnected screens have Hilbert spaces $\mathscr{H}_{\rm L}$ and $\mathscr{H}_{\rm L}$, and encode their respective static patches $P_{\rm L}$ and $P_{\rm R}$. Once we consider the full theory with Hilbert space $\mathscr{H}_{\rm L}\otimes\mathscr{H}_{\rm R}$, the entanglement between the two sets of degrees of freedom can be seen from the bulk point of view as the spatial connectivity between the two patches. In other words, the entanglement wedge of the union of the two screens is larger than the union of their independent entanglement wedges. In particular, it covers the full exterior region that bridges the two static patches. This enrichment of the bulk geometry emerges directly from entanglement in the dual theory. 

\begin{figure}[ht]
\begin{subfigure}[t]{0.48\linewidth}
	\centering
\includegraphics[width=0.52\linewidth]{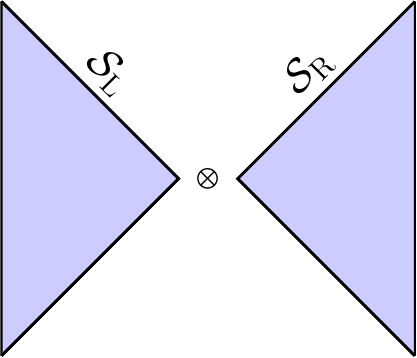}
\caption{}
\end{subfigure}
\begin{subfigure}[t]{0.48\linewidth}
	\centering
	\includegraphics[width=0.45\linewidth]{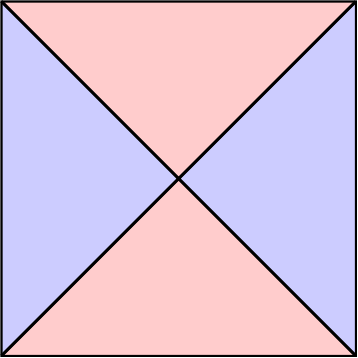}
\caption{}
\end{subfigure}
\caption{\footnotesize Extended static patch holography: (a) Two static patches, independently encoded on two holographic screens with Hilbert space $\mathscr{H}_{\rm L}$ and $\mathscr{H}_{\rm R}$. (b) The bulk dual to the Bunch-Davies state on a Hilbert space $\mathscr{H}_{\rm L}\otimes\mathscr{H}_{\rm R}$, which implements entanglement between the two screens, is global de Sitter spacetime. This picture is analogous to the AdS black hole presented in Figure~\ref{fig:ER=EPR}.\label{fig:bridge}}
\end{figure}

Similarly to the double-sided AdS black hole, the state of the dual quantum theory may be described by a thermofield-double state called the Bunch-Davies state~\cite{Goheer:2002vf}, constructed as follows. We consider a geodesic in de Sitter space, which is arbitrary and equivalent to any other one due to the isometry group of de Sitter space, and place the worldline of an observer along this geodesic. This worldline defines a static patch. Among the Killing vectors of de Sitter space, there is one that is timelike, future-directed, and defined everywhere in the static patch. This Killing vector maps the worldline to itself, by shifting the proper time on the worldline. The associated conserved charge $H$ is the Hamiltonian of the static patch.\footnote{In~\cite{Witten:2023xze}, this Hamiltonian is supplemented by the Hamiltonian of the observer itself.} Let us write $\ket{\Psi_i}$ and $E_i$ the eigenvectors and eigenvalues of $H$. The static patch is in a thermal state of inverse temperature $\beta$:
\begin{equation}
	\label{eq:th}
 \rho = \frac{1}{Z}e^{-\beta H}.
\end{equation}
We consider two copies of such a static patch, with Hilbert spaces $\mathscr{H}_{\rm L}$ and $\mathscr{H}_{\rm R}$ and associated eigenstates $\ket{\Psi_i}_{\rm L}$ and $\ket{\Psi_i}_{\rm R}$. Each of these theories independently describes a static patch, as in the left picture of Figure~\ref{fig:bridge}. 

We now construct the thermofield-double state whose Hilbert space is $\mathscr{H}_{\rm L}\otimes \mathscr{H}_{\rm R}$ and define the Hamiltonian
\begin{equation}
	H_{\rm BD}= H_{\rm L} - H_{\rm R},
\end{equation}
with $H_{\rm L}$ and $H_{\rm R}$ as two copies of the original Hamiltonian. Considering physical states of $\mathscr{H}_{\rm L}\otimes \mathscr{H}_{\rm R}$ with vanishing eigenvalues ensures that both static patches have the same time. With this constraint in mind, one constructs the thermofield-double state \eqref{eq:TFD} on the full screen $\mathcal{S}$,
\begin{equation}
 \ket{\Psi_{\rm BD}} = \frac{1}{\sqrt{Z}}\sum e^{-\frac{1}{2}\beta E_i}\ket{\Psi_i}_{\rm L}\otimes \ket{\Psi_i}_{\rm R},
\end{equation}
which satisfies the constraint $H_{\rm BD} \ket{\Psi_{\rm BD}}=0$. Tracing out the degrees of freedom from $\mathscr{H}_{\rm L}$ ($\mathscr{H}_{\rm R}$), one recovers the thermal density matrix \eqref{eq:th} associated with $P_{\rm L}$ ($P_{\rm R}$). The Bunch-Davies state describes the full de Sitter space as in the right panel of Figure~\ref{fig:bridge}. In this picture, the two static patches in blue are still encoded in $\mathcal{S}_{\rm L}$ and $\mathcal{S}_{\rm R}$, while the exterior region in red is encoded in $\ket{\Psi}_{\rm BD}$, emerging from the entanglement between $\mathcal{S}_{\rm L}$ and $\mathcal{S}_{\rm R}$.  It was argued in~\cite{Chandrasekaran:2022cip} that in a quantum description of de Sitter space, the static patches must be maximally entangled, which is associated with $\beta\rightarrow 0$.  This might seem counter-intuitive as we know that the de Sitter background can be associated with a finite temperature $1/\beta_{dS}=2\pi$. But it was argued in~\cite{Lin:2022nss} that it is possible to define a temperature-like quantity, which stays finite as the global temperature goes to infinity. Essentially, their argument is that a bulk spacetime with finite effective temperature can be described by a holographic quantum theory at infinite temperature. The fact that the DSSYK model satisfies such a property was one of the main motivations for developing the dS/DSSYK correspondence~\cite{Lin:2022nss}. 

Considering the screen in a Bunch-Davies state is consistent with the idea that the state of the whole universe should be unique and pure. The entropy of the pair of holographic screens is therefore expected to be pure not only at next-to-leading order but to all orders. Should one- doubt that the exterior region is encoded on the screens, one could argue that when the screens coincide at the bifurcate horizons, there is no exterior region, such that the two-screen system must be in a pure state. Indeed, there exist Cauchy slices $\Sigma'$ such that $\Sigma'_{\rm L }$ and $\Sigma'_{\rm R}$ join at the bifurcate horizon, and such that $\Sigma'_{\rm E}=\varnothing$. Moreover, the evolution operator along the screens should be linear (possibly unitary), which cannot transform a pure state into a mixed state. Hence, whether we believe or not that the exterior region is also encoded on the screens, the entropy of the two-screen system must vanish to all orders:
\begin{equation}
	S(\Sigma\vert_{\mathcal{S}})= 0.
\end{equation}

\subsection{An example of non-existence problem for extremal surfaces}
\label{subsec:arc}

We conclude this chapter by computing the entropy of a non-trivial subsystem of one of the two screens: $A\in\Sigma\vert_{\mathcal{S}_{\rm L}}$. We perform this computation in dS$_3$, where we use again $\theta_1=\theta$ and define $\theta_2=\varphi$. We show that there is no extremal surface $\gamma_{\rm e}(A;D_{\rm E})$ but that there always exists a C-extremal surface $\gamma_{\rm c}(A;D_{\rm E})$.

The left screen $\Sigma\vert_{\mathcal{S}_{\rm L}}$ and the right screen $\Sigma\vert_{\mathcal{S}_{\rm L}}$ are taken to be circles at fixed conformal times $\sigma_{\rm L}$ and $\sigma_{\rm R}$. They are parametrized by $\varphi$ and have a circumference of $2\pi$. As a subsystem $A$ of $\Sigma\vert_{\mathcal{S}_{\rm L}}\cup\Sigma\vert_{\mathcal{S}_{\rm R}}$, we take the arc of length $\theta_A$ on the screen $\Sigma\vert_{\mathcal{S}_{\rm L}}$, parametrized by $\varphi\in[0,\theta_A]$, for a given $\theta_A\in(0, 2\pi)$. 

With respect to each region, the subsystem is $A_{\rm L}=A_{\rm E}=A$ and $A_{\rm R}=\varnothing$. The empty surface is the minimal extremal surface in region $D_{\rm R}$. It is also C-extremal. To find the extremal surfaces in $D_{\rm L}$ and $D_{\rm E}$, we note that there are exactly two geodesics in global dS$_3$ that are anchored to the two endpoints of $A$, denoted $\partial A$. Here, we briefly review the computation of \cite{Franken:2023pni}. It is easier to find the geodesic using the definition \eqref{eq:dS_hyperb} of dS$_3$ as a hypersurface in four-dimensional Minkowski spacetime. The two geodesics anchored to $\partial A$ are parametrized by
\be
\left\{\!\begin{array}{ll}
X^0&\!\!\!=T_{\rm L} \big(\lambda_1^{\pm}(\lambda)+\lambda\big)\\
X^3&\!\!\!=|X^0|\esps\\
X^1&\!\!\!=\lambda\cos\theta_A + \lambda_1^{\pm}(\lambda)\esps\\
X^2&\!\!\!=\lambda\sin\theta_A\esps
\end{array}\right. ,
\label{geoL}
\ee
where $T_{\rm L}=\tan\sigma_{\rm L}$ and $\lambda$ is a running parameter. The function $\lambda_1^{\pm}(\lambda)$ describes an ellipse:
\begin{equation}
\lambda_1^{\pm}(\lambda)=-\lambda\cos \theta_A\pm \sqrt{1-\lambda^2\sin^2\theta_A},
\end{equation}
where $|\lambda|\le \frac{1}{|\sin\theta_A|}$. Restricted to the dS hyperboloid, these curves lie exclusively on the horizon. In particular, they are on the horizon of the pode when $\lambda+\lambda_1^{\pm}\geq0$ and on the horizon of the antipode when $\lambda+\lambda_1^{\pm}\leq 0$. The two geodesics connecting the endpoints of $A$ are bounded by the points $(0,1)$ and $(1,0)$ in the $(\lambda,\lambda_1)$ plane.

One of the two geodesics extends only on the horizon of the pode, and satisfies $|\sigma|\geq|\sigma_{\rm L}|$ at all of its points, see Figure \ref{fig:arc}. 
\begin{figure}[h!]
    \centering
    \includegraphics[width=0.35\linewidth]{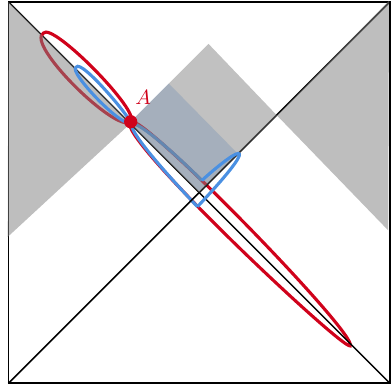}
    \caption{\footnotesize Codimension-two surfaces anchored to the subsystem $A$. The two extremal surfaces are depicted by red curves. The lower one is not contained in $D_{\rm E}$ and cannot be $\gamma_{\rm c}(A;D_{\rm E})$. Two examples of C-extremal surfaces that are not extremal are depicted in blue. The upper one is a candidate of $\gamma_{\rm c}(A;D_{\rm L})$. The second one is a candidate of $\gamma_{\rm c}(A;D_{\rm E})$ that leads to a non trivial entanglement wedge, shaded in blue. We made the curves slightly deviate from the horizon for clarity, but all of them are strictly contained in the union of the two cosmological horizons.}
    \label{fig:arc}
\end{figure}
Its length is $\inf(\theta_A,2\pi-\theta_A)$. The second geodesic extends on both horizons, and satisfies $|\sigma|\leq|\sigma_{\rm L}|$ at all of its points. In particular, $\sigma$ changes sign along the curve by intersecting the bifurcate horizon twice. Its length is $\sup(\theta_A,2\pi-\theta_A)$. The first geodesic is $D_{\rm L}$-homologous to $A$. It is therefore identified with $\gamma_{\rm e}(A;D_{\rm L})$. The second geodesic is an extremal surface. However, it spans both positive and negative values of $\sigma$. Thus, it is not contained in $D_{\rm E}$.\footnote{$D_{\rm E}$ is the causal diamond of $\Sigma_{\rm E}$ which is bounded by $\Sigma\vert_{\mathcal{S}_{\rm L}}$ and $\Sigma\vert_{\mathcal{S}_{\rm R}}$. These surfaces lie at fixed and positive $\sigma$. Thus, $D_{\rm E}$ does not extend beyond the bifurcate horizon.} There are no other geodesics anchored to $A$, so $\gamma_{\rm e}(A;D_{\rm E})$ does not exist.\footnote{Note that the empty surface is not a valid choice as it is not homologous to $A$.} This result also holds when the screen is located inside the static patch on a timelike hypersurface.

In \cite{Franken:2023pni}, it was shown that carefully applying a C-extremization procedure to the area functional leads to an infinite number of degenerate C-extremal surfaces, both in $D_{\rm L}$ and $D_{\rm E}$. This is due to the fact that any curve defined on the horizon, with $\partial A$ as its endpoints, and monotonic in $\varphi$, has length $\theta_A$ or $2\pi-\theta_A$. In $D_{\rm L}$, there is an infinite number of curves homologous to $A$ that extend along the horizon with $|\sigma|\geq |\sigma_A|$ and have a length of $\theta_A$ or $2\pi-\theta_A$. Among them, any curve that has length $\min(\theta_A,2\pi-\theta_A)$ is a C-extremal surface $\gamma_{\rm c}(A;D_{\rm L})$. In $D_{\rm E}$ there is also an infinite number of curves homologous to $A$ extending on the horizons, with length $\theta_A$.\footnote{Similar curves with length $2\pi-\theta_A$ exist, but they are not $D_E$-homologous to $A$.} These curves may extend only on the left horizon, between $A$ and the bifurcate horizon, but could also bifurcate at the bifurcate horizon to extend on the opposite horizon. Any of these curves can be $\gamma_{\rm c}(A;D_{\rm E})$.

Therefore, the entropy of $A$ is
\begin{align}
    S(A)    &= \left\{\!\begin{array}{ll}
       \dis \frac{2\mathrm{Area}(A)}{4G} +O(G) \, , &\mbox{if }0\le \theta _A\leq \pi\espD\\
       \dis  \frac{\pi}{2G}+O(G) \, , &\mbox{if } \pi<\theta_A \le2 \pi
    \end{array}\right. \!\!.
\end{align}
The entropy shows an interesting phase transition when $A$ becomes larger than half the size of the screen, as shown in Figure \ref{fig:plot}.
\begin{figure}[ht]
    \centering
    \includegraphics[width=0.5\linewidth]{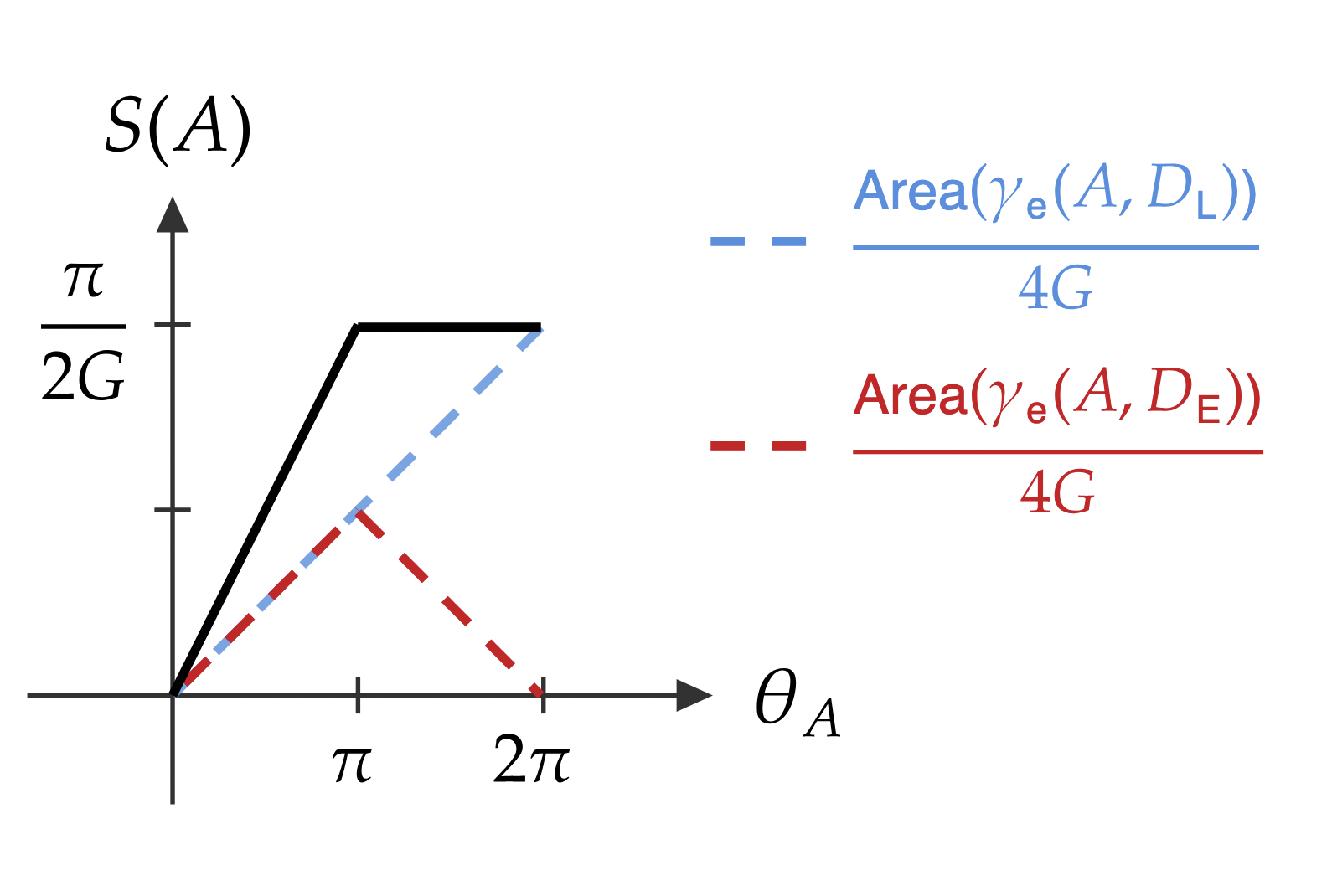}
    \caption{\footnotesize Plot of $S(A)$ as a function of $\theta_A$. The red and blue dashed curves correspond to the contribution from the exterior and left regions, respectively.}
    \label{fig:plot}
\end{figure}
Moreover, the leading contribution obeys a ``volume law'' (as opposed to an area law), suggesting that the holographic theory on the screens is nonlocal~\cite{Shaghoulian:2021cef}. 

The large degeneracy of C-extremal surfaces again leads to an ambiguity for the entanglement wedge of an arc. When $\theta_A\leq \pi$, quantum corrections lift this degeneracy. As first-order quantum corrections arise from the entropy on $\bigcup_I\mathcal{C}_I(A)$, choosing $\gamma_{\rm c}(A;D_{\rm L})$ and $\gamma_{\rm c}(A;D_{\rm E})$ to identify with $A$ reduces $\bigcup_I\mathcal{C}_I(A)$ down to a vanishing region. This choice trivially minimizes the quantum correction. Thus,
\begin{equation}
    W(A)\vert_{\theta_A\leq\pi} = \varnothing.
\end{equation}
Note that we only consider the bulk part of $W(A)$. Technically, $W(A)$ contains only $A$ itself. We discard the parts of the entanglement wedge that lie on the holographic screen itself.

When $\theta_A>\pi$, $A$ is not $D_{\rm E}$-homologous so that we cannot use the same argument. Extremizing the contribution from quantum fields in this case is a complicated problem that goes beyond the scope of this work. If $\gamma_{\rm c}(A;D_E)$ extends on the horizon between $A$ and $E$, the entanglement wedge does not extend into $D_E$ and $W(A)=\varnothing$. However, the C-extremization procedure does not forbid $\gamma_{\rm c}(A;D_E)$ to lie on the opposite cosmological horizon. In such a case, $W(A)$ would cover a non-trivial subregion of the inflationary region $D_{\rm E}$. In any case, $\gamma_{\rm c}(A,D_{\rm L})$ has length $2\pi-\theta_A$ when $\theta_A\geq \pi$, such that
\begin{equation}
    W(A)\vert_{\theta_A\geq\pi} \supset D_{\rm L}.
\end{equation}
Independently of the extent of $W(A)$ in the exterior region, there is a discontinuous phase transition at $\theta=\pi$, where the entanglement wedge in $D_{\rm L}$ goes from an empty set to the complete region $D_{\rm L}$.

\chapter{Generalization to FLRW holography}
\label{ch:FLRW}

In our endeavor to study the holographic principle in cosmological spacetimes, it is desirable to consider geometries beyond de Sitter space. In particular, the problem of the absence of asymptotic boundaries extends to all closed FLRW spacetimes. An additional motivation is that de Sitter space does not have a cosmological singularity. It is interesting both from a theoretical and cosmological perspective to explore cosmological singularities in the context of quantum gravity. In particular, a holographic understanding of Big Bang/Big Crunch cosmologies may be necessary to go beyond the effective description of semiclassical gravity, which breaks down in the vicinity of singularities. See \cite{Fischler:1998st, Hellerman:2001yi, Bak:1999hd, Bousso:1999cb, Diaz:2007mh, Sanches:2016sxy, Nomura:2016aww,Nomura:2016ikr,Caginalp:2019fyt,Noumi:2025lbb} for other attempts to formulate a holographic description of cosmological spacetimes.

In this chapter, we extend the static patch holography proposal and the bilayer holographic entropy prescription to generic closed FLRW cosmologies in arbitrary dimensions, which are characterized by a more involved horizon structure. We show that on Cauchy slices containing a region of trapped surfaces, there exists an infinite number of possible trajectories of the holographic screens, which can be grouped into equivalence classes. In each class, the effective holographic theory can be derived from a pair of “parent” screens on the apparent horizons. Analogously to the de Sitter case, we argue that a pair of holographic screens associated with antipodal observers can encode the geometry of the entire spacetime, with the regions of trapped and anti-trapped surfaces emerging from the entanglement between the two screens. Our results provide evidence for the consistency of C-extremization as a prescription to compute holographic entanglement entropy.

This chapter is mainly based on \cite{Franken:2023jas}. See also \cite{Rondeau:2024ftb}. Note that the inequivalence between extremal and C-extremal surfaces has strong consequences in the following analysis. As extremal surfaces are found not always to exist, we follow through with C-extremization, as presented in Section~\ref{sec:ext}. We conclude this chapter with a discussion on the possibility of considering maximin surfaces instead.

\section{Basics of closed FLRW cosmology}
\label{basics_FLRW}

In order to motivate our proposal for holographic systems dual to closed FLRW cosmological universes, we begin our discussion with a brief survey of basic features of closed FLRW cosmologies, the associated Penrose diagrams and horizon structures. 

We consider an $(d+1)$-dimensional \textit{closed FLRW cosmology}, where $d\ge 2$. In conformal gauge, the spacetime metric can be written as 
\begin{equation}
\label{meeta}
    d s^2= a^2(\eta) \left( -d\eta^2 + d\Omega^2_{d}\right)\!,
\end{equation}
where $\eta$ is the conformal time and $a(\eta)$ is the \textit{scale factor}.\footnote{In this chapter, we use a different notation than $\sigma$ for conformal time, as its domain differs from the de Sitter case, in general. Moreover, we rescale it so that $\eta$ ranges from $0$ to some end value $\eta_{\infty}$.} In this work, we restrict ourselves to an FLRW cosmology induced by a single perfect fluid of energy density $\rho$ and pressure $p$, satisfying the state equation
\be
p=w\rho, 
\label{seq}
\ee
where the constant $w\in[-1,1]$ is the perfect fluid index. As reviewed in Appendix~\ref{app:FLRW_cosmo}, the energy density satisfies 
\be
\rho=\frac{C}{a^{d(1+w)}},
\ee
where $C> 0$ is a constant. The evolution of the scale factor depends drastically on $w$, which admits a critical value 
\be
w_{\rm c}=-1+\frac{2}{d}\in(-1,0].
\ee
Throughout this chapter, we will focus on the generic case where $w\neq w_{\rm c}$,\footnote{The scale factor evolution for $w=w_{\rm c}$ is $a(\eta)=e^{\pm \eta \sqrt{{16\pi C\over n(d-1)}-1}}, ~~\quad \eta\in \R$.} which yields 
\be
\label{a(eta)}
a(\eta)=a_0\!\left(\sin{\frac{\eta}{ |\gamma|}}\right)^\gamma, ~~\quad \eta\in\big[0,|\gamma|\pi\big],
\ee
where we have defined
\be
a_0=\left(\frac{d(d-1)}{ 16\pi C}\right)^{\frac{1}{d(w_{\rm c}-w)}}, \qquad \gamma={2\over d(w-w_{\rm c})}={2\over d(1+w)-2}.
\label{a0gamma}
\ee
Qualitatively, the cosmological evolution is as follows:

\noindent $\bullet$ When $w$ increases from $-1$ to $w_{\rm c}$, the parameter $\gamma$ decreases from $-1$ to $-\infty$,
 \be
 -1\le w<w_{\rm c} \qquad \Longrightarrow \qquad -1\ge \gamma >-\infty.
 \label{g<-1}
 \ee 
Since $\gamma<0$, the cosmological evolution bounces. The scale factor decreases from an infinite value at $\eta=0$ and reaches its minimum $a_0$ at $\eta=|\gamma|\pi/2$. It then expands and becomes infinite at finite conformal time $\eta=|\gamma|\pi$. The evolution is nowhere singular. The particular case $w=-1$ ($\gamma=-1$) corresponds to a cosmological evolution induced by a positive cosmological constant, \textit{ie} a de Sitter spacetime. 

\noindent $\bullet$ When $w$ increases from $w_{\rm c}$ to 1, the parameter $\gamma$ decreases from $+\infty$ to $1/(d-1)$,
 \be
 w_{\rm c}< w\le 1 \qquad \Longrightarrow \qquad +\infty> \gamma \ge{1\over d-1}.
 \ee 
Since $\gamma>0$, the scale factor increases from a Big Bang singularity at $\eta=0$ where it vanishes, and reaches its maximum $a_0$ at $\eta=|\gamma|\pi/2$. It then decreases up to a Big Crunch singularity at $\eta=|\gamma|\pi$, where it vanishes. Semiclassical bulk computations can be trusted for a great part of the cosmological evolution as long as $a_0 >> l_P$. However, near the singularities, the geometrical description breaks down, and obtaining a holographic dual picture seems necessary to understand the physics. The case $w=1$ ($\gamma=1/(d-1)$) corresponds to a cosmological evolution induced by moduli fields. $w=1/n$ ($\gamma=2/(d-1)$) corresponds to an evolution induced by radiation. Finally, $w=0$ ($\gamma=2/(d-2)$ for $d\ge 3$) corresponds to an evolution induced by massive non-relativistic matter.

The spacetime metric~(\ref{meeta}) can be written in the explicitly SO$(d)$-symmetric form
\be
d s^2=a^2(\eta) \left( -d\eta^2 +d\theta^2+\sin^2(\theta)d\Omega_{d-1}^2\right)\!,
\label{so met}
\ee
where $\theta=\theta_1 \in [0,\pi]$ is a polar angle and $d\Omega^2_{d-1}$ is the metric of the unit $(d-1)$-dimensional sphere~$\mathbb{S}^{d-1}$. Slices of constant conformal time $\eta$ are spheres $\mathbb{S}^d$. Moreover, every generic point $(\eta,\theta)$ of the Penrose diagram corresponds to a sphere $\mathbb{S}^{d-1}$. In the particular cases where $\theta=0$ or $\theta=\pi$, $\mathbb{S}^{d-1}$ reduces to a point, which is the north pole or the south pole of the $\mathbb{S}^d$, respectively. Since $\theta$ ranges from $0$ to $\pi$ and $\eta$ ranges from $0$ to $|\gamma|\pi$, the diagrams are taller than wide when $|\gamma|\ge 1$. An example of the bouncing case is shown in \Figs{Penb}.
\begin{figure}[ht]
\centering
\begin{subfigure}[c]{0.3\linewidth}\centering
\!\!\!\!\!\!\!\!\!\!\!\!
\includegraphics[width=3.7cm]{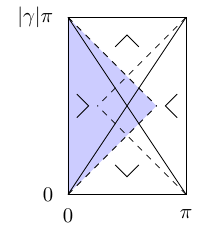}
    \caption{\footnotesize$-1>\gamma>-\infty$.\label{Penb}}
\end{subfigure}\hfill
\begin{subfigure}[c]{0.3\linewidth}\centering
\!\!\!\!\!\!\!\!\!\!\!\!
\includegraphics[width=3.7cm]{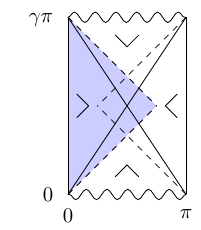}
    \caption{\footnotesize$1<\gamma<\infty$.\label{Penf}}
\end{subfigure}\hfill
\begin{subfigure}[c]{0.4\linewidth}\centering
\!\!\!\!\!\!\!\!\!\!\!\!
\includegraphics[width=6cm]{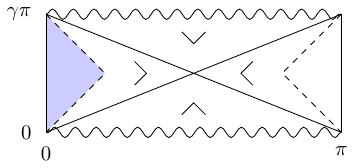}
    \caption{\footnotesize $1/(d-1)\le \gamma<1$.\label{PenT}}
\end{subfigure}
\caption{\footnotesize Penrose diagrams of closed FLRW spacetimes. Case (a) is an example of bouncing cosmology, while cases (b) and (c) correspond to examples of Big Bang/Big Crunch cosmologies. The causal patches of the pode and antipode are delimited by their cosmological horizons, shown in dashed lines. The causal patch of the pode appears in blue. The apparent horizons of the pode and antipode are the diagonal line segments. They divide the Penrose diagrams in four triangular domains in which the Bousso wedges are indicated. \label{Pen}}
\end{figure}
The diagrams are identical in the Big Bang/Big Crunch case when
$1\le \gamma<+\infty$, \textit{ie} $-1+4/d\ge w> w_{\rm c}$, up to wavy lines indicating the singularities in \Fig{Penf}. When $1/(d-1)\le \gamma<1$, \textit{ie} $1\ge w>-1+4/d$, the Penrose diagram is wider than tall, as shown in Figure \ref{PenT}.

Let us consider a pair of comoving observers sitting respectively at the pode and the antipode. Their worldlines on the Penrose diagrams are the left and right vertical lines. In the Penrose diagrams, the past and future horizons for the observers at the pode and the antipode correspond to segments on the lines:
\be
\begin{tabular} {c|c|c}
& pode &antipode\\ 
\cline{1-3}
past horizon & $\eta=\theta$ & $\eta=\pi-\theta$ \\ 
future horizon & $\eta=|\gamma|\pi-\theta$ & $\eta=|\gamma|\pi-\pi+\theta$ 
\end{tabular}
\ee
The horizons of the observers delimit their respective causal patches. These are shown in dashed lines in Figure~\ref{Pen}, while the causal patch of the pode is depicted in blue. When $|\gamma|\ge1$, the causal patches of the pode and antipode overlap, while for $1/(d-1)\le\gamma<1$ they are disconnected. 

As will be seen in the next section, another notion, which is that of \textit{apparent horizons} of the observers, enters naturally in the discussion of holographic dual descriptions of bulk cosmological evolutions. Strictly speaking, apparent horizons are the boundaries of the union of all trapped surfaces. However, in the present work, we extend the definition of apparent horizons to also include the boundaries of the union of all anti-trapped surfaces. Trapped (anti-trapped) surfaces are spacelike codimension~2 surfaces whose areas decrease (increase) locally along any future timelike direction. In the case of an FLRW cosmology, thanks to the SO$(d)$ symmetry of the metric~(\ref{so met}), the apparent horizons can be determined by looking for the domains of the Penrose diagram where all points $(\eta,\theta)$ correspond to trapped or anti-trapped spheres~$\mathbb{S}^{d-1}$. 

From Equation \eqref{so met}, the area of $\mathbb{S}^{d-1}$ located at $(\eta,\theta)$ is given by 
\be
\A(\eta,\theta)=v_{d-1}\,a_0^{d-1}\left(\sin{\eta\over |\gamma|}\right)^{\gamma(d-1)}(\sin\theta)^{d-1},
\label{AS}
\ee
where $v_{d-1}$ is the volume of the sphere of radius $1$. A sphere $\mathbb{S}^{d-1}$ is (anti-)trapped  if its expansion is negative (positive) along the two orthogonal null and future directions. The apparent horizons are the sets of points that saturate either of these bounds. As a result, they are the two diagonal lines of the Penrose diagrams,
\be
{\eta\over |\gamma|}=\theta,\qquad {\eta\over |\gamma|}=\pi-\theta.
\label{app}
\ee
The rectangular Penrose diagrams are thus divided into four triangular regions, as shown  in Figure~\ref{Pen}. In each triangle, the constant signs of the expansion are symbolized by so-called Bousso wedges~\cite{Bousso:1999xy}. The latter are 90$^{\circ}$ wedges, whose sides indicate the two null directions of negative expansion. The apparent horizons are timelike when  $|\gamma|> 1$, spacelike when $1/(d-1)\le \gamma<1$ and lightlike when  $|\gamma|= 1$. In the latter case, they coincide with the cosmological horizons of the pode and antipode. In all cases, we define the bifurcate horizon as the intersection of the apparent horizons of the pode and antipode. It is the $\mathbb{S}^{d-1}$ at $(\eta,\theta)=(\pi/2,|\gamma|\pi/2)$. 

\section{Holographic proposal}
\label{sec:FLRWholo}

The fundamental problems presented in Section~\ref{sec:holodS} associated with finding a holographic description of de Sitter space apply equally to any closed FLRW spacetime. In particular, their quantum mechanical description is expected to have a trivial Hilbert space. Notably, the dS/CFT correspondence \cite{Strominger:2001pn} does not apply to half of these cosmologies, whose null infinities are singularities. This provides an even stronger motivation to look for timelike holographic screens in the bulk.

Consider a foliation $\mathcal{F}$ of the geometry in Cauchy slices $\Sigma$ and a static observer on the pode. We aim to find the location of a holographic screen $\mathcal{S}_{\rm L}$ encoding the largest portion of each slice $\Sigma$ that contains the pode. This encoded spatial region is denoted $\Sigma_{\rm L}$. By abuse of language, we also refer to the intersection between $\mathcal{S}_{\rm L}$ and $\Sigma$, denoted $\Sigma\vert_{\mathcal{S}_{\rm L}}$, as the holographic screen. As in static patch holography, we associate a thermodynamic entropy to the holographic screen, given by its area divided by $4G$. We impose that the screen must lie in the causal patch of an observer, so that the associated information is accessible to this observer. In particular, the timelike tube theorem~\cite{Borchers1961,osti_4665531,Strohmaier:2023opz,Witten:2023xze} ensures that the algebra of observables on the screen is the same as the algebra of observables of this observer.

We repeat the strategy of Section~\ref{sec:holodS} and apply the spacelike projection theorem \ref{th:spacelike} to codimension-two surfaces. The crucial difference from the de Sitter case is that the cosmological horizon of an observer does not coincide with its apparent horizon in general. On each slice $\Sigma$, we look for the codimension-two surface $\Sigma\vert_{\mathcal{S}_{\rm L}}$ that maximizes $\Sigma_{\rm L}$. On any Cauchy slice, a codimension-two surface inside the apparent horizon satisfies all the assumptions of the spacelike projection theorem. See Figure \ref{fig:FLRWscreen}.

For $1/(d-1)\leq \gamma < 1$ cosmologies, see Figure \ref{PenT}, the causal patch lies inside the apparent horizon. Therefore, we saturate the constraint that the screen must be contained in the causal patch, and locate $\mathcal{S}_{\rm L}$ along the cosmological horizon.

For $|\gamma|\geq 1$ cosmologies, the apparent horizon lies inside the causal patch. At the very least, the holographic screen could be located on the apparent horizon of the pode. However, some slices contain surfaces that satisfy all the assumptions of the spacelike projection theorem, are inside the causal patch, but outside of the apparent horizon. Such a surface cannot lie beyond the opposite apparent horizon, so it must lie in the region bounded by the two apparent horizons. Cauchy slices being achronal, they contain at most one subsystem in the region of trapped or antitrapped surfaces. We call \textit{contracting slices} the slices that contain a region of trapped surfaces. We call \textit{expanding slices} the slices that contain a region of anti-trapped surfaces. There is one slice—the \textit{bifurcation slice}—in the foliation that intersects the bifurcate apparent horizon and is neither an expanding nor a contracting slice. Expanding slices do not have any surface outside of the apparent horizon of the pode that can be a holographic screen for the observer on the pode. Indeed, the region between the apparent horizons is spanned by surfaces that expand in all future directions. Therefore, in all expanding slices for $|\gamma|\geq 1$, the holographic screen is located on the apparent horizon of the pode. This is pictured in Figure \ref{fig:FLRWscreen}

\begin{figure}[ht]
\begin{subfigure}[t]{0.48\linewidth}
\centering
\includegraphics[width=0.5\linewidth]{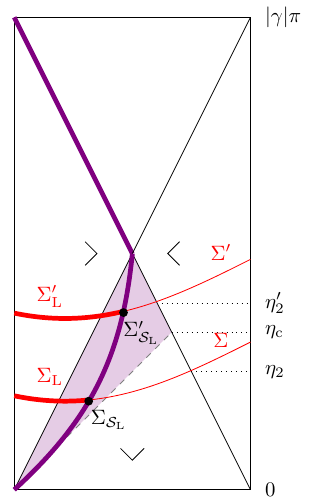}
\caption{\footnotesize Case of a bouncing cosmology, $\gamma\le -1$. \label{pushA}}
\end{subfigure}
\quad \,
\begin{subfigure}[t]{0.48\linewidth}
\centering
\includegraphics[width=0.5\linewidth]{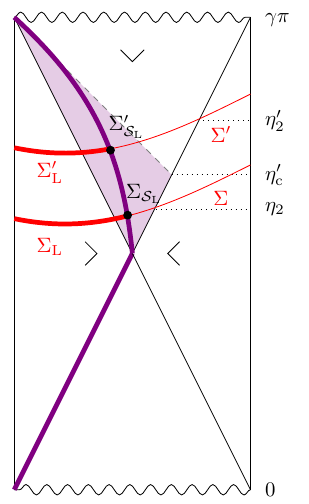}    
\caption{\footnotesize Case of a Big Bang/Big Crunch cosmology, \mbox{$\gamma\ge 1$}.\label{pushB}}
\end{subfigure}
    \caption{\footnotesize When the part of $\Sigma$ between the apparent horizons of the pode and antipode lies in the region of trapped surfaces, the screen $\Sigma\vert_{\mathcal{S}_{\rm L}}$ can follow any timelike (possibly locally lightlike) trajectory in the purple region. \label{fig:FLRWscreen}}
\end{figure}

The situation in contracting slices is more involved. On a contracting slice, any surface in the region between the apparent horizon and the cosmological horizon of the pode has a future-directed lightsheet directed towards the pode. Thus, any such surface is capable of encoding the part of the contracting slice that contains the pode and is bounded by that surface. Let $\eta_{\Sigma}$ be the conformal time at which $\Sigma$ intersects the apparent horizon of the antipode. Define the conformal time
\begin{equation}
    \eta_{c} =\frac{|\gamma|}{1+|\gamma|}\pi.
\end{equation}
For bouncing cosmologies $\gamma<-1$, the apparent horizon of the antipode intersects the past cosmological horizon of the pode at $\eta_c$. For Big Bang/Big Crunch cosmologies $\gamma>1$, the apparent horizon of the antipode intersects the future cosmological horizon of the pode at $\eta_c'=|\gamma|\pi-\eta_c$.  A few examples of $\eta_{\Sigma}$, $\eta_{c}$, and $\eta'_{c}$ are given in Figure \ref{fig:FLRWscreen} On contracting slices satisfying 
\begin{align}
    \eta_{\Sigma}&\in[\eta_c,|\gamma|\frac{\pi}{2}],\quad ~\gamma<-1,\\
    \eta_{\Sigma}&\in[\gamma\frac{\pi}{2},\eta_c'],\quad ~\gamma>1,
\end{align}
the cosmological horizon of the pode lies beyond its apparent horizon. Hence, on these slices the holographic screen may be located anywhere between the apparent horizon of the pode and the apparent horizon of the antipode. On contracting slices satisfying 
\begin{align}
    \eta_{\Sigma}&\in[0,\eta_c],\quad ~\gamma<-1,\\
    \eta_{\Sigma}&\in[\eta_c',\gamma\pi],\quad ~\gamma>1,
\end{align}
the cosmological horizon of the pode lies between the two apparent horizons and therefore constrains the location of the holographic screen. On these slices, the screen must lie between the apparent and cosmological horizons of the pode.

The hypersurface $\mathcal{S}_{\rm L}$ thus follows the apparent horizon for all expanding slices, and takes some timelike trajectory inside the purple region of Figure \ref{fig:FLRWscreen} defined from the constraints described above. $\mathcal{S}_{\rm L}$ intersects the bifurcation slice on the bifurcate horizon, which serves as a gluing surface between the screens of the expanding and contracting slices. The trajectory of the holographic screen must be continuous, but not necessarily differentiable. It is tempting to push the trajectory of the holographic screen in the purple region all the way to its boundary, to maximize the extent of all slices $\Sigma_{\rm L}$. We argue in the following that when a second screen is associated with an observer at the antipode, this is not necessary. 

An analogous argument can be followed for the holographic screen $\mathcal{S}_{\rm R}$ of the antipode. The different constraints found for the location of the screen $\mathcal{S}_{\rm L}$ can be applied to $\mathcal{S}_{\rm R}$ by applying a mirror symmetry $\theta\rightarrow\pi-\theta$ to our previous discussion. In principle, the screens $\mathcal{S}_{\rm L}$ and $\mathcal{S}_{\rm R}$ could intersect more than once to create an overlap between the bulk regions they encode. We avoid these situations by exchanging $\mathcal{S}_{\rm L}$ and $\mathcal{S}_{\rm R}$ between the intersections. Formally, the degrees of freedom on each screen may be exchanged when $\mathcal{S}_{\rm L}$ and $\mathcal{S}_{\rm R}$ intersect, which effectively exchanges their role as holographic screens.

On expanding slices, we have located the two screens on the apparent horizons of antipodal observers. We have argued that neither of those screens could be located in the expanding region of the slice. Thus, our setup maximizes the extent of $\Sigma_{\rm L}\cup \Sigma_{\rm R}$. On the other hand, there is an infinite number of screens $\Sigma\vert_{\mathcal{S}_{\rm L}}\cup\Sigma\vert_{\mathcal{S}_{\rm R}}$ for contracting slices that may be chosen equivalently. 

On contracting slices, the two-screen system may have the capacity to describe more than the states on $\Sigma_{\rm L}$ and $\Sigma_{\rm R}$. To see why, let us apply the spacelike projection theorem to the complement of $\Sigma_{\rm L}$ in $\Sigma$. From the Bousso wedges in the region of trapped surfaces and the right causal patch in Figure \ref{fig:FLRWscreen}, we see that the coarse-grained entropy on the complement of $\Sigma_{\rm L}$ is bounded by the area of $\Sigma\vert_{\mathcal{S}_{\rm L}}$ divided by $4G$. The same remark applies to the complement of $\Sigma_{\rm R}$ in $\Sigma$ and $\Sigma\vert_{\mathcal{S}_{\rm R}}$. Since $\Sigma$ is now equal to the union of these complementary slices (with some overlap), the additivity of coarse-grained entropy guarantees that the two-screen system has a sufficient number of degrees of freedom to describe the state on the entire $\Sigma$. This motivates us to consider arbitrary timelike trajectories for the screens in the region of trapped surfaces, as long as they remain in the causal patches of the respective observers and have the same endpoints as the segments of the respective apparent horizons in this region. This argument does not apply when $\Sigma\vert_{\mathcal{S}_{\rm L}}$ ($\Sigma\vert_{\mathcal{S}_{\rm R}}$) is inside the left (right) apparent horizon or if one tries to push $\Sigma\vert_{\mathcal{S}_{\rm L}}$ ($\Sigma\vert_{\mathcal{S}_{\rm R}}$) in the region of anti-trapped surfaces.

In addition to the non-uniqueness of holographic screens, the structure of the holographic dual presented here differs from that of static patch holography. Indeed, when one of the screens is in the causal patch of the other, they can exchange energy and information, possibly via time dependent interaction terms. The exchanges can be reciprocal when both screens are in each other's causal patch. This is in contrast to the example of the eternal AdS black hole (or de Sitter space), for which there are no interactions between the two copies of the dual CFT (or the two quantum theories on the horizons).  Additionally, the evolution of the holographic theory is not unitary, in the sense that it corresponds to a sequence of maps between Hilbert spaces of different dimensionalities. For the Big Bang/Big Crunch case, the dimensionality of the Hilbert space is maximum  when both screens are at the bifurcation point. Near the singularities, the dimensionality of the Hilbert space is very small, with the holographic dual having very few degrees of freedom. In the bouncing cases, the maximum dimensionality occurs when the screens are at the future or past null infinity. 

Despite these differences, the geometrical structure of our holographic proposal is identical to that of static patch holography. In particular, our discussion of holographic entanglement entropy and our motivations to introduce our entropy prescription \eqref{eq:FPRT} apply directly to this more general case. In the following, we discuss the application of this prescription to closed FLRW cosmologies.

\section{A time-dependent ER=EPR realization}

In this section, we study entanglement wedge reconstruction in FLRW holography, using the same holographic entanglement entropy prescription as in static patch holography. In particular, we show that the degeneracies found in de Sitter holography are lifted for general cosmologies, leading to a time-dependent entanglement structure. We first argue that, as in static patch holography, the complete spacetime for $|\gamma|\geq 1$ cosmologies can be reconstructed from the state of two holographic theories defined on $\mathcal{S}=\mathcal{S}_{\rm L}\cup \mathcal{S}_{\rm L}$. The contracting or expanding regions spatially connecting $\mathcal{S}_{\rm L}\cup \mathcal{S}_{\rm L}$ and $\mathcal{S}_{\rm L}\cup \mathcal{S}_{\rm R}$ emerge from the entanglement between the two screens. Interestingly, the time evolution of these regions can be related to the time evolution of the entanglement between the screens. We briefly discuss how $1/(d-1)\leq |\gamma|< 1$ cosmologies escape our discussion.

\subsection{The two-screen system in cosmologies with \texorpdfstring{\boldmath $|\gamma| \geq 1$}.}
\label{twoscreens}

Consider $A=\Sigma\vert_{\mathcal{S}}$. The analysis of extremal surfaces is identical to that of Section~\ref{sec:2screens}. In particular, the three C-extremal surfaces are the empty surfaces defined by an empty set of points. This results in a vanishing geometrical entropy. Moreover, the first order correction to this result is given by the entanglement entropy of fields contained in a Cauchy slice of the spacetime. Assuming that this quantity vanishes,
\begin{equation}
    S(\Sigma\vert_{\mathcal{S}}) = 0 + O(G).
\end{equation}
This is consistent with the assumption that the holographic dual is in a pure state. In particular, we expect this result to hold to all orders. However, in radiation-dominated cosmologies, the purity assumption does not hold and we expect corrections at order $G^0$. This case was discussed in detail in Section $6$ of \cite{Franken:2023jas}. The resulting entanglement wedge spans all Cauchy slices that contain $\Sigma\vert_{\mathcal{S}}$:
\begin{equation}
    W(\Sigma\vert_{\mathcal{S}}) = D_{\rm L} \cup D_{\rm E} \cup D_{\rm R}.
\end{equation}
Thus, we argue that conjecture \ref{conj:double} applies to all closed FLRW cosmologies with $|\gamma|\geq 1$. 

Interestingly, this result holds regardless of the choice of screen in the contracting region. As discussed in the previous section, an infinite number of screens can be chosen in the contracting regions. This result suggests that these configurations are equivalent, at least in their capacity to encode complete regions of spacetime.

Finally, notice that the state on $\Sigma\vert_{\mathcal{S}}$ remains pure if it starts pure at past infinity. As we noted in Section~\ref{sec:FLRWholo}, the evolution of the two-screen system, along with the bulk cosmology, is not unitary, in the sense that the number of the underlying degrees of freedom and the dimensionality of the Hilbert space changes. It could be that this evolution is isometric~\cite{Cotler:2022weg, Cotler:2023eza} and linear, mapping pure states to pure states. As we have seen, this possibility is indeed consistent with the bilayer proposal. 

\subsection{The single-screen system in cosmologies with \texorpdfstring{\boldmath $|\gamma|\geq 1$}.}
\label{inglescreenbouncing}

We now proceed to consider a single-screen system $\Sigma\vert_{\mathcal{S}_{\rm L}}$ or $\Sigma\vert_{\mathcal{S}_{\rm L}}$. In the expanding phase, the screens must be placed on the apparent horizons. In the initial contracting phase, they can be pushed farther into the region of trapped surfaces between the apparent horizons, as explained in Section~\ref{sec:FLRWholo} and shown in Figure \ref{fig:FLRWscreen}. We will analyze the left screen subsystem $\Sigma\vert_{\mathcal{S}_{\rm L}}$, but all results and conclusions can be adapted to apply equally to the right screen $\Sigma\vert_{\mathcal{S}_{\rm R}}$.

Consider $A=\Sigma\vert_{\mathcal{S}_{\rm L}}$. The C-extremal surfaces in $D_{\rm L}$ and $D_{\rm R}$ are the empty surface. The resulting entanglement wedges in the left and right patches are
\begin{equation}
    W_{\rm L}(\Sigma\vert_{\mathcal{S}_{\rm L}}) = D_{\rm L}, \quad~ W_{\rm R}(\Sigma\vert_{\mathcal{S}_{\rm L}}) = \varnothing.
\end{equation}
As expected, the left holographic screen encodes its interior. The leading order contribution to $S(\Sigma\vert_{\mathcal{S}_{\rm L}})$ arises solely from the exterior region.

Since region~$D_{\rm E}$ has a boundary, the extremization of the area functional involves Lagrange multipliers and auxiliary fields, in order to impose that all homologous extremal surfaces lie in this diamond, including its boundary. There is no extremal surface in the bulk, except the bifurcate horizon. When $\eta_{\rm L}=|\gamma|\frac{\pi}{2}$, $D_{\rm L}$ reduces to the bifurcate horizon and the C-extremization is trivial. In every other case, the bifurcate horizon is not contained in $D_{\rm E}$, and the C-extremal surface cannot be an extremum of the area functional. This is another example  of a non-existence problem for extremal surfaces. Thus, from definition \ref{def:Rext}, the C-extremal surface must be a surface on $\partial D_{\rm E}$ that has the smallest area in $D_{\rm E}$.\footnote{In the general definition of C-extremal surfaces, we allow surfaces to only partially lie on the boundary. Here we assume that the spherical symmetry is not spontaneously broken by the extremization. Perfect spheres either lie in the interior of $D_{\rm E}$ or on its boundary.} The surface with smallest area in $D_{\rm E}$, denoted $M$, can be shown to lie on the top or bottom of the causal diamond $D_{\rm E}$, depending on whether $D_{\rm E}$ is in an expanding or contracting region. 
In both cases, the resulting homology region $\mathcal{C}_{\rm E}(\Sigma\vert_{\mathcal{S}_{\rm L}})$ is a null codimension-one hypersurface.  This is depicted in Figure~\ref{Penrose_gamma<-1}. In Appendix B of \cite{Franken:2023jas}, $M$ is shown to be the C-extremal surface using the method of Lagrange multipliers. As an example, the computation is done in de Sitter space ($\gamma=-1$) in Appendix \ref{app:Lagrange}. The computation easily generalizes to other cosmologies. 
\begin{figure}[ht]
\centering
\begin{subfigure}[t]{0.48\linewidth}
\centering
\includegraphics[width=0.4\linewidth]{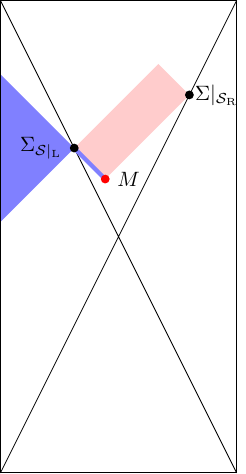}
\caption{\footnotesize $\gamma<-1$: At large times, the entropy of $\Sigma\vert_{\mathcal{S}_{\rm L}}$ saturates at a finite value.\label{fig:MExp}}
\end{subfigure}\hfill
\begin{subfigure}[t]{0.48\linewidth}
\centering
\includegraphics[width=0.43\linewidth]{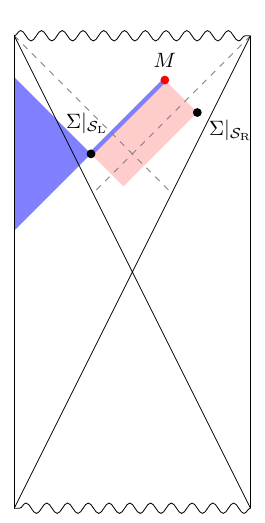}
\caption{\footnotesize $\gamma>1$: The leading order entropy of $\Sigma\vert_{\mathcal{S}_{\rm L}}$ goes to zero at a finite time.\label{fig:MCont}}
\end{subfigure}\hfill
    \caption{\footnotesize Penrose diagram for $\gamma<-1$ and $\gamma>1$. The dark blue region corresponds to the entanglement wedge of the single-screen system $\Sigma\vert_{\mathcal{S}_{\rm L}}$. $D_{\rm E}$ is red shaded. The C-extremal sphere, denoted by the red dot $M$, corresponds to the lower or upper vertex.}
    \label{Penrose_gamma<-1}
\end{figure}
In the expanding phase of bouncing cosmologies, as depicted in Figure~\ref{fig:MExp}, the entropy increases as $\eta_{\rm L}$ and $\eta_{\rm R}$ increase, and eventually saturates to a finite upper bound, as $\eta_{\rm L}, \eta_{\rm R} \to |\gamma|\pi$. Notice that this result holds despite the area of the screen $\Sigma\vert_{\mathcal{S}_{\rm L}}$  growing to infinity. At later times, the entanglement entropy is a small fraction of the maximal possible value.\footnote{The maximum value is the coarse grained entropy, equal to the area of the screen itself.} The extra degrees of freedom added to the screens as these evolve along the apparent horizons remain disentangled.  

In the contracting phase of Big Bang/Big Crunch cosmologies, as depicted in Figure \ref{fig:MCont}, the leading contribution to the entropy vanishes at a finite time. When $(\eta_{\rm L}+\eta_{\rm R})/2 < \gamma\pi$, the entanglement entropy is significantly large, of order $G^{-1}$, and decreasing. At $(\eta_{\rm L}+\eta_{\rm R})/2= \gamma\pi$ the minimal extremal surface $M$ hits the Big Crunch singularity. As a result, the part of the entanglement wedge of $\Sigma\vert_{\mathcal{S}_{\rm L}}$ in $D_{\rm E}$ is pinched to the singularity. The effective bridge connecting the screens closes off, leading to disentanglement (or a significant decrease in the entanglement entropy), well before the screens reach the Big Crunch singularity.  The entanglement entropy becomes small, of order $G^0$, even if the Universe has not yet collapsed. The leading classical contributions to the entropy remain vanishing for the rest of the evolution of the screens along their trajectories. Thus, the entropy at later times is dominated by the semiclassical contributions. 

The disentanglement phenomenon cannot occur when the two screens are in causal contact, namely when they lie in the overlap region of the pode and antipode causal patches. It is easy to see that in this case $M$ has to lie below the intersection of the event horizons delimiting the pode and antipode causal patches.

\subsection{The case of \texorpdfstring{\boldmath $1/(d-1)\leq |\gamma|< 1$}~ cosmologies}

Cosmologies with $1/(d-1)\leq |\gamma|< 1$ seem to escape our discussion. Indeed, the bulk picture suggests that the two screens cannot be in a pure state. The region between the cosmological horizons remains inaccessible to the observers at the pode and antipode at all times. Particles created at the Big Bang singularity in the past may reach the Big Crunch singularity without entering the pode and antipode causal patches and without crossing either of the screen trajectories along the cosmological horizons.  In order to describe the exterior region, we would need additional observers following trajectories in the exterior region and the corresponding holographic screens. Therefore, we do not have evidence that the degrees of freedom in the exterior region can be encoded holographically on a two-screen system at the cosmological horizons. Moreover, $\Sigma_{\rm L}\cup \Sigma_{\rm R}$ never amounts to a complete bulk Cauchy slice. It seems plausible that the bilayer prescription, which requires both exterior and interior contributions, is not applicable in these cosmological cases, even though we do not have direct evidence for this from a replica bulk path integral computation. Indeed, applying this proposal blindly for these cases leads to paradoxical results from the point of view of bulk causality, as it would seem to imply that the entanglement wedge of the two screen system comprises complete bulk Cauchy slices. 

Rather, the holographic entanglement entropy prescription could involve contributions from the two interior regions only. For the single screen subsystem, the C-extremal surface in region L is the empty surface, implying that the leading classical geometrical contribution to the entanglement entropy vanishes. The entanglement entropy should arise from semiclassical entropy contributions associated with bulk field degrees of freedom on Cauchy slices connecting the pode and the screen. Indeed, we can imagine entangled spin pairs, which start in the early past and then separate further, entering the pode and antipode causal patches, respectively. As a result, there can be non-trivial quantum entanglement between the two causal patches. Notice that this entanglement entropy is significantly smaller than the thermodynamic entropy for a single screen subsystem, which is $O(1/G)$. 
Similar conclusions hold for the two-screen system. The leading geometrical entropy is zero but there are non-vanishing contributions at order $G^0$. Since the entanglement wedge does not extend into the exterior region, the semiclassical entropy associated with the two-screen system cannot be zero, showing that the screens are not in a pure state. 

\subsection{Summary}

To summarize, we argued that closed FLRW cosmologies can be holographically described by a pair of holographic screens located on timelike hypersurfaces in the causal patches of two antipodal observers. In the expanding phase of the geometry, the screens are located along the apparent horizons of the two observers. In the contracting phase, there is an infinite number of alternative configurations where the screen of each observer lies between its apparent and cosmological horizons. We argue in the next section that spacelike slices of these different screens can be grouped into equivalence classes. In each class, the entanglement wedge and fine-grained entropy of every screen are identical. Moreover, each class contains a ``parent screen'' located on the apparent horizons, that has the most degrees of freedom and contains information unnecessary to bulk reconstruction.

The entanglement entropy between the two screens can be directly related to the geometry of an effective geometrical bridge in the exterior region. The effective geometrical bridge is the union of the two lightlike homology regions $\mathcal{C}_{\rm E}(\Sigma\vert_{\mathcal{S}_{\rm L}})$ and $\mathcal{C}_{\rm E}(\Sigma\vert_{\mathcal{S}_{\rm R}})$, which connect the screens with the C-extremal surface $\gamma_{\rm c}(\Sigma\vert_{\mathcal{S}_{\rm L}},D_{\rm E})=M$. It corresponds to the limiting slice on $D_{\rm E}$, which contains the smallest bottleneck $M$. In the Big Bang/Big Crunch cases, the bottleneck pinches and closes off, precisely when the two screen subsystems effectively disentangle. The presence of a complete geometrical bridge between the screen subsystems is a manifestation of the ER=EPR paradigm in a cosmological, time-dependent setting.

The entropy of the left screen is plotted in Figure~\ref{S_AH}, for three-dimensional cosmological spacetimes ($d=2$), and for several values of $w$ in the range $[-1,1]$. For simplicity, we consider the case where the screen trajectories coincide with the apparent horizons, and take the screen conformal times to be equal, $\eta_{\rm L}=\eta_{\rm R}$.
\begin{figure}[ht]
    \centering
\includegraphics[width=0.7\linewidth]{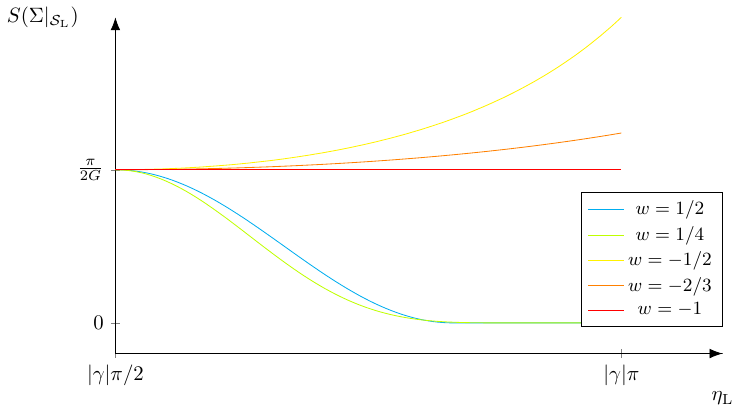}
\caption{\footnotesize Entropy of a single screen subsystem as a function of the conformal time, for $d=2$ and various values of $w$. The two screens are placed on the apparent horizons and taken to lie at equal conformal times. The entropy is rescaled with respect to $|\gamma|\pi$.}
    \label{S_AH}
\end{figure}

As we already remarked, the coarse-grained entropy of a single screen subsystem is given by the area of the screen in Planck units. This suggests that the evolution of the holographic dual to a generic closed FLRW universe is not unitary, a very interesting possibility that has been studied recently in \cite{Cotler:2022weg, Cotler:2023eza}. In this chapter, we found that the entanglement entropy to leading order is less than the coarse-grained entropy (given by the area divided by $4G$). In particular, the difference between the two values increases as $|\eta_{\rm L}+\eta_{\rm R}|$ increases, and goes to zero as $\eta_{\rm L}+\eta_{\rm R}\rightarrow 0$. This suggests that the state of a screen $\Sigma\vert_{\mathcal{S}_{\rm L}}$ (or $\Sigma\vert_{\mathcal{S}_{\rm R}}$) is thermal on a time-symmetric slice and the states evolve away from the thermal state as $|\eta_{\rm L}+\eta_{\rm R}|$ increases. It would be interesting to test this in explicit conjectures of holographic model for FLRW spacetimes.

\section{Equivalence classes of holographic screens}

As we already remarked in Section~\ref{sec:FLRWholo}, there are infinitely many possible choices of trajectories for the two screens $\Sigma\vert_{\mathcal{S}_{\rm L}}$ and $\Sigma\vert_{\mathcal{S}_{\rm R}}$ in the contracting phase of the cosmologies with $\gamma<-1$, $\gamma > 1$, leading to different holographic constructions. However, by examining the full two-screen and single screen subsystems, we find evidence suggesting that these screen configurations can be grouped into equivalence classes. In each class, the effective holographic theory on the screens can be derived from a pair of ``parent'' screens on the apparent horizons of the pode and antipode, yielding identical predictions for the entropy of certain gravitational bulk systems, which can be reconstructed from the single screen subsystems. To arrive at these conclusions, consider a Cauchy slice $\Sigma$ of the foliation $\mathcal{F}$, where $\Sigma\vert_{\mathcal{S}_{\rm L}}$ is in the purple region in Figure \ref{pushback}.
\begin{figure}[ht]
\begin{subfigure}[t]{0.48\linewidth}
\centering
\includegraphics[width=3cm]{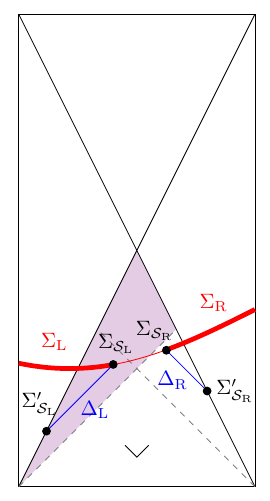}
\caption{\footnotesize Case of a bouncing cosmology, $\gamma\le -1$.\label{puschbackA}}
\end{subfigure}
\quad \,
\begin{subfigure}[t]{0.48\linewidth}
\centering
\includegraphics[width=3cm]{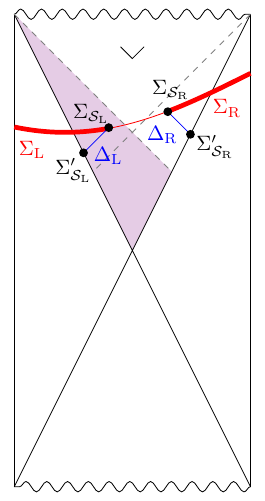}    
\caption{\footnotesize Case of a Big Bang/Big Crunch cosmology, \mbox{$\gamma\ge 1$}.\label{puschbackB}}
\end{subfigure}
    \caption{\footnotesize  The screens $\Sigma\vert_{\mathcal{S}_{\rm L}}$ and $\Sigma\vert_{\mathcal{S}_{\rm R}}$ can be moved along the lightlike slices $\Delta_{\rm L}$ and $\Delta_{\rm R}$, which are parallel to the $x^+$ and $x^-$ axes in the region of trapped surfaces. The ``parent screens'' $\Sigma'\vert_{\mathcal{S}_{\rm L}}$, $\Sigma'\vert_{\mathcal{S}_{\rm R}}$ are located on the spheres of greatest areas along $\Delta_{\rm L}$, $\Delta_{\rm R}$, respectively, which lie on the apparent horizons.\label{pushback}}
\end{figure}

Let us define $\Delta_{\rm L}$ as the null hypersurface that emanates orthogonally from $\Sigma\vert_{\mathcal{S}_{\rm L}}$ towards the past and the apparent horizon of the pode. Likewise, $\Delta_{\rm R}$ is a null hypersurface emanating orthogonally from $\Sigma\vert_{\mathcal{S}_{\rm R}}$ towards the past and the apparent horizon of the antipode. All slices of $\Delta_{\rm L}$ and $\Delta_{\rm R}$ are S$^{n-1}$ spheres with different areas. If we displace the screens $\Sigma\vert_{\mathcal{S}_{\rm L}}$ and $\Sigma\vert_{\mathcal{S}_{\rm R}}$ to any spheres of $\Delta_{\rm L}$ and $\Delta_{\rm R}$, respectively, we obtain the same result for the fine-grained entanglement entropy between the two screens, irrespective of their locations on $\Delta_{\rm L}$ and $\Delta_{\rm R}$. Indeed, the exterior region associated with any surface on $\Delta_{\rm L}$ contains the exterior region $D_{\rm E}$ associated with $\Sigma\vert_{\mathcal{S}_{\rm R}}$. In particular, the minimal surface $M$ is contained in the exterior region associated with any choice of screen along $\Delta_{\rm L}$. Hence, the C-extremal surface is $M$ for any such screen. Moreover, including higher order corrections requires adding the semiclassical entropy on the union of $\Sigma_{\rm L}$ with the null hypersurface $\Delta_{\rm L}$ between the screen and $M$. The resulting slices for any two screens in $\Delta_{\rm L}$ are unitarily related, so that their entropies are equal. Hence, the equivalence holds at the level of quantum corrections. Based on these observations, we define the following equivalence class.

\textit{ In the contracting phase of a cosmology, two screens $\Sigma\vert_{\mathcal{S}}$ and $\tilde{\Sigma}\vert_{\mathcal{S}}$ are equivalent if there exist $\Sigma'\vert_{\mathcal{S}}$ on the apparent horizon such that $\Sigma\vert_{\mathcal{S}_{\rm L}}$ and $\tilde{\Sigma}\vert_{\mathcal{S}_{\rm L}}$ belong to the hypersurface $\Delta_{\rm L}$ emanating from $\Sigma'\vert_{\mathcal{S}_{\rm L}}$, and similarly for $\Sigma\vert_{\mathcal{S}_{\rm R}}$ and $\tilde{\Sigma}\vert_{\mathcal{S}_{\rm R}}$.}

However, $\Delta_{\rm L}$ and $\Delta_{\rm R}$ have positive expansion.\footnote{Note that these null hypersurfaces may be past or future directed. We define the affine parameter along the hypersurface to grow away from $\Sigma\vert_{\mathcal{S}_{\rm L}}$ or $\Sigma\vert_{\mathcal{S}_{\rm R}}$.} Therefore, the intersections of $\Delta_{\rm L}$ and $\Delta_{\rm R}$ with the apparent horizons have the greatest areas, and thus the greatest coarse-grained entropy. We locate alternative screens $\Sigma'\vert_{\mathcal{S}_{\rm L}}$ and $\Sigma'\vert_{\mathcal{S}_{\rm R}}$ on these two surfaces, as in Figure \ref{pushback}. This construction suggests that the effective holographic theory on $\Sigma\vert_{\mathcal{S}_{\rm L}}\cup \Sigma\vert_{\mathcal{S}_{\rm R}}$ can be obtained by integrating out some degrees of freedom of the ``parent'' system $\Sigma\vert_{\mathcal{S}_{\rm L}}'\cup\Sigma\vert_{\mathcal{S}_{\rm R}}'$. This is reminiscent of an RG flow in local quantum field theories.\footnote{However, as in the de Sitter case $\gamma=-1$, we do not expect the holographic dual theory to these closed FLRW cosmologies to be local \cite{Shaghoulian:2021cef, Shaghoulian:2022fop, Franken:2023pni}.} 

Having argued that the screens may be grouped into equivalence classes, characterized by an identical prediction of the holographic entanglement entropy between the screens, we now show that the bulk region encoded in screens belonging to the same equivalence class is identical. Let us focus on $\Sigma\vert_{\mathcal{S}_{\rm L}}$ and the parent screen $\Sigma'\vert_{\mathcal{S}_{\rm L}}$. In the exterior region, the entanglement wedge of $\Sigma\vert_{\mathcal{S}_{\rm L}}$ consists of the lightlike segment between $\Sigma\vert_{\mathcal{S}_{\rm L}}$ and $M$, denoted by $\Delta_{\Sigma\vert_{\mathcal{S}_{\rm L}}M}$. See Figure~\ref{entanglementwedgesb}. The part of the entanglement wedge of $\Sigma'\vert_{\mathcal{S}_{\rm L}}$ in $D_{\rm E}$ is the union of $\Delta_{\rm L}$ and $\Delta_{\Sigma\vert_{\mathcal{S}_{\rm L}}M}$ -- see Figure~\ref{entanglementwedgesa}.

\begin{figure}[ht]
\centering
\begin{subfigure}[t]{0.3\linewidth}
\centering
\includegraphics[width=3cm]{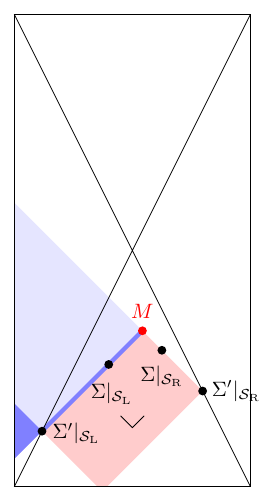}
\caption{\footnotesize ``Parent'' screen configuration. \label{entanglementwedgesa}}
\end{subfigure}\hfill
\begin{subfigure}[t]{0.3\linewidth}
\centering
\includegraphics[width=3cm]{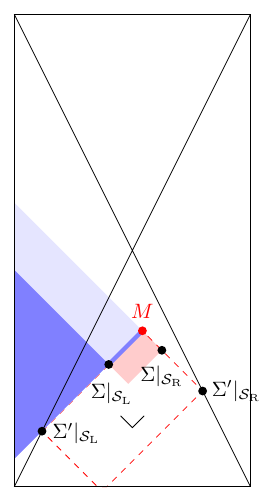}
\caption{\footnotesize A ``daughter'' screen configuration.\label{entanglementwedgesb}}
\end{subfigure}\hfill
\begin{subfigure}[t]{0.3\linewidth}
\centering
\includegraphics[width=3cm]{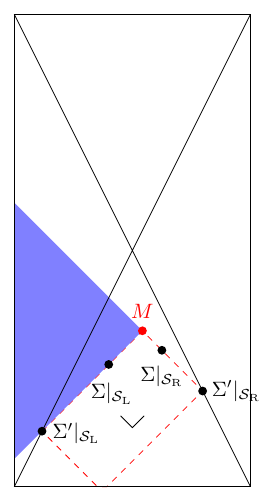}
\caption{\footnotesize Minimal screen configuration.\label{entanglementwedgesc}}
\end{subfigure}
\caption{\footnotesize Several screen configurations in a given equivalence class, defined by a pair of screens $\Sigma'\vert_{\mathcal{S}_{\rm L}}$ and $\Sigma'\vert_{\mathcal{S}_{\rm R}}$ on the apparent horizons of the pode and the antipode. The light red shaded region is $D_{\rm E}$, and the dark blue region is the entanglement wedge of the single left-screen system. The light blue region is the causal completion of the entanglement wedge.}\label{entanglementwedges}
\end{figure}

In fact, as shown in Figure~\ref{entanglementwedges}, the full entanglement wedge of $\Sigma\vert_{\mathcal{S}_{\rm L}}$ is larger and contains the entanglement wedge of the parent screen $\Sigma'\vert_{\mathcal{S}_{\rm L}}$. However, since there is unitary evolution among the slices in the entanglement wedge of $\Sigma\vert_{\mathcal{S}_{\rm L}}$, the state on them can be determined from the state on a slice in the entanglement wedge of the parent screen $\Sigma'\vert_{\mathcal{S}_{\rm L}}$ (since the latter lies in the entanglement wedge of $\Sigma\vert_{\mathcal{S}_{\rm L}}$). The converse is also true. The state on slices in the entanglement wedge of the parent screen can be determined from the state on the slices in the entanglement wedge of the ``daughter'' screen.\footnote{This remark is reminiscent of the comment below the definition of entanglement wedge \ref{def:EW}, where we noted that the bulk region dual to a subsystem $A$ of the screen should technically be the causal completion of $W(A)$.} In other words, the causal completion of the entanglement wedge is the same for all screens in the same equivalence class. Therefore, the extra degrees of freedom that are integrated out in order to obtain the effective holographic theory on the ``daughter'' screen are not really needed for the reconstruction of the bulk state on these slices. In particular, in contracting regions, the spacelike projection theorem implies that the area of the sphere $M$ divided by $4G$ already provides an upper bound on the coarse-grained entropy on Cauchy slices in the entanglement wedge of the ``parent'' and (any of) the ``daughter'' screen(s). 

One may choose to place both screens at $M$. The entanglement wedge of the left screen at $M$ now coincides with the causal completion of the entanglement wedge of alternative screens in the equivalence class of $M$. Furthermore, this is the minimal configuration with the smallest possible number of degrees of freedom that are necessary to reconstruct the dual bulk subsystems. The density matrix in this minimal dual holographic theory thermalizes since the von Neumann entropy is of the same order as the area of the screen in Planck units—the latter sets the number of degrees of freedom in the holographic theory. For the rest of the configurations, the larger density matrix should have many zero eigenvalues, with the non-trivial block of non-zero eigenvalues corresponding to the subspace associated with the minimal screen $M$, reflecting the fact that the additional number of degrees of freedom do not participate in the entanglement pattern with the right screen.

\section{C-extremization vs maximin}
\label{sec:discussion}

The reader familiar with holographic entanglement entropy might be bothered by our result that constrained-extremization leads to the entropy being computed by a minimin surface.\footnote{See Lemma \ref{lem:minimin}: A C-extremal surface that is not extremal in the usual sense is a minimin surface, obtained by finding the minimal area surface on every slice, and taking the minimal area surface among the minimal area surfaces of every slice.} This is quite unconventional. In particular, one might ask why the entropy is not computed by maximin surfaces. Note that the screens $\Sigma\vert_{\mathcal{S}_{\rm L}}$ and $\Sigma\vert_{\mathcal{S}_{\rm R}}$ are themselves maximin surfaces in $D_{\rm E}$, and saddle points along the boundary $\partial D_{\rm E}$. The entropy of the screen being computed by their area would be consistent with the de Sitter case, making this alternative proposal appealing.

First of all, recall that while $\Sigma\vert_{\mathcal{S}_{\rm L}}$ and $\Sigma\vert_{\mathcal{S}_{\rm R}}$ are maximin surfaces, they are not extremal. Thus, there cannot be a complete analogy with AdS extremal surface prescriptions. Moreover, the results of \cite{Hao:2024nhd} show that explicit computations in JT gravity find that entanglement entropy in de Sitter space is not computed by finding extremal surfaces but by globally minimizing the area functional over a domain. This is consistent with C-extremization (although the setups differ) and we expect this result to remain valid in more general closed geometries, such as closed FLRW metrics.

Second, a maximin procedure leads to unexpected results. Indeed, following a maximin procedure would identify the entropy of a screen with its area divided by $4G$. In pure de Sitter space, this is expected. Indeed, the Gibbons-Hawking entropy \eqref{eq:GH} associates an entropy to the horizon, equal to its area divided by $4G$. The de Sitter static patch is in a thermal state, where the fine-grained entropy saturates its upper bound, which is given by the coarse-grained entropy. Thus, when locating a holographic screen on the horizon, one expects that its fine-grained entropy—computed by a holographic entanglement entropy prescription—equals its coarse-grained entropy—computed by its area divided by $4G$. This is not true for other closed cosmologies.\footnote{Associating an entropy with the apparent horizon is supported by \cite{Cai:2006rs}, where the first law of black hole thermodynamics can be generalized to FLRW apparent horizons, with an entropy equal to the area divided by $4G$.} In particular, the interior of the apparent horizon of an observer is not time-independent and not in a thermal state. Therefore, the fine-grained entropy of $\Sigma\vert_{\mathcal{S}_{\rm L}}$ should be strictly smaller than its coarse-grained entropy.

\chapter{Causality on the holographic screen}
\label{ch:CWT}

To understand a holographic spacetime from a ultraviolet (UV) dual quantum theory, a natural approach is to identify a location for the dual degrees of freedom to live. In static patch holography, the holographic screen is placed away from the asymptotic boundary, on the cosmological horizon. Generally, a theory on such a screen is expected to be nonlocal, which is supported by the considerations of Section~\ref{sec:EWdS}, so causality must be treated carefully. Furthermore, a precise understanding of causality on the boundary/screen consistent with the bulk causality is essential to resolve various puzzles related to holographic entanglement entropy such as a violation of subadditivity~\cite{Kawamoto:2023nki,Mori:2023swn,Grado-White:2020wlb,Lewkowycz:2019xse,Noumi:2025lbb}.

In Chapter~\ref{ch:CWTAdS}, we introduced the connected wedge theorem and argued that it must be valid in any holographic theory. In this chapter, we consider holographic scattering in the context of static patch holography and its extension to asymptotically de Sitter spacetime and bouncing FLRW cosmologies. To maintain consistency with the connected wedge theorem, we introduce the notion of induced causality on the screen and prove the connected wedge theorem in de Sitter holography using this notion of causality. We show that induced causality resolves the previous apparent contradiction and proves to be the minimal proposal in at least a number of non-trivial examples. This resolution leads to three important consequences: 1) An insight for the dS connected wedge theorem, presented and proven in Section~\ref{sec:theorem} for three-dimensional asymptotically dS spacetime. 2) Bulk local excitations in the interior of an observer's holographic screen, emanating from the screen, are not described by local operators on the screen. 3) A local excitation on the screen should be mapped to a local operator on $\mathcal{I}^{\pm}$, hinting at a relation between static patch holography and the dS/CFT correspondence. We focus our examples and discussions on asymptotically dS$_3$ space, although we expect our results to apply to bouncing ($\gamma<-1$) three-dimensional FLRW cosmologies.\footnote{The results of the following Chapter require spacetime to have an expanding phase in its future and a contracting phase in its past.}

Note that, in Chapter~\ref{ch:CWTAdS}, we implicitly assume unitarity on the screen/boundary in the quantum information argument for the connected wedge theorem. One may think this is not the case for a static patch of dS as it is a part of the global dS. This would be the case at least in the Bunch-Davies vacuum. We stress that, however, a causal patch may not have a timelike Killing vector in asymptotically dS. Nevertheless, the gravitational proof that we present later in Section~\ref{sec:theorem} works even in these cases by considering the causal region (timelike envelope) of the observer's worldline, as in Chapter~\ref{ch:FLRW}.

This chapter solely focuses on the causal patch of an observer, say the right (antipode) one. For simplicity, we denote by $\mathcal{S}$ the screen associated with the antipode observer. In paticular, we generalize Definition \ref{def:ap-sp} to include general asymptotically dS spacetimes and bouncing FLRW cosmologies.
\begin{definition}[Holographic screen]
    \label{def:ap}
    The holographic screen $\mathcal{S}$ associated with an observer is the codimension-one convex timelike boundary of a region in which all closed codimension-two surfaces for which null geodesics orthogonal to the surface and directed towards the observer's worldline are of non-positive expansion.
\end{definition}
This definition locates the holographic screen inside the apparent horizon of the observer.

\section{An apparent violation of the connected wedge theorem}
\label{sec:violation}

Let us consider the limiting case where the holographic screen $\mathcal{S}$ of an observer in pure dS${}_3$ is located on the cosmological horizon. We consider here an example of a $2$-to-$2$ scattering $c_1,c_2\rightarrow r_1,r_2$ with
\begin{align}
\label{eq:Pscat}
\begin{alignedat}{2}
    c_1 &= (-\pi/4,3\pi/4,\pi/2), &\quad c_2 &= (-\pi/4,3\pi/4,-\pi/2), \\
    r_1 &= (\pi/4,3\pi/4,0), &\quad r_2 &= (\pi/4,3\pi/4,\pi),
\end{alignedat}
\end{align}
in conformal coordinates $(\sigma,\theta,\varphi)$.  See Figure~\ref{fig:scat-scr} for a schematic picture. 
\begin{figure}[ht]
    \centering
    \includegraphics[width=0.6\linewidth]{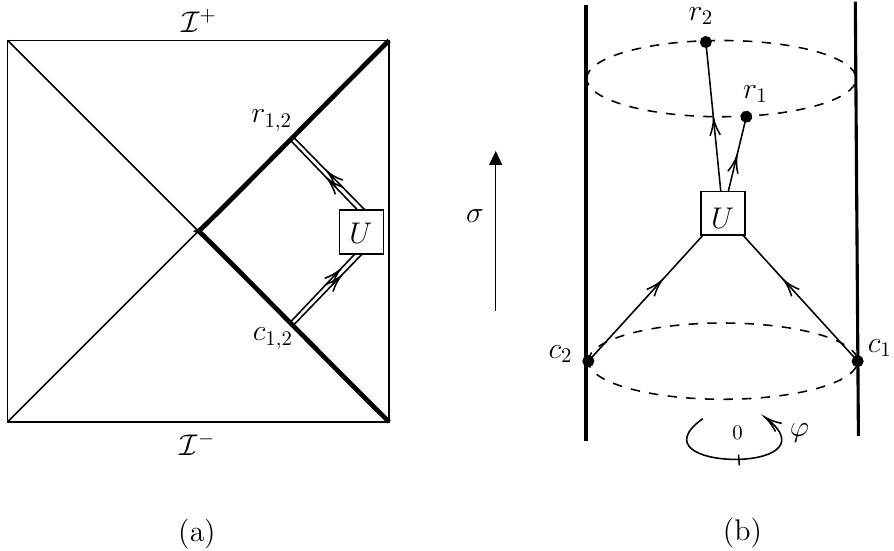}
    \caption{\footnotesize A schematic picture of a 2-to-2 scattering in the static patch of pure dS${}_3$, from input points $c_1,c_2$ to output points $r_1,r_2$ on the cosmological horizon. 
    The white box denoted $U$ corresponds to a unitary operation acting on the signal qubits in the bulk. In both pictures, $\sigma$ {indicates} the time direction. (a) Picture of the scattering in the Penrose diagram of de Sitter space. (b) Picture of the scattering in the static patch. 
    Each cosmological horizon (or stretched horizon) has the topology of a cylinder and the bulk (static patch) fills inside.}
    \label{fig:scat-scr}
\end{figure}
This scattering is possible in the bulk. In particular, the scattering can only occur at one point:
\begin{equation}
    J_{12\rightarrow 12} = J^+(c_1)\cap J^+(c_2) \cap J^-(r_1) \cap J^-(r_2) = (0,\pi/2,0).
\end{equation}
We would like to test if the connected wedge theorem (Theorem~\ref{th:CWT}) applies to this example of scattering in dS spacetime. For this, one computes the decision regions, where each decision region $\mathcal{R}_i$ is defined as an intersection on the screen $\mathcal{S}$ between the future domain of dependence of $c_i$ and the past domains of dependence of $r_1$ and $r_2$ for $i=1,2$.

To this end, one needs to identify the appropriate causality on the screen $\mathcal{S}$. 
One possibility would be the causality based on the induced metric on $\mathcal{S}$. In this case, the screen causality is completely blind to the holographic bulk. In the current limiting case, the discrepancy between the bulk causality and boundary causality is the highest, as $\mathcal{S}$ is located on the cosmological horizon, \textit{ie} the induced metric is
\begin{equation}
    ds^2_\mathcal{S}=d\varphi^2.
\end{equation}
Thus, under this causality, only light can propagate at a fixed angle $\varphi$. This suggests that the decision region is empty as the spatial location of each input and output point differs.\footnote{There is a subtlety at $\sigma=0$ as the cosmological horizon bifurcates. However, this subtlety can be avoided without significantly modifying the metric by replacing $\mathcal{S}$ with the stretched horizon. Thus, the conclusion should be the same.} Note that the decision region becomes nonempty only when the spatial location of one of the input points and both output points coincide. Even if this happens, the decision region is pointlike, so it leads to a significant violation of the connected wedge theorem. 

Another possibility is to consider the bulk causality restricted on the screen. In this case, each decision region is given by $J^+(c_{i})\vert_\mathcal{S}\cap J^-(r_1)\vert_\mathcal{S} \cap J^-(r_2)\vert_\mathcal{S}$ for $i=1,2$, where we define the lightcone of a point $p$ on the screen as
\begin{equation}
\label{eq:deflc}
    J^{\pm}\vert_{\mathcal{S}}(p) = J^{\pm}(p) \cap \mathcal{S},\quad p\in\mathcal{S}.
\end{equation}
To find the decision region, we compute the intersection between the boundary of the bulk lightcone of a point on the horizon and the cosmological horizon itself. The solution in conformal coordinates is
\begin{equation}
    \begin{cases}
        \begin{alignedat}{2}
            \varphi &= \varphi_0 &\quad &(\sigma\sigma_0 \geq 0)\\
            \tan\sigma &= -\cot\sigma_0\sin^2\left(\frac{\varphi-\varphi_0}{2}\right) &\quad &(\sigma\sigma_0 \leq 0)
        \end{alignedat}
    \end{cases}
    ,
    \label{eq:embed-sol}
\end{equation}
where $(\sigma_0,\varphi_0)$ are the coordinates of $p$ on the horizon. An example lightcone on the horizon for $\sigma_0> 0$ is pictured in Figure~\ref{fig:LC}.
\begin{figure}[ht]
    \centering
    \begin{subfigure}[t]{0.3\textwidth}
    \centering
    \includegraphics[width=5cm]{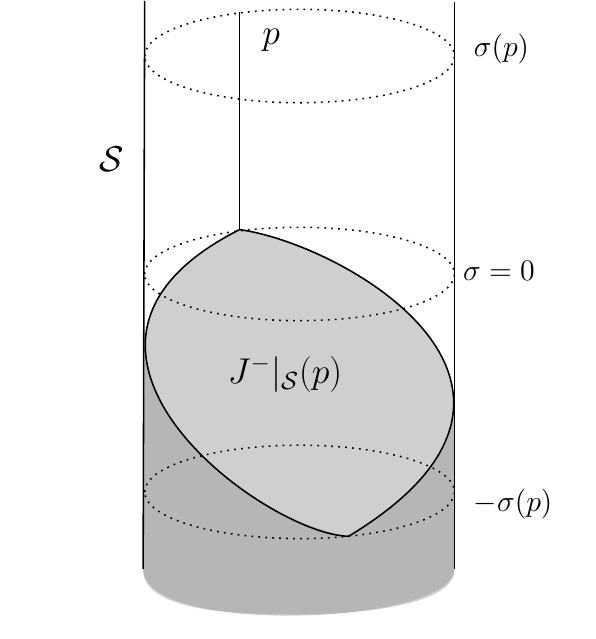}
    \caption{}
    \label{fig:LC}
    \end{subfigure} 
    \hfill
    \begin{subfigure}[t]{0.6\textwidth}
    \centering
    \includegraphics[width=10cm]{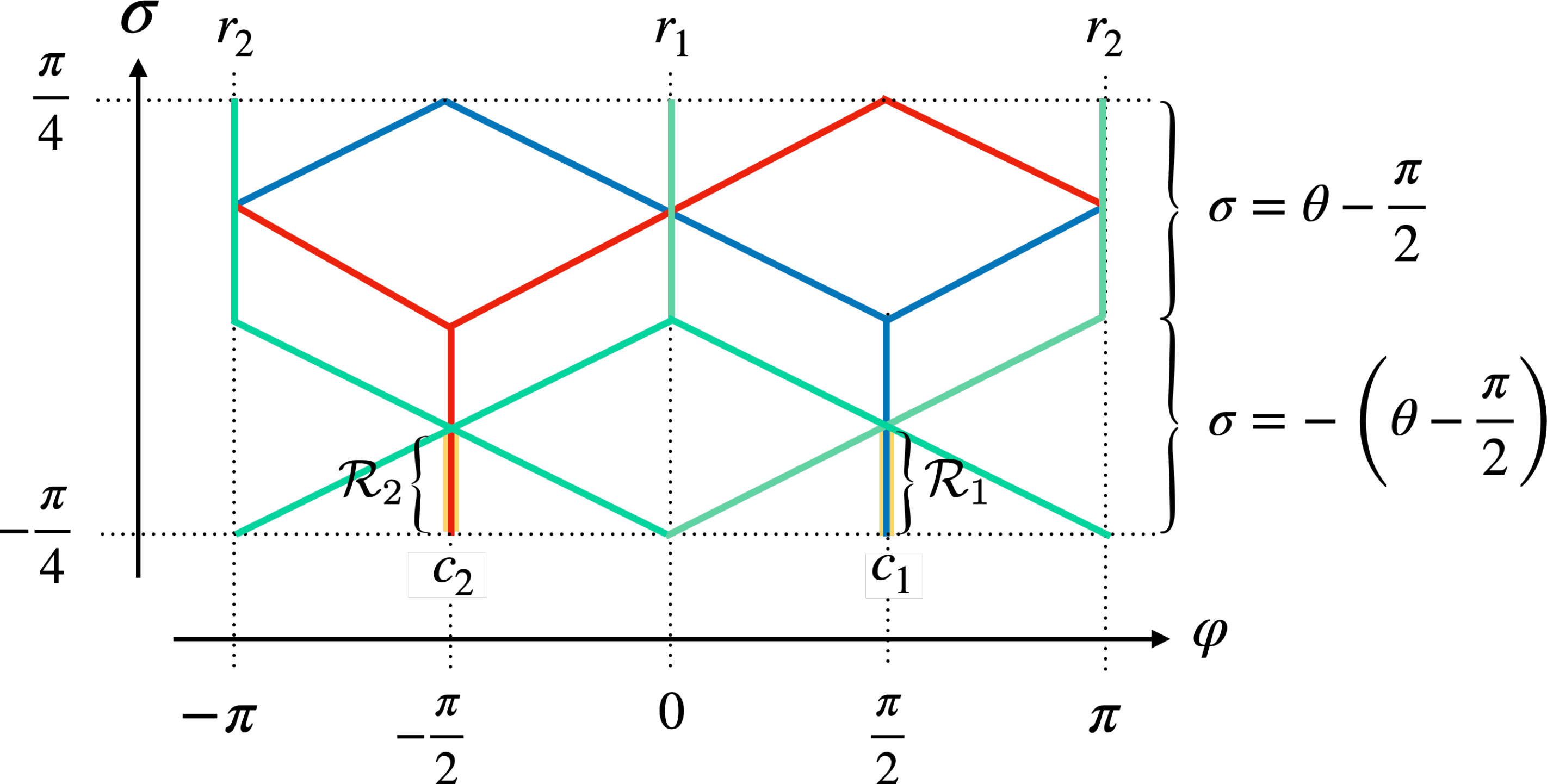}
    \caption{}
    \label{fig:decR}
    \end{subfigure}
    \caption{\footnotesize (a) Past lightcone (shaded in grey) of a point $p$ on the screen $\mathcal{S}$. The screen is located on the cosmological horizon, which has the topology of a cylinder $X_1^2+X_2^2=1$. The boundary of the lightcone consists of the union of the segment in the region $\sigma > 0$ with the ellipsoid curve in the region $\sigma \leq  0$. (b) Decision regions $\mathcal{R}_1$ and $\mathcal{R}_2$ for the scattering \eqref{eq:Pscat} constructed from lightcones on the screen $\mathcal{S}$. $\partial J\vert_{\mathcal{S}}^+(c_1)$ is denoted by a red line, $\partial J\vert_{\mathcal{S}}^+(c_2)$ is denoted by a blue line, and $\partial J\vert_{\mathcal{S}}^-(r_{1,2})$ are denoted by green lines. The decision regions have pointlike spatial sections, and therefore vanishing entanglement.}
\end{figure}
Using Equation \eqref{eq:embed-sol}, we find that the decision regions associated with the input and output points~\eqref{eq:Pscat} are two timelike segments (Figure~\ref{fig:decR}). If the connected wedge theorem is true in dS, the decision regions should be connected by an entanglement wedge, but this is not the case here. A spacelike slice of the decision regions reduces to a set of two points on the horizon. The holographic entanglement entropy of two pointlike regions is zero and the associated entanglement wedge vanishes. This is an explicit apparent counterexample to the connected wedge theorem in dS spacetime.
This presents a puzzle as the connected wedge theorem (Theorem~\ref{th:CWT}) is expected to hold in any holographic setup.

\section{Induced causality}
\label{sec:ind_LC}

{What needs to be modified for
the connected wedge theorem to hold?} Since we start from the semiclassical limit of the holographic duality, the existence of bulk scattering from bulk causality should be taken for granted.
On the other hand, there is room to change the causality and decision regions fixed from it on the screen where the dual theory lives. We implicitly assumed through the definition of the decision regions that a bulk signal emanating from or reaching a point on the screen is given by an excitation of a local operator on the screen. This is the case in AdS/CFT, where a bulk excitation can be created by a local operator on the asymptotic boundary~\cite{Nozaki:2013wia,Terashima:2023mcr,Tanahashi:2025fqi}.\footnote{In such a case, the lightcone associated with a localized perturbation propagating in the bulk and emanating from the screen can be shown to be the lightcone in induced coordinates.} It was noted in~\cite{Mori:2023swn} that this is not the case in AdS holography with braneworld or cutoff surfaces. 
They suggest that a correct boundary dual of a localized wave packet in the bulk is given by a local excitation on a fictitious boundary, so that it is effectively smeared and nonlocal on the brane/cutoff surface.\footnote{Quantum mechanically, the fictitious evolution is interpreted as a state preparation on the hypersurface by a finite Lorentzian time evolution. Alternatively, a fine-tuned causal structure from microscopic aspects of the dual theory may resolve the puzzle~\cite{Omiya:2021olc}. In this paper, we seek a resolution that only relies on the geometry, without such a fine tuning of the theory.}
This hints that we need to define another notion of causality on the non-asymptotic boundary, namely, the induced causality.

\subsection{Lightcones induced from the dS boundary \texorpdfstring{\boldmath $\mathcal{I}^{\pm}$}{} }

In this work, we follow the strategy of~\cite{Mori:2023swn} and consider `fictitious' local perturbations at the conformal boundaries of asymptotically de Sitter spacetime. We quote the word `fictitious' because in dS the screen is not a boundary, where spacetime terminates, so the conformal boundary is not fictitious, contrary to the previous work in AdS.
This is first motivated by the analogy with the case studied in~\cite{Mori:2023swn}, reviewed in Section~\ref{ch:CWTAdS}, and related to geometric properties of causal horizons, which will be made precise in Section~\ref{sec:theorem}. The second motivation is the dS/CFT correspondence~\cite{Strominger:2001pn}, which could provide a physical interpretation of the fictitious points providing the induced lightcones of perturbations on the holographic screen. 

Considering a $2$-to-$2$ scattering connecting points on the screen, $c_1,c_2,\in \mathcal{S} \rightarrow r_1,r_2 \in \mathcal{S}$, we define `fictitious' input and output points as follows.
\begin{definition}[`Fictitious'/tilded point]
\label{def:R}
    We define points on the conformal boundaries $\Tilde{c}_1,\Tilde{c}_2\in \mathcal{I}^-,\Tilde{r}_1,\Tilde{r}_2\in \mathcal{I}^+$ as points that are causally connected to the points $c_1,c_2,r_1,r_2$ on the screen $\mathcal{S}$. That is,
    \begin{equation}
        \tilde{c}_i \in J^-(c_i) \cap \mathcal{I}^- \quad,\quad \tilde{r}_i \in J^+(r_i) \cap \mathcal{I}^+.
    \end{equation}
\end{definition}
The tilded point associated with a point on the screen leads to the definition of an induced lightcone:
\begin{definition}[Induced causality]
\label{def:ind}
    The induced lightcone of a point $p\in\mathcal{S}$ induced from $\tilde p$ is
    \begin{equation}
        \hat J^{\pm}(p) = J^{\pm}(\tilde p) \cap \mathcal{S}.
    \end{equation}
\end{definition}
See Figure~\ref{fig:ind_lc} for a schematic example of the induced past lightcone.
\begin{figure}
    \centering

        \includegraphics[width=8cm]{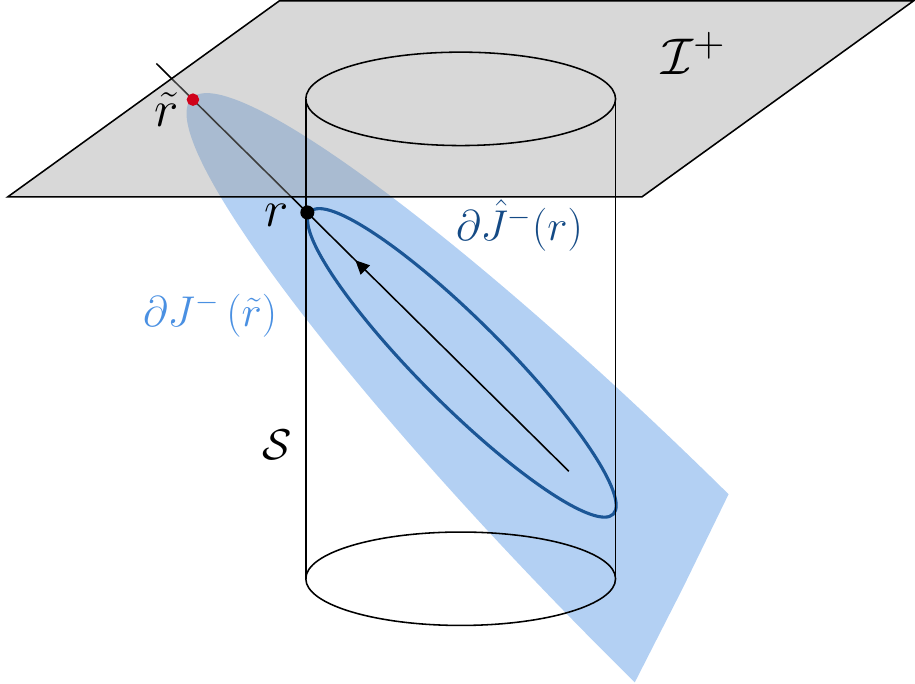}
    \caption{\footnotesize A bulk perturbation propagating towards the boundary is pictured by an arrow. Given an output $r_1$ on the holographic screen $\mathcal{S}$ (cylinder here), a point $\tilde r$ is defined on the future conformal boundary $\mathcal{I}^+$ (in grey). The boundary of the past induced lightcone of $r$ (in dark blue) is given by the intersection of the past lightcone of $\tilde{r}$ (light blue) with the screen. Similarly, by exchanging $+$ and $-$, future and past, one obtains a future induced lightcone.}
            \label{fig:ind_lc}
\end{figure}
For convenience, we define the induced lightcone in this way, however, due to the arbitrariness in choosing $\tilde{p}$, there are multiple alternative induced lightcones for a given point $p\in\mathcal{S}$. Most of these induced lightcones may look counterintuitive as these future/past induced lightcones can contain a point causally in the past/future of $p$ (in terms of the bulk causality), respectively. In general, $\hat{J}^{\pm}(p)$ may refer to any of them arbitrarily, and the proof of the connected wedge theorem is valid for any choice of $\tilde{p}$, as we will see in the next section. However, in some cases, it is useful to select a special induced lightcone such that it agrees with the conventional expectation for a definition of a lightcone, that is, no point causally in the past/future relative to a point $p$ should be present in the future/past lightcone of $p$. For this, it is useful to think of $\hat{J}^{\pm}$ as the most restrictive one. By the most restrictive, we mean the choice of point $\tilde p$ which minimizes the span of $\hat{J}^{\pm}$, in particular, such that the tip of the cone lies exactly at $p$, or as close as possible to it. In particular, one may replace Definition~\ref{def:ind} by
    \begin{equation}
        \hat{J}^{\pm}(p) = \min_{\tilde{p}} [J^{\pm}(\tilde{p})\cap \mathcal{S}],
    \end{equation}
    where we have defined the minimization so that the point $\tilde{p}=\tilde{p}_\ast$ found by the minimization satisfies $\hat J^{\pm}(\tilde{p}_\ast)\subseteq \hat J^{\pm}(\tilde{p}^\prime)$ for any $\tilde{p}^\prime$ satisfying Definition~\ref{def:R}.
    
We construct the induced lightcones on a screen $\mathcal{S}$ located on the stretched horizon of pure dS$_3$ defined by a fixed $r\leq 1$ coordinate, and induced from an arbitrary point $\tilde c\in \mathcal{I}^-$ or $\tilde r \in \mathcal{I}^+$. We denote their conformal coordinates as
\begin{equation}
    (\sigma_i,\theta_i,\varphi_i)=(k(\pi/2 - \epsilon_{dS}), \theta_i , \varphi_i),
\end{equation}
where $k=+1$ for $\tilde r$ and $k=-1$ for $\tilde{c}$, and $\epsilon_{dS}$ is the UV cutoff of $\mathcal{I}^{\pm}$. The boundary of the induced lightcone is given by the intersection of this hypersurface with the stretched horizon:
\begin{equation}
\label{eq:ind_lc}
    \tan\sigma = kr\frac{\cos(\varphi-\varphi_i)}{\sin\theta_i} \pm \cot\theta_i\sqrt{1-r^2\sin^2(\varphi-\varphi_i)}.
\end{equation}
This gives the formula for the induced lightcone on a screen at a constant $r$ in the static patch in the conformal coordinates $(\sigma,\theta=\pi-\arcsin(r\cos\sigma),\varphi)$ as a function of the {tilded} point $(\sigma_i=k(\pi/2-\epsilon_{dS}),\theta_i,\varphi_i)$ at $\mathcal{I}^\pm$.\footnote{We define the domain of the inverse sine function as $\arcsin x \in [-\pi/2,\pi/2]$.}

\subsection{Induced causality vs local causality}
\label{sec:stretched}

We highlighted in Section~\ref{sec:violation} that a naive approach to causality on the screen leads to results that contradict the connected wedge theorem. In Section~\ref{sec:ind_LC}, we introduced the notion of the induced lightcone, which we conjecture to be the right object to consider when computing the causal region of the screen operator encoding a localized wavepacket in the causal patch. We will further motivate this conjecture by showing that this prescription leads to a well-defined connected wedge theorem. In this section, we consider the effect of the location of the holographic screen on induced lightcones.

We construct the lightcone of a point $p$ on the holographic screen located at fixed $r$ in pure dS$_3$. The result for $r=1$ was given in Equation \eqref{eq:embed-sol}. The computation for $r<1$ is analogous, using the parametrization $X_3^2-X_0^2=1-r^2$, $X_1=r\cos\varphi$ and $X_2=r\sin\varphi$ of the fixed $r$ hypersurface in embedding coordinates. Combined with the bulk lightcone equation, one obtains the general solution for the lightcone of a point
\begin{equation}
    p:\, (\sigma_0,\theta_0=\pi-\arcsin(r\cos\sigma_0),\varphi_0)
\end{equation}
(in the conformal coordinates) on a screen $\mathcal{S}$ at a fixed $r$:
\begin{align}
\begin{split}
    \tan\sigma &= \frac{1}{1-r^2}\biggl[-\tan\sigma_0(1-r^2\cos(\varphi-\varphi_0))\biggr.\\
    &+\biggl.\sqrt{r^2(1-\cos(\varphi-\varphi_0))(1-r^2\cos(\varphi-\varphi_0)+1-r^2)(\tan^2\sigma_0+1-r^2)}\biggr].
    \end{split}
\end{align}
To compare this with the induced lightcone, we calculate the conformal coordinates of the induced point $\tilde{p}$ associated with $p$:
\begin{align}
\begin{split}
    \tilde{p}: \, (\pm(\pi/2-\epsilon_{dS}),\sigma_0+\arccos(r\cos\sigma_0),\varphi_0)).
    \end{split}
\end{align}
Applying Equation \eqref{eq:ind_lc} to $\tilde{p}$, we find the induced lightcone of $p$:
\begin{equation}
\begin{split}
    \tan\sigma =&
    kr\frac{\cos(\varphi-\varphi_0)}{\sin(\sigma_0+\arccos(r\cos\sigma_0))} \\
    &\pm \cot(\sigma_0+\arccos(r\cos\sigma_0))\sqrt{1-r^2\sin^2(\varphi-\varphi_0)}.
    \end{split}
\end{equation}
The lightcones from a point $p$ on $\mathcal{S}$ and the induced lightcones for the point $p$ are pictured in Figure~\ref{fig:comp} for different values of $r$. 

\begin{figure}
    \centering
      \includegraphics[width=0.9\linewidth]{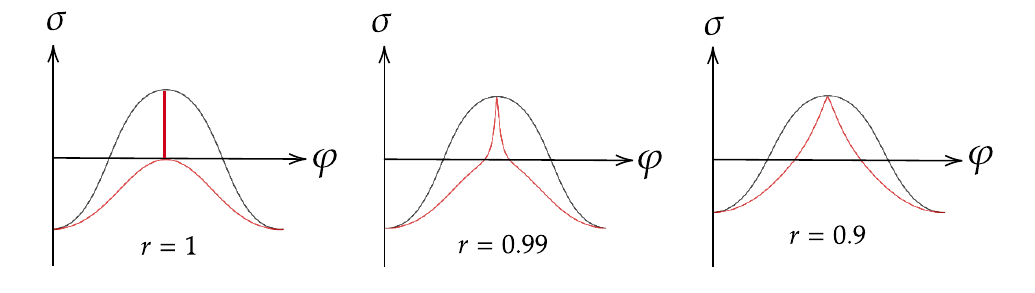}

    \includegraphics[width=0.9\linewidth]{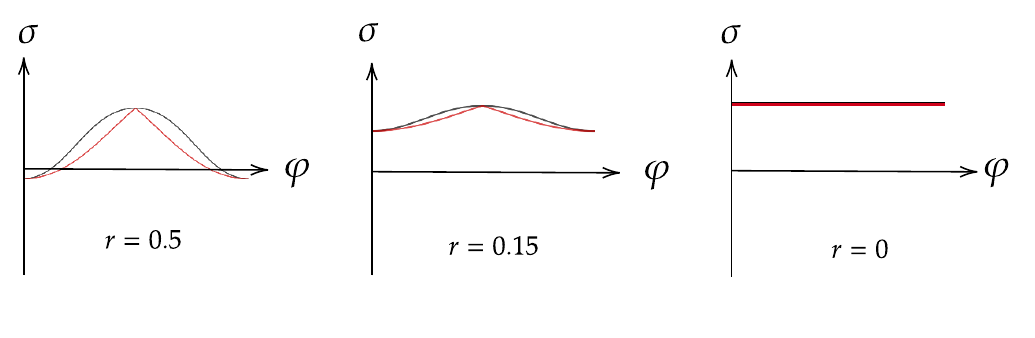}
    \caption{\footnotesize Induced lightcone $\hat{J}^-(p)=\min_{\tilde{p}} J^-(\tilde{p})\cap \mathcal{S}$ (in black) and lightcone from a point $p$ on the screen $J^-\vert_{\mathcal{S}}(p)=J^-(p)\cap\mathcal{S}$ (in red) located at the fixed $r=1,0.99,0.9,0.5,0.15,0$ hypersurface. As $r\rightarrow 0$, the lightcone increases in size, becoming a better approximation of the induced lightcone. Note that at $r=0$ there is no spatial extent in the $\varphi$ direction because it is a pode in dS. }
    \label{fig:comp}
\end{figure}
The induced lightcone $\hat{J}$ always contains the lightcone $J\vert_{\mathcal{S}}$. Moreover, the induced lightcone covers a considerably larger portion of the screen than the lightcone of a point when the screen is located close to the cosmological horizon. As $r\rightarrow 0$, the lightcone of a point spreads and approaches the induced lightcone. However, they do not converge, even to the first order in~$r$.

The induced lightcone is interpreted as the causal region associated with the nonlocal operator encoding the local perturbation in the bulk at $p$. The fact that it contains trajectories that are apparently superluminal is due to the nonlocality of this operator. 
On the other hand, the lightcone of a point $p$ is the set of points on the screen that are causally connected to a local operator at $p\in\mathcal{S}$ through the bulk. The difference between the induced lightcone and the lightcone of the point $p$ becomes smaller as $r\rightarrow 0$. We interpret this as a localization of the operator on the screen encoding the perturbation at $p$.

\section{Resolution of the apparent contradiction}
\label{sec:resolution}

We reconsider the example of the 2-to-2 scattering of Section~\ref{sec:violation}. We now apply the prescription of induced lightcones from $\mathcal{I}^{\pm}$ for a screen located at the cosmological horizon $r=1$. The equation of $\partial \hat{J}^{\pm}(p)$ \eqref{eq:ind_lc} simplifies to
\begin{equation}
\label{eq:ind}
    \tan\sigma = \left\{
    \begin{array}{ll}
        k\tan\frac{\theta_i}{2}\cos(\varphi-\varphi_i) & \mbox{if } \sigma k \geq 0 \\
        k\cot\frac{\theta_i}{2}\cos(\varphi-\varphi_i) & \mbox{if } \sigma k \leq 0 
    \end{array}
\right..
\end{equation}
We choose `fictitious' points $\tilde{c}_i$ and $\tilde{r}_i$ by merely extrapolating the lightlike signals emanating from $c_i$ or reaching $r_i$. These points are 
\begin{align}
\label{eq:fic_coord}
\begin{alignedat}{2}
    \Tilde{c}_1 &=(-\pi/2+\epsilon_{dS},\pi/2+\epsilon_{dS},\pi/2), &\quad \Tilde{c}_2 &=(-\pi/2+\epsilon_{dS},\pi/2+\epsilon_{dS},-\pi/2), \\
    \Tilde{r}_1 &=(\pi/2-\epsilon_{dS},\pi/2+\epsilon_{dS},0), &\quad \Tilde{r}_2 &=(\pi/2-\epsilon_{dS},\pi/2+\epsilon_{dS},\pi).
\end{alignedat}
\end{align}
It is easy to check that this choice of `fictitious' points is minimal, as $J^+(\tilde c_1)\cap J^+(\tilde c_2) \cap J^-(\tilde r_1) \cap J^-(\tilde r_2) = (0,\pi/2,0) = J_{12\rightarrow 12}$. We now modify the previous definition of decision regions to accommodate the notion of induced lightcones.
\begin{definition}
\label{def:decision}
    The (induced) decision regions $\mathcal{R}_1$ and $\mathcal{R}_2$ associated with the $2$-to-$2$ scattering $c_1,c_2\rightarrow r_1,r_2 \in \mathcal{S}$ are defined as
    \begin{equation}
    \mathcal{R}_i = \hat{J}^+(c_i)\cap \hat{J}^-(r_1) \cap \hat{J}^-(r_2).
    \end{equation}
Let us denote the largest spatial section of $\mathcal{R}_i$ by $\mathcal{V}_i$, such that $\mathcal{R}_i=\hat{D}(\mathcal{V}_i)$ where we defined $\hat{D}(\mathcal{V}_i)$ as the set of points $p\in\mathcal{S}$ such that $\mathcal{V}_i\in \hat{J}^-(p)\cup\hat{J}^+(p)$. 
\end{definition}
For the example \eqref{eq:Pscat}, we get the following edges of decision regions $\mathcal{R}_1$ and $\mathcal{R}_2$, as illustrated in Figure~\ref{fig:comparison} for an illustration.
\begin{figure}[ht]
    \centering
    \includegraphics[width=0.9\linewidth]{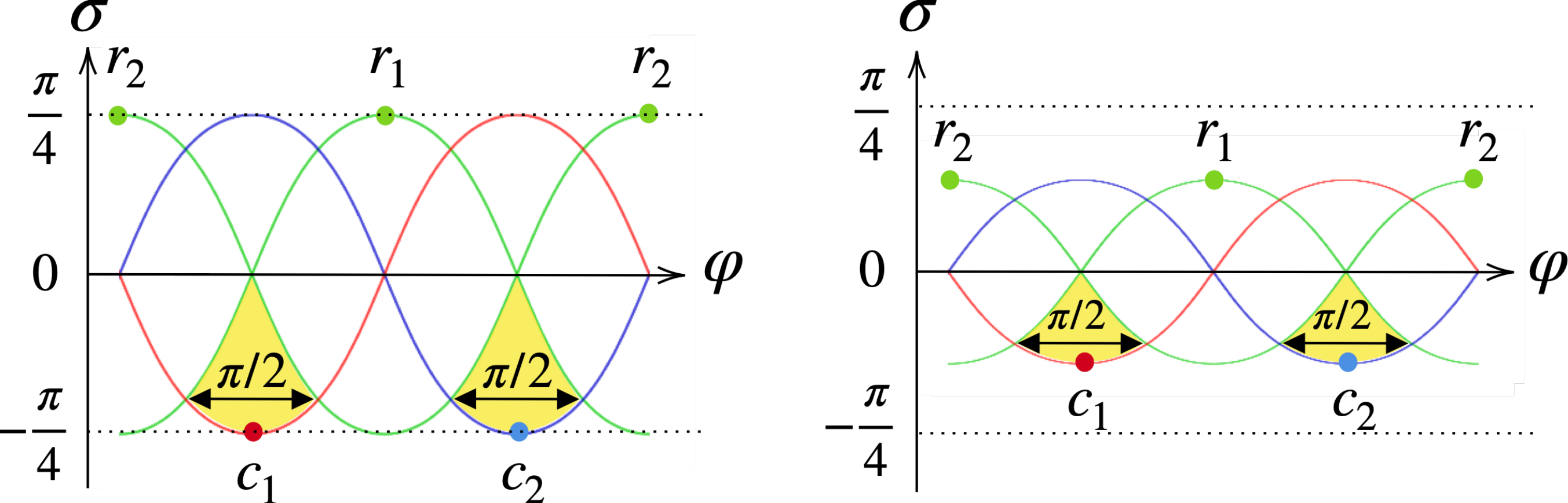}
    \caption{\footnotesize The decision regions $\mathcal{R}_{1,2}$ for the scattering \eqref{eq:Pscat} on a screen located on the cosmological horizon (left diagram) and on a stretched horizon located at $r=0.5$ (right diagram) extend over $\pi/2$ when their lightcones are induced from the input/output points on the null infinities. $\partial \hat{J}^+(\tilde{c}_1)$ is denoted by a red line, $\partial \hat{J}^+(\tilde{c}_2)$ is denoted by a blue line, and $\partial \hat{J}^-(\tilde{r}_{1,2})$ are denoted by green lines.}
    \label{fig:comparison}
\end{figure}
\begin{align}
\begin{split}
\label{eq:V1V2}
  \mathcal{V}_1:\quad & \sigma=-\arctan(1/\sqrt{2})~\quad \varphi\in [\pi/4,3\pi/4],\\
  \mathcal{V}_2: \quad& \sigma=-\arctan(1/\sqrt{2}) ~\quad \varphi\in [-3\pi/4,-\pi/4].
  \end{split}
\end{align}
Under the induced screen causality, a direct scattering on $\mathcal{S}$ is not possible, as $\mathcal{R}_1\cap\mathcal{R}_2=\varnothing$. Now that $\mathcal{V}_1$ and $\mathcal{V}_2$ are specified, their entanglement entropy can be computed. As reviewed in Chapter~\ref{ch:bridge}, the entanglement entropy of a spatial subsystem of the screen $\mathcal{S}$ has three contributions. The entropy is given by the sum of the areas of the homologous minimal extremal surfaces 1) in the interior of $\mathcal{S}$, 2) in the interior of a complementary screen associated with an antipodal observer, and 3) in the exterior region bounded by the union of these two screens. These three regions are illustrated in gray in Figure~\ref{fig:qucorr}. As we focus on  a specific patch, we simplify the notation and denote by $D$ the causal diamond inside $\mathcal{S}$ in which we extremize the area.\footnote{In the interior region $D$, extremal surfaces are C-extremal surfaces and conversely, see Appendix \ref{app:proofs}.}

There are two candidate extremal surfaces of $\mathcal{V}_1\cup\mathcal{V}_2$ in the interior of $\mathcal{S}$. The first one, associated with a disconnected entanglement wedge, is the union of the extremal surfaces for $\mathcal{V}_1$ and $\mathcal{V}_2$, namely, 
\begin{equation}
    \gamma_{\rm e}^{dis}(\mathcal{V}_1\cup\mathcal{V}_2;D) = \gamma_{\rm e}(\mathcal{V}_1;D) \cup \gamma_{\rm e}(\mathcal{V}_2;D),
\end{equation}
where $\gamma_{\rm e}(\mathcal{V}_i;D)$ is the geodesic anchored to $\mathcal{V}_i$ in the region inside $\mathcal{S}$, whose length is $\min(\Delta\varphi,2\pi-\Delta\varphi)$ where $\Delta\varphi$ is the angle of the arc.\footnote{See Section~\ref{subsec:arc} or appendix B.1 of~\cite{Franken:2023pni} for a detailed computation.} 

The second candidate in the interior of $\mathcal{S}$ is the extremal surfaces associated with the complement of $\mathcal{V}_1\cup\mathcal{V}_2$ on $\mathcal{S}$. These extremal surfaces are homologous to $\mathcal{V}_1\cup\mathcal{V}_2$, since the union of $\mathcal{V}_1\cup\mathcal{V}_2$ with its complement defines a slice of~$\mathcal{S}$. The associated entanglement wedge is connected through the interior of $\mathcal{S}$, namely,
\begin{equation}
    \gamma_{\rm e}^{con}(\mathcal{V}_1\cup\mathcal{V}_2;D) = \gamma_{\rm e}^{dis}((\mathcal{V}_1\cup\mathcal{V}_2)^c\vert_{\mathcal{S}};D),
\end{equation}
where $(\cdot)^c\vert_{\mathcal{S}}$ represents a codimension-two complementary subregion on an achronal surface on the (single) screen containing the subregion.

The exterior contribution comes from the arcs themselves, as the existence of the complementary screen prevents $\mathcal{V}_1\cup\mathcal{V}_2$ from being homologously connected to their complement $(\mathcal{V}_1\cup\mathcal{V}_2)^c\vert_{\mathcal{S}}$. Consequently, due to the homology condition, $W_{\rm E}(\mathcal{V}_1\cup\mathcal{V}_2)$ is always disconnected. 

Finally, $\mathcal{V}_1\cup\mathcal{V}_2$ does not extend on the complementary screen, so there is no entropy contribution from the complementary patch. Thus, the entanglement wedge of $\mathcal{V}_1\cup\mathcal{V}_2$ can only be connected through the interior of $\mathcal{S}$, if $\gamma_{\rm e}^{con}$ is smaller than $\gamma_{\rm e}^{dis}$. 

Considering the scattering \eqref{eq:Pscat} and its associated induced decision regions \eqref{eq:V1V2}, we find
\begin{equation}
    {\rm Area}(\gamma_{\rm e}^{dis}(\mathcal{V}_1\cup\mathcal{V}_2;D_L)) -{\rm Area}(\gamma_{\rm e}^{con}(\mathcal{V}_1\cup\mathcal{V}_2;D_L))= 0.
\end{equation}
In other words, the scattering we considered corresponds to the transition case where the disconnected and connected entanglement wedges are equivalent. Even though we considered input and output points on a screen located on the cosmological horizon, this result would have been identical for any set of input/output points on an arbitrary stretched horizon at fixed radius $r$ such that $\tilde{c}_i$ and $\tilde{r}_i$ of Equation \eqref{eq:fic_coord} are their associated set of {tilded} points on $\mathcal{I}^{\pm}$. Indeed, for these {tilded} points, Equation \eqref{eq:ind_lc} becomes linear in $r$, such that the size of $\mathcal{V}_1$ and $\mathcal{V}_2$ is independent of $r$. This is depicted in Figure~\ref{fig:comparison}. Hence, the fact that we found an exact match between the lengths of the connected and disconnected geodesics for these induced points is not specific to a screen on the horizon, but a general property of these `fictitious' points that are associated with a $J_{12\rightarrow 12}$ reducing to a point. This is consistent with the connected wedge theorem, and the typical behavior of pointlike holographic scatterings where the lengths of the connected and disconnected geodesics are equivalent~\cite{May:2019odp, May:2019yxi, May:2021nrl, May:2022clu}. This example thus provides evidence that the induced lightcones defined in Section~\ref{sec:ind_LC} are the right objects to consider when studying causality on the holographic screen. In particular, this motivates the idea that the induced lightcone is the minimal definition of lightcone consistent with the connected wedge theorem.\footnote{Proving this would be an interesting future direction.}

\section{Proof of the de Sitter connected wedge theorem}
\label{sec:theorem}

Inspired by the definition of the induced lightcone from points at $\mathcal{I}^{\pm}$, we generalize the result obtained in the previous section. The proof of our statement closely follows that of~\cite{May:2019odp, Mori:2023swn} and relies on the static patch holographic conjecture (Conjecture~\ref{conj:SPH}) and its extension to asymptotically de Sitter space and closed FLRW cosmologies, as well as the focusing theorem and the second law for causal horizons.\footnote{When we also include quantum matter corrections, these theorems are replaced by the restricted quantum focusing conjecture and the generalized second law.} The statement of the connected wedge theorem in the context of static patch holography is as follows.

\begin{theorem}[de Sitter connected wedge theorem]
 \label{th:CWTpatch}
Let $\mathcal{S}$ be the holographic screen of an observer in an asymptotically dS$_3$ spacetime. Assuming static patch holography (Conjecture~\ref{conj:SPH}), if the $2$-to-$2$ scattering $c_1,c_2 \in \mathcal{S} \rightarrow r_1,r_2\in\mathcal{S}$ is possible in the bulk,
\begin{equation}
    J_{12\rightarrow 12} \neq \varnothing,
\end{equation}
and not on $\mathcal{S}$, 
\begin{equation}
\label{eq:nosc}
    \hat{J}_{12\rightarrow 12} = \hat{J}^+(c_1)\cap \hat{J}^+(c_2)\cap \hat{J}^-(r_1)\cap \hat{J}^-(r_2) =\varnothing,
\end{equation}
then $\mathcal R_1$ and $\mathcal R_2$ have a mutual information $O(1/G)$, and their entanglement wedge is connected in the interior of $\mathcal{S}$. We assume that $\mathcal R_1$ and $\mathcal R_2$ consist of connected regions.\footnote{We assume this because some obstacles with a causal horizon in $J^-(\tilde{r}_{1,2})\cap J^+(\tilde{c}_{1,2})$ may cause the decision regions to split apart, invalidating our proof below~\cite{Mori:2023swn}.}
\end{theorem}

\begin{proof}
    We will use three important properties here. 
    \begin{itemize}
        \item In the bulk region inside $\mathcal{S}$, minimal extremal surfaces are maximin surfaces and conversely. We develop this statement in Theorem~\ref{th:patch} of Appendix~\ref{app:proofs}.
        \item A congruence of lightrays emanating orthogonally from an extremal surface is of non-positive expansion. This follows from the focusing theorem \ref{thm:focus}.
        \item The second law of causal horizons~\cite{Jacobson:2003wv}: A causal horizon is the boundary of the causal past of a timelike worldline ending at time infinity $\mathcal{I}^+$. The second law states that the area of such a horizon cannot decrease in time.
    \end{itemize}

Let $\mathcal{V}_i$ be the achronal codimension-two surface on $\mathcal{S}$ such that $\hat{D}(\mathcal{V}_i)=\mathcal{R}_i$, as in Definition~\ref{def:decision}. $\Sigma\vert_\mathcal{S}$ denotes an achronal slice of $\mathcal{S}$ and $\Sigma$ denotes an achronal slice containing the observer of interest in the bulk such that $\partial\Sigma=\Sigma\vert_\mathcal{S}$.
$D=D(\Sigma)$ denotes the causal diamond of slices in the causal patch that is bounded by $\Sigma\vert_\mathcal{S}$. Note that the definition of $\Sigma$ here is different from the convention used in Chapter~\ref{ch:bridge}.
The exact location of the complementary screen $\mathcal{S}_{\rm L}$ is not important here. 
Additionally, the exterior contribution to the entropy is irrelevant here as the exterior entanglement wedge is always disconnected. The contribution from the complementary patch is absent because the decision regions do not lie on the complementary screen. For a more detailed discussion, see Section~\ref{sec:resolution}.
Let us denote $\gamma_i=\gamma_{\rm e}(\mathcal{V}_i;D)$ with $i=1,2$ and $\gamma_{1\cup 2}=\gamma_{\rm e}(\mathcal{V}_1\cup\mathcal{V}_2;D)$ for simplificity. To show Theorem~\ref{th:CWTpatch}, we follow the strategy of~\cite{May:2019odp}. We need to prove that 
\begin{equation}
    {\rm Area}(\gamma_1\cup\gamma_2) \leq {\rm Area}(\gamma_{1\cup 2}).
\end{equation}
If this is true, the entanglement wedge is connected and mutual information is $O(1/G)$. 

For any $\Sigma$, let us construct a codimension-one surface, called the null membrane $\mathscr{N}_{\Sigma}$. See Figure~\ref{fig:membrane} for a sketch.
\begin{figure}[ht]
    \centering
    \includegraphics[width=0.5\linewidth]{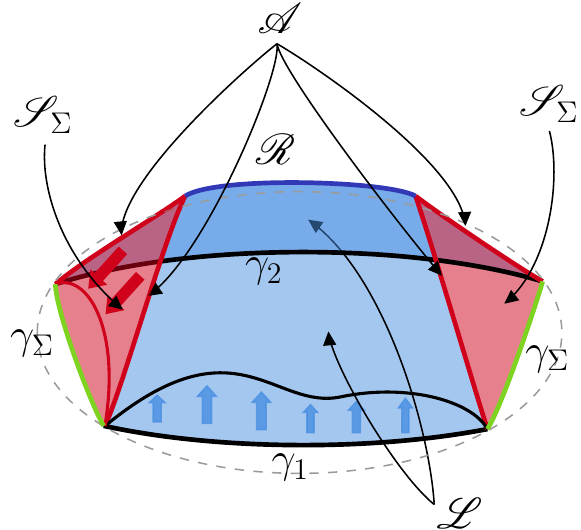}
    \caption{\footnotesize Sketch of the null membrane $\mathscr{N}_{\Sigma}=\mathscr{L}\cup\mathscr{S}_{\Sigma}$ created from $\gamma_1\cup\gamma_2$, depicted by two thick black lines. The slice $\Sigma$ is a disc-like surface bounded by the dashed line representing $\Sigma\vert_{\mathcal{S}}$. The lift $\mathscr{L}$ is depicted by a light blue surface, while the slope $\mathscr{S}_\Sigma$ is depicted by two light red surfaces. The ridge $\mathscr{R}$ is the thick blue line, and the four spacelike cusps $\mathscr{A}=\mathscr{L}\cap\mathscr{S}_{\Sigma}$ are depicted by thick red lines. The surface $\gamma_{\Sigma}$ leading to the contradiction is depicted by two green lines. A deformation of $\gamma_1$ on the lift is depicted by a thin black curve, with the direction of contraction of lightrays depicted by blue arrows. Similarly, a deformation of a set of two connected cusps is depicted by a thin red curve, with the direction of contraction of lightrays depicted by red arrows.}
    \label{fig:membrane}
\end{figure}
$\mathscr{N}_{\Sigma}$ is constructed from the union of null surfaces. The first one, called the lift is
\begin{equation}
    \mathscr{L} = \partial J^+_{in}(\gamma_1\cup\gamma_2) \cap J^-(\tilde{r}_1) \cap J^-(\tilde{r}_2),
\end{equation}
where $J^+_{in}(\gamma_1\cup\gamma_2)$ refers to the congruence of lightrays emanating orthogonally from $\gamma_1\cup\gamma_2$ and directed towards the interior of $D$. The lift $\mathscr{L}$ consists of the portion of the union of the lightsheets emanating from $\gamma_1$ and $\gamma_2$ that lies in the past of both points $\tilde{r}_1$ and $\tilde{r}_2$, and ends at their meeting points. These meeting points form a spacelike codimension-two surface called the ridge $\mathscr{R}$.

Let us show that the ridge is non-empty. It was shown in~\cite{Wall:2012uf} that whenever an extremal surface is also a maximin surface, the surface must lie outside the causal wedge. This implies that
\begin{equation}
\label{eq:P}
    J_{12\rightarrow 12} \subseteq J^+(\mathcal{C}_1\cup \mathcal{C}_2) \cap J^-(\tilde{r}_1) \cap J^-(\tilde{r}_2),
\end{equation}
where $\mathcal{C}_1,\mathcal{C}_2\in \Sigma$ are the homology regions associated with $\gamma_1$ and $\gamma_2$ on $\Sigma$. $J_{12\rightarrow 12}$ is assumed to be non-empty, such that the right-hand side of Equation \eqref{eq:P} is also non-empty. This region being not empty implies that a subset of $J^+(\mathcal{C}_1\cup \mathcal{C}_2)$ exists in the past of $\tilde{r}_1$ and $\tilde{r}_2$. Because $J^+(\mathcal{C}_1\cup \mathcal{C}_2) = J^+(\mathcal{C}_1) \cap J^+(\mathcal{C}_2)$, $J^+(\mathcal{C}_1)$ and $J^+(\mathcal{C}_2)$ must meet in the past of  $\tilde{r}_1$ and $\tilde{r}_2$. The ridge is therefore non-empty.

The other constituent of $\mathscr{N}_{\Sigma}$, called the slope $\mathscr{S}_{\Sigma}$, is defined as
\begin{equation}
    \mathscr{S}_{\Sigma} = \partial[J^-(\tilde{r}_1)\cup J^-(\tilde{r}_2)] \cap J^-[\partial J^+_{in}(\gamma_1\cup\gamma_2)] \cap J^+(\Sigma).
\end{equation}
The slope is a subsystem of two causal horizons associated with $\tilde{r}_1$ and $\tilde{r}_2$ on $\mathcal{I}^+$, which is in the future of $\Sigma$ and the past of the lightsheets emanating from $\gamma_1$ and $\gamma_2$. We now construct the null membrane as
\begin{equation}
    \mathscr{N}_{\Sigma} = \mathscr{S}_{\Sigma} \cup \mathscr{L}.
\end{equation}
Let $\gamma_{\Sigma}$ be the past boundary of the slope, that is
\begin{equation}
     \gamma_{\Sigma} = \partial[J^-(\tilde{r}_1)\cup J^-(\tilde{r}_2)] \cap J^-[\partial J^+_{in}(\gamma_1\cup\gamma_2)] \cap \Sigma.
\end{equation}
The boundary of $\gamma_{\Sigma}$ must be the same as the one of $\gamma_1\cup\gamma_2$, as the latter is defined to be on the past lightcone of $\tilde{r}_1$ and $\tilde{r}_2$.\footnote{Note that the boundary of $\gamma_1\cup\gamma_2$ is the same as the boundary of $\gamma_{1\cup 2}$ by the homology condition.} In particular, the intersection of these past lightcones with $\partial\Sigma=\Sigma\vert_\mathcal{S}$ must therefore coincide with $\partial \mathcal{V}_i = \partial \gamma_i$. Hence, $\gamma_{\Sigma}\cup\gamma_1\cup\gamma_2$ is a closed codimension-two surface on $\Sigma$.\footnote{The connectivity of $\gamma_{\Sigma}\cup\gamma_1\cup\gamma_2$ is ensured by hyperbolicity of asymptotically de Sitter spacetime.}

The slope and the lift intersect on a set of four spacelike cusps, denoted by $\mathscr{A}$. These cusps must exist, otherwise, it would imply that the past lightcones of $\tilde{r}_1$ and $\tilde{r}_2$ have intersected and hit the boundary before reaching the ridge $\mathscr{R}$, meaning $\mathcal R_1$ and $\mathcal R_2$ have an intersection, such that the theorem is trivially satisfied. The lift is a subsystem of a lightsheet (since it emanated from an extremal surface), \textit{ie} it is of negative expansion. Therefore,
\begin{equation}
\label{eq:lift}
    \text{Area}(\gamma_1\cup\gamma_2) \geq \text{Area}(\mathscr{A}) + 2\text{Area}(\mathscr{R}).
\end{equation}
The second law of causal horizons implies that $\text{Area}(\mathscr{A})\geq \text{Area}(\gamma_{\Sigma})$. Combining this with Equation \eqref{eq:lift}, we get
\begin{equation}
\label{eq:slope}
    \text{Area}(\gamma_1\cup\gamma_2) \geq \text{Area}(\gamma_{\Sigma}).
\end{equation}
We have constructed on every slice $\Sigma$ a codimension-two surface homologous to $\mathcal{V}_1\cup\mathcal{V}_2$ with a smaller area than the extremal area surface $\gamma_1\cup\gamma_2$ with a disconnected entanglement wedge, implying that $\gamma_1\cup\gamma_2$ is not a true maximin surface,
since it is not minimal on any slice $\Sigma$. Minimal extremal surfaces in the interior of $D$ are maximin surfaces (Theorem~\ref{th:patch}) so $\gamma_1\cup\gamma_2$ cannot be the smallest extremal surface, concluding the proof.
\end{proof}

    Theorem~\ref{th:CWTpatch} and its proof may be generalized to semiclassical spacetimes, where holographic entanglement entropy includes quantum matter corrections, by replacing everywhere the area of surfaces by their generalized entropy~\eqref{eq:Sgen}, and assuming the restricted quantum focusing conjecture~\cite{Shahbazi-Moghaddam:2022hbw, Bousso:2015mna} as well as the generalized second law of causal horizons~\cite{Wall:2009wm,Wall:2011hj}.

\addtocontents{toc}{\protect\end{adjustwidth}}

\chapter*{Conclusion}
\addcontentsline{toc}{part}{Conclusion} 

In this thesis, we presented results in semiclassical gravity and holography that stem from quantum information.
We reviewed how black hole thermodynamics inspired the formulation of general entropy bounds in semiclassical gravity. One of the endpoints of this thought process is the holographic principle, which puts an upper-bound on the information content of spacetime subregions. These information-theoretic constraints are well-motivated in classical gravity where the matter content satisfies the null energy condition.

As is well-known, this condition is universally broken in semiclassical gravity, where matter is studied at the quantum level on a curved background. Motivated by the generalized second law of thermodynamics, we define semiclassical gravity conjectures by taking classical statements and promoting the usual geometric entropy $\mathrm{Area}/4G$ to the generalized entropy $\mathrm{Area}/4G+S_{\rm out}$. This leads to a number of fundamental statements in gravity, quantum field theory, and holography. In particular, the focusing theorem of general relativity is promoted to a quantum focusing conjecture which underlies most of the known entropy bounds in semiclassical gravity, as well as the existence of holographic entanglement entropy and associated constraints at the semiclassical level. Additionally, the holographic principle is modified to take into account entanglement between quantum fields. The resulting statement is called the quantum Bousso bound.

Proving or testing these conjectures in semiclassical gravity is a complex endeavor, as accounting for the entanglement between quantum fields is non-trivial even in very simple models. An efficient route to overcome this computational difficulty is to consider toy-models of quantum gravity. Among them, JT gravity has emerged as an extremely powerful tool in recent years. We leverage this framework to prove the quantum Bousso bound in semiclassical JT gravity. While closely related, the quantum focusing conjecture does not follow from analogous computations. We discuss a weaker version of quantum focusing which has been shown to be sufficient for all known applications of quantum focusing. We prove this weaker statement in semiclassical JT gravity. Interestingly, we find violations of quantum focusing in the vicinity of black hole and cosmological horizons in JT gravity. This strongly motivates us to reconsider the restricted formulation of quantum focusing as the fundamental statement in quantum gravity, rather than quantum focusing which is violated.

We then considered applications of these developments to quantum cosmology and in particular to holography in closed universes.

The holographic principle naturally suggests a holographic setup in de Sitter spacetime, called static patch holography, where the causal patch of an observer is described by a quantum theory defined on the cosmological horizon. Motivated by the success of holographic entanglement entropy in AdS/CFT, we considered the problem of computing entanglement of subregions of the horizon theory in terms of geometric bulk quantities. We defined a covariant prescription inspired by the bilayer proposal of Shaghoulian, while arguing that the monolayer proposal is inconsistent. It appears that the usual extremization problem considered in AdS/CFT suffers from existence problems in de Sitter space. We motivate an alternative extremization problem where the area functional is supplemented by Lagrange multipliers that ensure the extremal surface sits in a specific region of the bulk. This leads to large degeneracies that can be lifted by quantum corrections. Based on the computation of extremal surfaces in this context, we argue that global de Sitter spacetime emerges from a quantum theory defined on the two holographic screens of antipodal observers. The entanglement between the two holographic screens leads to the description of the inflationary region.

These developments can be generalized to closed FLRW cosmologies, with the holographic screen positioned on the apparent horizons of two antipodal observers. Interestingly, in the contracting phase of these geometries, the screens can be pushed outside of the apparent horizon, leading to an infinite number of alternative holographic screens. We group these holographic screens into equivalence classes, where each equivalence class contains a "parent screen" located at the apparent horizon. Every other screen in the equivalence class is obtained by integrating out degrees of freedom from the parent screen. In each class, there is a minimal screen with the smallest number of degrees of freedom, that thermalizes close to the time-symmetric slice. This generalization of our results to closed FLRW cosmologies provides a time-dependent realization of the ER=EPR conjecture.

Finally, we considered the connected wedge theorem that can be proven from quantum information theory, which states that if four points on the holographic screen are causally related in the bulk, but not on the screen, then there must be a high amount of mutual information in the screen theory. This theorem was initially motivated by the AdS/CFT correspondence but we argue that it must be valid in any holographic theory. In particular, static patch holography and our holographic entanglement entropy proposal should be consistent with the connected wedge theorem. Applying this theorem demands an explicit description of causality on the holographic screen, which is not known in static patch holography. We propose a definition of causality on the screen, induced from points at null infinity, and prove the connected wedge theorem in static patch holography from geometric arguments using this definition.
We conclude this thesis with some remarks and possible future directions.

\textbf{Physical interpretation of the Strominger-Thompson quantum Bousso bound}

The Strominger-Thompson bound puts an upper bound on the hydrodynamic entropy of a region of spacetime, which depends on the generalized entropy. As it mixes different definitions of entropy, the correct physical interpretation of this bound is not clear. In particular, removing the hydrodynamic entropy from the expression leads to the BFLW quantum Bousso bound that one obtains from the quantum focusing conjecture. Based on the hydrodynamic conditions that are used to prove the Strominger-Thompson bound, we believe that it is only valid in a regime where matter can be decoupled into two independent systems. The first one is classical and well-approximated by a hydrodynamic description. The second is quantum and contributes to the generalized entropy. Considering the BFLW bound and coarse-graining over the degrees of freedom of the classical subsystem should lead to the Strominger-Thompson bound. It would be interesting to explore this idea further in order to obtain a fundamental derivation of this bound, that does not rely on the hydrodynamic regime. Moreover, it is also not clear how to interpret the spacelike projection theorem that one would obtain from this semiclassical bound. It would suggest that the entropy of classical matter in a spacetime region is bounded from above by the generalized entropy of its boundary. In this setup, the encoding of quantum matter is ignored, and one wonders how to properly generalize the holographic principle of Bousso, which was derived from the classical Bousso bound.

\textbf{General proof of the restricted QFC and violations of the QFC}

Our results suggest that the QFC might be broken in semiclassical gravity in more than two dimensions. It would be interesting to look for explicit QFC violations beyond JT gravity. Such violations have already been argued for in \cite{Fu:2017lps}, but were contradicted in \cite{Leichenauer:2017bmc}. Additionally, the restricted QFC has now been proven in semiclassical JT gravity and braneworld holography. A general proof is still lacking. The powerful tools of algebraic quantum field theory, that led to general proofs of some corollaries of the QFC such as the QNEC \cite{Balakrishnan:2017bjg,Kudler-Flam:2023hkl} and the GSL \cite{Faulkner:2024gst,Kirklin:2024gyl}, might be a promising direction.

\textbf{Derivation of holographic entropy in dS holography}

A great part of this work consisted of constructing a self-consistent prescription to compute holographic entanglement entropy in static patch holography. The fact that we find such a definition suggests that there must be an entropy-like object in the dual theory that is computed by it. An exact dictionary would require some fundamental derivation. In particular, a derivation of our proposal from a path-integral computation would be a crucial breakthrough, especially to settle the ambiguity between the inequivalent definitions of extremal surfaces discussed in Section \ref{sec:const} and Appendix \ref{app:proofs}. We expect such a derivation to rely on replica path integrals, as in~\cite{Lewkowycz:2013nqa,Dong:2016hjy}. A difficulty that we expect to encounter is that the holographic screen is not located in an asymptotic, weakly gravitating, region. In AdS/CFT, the location of the holographic screen provides good control over fluctuations in the bulk. While bulk fluctuations can be included in the computation, asymptotic coldness ensures that observables and subregions can be defined properly on the boundary theory. However, it is not even clear how to define a screen subregion beyond the semiclassical limit in static patch holography.\footnote{We thank Luca Ciambelli for bringing this to our attention.}  Moreover, the similarity of the results of \cite{Hao:2024nhd} with our definition of C-extremization suggests that embedding dS in a higher-dimensional AdS spacetime could be used.

\textbf{Boundary conditions}

A related issue is the need for boundary conditions associated with the holographic screen in a fluctuating background. Dirichlet boundary conditions are widely used in the literature. However, it has been shown that these conditions do not lead to a well-defined initial boundary-value problem in general relativity \cite{An_2021,Anderson_2008}. Instead, it has been shown that conformal boundary conditions, where the conformal class of the induced metric and its extrinsic curvature are fixed, lead to a well-defined problem \cite{An_2021,An:2025rlw}. These boundary conditions have been studied in de Sitter space in \cite{Anninos:2024wpy}, leading to promising results. 

\textbf{The algebraic approach}

We have briefly discussed the algebraic approach in this thesis. In particular, we mentioned that the algebra of observables of the static patch can only be non-trivial if one includes an observer in the theory through a crossed-product construction \cite{Chandrasekaran:2022cip}. While we argued that a holographic description of de Sitter space should be based on the description of static patches, which are observer dependent, our discussion does not explicitly contain an observer. An interesting direction would be to connect some of our results to algebraic considerations. For example, Engelhard and Liu \cite{Engelhardt:2023xer} have presented an algebraic formulation of the ER=EPR conjecture in AdS/CFT, in terms of entanglement wedges and the algebra of observables of boundary regions. In particular, they relate the type of algebra associated with disconnected holographic screens with specific conditions on the entanglement wedge of these screens. Among these conditions, it appears that a connected entanglement wedge with large quantum fluctuations can be related to type II algebras. This is consistent with the conclusions of this thesis, and it would be interesting to develop the analysis of \cite{Engelhardt:2023xer} in the context of static patch holography. Another interesting direction would be to study the algebra of observables of the causal patch of an observer in FLRW spacetime. In this context, we do not expect the algebra to be type II$_1$,\footnote{Indeed, Hilbert spaces that are associated with a type II$_1$ algebra have the property of containing a maximal entropy state. This is a well-known property of the static patch but we do not know whether this property easily generalizes to other cosmologies.} and the overlap between complementary causal patches is expected to have crucial consequences.

\textbf{Holographic screens for open regions}

The holographic description of the de Sitter static patch is very peculiar, as it describes an open region of de Sitter which is in thermal equilibrium. In particular, the existence of a timelike Killing vector ensures that the static patch evolves under unitary evolution. An interesting direction would be to study holographic screens in other states than the Bunch-Davies state, such as the Unruh-de Sitter state. For example, one may wonder whether it is possible to match the evolution of the size of the apparent horizon—where the screen should be located—with the size of the Hilbert space associated with the effective description of its interior. This comment also applies to the holographic screen of closed FLRW spacetimes, which are expected to evolve non-unitarily. Non-isometric evolution in quantum cosmology has been discussed in \cite{Cotler:2022weg,Cotler:2023eza}. Additionally, an implicit assumption made in the connected wedge theorem is that the evolution on the holographic screen(s) is local. Otherwise, the time evolution manifestly entangles two disconnected regions so it cannot be viewed as a nonlocal quantum computation.\footnote{We thank Beni Yoshida for pointing this out.} The screen dual of the evolution in the Milne patch toward $\mathcal{I}^+$ may require coupling the left and the right screens. Thus, it is essential to understand the locality of time evolution on the screens to justify the connected wedge theorem in the exterior region.

\textbf{Pushing the screen inside the static patch}

We briefly commented in Section~\ref{sec:stretched} on the effect of the location of the screen on causality. We found that pushing the screen deeper into the static patch tends to localize the effective holographic theory, as the causal region tied to a bulk perturbation approaches that of a local operator near the observer's worldline. In general, we expect that the effective holographic theory defined on the screen follows some kind of renormalization group flow as one pushes the screen closer to the observer's worldline. It would be interesting to describe the details of the coarse-graining associated with moving the holographic screen in the static patch. In parallel, one might consider holographic scattering in alternative static patch holographic setups, such as worldline holography or half-de Sitter holography~\cite{Anninos:2011af,Banihashemi:2022htw,Kawamoto:2023nki}.

\textbf{Precise nature of excitations on a screen}

In Chapter~\ref{ch:CWT}, we did not discuss the detailed nature of a local excitation at null infinity or a 'smeared' excitation on the screen. In fact, knowing what types of excitations are allowed is very important to support our proof of the connected wedge theorem~\cite{Mori:2023swn}. Nevertheless, at this point the precise holographic dual of de Sitter, in particular within static patch holography, is unclear. In the proposed dS/DSSYK duality~\cite{Susskind:2021esx,Lin:2022nss,Rahman:2022jsf,Goel:2023svz,Narovlansky:2023lfz,Verlinde:2024znh,Verlinde:2024zrh,Blommaert:2023opb, Blommaert:2023wad,Rahman:2024iiu}, Susskind has argued that only a small number of special collective excitations of the fermions in the DSSYK model would propagate deep in the bulk \cite{Susskind:2022bia}. It is interesting to see what types of excitations in the theory realize holographic scattering, if there are any, and to see if the model shows signs of induced causality. We can already point out that the induced lightcones constructed in this paper showcase a difference between perturbations propagating into the bulk and perturbations that are confined close to the stretched horizon. Indeed, the induced lightcone of a bulk perturbation is apparently superluminal. This means such an excitation is nonlocal and obeys a nonlocal evolution from the screen point of view. On the other hand, the lightcone of a perturbation confined on the screen is a lightcone based on the induced metric, and it does not exhibit superluminality, suggesting a local operator. A further investigation of these perturbations and their dual in a concrete theory could help us to identify the true UV boundary, from which holographic dS emerges.

\textbf{Scattering between holographic screens}

In this work, we only considered holographic scattering in the static patch. Since global de Sitter spacetime is conjectured to be encoded on the holographic screens associated with two antipodal observers, an interesting direction would be to consider scatterings connecting the two screens to probe the exterior region. One clear obstruction is that no direct scattering is possible in the static patch, due to cosmological horizons between the screens. One way to overcome this issue would be to consider explicit solutions of asymptotically de Sitter spacetimes in which the static patches always overlap~\cite{Gao:2000ga,Leblond:2002ns}. Another possibility would be to use the mapping between perturbations on the screen and points at null infinity. In addition to providing evidence for the connection between static patch holography and dS/CFT, scattering in the exterior region may provide a useful tool in determining the correct entanglement entropy prescription in the region between the screens.

\textbf{Relating static patch holography to dS/CFT}

\begin{figure}[ht]
    \centering
    \includegraphics[width=0.3\linewidth]{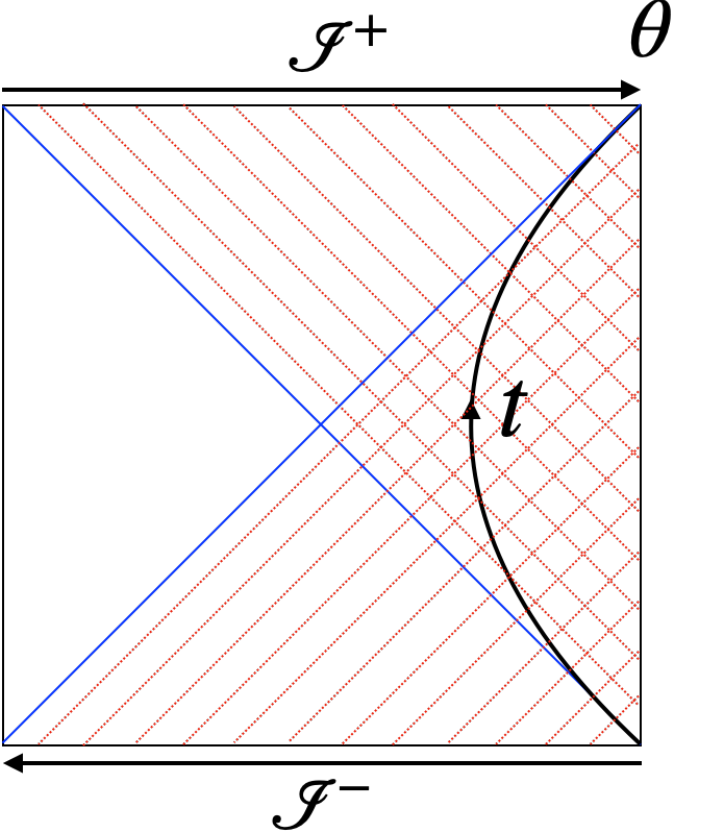}
    \caption{\footnotesize The mapping between points on a holographic screen $\mathcal{S}$ and points on the null infinities $\mathcal{I}^\pm$. Left: The proposed mapping is based on the induced causality in the de Sitter Penrose diagram. The time translation in the static patch is mapped to a spatial translation on $\mathcal{I^\pm}$. }
    \label{fig:mapping}
\end{figure}

In the de Sitter version of the connected wedge theorem, we related input and output points to points at past and future null infinities. This maps a bulk scattering from the static patch holographic screens to a bulk scattering from conformal boundaries. It would be interesting if the dS/CFT correspondence could provide a precise framework to describe this type of scattering. In particular, can we prove a connected wedge theorem in dS/CFT? Doing so would require precise notions of causality on the Euclidean CFT, which is problematic as there is no notion of time evolution.
Our construction may serve as an alternative definition of ``time'' in dS/CFT as it provides a mapping between points on the static patch, which follows the real time evolution, to the points on the conformal boundary.\footnote{See \cite{Parikh:2002py} for a related mapping.} See Figure~\ref{fig:mapping} for a pictorial representation.
Additionally, studying connected wedges in dS/CFT would require a precise prescription to compute the entanglement entropy of CFT subregions. This has been studied in~\cite{Doi:2022iyj,Narayan:2015vda,Narayan:2020nsc,Narayan:2022afv, Narayan:2023zen,Doi:2023zaf, Das:2023yyl,Doi:2024nty}.
Succeeding in this challenge would open a path to a better understanding of time, entanglement, and causality in dS/CFT.

\newpage

\appendix
\addcontentsline{toc}{part}{Appendices}

\addtocontents{toc}{\protect\begin{adjustwidth}{1cm}{0cm}}

\setcounter{chapter}{0}
\renewcommand{\thechapter}{\Alph{chapter}}

\chapter{Entanglement entropy in a box}
\label{app:entanglement_entropy}

In this Appendix, adapted from Appendix A of~\cite{Franken:2023ugu}, we provide a general entanglement entropy formula for a CFT$_2$ in a two-dimensional curved background, with two reflective boundaries. We assume that matter fields can be expanded in right- and left-moving modes that are uncorrelated at $\mathcal{I}^-$\cite{Fiola:1994ir}. 

\section{Entanglement entropy with reflecting boundaries}
\label{app:reflec_bound}

In a model with reflecting boundaries like the half reduction model of de Sitter JT gravity, the reflection of right-moving modes into left-moving modes and vice-versa induces correlations between them. A study of entanglement entropy in the presence of a single moving mirror following the timelike trajectory $x^+ - x^- =$ constant has been provided in~\cite{Fiola:1994ir}. In this section, we discuss a more general case where the spacetime contains two reflecting boundaries following more general trajectories.

Let us consider the Kruskal coordinates $(x^+,x^-)$ and associated conformal factor \eqref{eq:Kruskal}. This set of coordinates naturally splits the spacetime into four distinct regions. Inspired by the half reduction model of de Sitter JT gravity, we consider the case where the spacetime contains two spatial reflecting boundaries lying entirely in the regions $x^+ x^-\leq 0$. We parameterize their trajectories through a function\footnote{One could find a change of coordinates $y^{\pm}(x^{\pm})$ such that the boundary is described by $y^+ - y^- =$ constant. However, choosing a coordinate system in the entropy formula corresponds to selecting a specific vacuum, defined in these coordinates. To generalize our discussion to arbitrary vacua, we must consider an arbitrary function $f$.} $x^-=f(x^+)$, hence satisfying $x^+ f(x^+)\leq 0$, and impose in addition the following conditions:
\begin{enumerate}
    \item When introducing an infrared cutoff $L$, the left boundary is bounded by the points of $x^{\pm}$ coordinates $(-L,0)$ and $(0,L)$, while the right boundary is bounded by the points $(0,-L)$ and $(L,0)$. We thus have $f:I \rightarrow I$, with  $I=(-L,0)\cup (0,L)$.
    \item All modes emanating from any spacelike slice bounded by the points $(-L,0)$ and $(0,-L)$ reflect once and only once on a boundary, which is equivalent to having $f$ be a bijection from $I$ to $I$.
    \item The trajectory must be continuous and differentiable, which is equivalent to imposing $f$ to be a $C^1$ function.
    \item The trajectory must be timelike to define a spatial boundary, which is equivalent to having $0<f'<+ \infty$, where $ f'\equiv d f/d x^+$.
    \item We restrict ourselves to states that have symmetric left- and right-moving radiation, and in general to systems symmetric under $x^+\leftrightarrow x^-$. This implies that $f$ is an involution, \emph{ie} $f=f^{-1}$.
\end{enumerate}
This construction ensures that, when a mode reflects on a boundary:
\begin{itemize}
    \item Left-moving modes propagating along $x^-$ at fixed $x^+=x_0^+$ transform into right-moving modes propagating along $x^+$ at fixed $x^-=f(x_0^+)$.
    \item Right-moving modes propagating along $x^+$ at fixed $x^-=x_0^-$ transform into left-moving modes propagating along $x^-$ at fixed $x^+=f(x_0^-)$.
\end{itemize}
In particular, we are able to express all modes either as right-moving modes emanating from the interval $[-L,L]$ on the $x^-$ axis, or as left-moving modes emanating from the interval $[-L,L]$ on the $x^+$ axis.

Similarly to the spatial boundaries, we parameterize $\mathcal{I}^{\pm}$ as a spacelike curve described by $x^- = g(x^+)$, with $g:I\rightarrow I$ a $C^1$ involution such that $g(x^-)x^-\geq 0$ and $-\infty <  g' < 0$. From now on, we take the limit $L\rightarrow \infty$, which will be justified at the end of this section. In Figure~\ref{fig:LR_modes},
\begin{figure}[ht]
\begin{subfigure}[t]{0.48\linewidth}
\centering
\includegraphics[width=1\linewidth]{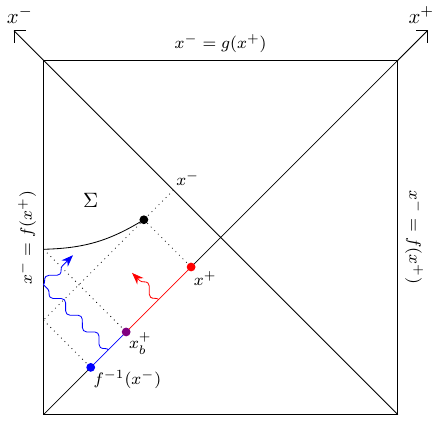}
\caption{\footnotesize Case $x^-\geq 0$. The left-moving modes passing through $\Sigma_p$ are the left-moving modes emanating from the interval $[x_b^+,x^+]$ depicted in red. The right-moving modes passing through $\Sigma_p$ are the left-moving modes emanating from the interval $[f^{-1}(x^-),x_b^+]$ depicted in blue. Tracing over every mode on $\Sigma_p$ is therefore equivalent to tracing over left-moving modes on the interval $[f^{-1}(x^-),x^+]$. \label{fig:LR_modes_case1}}
\end{subfigure}
\quad \,
\begin{subfigure}[t]{0.48\linewidth}
\centering
\includegraphics[width=1 \linewidth]{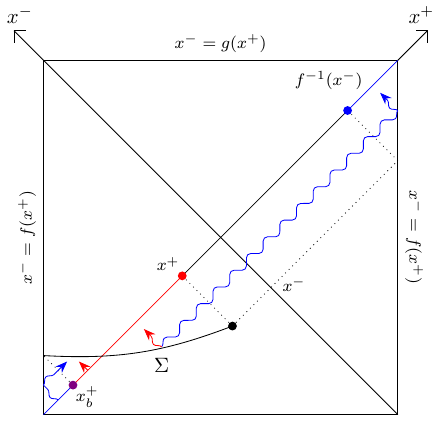}   
\caption{\footnotesize Case $x^-\leq 0$. The left-moving modes passing through $\Sigma_p$ are the left-moving modes emanating from (or which will cross) the interval $[x_b^+,x^+]$ depicted in red. The right-moving modes passing through $\Sigma_p$ are the left-moving modes emanating from (or which will cross) the interval $(-\infty,x_b^+] \cup [f^{-1}(x^-),+\infty)$ depicted in blue. Tracing over every mode on $\Sigma_p$ is therefore equivalent to tracing over left-moving modes on the interval $(-\infty,x^+]\cup [f^{-1}(x^-),+\infty)$ or on the complementary interval $[x^+,f^{-1}(x^-)]$. \label{fig:LR_modes_case2}}
\end{subfigure}
    \caption{\footnotesize Spacelike slice $\Sigma_p$ bounded by a point $(x_b^+,f(x_b^+))$ on the left boundary of the spacetime and a point $(x^+,x^-)$ in the bulk. Considering reflective boundary conditions at the timelike boundaries of the spacetime induces correlations between left- and right-moving modes passing through $\Sigma_p$. \label{fig:LR_modes}}
\end{figure}
the timelike and spacelike boundaries are depicted with examples of reflecting lightrays, and we introduce the following notation to describe segments of the cosmological horizons:
\begin{align}
  \begin{aligned}
   H_-^- &= \{(x^+,x^-)|x^+=0,x^-\leq 0\}, \\   H_+^- &= \{(x^+,x^-)|x^-=0,x^+\leq 0\},
  \end{aligned}
  &&
  \begin{aligned}
   H_-^+ &= \{(x^+,x^-)|x^+=0,x^-\geq 0\}, \\   H_+^+ &= \{(x^+,x^-)|x^-=0,x^+\geq 0\},
  \end{aligned}
 \end{align}
so that $H_- = H_-^- \cup H_-^+$ is the horizon along $x^-$ and $H_+=H_+^-\cup H_+^+$ is the horizon along $x^+$.

From our construction, we consider right- and left-moving modes at $\mathcal{I}^-$ and note that all the right-moving modes pass through the half line $H_-^-$ while all the left-moving modes pass through the half line $H_+^-$. Because of the reflecting boundary conditions, all right-moving (left-moving) modes passing through $H_-^-$ ($H_+^-$) also cross the half line $H_+^+$ ($H_-^+$) as left-moving (right-moving) modes. In other words, all modes coming from $\mathcal{I}^-$ cross $H_-$ as right-movers and $H_+$ as left movers, such that we can express the quantum state of the fields as left-moving (right-moving) states at $H_-$ ($H_+$). From here we always refer to the state of the fields as left-moving states on $H_+$. The condition that $f$ is an involution ensures that the following procedure is independent of the reference axis.

Now, consider a subsystem $\Sigma_p$ of some Cauchy slice $\Sigma$, bounded by the left boundary at $b=(x_b^+,f(x_b^+))$ and some point $p=(x^+,x^-)$. As mentioned above, right- and left-moving modes on $\Sigma_p$ are correlated and tracing over right- and left-movers on $\Sigma$ is equivalent to tracing over left-movers on some interval of $H_+$. First, all left-moving modes on $\Sigma$ are the same modes as the left-moving modes on the interval $[x_b^+,x^+]$ of $H_+$. As for right-movers on $\Sigma_p$, the projection slightly varies depending on the position of the endpoint $(x^+,x^-)$:
\begin{itemize}
\item When $x^- \geq 0$, the right-moving modes on $\Sigma_p$ are the left-moving modes in the interval $[f^{-1}(x^-),x_b^+]$ of $H_+$.\footnote{We keep the notation $f^{-1}$ for clarity, although we imposed $f=f^{-1}$.} Therefore, tracing over every mode on $\Sigma_p$ is equivalent to tracing over left-moving modes on the interval $[f^{-1}(x^-),x^+]$ of $H_+$, see Figure~\ref{fig:LR_modes_case1}. 
\item When $x^-\leq 0$, the right-moving modes on $\Sigma_p$ are the same as the left-moving modes in the subset $(-\infty,x_b^+] \cup [f^{-1}(x^-),\infty)$ of $H_+$. Therefore, tracing over all modes on $\Sigma_p$ is equivalent to tracing over left-moving modes on the interval $(-\infty,x^+] \cup [f^{-1}(x^-),\infty)$ of $H_+$, see Figure~\ref{fig:LR_modes_case2}. Since the left-movers on $H_+$ should be in a pure state, this is equivalent to tracing over the complement of this subset, \emph{ie} $[x^+,f^{-1}(x^-)]$.
\end{itemize}
Assuming $|x^+-f^{-1}(x^-)|\ll L$, we use \eqref{eq:entropy_left} to get
\begin{equation}
    S(\Sigma_p) = \frac{c}{12}\ln\left[\frac{(x^+ -f^{-1}(x^-))^2}{\delta x^+~\delta f^{-1}(x^-)}\right],
\end{equation}
where $\delta x^+$ and $\delta f^{-1}(x^-)$ are the UV cutoffs at the endpoints of the interval of $H_+$. In other words, we have smeared over the regions $[x^+ -\delta x^+,x^+]$ and $[f^{-1}(x^-), f^{-1}(x^-)+ \delta f^{-1}(x^-)]$ on $H_+$ in the case $x_-\geq 0$, and similarly when $x_-\leq 0$. We would like to relate these quantities to the invariant cutoff $\delta^2$, \emph{ie} to the UV cutoff associated with the position of $(x^+,x^-)$. Since the left-moving modes on $[x^+ -\delta x^+,x^+]$ are the same as the left-moving modes on the same interval at $x^+$ fixed, we write $\delta x^+=\varepsilon_{\rm L}(x^+,x^-)$. To relate the interval $[f^{-1}(x^-), f^{-1}(x^-)+ \delta f^{-1}(x^-)]$ on $H_+$ to an interval around $(x^+,x^-)$, we need to take into account the reflection on the left boundary. Considering the reflection of two lightrays coming from the two endpoints of the interval, smearing over the left-moving modes on $[f^{-1}(x^-), f^{-1}(x^-)+ \delta f^{-1}(x^-)]$ on $H_+$ is equivalent to smearing over the right-moving modes on $[f(f^{-1}(x^-)), f(f^{-1}(x^-)+ \delta f^{-1}(x^-))]$ at fixed $x^+$ around the point $(x^+,x^-)$. At first order in $\delta f^{-1}(x^-)$, this interval is $[x^-,x^- +\delta f^{-1}(x^-) f'(f^{-1}(x^-))]$, from which we identify the cutoff for the right movers at the point $(x^+,x^-)$ to be:
\begin{equation}
    \varepsilon_{\rm R}(x^+,x^-) = \delta f^{-1}(x^-)~f'(f^{-1}(x^-)).
\end{equation}
We conclude that
\begin{equation}
\label{eq:ent_bdy}
    S(\Sigma_p) = \frac{c}{6}\omega(x^+,x^-)+\frac{c}{12}\ln\left[f'(f^{-1}(x^-))\frac{(x^+ -f^{-1}(x^-))^2}{\delta^2}\right],
\end{equation}
where $\delta^2=\delta_{\rm L}\delta_{\rm R}$ is the cutoff in inertial coordinates at the endpoint of $\Sigma$.

For this formula to be well defined, we need to motivate the assumption $|x^+-f^{-1}(x^-)|\ll L$. As $L\rightarrow \infty$, this is equivalent to imposing $-L\ll f^{-1}(x^-)\ll L$. For $f$ to break this condition, one would need to localize $x^-$ with an infinite precision as $L\rightarrow \infty$, which cannot be done in the presence of the UV cutoff $\delta$.

\section{Change of vacuum}
\label{sec:changevac}

As noted in~\cite{Fiola:1994ir} and in Section~\ref{sec:SCFT}, the entropy formula \eqref{eq:entropy_left} is only valid in the vacuum state defined with respect to the coordinates $x^{\pm}$, with the associated metric $ds^2 = -e^{2\omega_x(x^+,x^-)}dx^+dx^-$, where we have made the coordinate system explicit in the notation $\omega_x$. In general, the vacuum state may be defined with respect to an arbitrary set of lightcone coordinates $y^{\pm}(x^{\pm})$, in which the metric is $ds^2 = -e^{2\omega_y(y^+,y^-)}dy^+dy^-$. Under this change of coordinates, the endpoints of $\Sigma$ become $(y^+(x_1^+),y^-(x_1^-))$ and $(y^+(x_2^+),y^-(x_2^-))$. By diffeomorphism invariance, the Cauchy slice defined by this interval in coordinates $y^{\pm}$ is physically identical to the Cauchy slice defined by the endpoints $(x_1^+,x_1^-)$ and $(x_2^+,x_2^-)$ in the original coordinates. For this reason, we refer to them both as $\Sigma$, independently of the coordinate system in which it is described.

To write the entropy contribution from the left-moving modes in a vacuum defined in coordinates $y^{\pm}$, one writes the size of the interval $[x_1^+,x_2^+]$ in $y^{\pm}$ coordinates~\cite{Fiola:1994ir},
\begin{equation}
    S_y(\Sigma) = \frac{c}{12}\ln\left[\frac{(y^+(x_2^+)-y^+(x_1^+))^2}{\varepsilon_{\rm L}(y(x_1))\varepsilon_{\rm L}(y(x_2))}\right],
\end{equation}
where the index $y$ denotes the vacuum in which the entropy is computed. As we change the coordinate system, the cutoffs $\varepsilon_{\rm L( \mathrm{R})}$ must change according to:
\begin{equation}
    \varepsilon_{\rm L( \mathrm{R})}(y^+,y^-) = e^{-\omega_y(y^+,y^-)}~ \delta_{\rm L( \mathrm{R})}.
\end{equation}
Therefore, the entropy $S_y(\Sigma)$ of $\Sigma$ in the vacuum state defined in coordinates $y^{\pm}$ is given by~\cite{Fiola:1994ir}:
\begin{align}
\begin{split}
\label{transf_entropy}
    S_{y}(\Sigma) &= \frac{c}{6}\left(\omega_y(y(x_1))+\omega_y(y(x_2))\right)+\frac{c}{12}\ln\left[\frac{(y^-(x^-_2)-y^-(x^-_1))^2(y^+(x^+_2)-y^+(x^+_1))^2}{\delta_1^2\delta_2^2}\right] \\
    &=\frac{c}{12}\left[2\left(\omega_x(x_1)+\omega_x(x_2)\right)+\ln\left[\left.\frac{dx^+}{dy^+}\right|_{x_1^+}\left.\frac{dx^+}{dy^+}\right|_{x_2^+}\left.\frac{dx^-}{dy^-}\right|_{x_1^-}\left.\frac{dx^-}{dy^-}\right|_{x_2^-}\right] \right.\\
    &+\left.\ln\left[\frac{(y^+(x^+_2)-y^+(x^+_1))^2(y^-(x^-_2)-y^-(x^-_1))^2}{\delta_1^2\delta_2^2}\right]\right],
\end{split}
\end{align}
where we have used 
\begin{equation}
    \omega_{y}(y^+(x^+),y^-(x^-)) = \omega_{x}(x^+,x^-) -\frac{1}{2}\ln\left[\frac{dy^+}{dx^+}\frac{dy^-}{dx^-}\right],
\end{equation}
as explained above Equation \eqref{eq:transfo_conf_factor}. The formulas above should be applied to spacetimes that have infinite extends in both $x^+$ and $x^-$ (or $y^+$ and $y^-$) directions. In particular they do not take into account any possible boundaries. 

A similar transformation law can be derived in the presence of spatial boundaries. First, assuming that the boundary in coordinates $x^{\pm}$ follows a trajectory $x^-=f_x(x^+)$, the spatial boundary in coordinates $y^{\pm}$ is described by
\begin{equation}
    y^- = f_y(y^+) =y^-(f_x(x^+)).
\end{equation}
From this, we compute the transformation law of the term $f'_x(f_x^{-1}(x^-))$:
\begin{equation}
    f'_x(f_x^{-1}(x^-)) \rightarrow f'_y(f_y^{-1}(y^-)) = f'_x(f_x^{-1}(x^-))\frac{\frac{dy^-}{dx^-}(x^-)}{\frac{dy^+}{dx^+}(f_x^{-1}(x^-))}.
\end{equation}
Hence, the entropy of a Cauchy slice $\Sigma$ bounded by a point $(x^+,x^-)$ and the boundary \linebreak $x^-=f_x(x^+)$, on a curved background with the metric $ds^2 =-e^{2\omega_x}dx^+dx^-$ and in a vacuum state defined in coordinates $y^{\pm}(x^{\pm})$, is given by:
\begin{align}
\begin{split}
    S_{y}(\Sigma) &= \frac{c}{6}\omega_x(x^+,x^-) + \frac{c}{12}\ln\left[f'_x(f_x^{-1}(x^-))\frac{\left(y^+(x^+) -y^+(f_x^{-1}(x^-))\right)^2}{\delta^2}\right]\\
    &-\frac{c}{12}\ln\left[\frac{dy^+}{dx^+}(x^+)\frac{dy^+}{dx^+}\left(f_x^{-1}(x^-)\right)\right].
\end{split}
\end{align}

\chapter{Proof of sufficient condition \ref{prop:sufc}}
\label{app:proofsufc}
\numberwithin{equation}{chapter}

For any null vector $k^{\pm}$, the dilaton equation of motion \eqref{eq:dilaton_eom} gives:
\begin{equation}
8\pi G k^{\pm} k^{\pm}\left(\braket{\tau_{\pm\pm}}-\frac{c}{24\pi}t_{\pm}\right)=-\Phi''=-\partial_{\pm}\left(\partial_{\pm}\Phi\frac{\partial x^{\pm}}{\partial\lambda}\right)\frac{\partial x^{\pm}}{\partial\lambda},
\end{equation}
independently of the sign of the cosmological constant $\Lambda$. Inserting this equation into the first entropy condition \eqref{eq:cond1_QNEC} leads to the inequality:
\begin{equation}
\partial_{\pm}s~\frac{\partial x^{\pm}}{\partial\lambda}\leq -\frac{\Phi''}{4G}+\frac{c}{12}k^{\pm} k^{\pm} t_{\pm},
\end{equation}
or equivalently, using the definition of the quantum area of a point \eqref{eq:quant_area_point}:
\begin{equation}
\partial_{\pm} s~\frac{\partial x^{\pm}}{\partial\lambda}\leq -\frac{\text{Area}_{\rm qu}''}{4G}+S''+\frac{c}{12}k^{\pm}k^{\pm} t_{\pm}.
\end{equation}
We now show that
\begin{equation}\label{eq:sufficient_cond}
S''+\frac{c}{12}k^{\pm} k^{\pm} t_{\pm}\leq 0.
\end{equation}
is a sufficient condition for the Strominger-Thompson bound \eqref{eq:STBB}. Written in terms of the normal ordered stress tensor, this condition is
\begin{equation}
    2\pi k^{\pm}k^{\pm}\braket{:T_{\pm\pm}:}\geq S''.
\end{equation}
Assuming inequality \eqref{eq:sufficient_cond}, one finds
\begin{equation}\label{eq:starting_inequality}
\partial_{\pm}s~\frac{\partial x^{\pm}}{\partial\lambda}\leq -\frac{\text{Area}_{\rm qu}''}{4G},
\end{equation}
Let us consider, without loss of generality, the case of a past directed lightsheet along $x^+$. Since $\partial x^+/\partial\lambda<0$, \eqref{eq:starting_inequality} gives:
\begin{equation}
\partial_+s \geq -\frac{1}{4G}\partial_+\left(\partial_+\text{Area}_{\rm qu}\frac{\partial x^+}{\partial\lambda}\right).
\end{equation}
Integrating this inequality between $x^+$ and $x_1^+>x^+$ yields
\begin{equation}
s(x_1^+)-s(x^+)\geq-\frac{1}{4G}~\partial_+\text{Area}_{\rm qu}(x_1^+)~\frac{\partial x^+}{\partial\lambda}(x_1^+)+\frac{1}{4G}~\partial_+\text{Area}_{\rm qu}(x^+)~\frac{\partial x^+}{\partial\lambda}(x^+),
\end{equation}
which can be further simplified using the second entropy condition \eqref{eq:cond2_QNEC} into
\begin{equation}
-s(x^+) \geq \frac{1}{4G}\partial_+\text{Area}_{\rm qu}(x^+)~\frac{\partial x^+}{\partial\lambda}(x^+).
\end{equation}
From $s=-k^+ s_+$ and $k^+=-\frac{\partial x^+}{\partial\lambda} > 0$ for a past-directed lightsheet, this gives:
\begin{equation}
-s_+(x^+)\leq \frac{1}{4G}\partial_+\text{Area}_{\rm qu}(x^+).
\end{equation}
Hence,
\begin{equation}
\int_0^1 d\lambda~s(\lambda)=\int_{x_1^+}^{x_2^+}d x^+\frac{\partial\lambda}{\partial x^+}(-k^+s_+) =\int_{x_2^+}^{x_1^+}d x^+(-s_+) \leq \frac{1}{4G}\int_{x_2^+}^{x_1^+}d x^+\partial_+\text{Area}_{\rm qu},
\end{equation}
from which we derive the Strominger-Thompson quantum Bousso bound:
\begin{equation}
\int_0^1 d\lambda~s(\lambda)\leq\frac{1}{4G}(\text{Area}_{\rm qu}(x_1)-\text{Area}_{\rm qu}(x_2)).
\end{equation}

\chapter{Area extremization in a causal diamond}
\label{app:Lagrange}

At fixed $(\sigma,\theta_1)$, the metric of dS$_{n+1}$ given in Equation \eqref{eq:conf} is explicitly  ${\rm SO}(d)$ symmetric, as it reduces to the metric of an $(d-1)$-dimensional sphere. The area of this $\mathbb{S}^{d-1}$ is 
\begin{equation}
\A=v_{d-1}\left(\frac{\sin\theta_1}{\cos\sigma}\right)^{d-1},
\end{equation}
where $v_{d-1}$ is the volume of the sphere of radius $1$. Our first aim in this appendix is to show that this function has a unique extremum, which is a minimum, when the domain of definition of the function is the causal diamond of the Cauchy slices $\Sigma'_{\rm E}$. In a second step, we will show that extra minima and maxima exist on the boundary of the diamond, and can be seen as solutions of an extremization problem. We consider spherically symmetric slices $\Sigma$. The conformal times $\sigma_1$ and $\sigma_2$ at which the screens $\Sigma\vert_{\mathcal{S}_{\rm L}}$ and $\Sigma\vert_{\mathcal{S}_{\rm R}}$ are located are considered to have the same signs. 
In the following, for simplicity, we restrict our discussion to positive $\sigma_1$ and $\sigma_2$, in the generic case where 
\be
0<\sigma_1<\frac{\pi}{2}, \qquad 0<\sigma_2<\frac{\pi}{2}.
\ee


\section{Extremal points of the area function}
\label{A1}
\numberwithin{equation}{section}

To describe the causal diamond, it is convenient to define null coordinates  
\be
x^+=\sigma+\theta_1,\qquad x^-=\sigma-\theta_1.
\label{x+-}
\ee
When $0<\sigma_1+\sigma_2\le \pi/2$, the diamond corresponds to the domain
\be
\left\{\!\!\begin{array}{l}
\dis \phantom{-}{\frac{\pi}{2}}\le x^+\le 2\sigma_2+{\frac{\pi}{2}}\\
\dis -{\frac{\pi}{2}}\le x^-\le 2\sigma_1-{\frac{\pi}{2}}\esp
\end{array}\right..
\label{diam}
\ee
Its boundary is composed of 4 segments labelled as $1,\dots,4$ in Figure~\ref{dL}. 
%
\begin{figure}[ht]
\begin{subfigure}[t]{0.48\linewidth}
\centering
\includegraphics[width=0.9\linewidth]{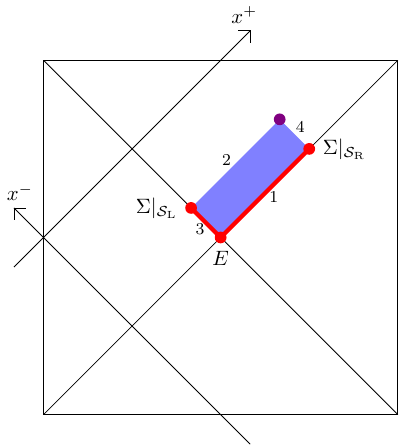}
\caption{\footnotesize When $0<\sigma_1+\sigma_2\le \pi/2$, the diamond has 4 boundary line segments. \label{dL}}
\end{subfigure}
\quad \,
\begin{subfigure}[t]{0.48\linewidth}
\centering
\includegraphics[width=0.9\linewidth]{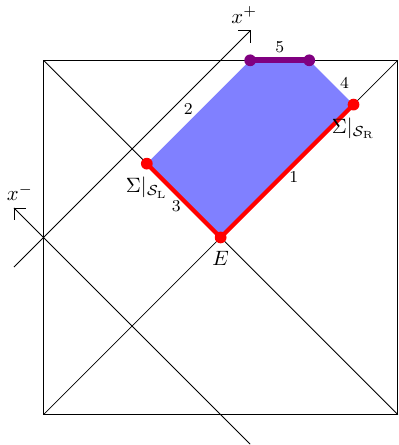}
\caption{\footnotesize When $\pi/2<\sigma_1+\sigma_2<\pi$,  the diamond has 5 boundary line segments.\label{dR}}
\end{subfigure}
    \caption{\footnotesize Minima (in red) and maxima (in purple) of the area of the sphere $\mathbb{S}^{d-1}$ as its position varies in the causal diamond of all Cauchy slices $\Sigma_{\rm E}$.}
    \label{fig:homologous_regions}
\end{figure}
%
Conversely, when $\pi/2<\sigma_1+\sigma_2<\pi$, the domain is further restricted to $\sigma\le \pi/2$, \textit{ie}
\be
x^+ + x^-\le \pi .
\label{extra}
\ee
Its boundary consists of 5 segments labelled as $1,\dots,5$ in Figure~\ref{dR}. 

In both cases, the derivatives of the area, 
\begin{align}
\frac{\partial \!\A}{\partial x^+}&=\phantom{-}v_{d-1}\,\frac{d-1}{2}\,\cos (x^-)\,\frac{(\sin\theta_1)^{d-2}}{(\cos\sigma)^d},\nonumber\\
\frac{\partial \!\A}{\partial x^-}&=-v_{d-1}\,\frac{d-1}{2}\,\cos (x^+)\,\frac{(\sin\theta_1)^{d-2}}{(\cos\sigma)^d},
\label{eq:deriv}
\end{align}
vanish simultaneously in the diamond only at $(x^+,x^-)=(\pi/2,-\pi/2)$ \textit{ie} $(\theta_1,\sigma)=(\pi/2,0)$, which corresponds to the bifurcate horizon, $E$. Hence, the $\mathbb{S}^{d-1}$ at this location is the only one with extremal area, in the sense that it satisfies $\delta\!\A=0$ as the position of the sphere varies infinitesimally in the diamond. This area is minimal, since it increases when $\sigma$ increases from $E$. However, we know that all spheres located on the boundaries~1 and~3 of the diamond  are degenerate. They are therefore other minima, even though they are not extrema, if we mean by this that $\delta\!\A\neq 0$ when the sphere moves towards the bulk of the diamond. Conversely, none of the spheres lying on the remaining boundaries of the diamond is a minimum or a maximum, unless it is  located at the upper tip of the diamond in Figure~\ref{dL}, or along the boundary~5 in Figure~\ref{dR}. The reason for this is that the area increases as we move upward along the boundaries~2 and~4. The upper tip (boundary~5) thus corresponds to a maximum  (degenerate maxima) of the area, which again is  not an extremum (are not extrema). 


\section{Minima \& maxima as solutions of an extremization problem}
\label{A2}

In the following, we show how all minima and maxima of the area present in the exterior causal diamond can be recovered by solving an extremization problem.

\noindent $\bullet$ Let us  first consider the case where $0<\sigma_1+\sigma_2\le \pi/2$. The inequalities~(\ref{diam}) can be imposed by supplementing the area functional with terms proportional to Lagrange multipliers~$\nu_I$, $I\in\{1,2,3,4\}$,
\begin{align}
\hA(x^+,x^-,\nu_I,a_I)=\A(x^+,x^-)&+\nu_1\left(x^-+\frac{\pi}{2}-a_1^2\right)+\nu_2\left(2\sigma_1-\frac{\pi}{2}-x^--a_2^2\right)\nonumber\\
&+\nu_3\left(x^+-\frac{\pi}{2}-a_3^3\right)+\nu_4\left(2\sigma_2+\frac{\pi}{2}-x^+-a_4^2\right)\!.
\label{hA}
\end{align}
Besides the multipliers, $a_I$ are extra variables whose squares are the ``positive distances from the boundary segment $I$ in the Penrose diagram.''\footnote{One may think that the distances $a_1^2$ and $a_2^2$ (and similarly $a_3^2$ and $a_4^2$) should be related from the onset. However, it is important to recover their relation only ``on-shell'' since otherwise $\nu_1$ and $\nu_2$ ($\nu_3$ and $\nu_4$) would simply add in Equation \eqref{hA} and we would lose half of the constraints.} 
The key point is that $\hA$ is defined in a domain without boundary, as $x_\pm$, $\nu_I$, $a_I$ are all spanning $\R$. To find its extrema, \textit{ie} the points in $\R^{10}$ satisfying $\delta\!\hA=0$, we first vary the function $\hA$ with respect to $x_\pm$, which yields\!
\begin{subequations}
\label{x}
 \begin{align}
\nu_4-\nu_3&= \frac{\partial{\A}}{\partial x^+},\label{x+}\\
\nu_2-\nu_1&= \frac{\partial{\A}}{\partial x^-}.\label{x-}
\end{align}
\end{subequations} 
Varying with respect to the Lagrange multipliers leads to the equations 
\begin{subequations}
\label{n}
 \begin{align}
\label{n1}
a_1^2&= x^-+\frac{\pi}{2},\\
a_2^2&=2\sigma_1-\frac{\pi}{2}-x^-,\label{n2}\\
a_3^2&=x^+-\frac{\pi}{2},\label{n3}\\
a_4^2&=2\sigma_2+\frac{\pi}{2}-x^+,\label{n4}
\end{align}
\end{subequations} 
which imply $\hA=\A$ when satisfied.
Finally, the variations with respect to $a_I$ give
\begin{subequations}
\label{a}
\begin{align}
\nu_1a_1&=0,\label{a1}\\
\nu_2a_2&=0,\label{a2}\\
\nu_3a_3&=0,\label{a3}\\
\nu_4a_4&=0.\label{a4}
\end{align}
\end{subequations}
To find all solutions of the system of 10 equations, we organize our discussion based on the location of $(x^+,x^-)\in\R^2$:
 
- When $(x^+,x^-)$ is not on the boundary of the diamond, \Eqs{n} impose it to lie in the bulk of the diamond and determine the values of $a^2_I> 0$, $I\in\{1,2,3,4\}$. As a result, $\nu_I=0$ from \Eqs{a}. However, \Eqs{x} are not satisfied, as seen below Equation \eqref{eq:deriv}. 

- Now consider $(x^+,x^-)$ in the bulk of the boundary segment 1, \textit{ie} not at its endpoints. We have $x^-+\pi/2=0$ and thus $a_1=0$ from Equation \eqref{n1}. This implies that Equation \eqref{a1} is satisfied. \Eqs{n2}--(\ref{n4}) determine $a^2_{2,3,4}> 0$, which imposes $\nu_{2,3,4}=0$ from \Eqs{a2}--(\ref{a4}). In that case, Equation \eqref{x+} is solved for any $x^+$ and Equation \eqref{x-} fixes $\nu_1$. We thus have found degenerate extrema of $\hA$. 

- For $(x^+,x^-)$ in the bulk of the boundary segment 3, a similar analysis can be carried out,  with the same conclusion.

- When $(x^+,x^-)$ is in the bulk of the boundary segment 2, we have $x^-=2\sigma_1-\pi/2$. \Eqs{n3}, (\ref{n4}) determine $a^2_{3,4}> 0$, which implies $\nu_{3,4}=0$ from  \Eqs{a3}, (\ref{a4}). However, since $\cos x^-=\sin(2\sigma_1)\neq 0$, Equation \eqref{x+} is not satisfied. 

- Similarly, when $(x^+,x^-)$ is in the bulk of the boundary segment 4, Equation \eqref{x-} cannot be satisfied. 

- If $(x^+,x^-)$ is at the upper tip of the diamond, \textit{ie}  the intersection of the boundary segments 2 and 4, we have $2\sigma_1-\pi/2-x^-=0$, $2\sigma_2+\pi/2-x^+=0$ and thus $a_{2,4}=0$ from \Eqs{n2}, (\ref{n4}). \Eqs{a2}, (\ref{a4}) are thus satisfied. We also have $a^2_1=2 \sigma_1> 0$, $a^2_3=2 \sigma_2> 0$ from \Eqs{n1}, (\ref{n3}), which implies $\nu_{1,3}=0$ from \Eqs{a1}, (\ref{a3}). \Eqs{x+},(\ref{x-}) then determine $\nu_{4,2}$. We thus have found an extremum of $\hA$. 

- Similarly, when $(x^+,x^-)$ is at another tip $\Sigma\vert_{\mathcal{S}_{\rm L}}$, $\Sigma\vert_{\mathcal{S}_{\rm R}}$ or $E$ of the diamond, one finds an extremum of $\hA$. 

To summarize, we have shown that the minima and the maximum of the function $\A$ in the diamond can be seen as extrema of $\hA$ in a domain of definition of higher dimension but without boundary. Conversely, all extrema of $\hA$ correspond to minima or the maximum of $\A$ in the diamond. However, it should be noticed that the description in terms of $\hA$ introduces an extra degeneracy, since any on-shell value of $a_I^2>0$ admits two roots, $a_I=\pm\sqrt{a_I^2}$. 
In Figure~\ref{dL}, the maximum is depicted as the purple dot, while the degenerate minima correspond to the red segments.


\noindent $\bullet$ In the case where  $\pi/2<\sigma_1+\sigma_2< \pi$, we introduce extra variables $\nu_5$, $a_5$ in $\R$ to impose the condition~(\ref{extra}) and define
\be
\tilde{\text{Area}}(x^+,x^-,\nu_I,\nu_5,a_I,a_5)=\hA(x^+,x^-,\nu_I,a_I)+\nu_5\big(\pi-x^+-x^--a_5^2 \big).
\ee
Solving the equation $\delta\!\tilde{\text{Area}}=0$ in $\R^{12}$ amounts to satisfying 
\begin{subequations}
\label{x'}
 \begin{align}
\nu_5+\nu_4-\nu_3&= \frac{\partial{\A}}{\partial x^+},\label{x+'}\\
\nu_5+\nu_2-\nu_1&= \frac{\partial{\A}}{\partial x^-},\label{x-'}
\end{align}
\end{subequations} 
\Eqs{n} and 
\be
a_5^2=\pi-x^+-x^-,
\label{n5}
\ee
along with \Eqs{a} and   
\be
\nu_5\,a_5=0.
\label{a5}
\ee
When \Eqs{n} and~(\ref{n5}) are fulfilled, we have $\tilde{\text{Area}}=\A$. As before, we organize our discussion based on the location of $(x^+,x^-)\in\R^2$:

- When $(x^+,x^-)$ is away from the line $\sigma=\pi/2$ (see Figure~\ref{dR}), Equation \eqref{n5} imposes it to be below this line and determines $a_5^2> 0$. Therefore,  $\nu_5=0$ from Equation \eqref{a5} and we have $\tilde{\text{Area}}=\hA$. We can then apply all steps of the analysis below Equation \eqref{a4} to the full rectangular diamond defined in Equation \eqref{diam}, where now $\pi/2<\sigma_1+\sigma_2<\pi$, and simply omit the solution located above the line $\sigma=\pi/2$, namely the upper tip of the rectangular diamond. We have thus found degenerate extrema of $\tilde{\text{Area}}$ located along the boundaries 1 and 3 of the diamond depicted in Figure~\ref{dR}.

- Now consider $(x^+,x^-)$ along the line $\sigma=\pi/2$, but not at the boundary endpoints of the segment 5. \Eqs{n} impose it to lie in the bulk of segment 5 and determine $a_I^2>0$, $I\in\{1,2,3,4\}$. Hence, $\nu_I=0$ from Equation \eqref{a}. We also have $x^++x^-=\pi$ and thus $a_5=0$ from Equation \eqref{n5}, while Equation \eqref{a5} is satisfied. Finally, \Eqs{x+'} and~(\ref{x-'}) are equivalent, giving
\be
\nu_5=\lim_{\sigma\to {\frac{\pi}{2}}^-}v_{d-1}\,\frac{d-1}{2}\,\frac{(\sin\theta_1)^{d-1}}{(\cos\sigma)^n}=+\infty.
\ee 
Indeed, since homologous surfaces at infinite distances (and thus infinite sizes) should be allowed in limiting cases, other variables such as Lagrange multipliers should also be allowed to be infinite. We have thus found degenerate extrema of $\tilde{\text{Area}}$. 

- If $(x^+,x^-)$ is at the intersection of the boundary segments 2 and 5, we have $2\sigma_1-\pi/2-x^-=0$, $\pi-x^+-x^-=0$, which determine $x^+$, $x^-$, and thus $a_{2,5}=0$ from \Eqs{n2},~(\ref{n5}). \Eqs{a2},~(\ref{a5}) are thus satisfied. 
\Eqs{n1}, (\ref{n3}), (\ref{n4}) also fix $a^2_{1,3,4}> 0$, which implies $\nu_{1,3,4}=0$ from \Eqs{a1}, (\ref{a3}), (\ref{a4}). \Eqs{x+'}, (\ref{x-'}) then determine $\nu_5=+\infty$, $\nu_2=0$. We thus have found an extremum of $\tilde{\text{Area}}$.  

- Similarly, $(x^+,x^-)$ at the intersection of the boundary segments 4 and 5 leads to an extremum of $\tilde{\text{Area}}$. 

In conclusion, all minima and maxima of the function $\A$ are recovered as extrema of the function $\tilde{\text{Area}}$.

\chapter{Extremal, C-extremal, and maximin surfaces}
\label{app:proofs}

\section{Maximin surfaces in static patch holography}

In the context of AdS/CFT, the HRT proposal is proven to be equivalent to a maximin procedure~\cite{Wall:2012uf}. Here, we consider the restricted maximin procedure~\cite{Marolf:2019bgj, Grado-White:2020wlb}.
\begin{definition}
    On any achronal slice $\Sigma_I$ such that $A\in\partial\Sigma_I$, which is not necessarily complete, $\gamma_{\min}(A;\Sigma_I)$ is the codimension-two surface $D(\Sigma_I)$-homologous to $A$ with minimal area.
\end{definition}
This definition differs from that of~\cite{Wall:2012uf} by the fact the achronal slices considered here are not complete. This leads to the possibility that $\gamma_{\min}(A;\Sigma)$ is not an extremum with respect to variations along $\Sigma_I$. 

\begin{definition}
\label{def:maximin}
\textbf{[Maximin]}
    Let $\Sigma\vert_{\mathcal{S}}$ be a spacelike slice of some holographic screen $\mathcal{S}$ and $A\in \Sigma\vert_{\mathcal{S}}$ some subsystem. Let $D$ denote the bulk region whose spatial boundary is $\Sigma\vert_{\mathcal{S}}$. The maximin surface associated with $A$, denoted $\gamma_{m}(A;D)$, is the surface $\gamma_{\min}(A;\Sigma')$ which is maximal when varying over all slices $\Sigma'$ such that $\partial\Sigma' = \Sigma\vert_{\mathcal{S}}$. The achronal slice on which $\gamma_{m}(A;D)$ is minimal is denoted $\Sigma_{m}(A;D)$. Additionally, $\gamma_{m}(A;D)$ must be stable in the sense that considering a small variation of $\Sigma_{m}(A;D)$, the new slice must have a local minimum (possibly non-extremal) in the neighborhood of $\gamma_{m}(A;D)$, with no greater area.
\end{definition}

We will not prove the existence of maximin surfaces, as it goes well beyond the scope of this work. Note that this was proven for certain classes of asymptotically AdS spacetimes~\cite{Wall:2012uf, Marolf:2019bgj}. We will always that such surfaces exist, which does not seem to be an extravagant assumption for at least the simple examples we consider in this thesis.

\section{Proofs of (in)equivalence}

First, we show a few properties to specify the origin of the inequivalence between extremization, C-extremization, and maximin. We then show the equivalence of the three prescriptions in the interior of the holographic screen of an observer.
\begin{lemma}
\label{lem:minimin}
    A codimension-two spacelike surface $\gamma_{\min}(A;D_I)$ that has the smallest area in bulk region $D_I$ and is $D_I$-homologous to $A$ is the minimin surface of $A$ in region $D_I$, and conversely.\footnote{Here minimin is defined by changing the maximization of Definition~\ref{def:maximin} to a minimization.}
\end{lemma}
\begin{proof}
    The smallest area surface $\gamma_{\min}(A;D_I)$ must be $\gamma_{\min}(A;\Sigma_I)$ for any $\Sigma_I$ containing it. The minimal area surface on every other slice cannot be smaller than $\gamma_{\min}(A;D_I)$ otherwise it would itself be the smallest area surface. Hence $\gamma_{\min}(A;D_I)$ is the minimin surface.
\end{proof}

\begin{proposition}[Maximin+Extremality along $\Sigma_m$ $\Rightarrow$ Extremality]
\label{th:ext=maximin}
    A surface $\gamma_I\in D_I$ is the extremal surface of $A$ if it is the maximin surface of $A$ and is extremal along $\Sigma_{m}(A;D_I)$. The converse also holds only if there exists such a maximin surface extremal along $\Sigma_m$.
\end{proposition}
\begin{proof}
    The proof of Theorem~$15$ of~\cite{Wall:2012uf} applies directly here when the maximin surface is extremal along $\Sigma_m$. The main assumptions in~\cite{Wall:2012uf} are that the bulk spacetime is classical, smooth, satisfies the null energy condition, and is globally hyperbolic. All these conditions are considered to be satisfied here. The proof does not depend on the AdS geometry and the only assumption that is nontrivial here is the existence of an extremum on every slice since the slices are not global. To avoid this difficulty, we impose explicitly the extremality along $\Sigma_{m}(A;D_I)$. However, even though we assume the maximin surface exists, there might be no maximin surface that is extremal along the corresponding $\Sigma_m(A;D_I)$. In such cases, the converse of the proposition is trivially not satisfied, and the extremal surface is a minimax surface.
\end{proof}
An example of the converse of Proposition~\ref{th:ext=maximin} being false is the extremal surface in the exterior region associated with one of the screens. The extremal surface homologous to a screen in the exterior is the bifurcate horizon. However, the stable maximin surface associated with the screen is always the screen itself or the complementary screen. The bifurcate horizon is an unstable maximin surface, but a stable minimax surface.

\begin{theorem}
\label{ineq}
     Maximin, extremization, and C-extremization are inequivalent. Moreover, none of them implies another.
\end{theorem}
\begin{proof}
    Minimin surfaces are not maximin surfaces in general, so a C-extremal surface is not always a maximin surface. Conversely, if the maximin surface is not extremal along $\Sigma_m$ and does not lie on $\partial D$, it cannot be C-extremal. Moreover, as we already noted, maximin surfaces can fail to be extremal along $\Sigma_{m}$, so being a maximin surface does not imply extremality. Conversely, Proposition~\ref{th:ext=maximin} states that extremality does not imply being a maximin surface. Finally, C-extremality does not imply extremality, and conversely, by Definition~\ref{def:Rext}.
\end{proof}
Theorem~\ref{ineq} implies that the C-extremization and maximin prescriptions are not appropriate analogs of their AdS counterparts, as they fail to be equivalent. This is not necessarily a problem in the context of the connected wedge theorem of Chapter~\ref{ch:CWT}, since the proof only uses the fact that the extremal surface must be a minimum on at least one slice. This is the case for all C-extremal surfaces and maximin surfaces. The only important issue is that C-extremal surfaces are not always extremal, meaning that we cannot always apply the focusing theorem. In the rest of this section, we show that the tension between the prescriptions is absent in the interior of the screen associated with any observer. The analysis of the exterior region is deferred to future work.

For the rest of this section, we denote by $\Sigma$ any achronal slice intersecting an observer's worldline and such that $\partial\Sigma=\Sigma\vert_{\mathcal{S}}$ with $\mathcal{S}$ the holographic screen associated with this observer.
\begin{lemma}
\label{prop:maximax}
    The surface $\Sigma\vert_{\mathcal{S}}$ is not a non-extremal minimal surface on any $\Sigma$.
\end{lemma}
\begin{proof}
    Let $\Sigma$ be an achronal slice with $\partial\Sigma=\Sigma\vert_{\mathcal{S}}$. Let us construct an arbitrarily closed codimension-two surface $\tilde\chi \in \Sigma$ arbitrary close to $\Sigma\vert_{\mathcal{S}}$, such that every point of $\tilde\chi$ is infinitesimally close to $\Sigma\vert_{\mathcal{S}}$. Consider a null hypersurface $\mathcal{L}(\Sigma\vert_{\mathcal{S}})$ emanating from $\Sigma\vert_{\mathcal{S}}$ and directed towards the interior of $\mathcal{S}$. From Definition~\ref{def:ap-sp}, $\mathcal{L}(\Sigma\vert_{\mathcal{S}})$ is of non-positive expansion—is a lightsheet—since it is in the interior of the screen. The lightsheets discussed here are depicted in Figure~\ref{fig:proof}. Consider an arbitrary codimension-two surface $\chi\in\mathcal{L}(\Sigma\vert_{\mathcal{S}})$, such that on each generator of the lightsheet $\mathcal{L}(\Sigma\vert_{\mathcal{S}})$, there is exactly one point of $\chi$. Construct the congruence of lightrays $\mathcal{L}(\chi)$ emanating orthogonally from $\chi$ and directed towards the interior of $\mathcal{S}$. This congruence is also of non-positive expansion from Definition~\ref{def:ap-sp}. We define $\tilde\chi = \Sigma \cap \mathcal{L}(\chi)$, which is a closed codimension-two surface on $\Sigma$. $\mathcal{L}(\Sigma\vert_{\mathcal{S}})$ and $\mathcal{L}(\chi)$ being of non-positive expansion, $\tilde\chi$ must have a smaller area than $\Sigma\vert_{\mathcal{S}}$. The procedure could be followed the other way around by starting from $\tilde{\chi}$. We showed that $\Sigma\vert_{\mathcal{S}}$ has an area greater than or equal to that of any of its infinitesimal deformation on $\Sigma$. Hence, if it is a minimal surface on some slice, it must saturate the bound and be an extremum along this slice.
\end{proof}

\begin{figure}[ht]
    \centering
    \includegraphics[width=0.5\linewidth]{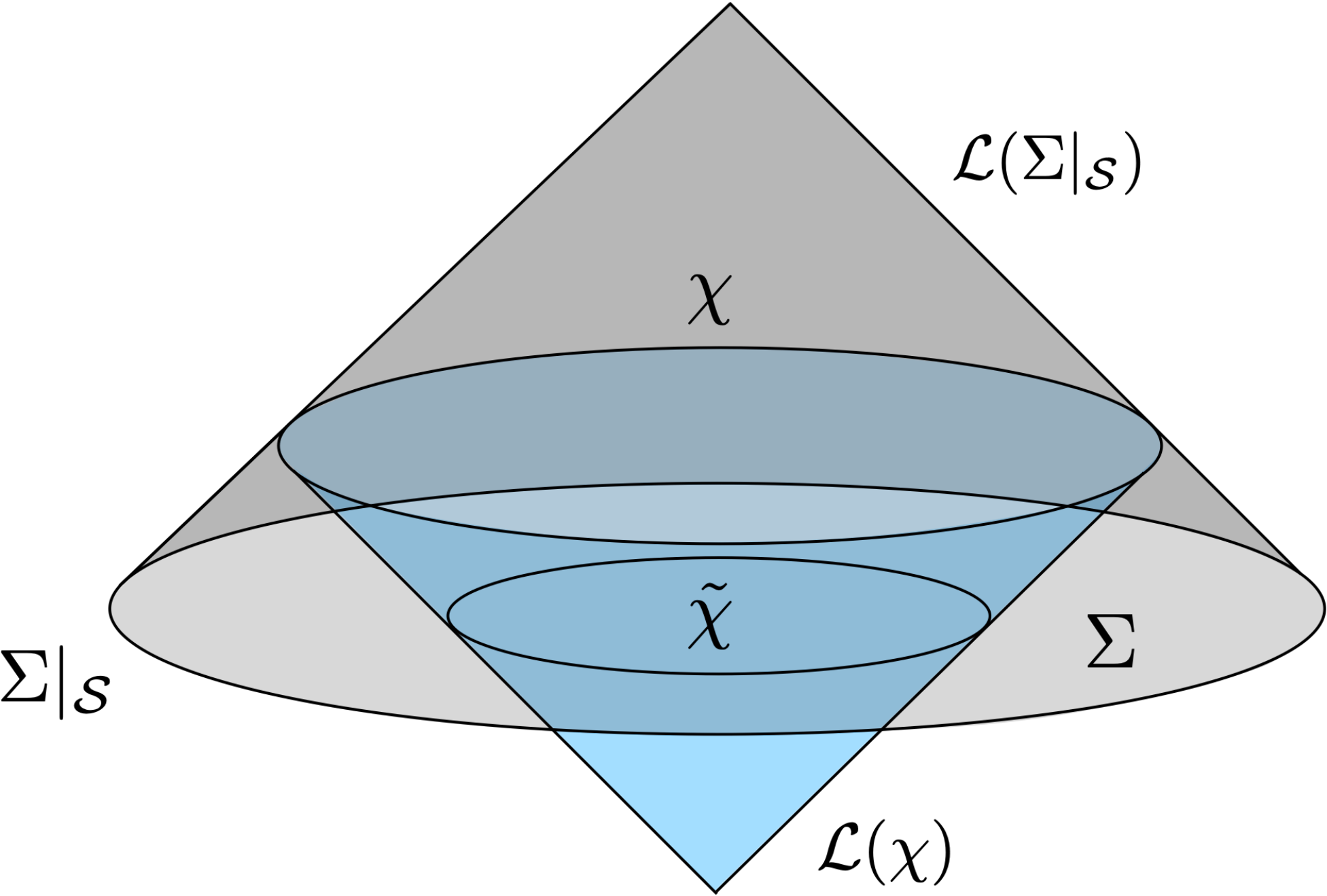}
    \caption{\footnotesize Schematic picture of the proof of Lemma~\ref{prop:maximax}. One constructs an arbitrary closed surface infinitesimally close to $\Sigma\vert_{\mathcal{S}}$ on $\Sigma$ by constructing the lightsheet $\mathcal{L}(\Sigma\vert_{\mathcal{S}})$ (in gray), and then a second lightsheet $\mathcal{L}(\chi)$ (in blue) emanating from an arbitrary closed surface $\chi \in \mathcal{L}(\Sigma\vert_{\mathcal{S}})$ infinitesimally close to $\Sigma\vert_{\mathcal{S}}$. The intersection $\tilde\chi = \mathcal{L}(\chi)\cap\Sigma$ is always smaller than $\Sigma\vert_{\mathcal{S}}$.}
    \label{fig:proof}
\end{figure}

\begin{theorem}
\label{th:patch}
    Maximin surfaces, extremal surfaces, and C-extremal surfaces are equivalent in the interior of the holographic screen of an observer.
\end{theorem}
\begin{proof}
    In this proof, we denote $D=D(\Sigma)$ and $A\in\Sigma\vert_{\mathcal{S}}$.
    \begin{itemize}
        \item \textbf{Maximin $\Rightarrow$ Extremal:} From Proposition~\ref{th:ext=maximin}, it is sufficient to show that all maximin surfaces $\gamma_{\rm m}(A;D)$ are extremal along $\Sigma_{m}(A;D)$. Consider a surface $\gamma$, $D$-homologous to $A$, and infinitesimally close to it. Definition~\ref{def:homol} implies that $\partial A = \partial \gamma$, and $\partial A^c= \partial A$ where $A^c$ is the complement of $A$ on $\Sigma\vert_{\mathcal{S}}$. Hence, $A^c\cup \gamma$ is a closed surface infinitesimally close to $\Sigma\vert_{\mathcal{S}}$, and it lies on some Cauchy slice $\Sigma$. By Lemma~\ref{prop:maximax},
    \begin{equation}
    \label{eq:ineq}
        \text{area}(A^c\cup \gamma) \leq \text{area}(\Sigma\vert_{\mathcal{S}}) = \text{area}(A^c \cup A),
    \end{equation}
    which implies $\text{area}(\gamma)\leq \text{area}(A)$. Hence, $A$ cannot be a non-extremal minimal surface on any $\Sigma$. An analogous argument applies to a surface $B$ infinitesimally close to $A^c$. $A$ and $A^c$ are therefore not the non-extremal minima, $D$-homologous to $A$ on any slice $\Sigma$. This extends to any surface $\gamma_A$, $D$-homologous to $A$, that has a positive measure subset on $A$ and/or $A^c$. The above arguments show that pushing $\gamma_A \cap \Sigma\vert_{\mathcal{S}}$ in the interior decreases the area, such that $\gamma_A$ is not a non-extremal minimal surface on any slice $\Sigma$.

    A surface that is minimal on some slice $\Sigma$ can only be non-extremal if it has a measurable subsystem on $\mathcal{S}$. However, we showed that such a surface cannot be a non-extremal minimum. Hence, any minimal surface on a slice $\Sigma$ must be extremal along $\Sigma$.\footnote{The minimal surface always exists by the extreme value theorem, see proof of Theorem~$9$ in~\cite{Wall:2012uf}, and if it is not on the boundary of $\Sigma$, it is an extremum along $\Sigma$.} This includes the maximin.

    \item \textbf{Extremal $\Rightarrow$ Maximin} is implied by Proposition~\ref{th:ext=maximin} and the above proof that all maximin surfaces are extremal along $\Sigma_m$.
    \item \textbf{Extremal $\Leftrightarrow$ C-extremal:} We proceed by contradiction and assume there exists a surface $\gamma_{\rm min}(A;D)$ $D$-homologous to $A$ with a positive measure subsystem on $\partial D$ while being the smallest area surface in $D$. By Lemma~\ref{lem:minimin}, this surface is a minimin. Any infinitesimal deformation of $\gamma_{\rm min}(A;D)$ along the null direction directed towards the interior of $D$ must decrease the area by Definition~\ref{def:ap-sp}. This is in contradiction with the assumption that $\gamma_{\rm min}(A;D)$ is a minimin.
    \end{itemize}
\end{proof}
These results imply that non-extremal maximin surfaces always arise in the exterior region $D_E$. An example of this is the maximin surface associated with $\Sigma\vert_{\mathcal{S}_{\rm L}}$ or $\Sigma\vert_{\mathcal{S}_{\rm R}}$ in the exterior. It is easy to see that this surface is either $\Sigma\vert_{\mathcal{S}_{\rm L}}$ or $\Sigma\vert_{\mathcal{S}_{\rm R}}$. In the case where the holographic screens are located on the cosmological horizons of pure de Sitter space, there is an infinite number of maximin surfaces on the horizon, see~\cite{Franken:2023pni}, but the only ones that satisfy the stability conditions are $\Sigma\vert_{\mathcal{S}_{\rm L}}$ and $\Sigma\vert_{\mathcal{S}_{\rm R}}$. The only closed extremal surfaces in de Sitter spacetime are the empty surface and the bifurcate horizon, such that the maximin surface is not extremal in this example.\footnote{Except if we consider achronal slices that intersect the horizons at the bifurcate horizon when the holographic screen is the cosmological horizon.} In particular, these surfaces are not extremal along any spacelike Cauchy slice. A similar analysis applies to spatial subsystems of the holographic screens in dS$_3$, which are one-dimensional arcs or a union of them. In particular, the extremal surface $\gamma_{\rm e}(A;D_E)$ associated with such a subsystem does not exist in pure dS$_3$, requiring the use of the maximin or the C-extremization procedure~\cite{Franken:2023pni}.

\chapter{Scale factor in closed FLRW cosmology}
\label{app:FLRW_cosmo}
\numberwithin{equation}{chapter}

In this appendix, we compile salient results concerning the FLRW cosmological evolution of a closed universe filled with a perfect fluid. 

For a closed universe, the metric of an $(n+1)$-dimensional FLRW cosmology takes the form 
\be
d s^2=-N^2(x^0)(d x^0)^2+a^2(x^0)d\Omega_n^2,
\label{me}
\ee
where $N(x^0)$ is the lapse function, $a(x^0)$ is the scale factor, and $d\Omega_n^2$ is the metric of the sphere S$^n$ of radius 1. For $n\ge 2$, the metric of S$^n$ is related to that of S$^{n-1}$ as follows:
\be
d\Omega_n^2=\frac{d r^2}{1-r^2}+r^2 d\Omega_{n-1}^2.
\ee
After performing the change of coordinate
\be
d\theta={d r\over \sqrt{1-r^2}},
\ee
\Eq{me} takes the alternative form
\begin{equation}
   d s^2=-N^2(x^0)(d x^0)^2+a^2(x^0)\big[d\theta^2+\sin^2(\theta)d\Omega_{n-1}^2\big].
\end{equation}

The Friedmann equations of motion for $N$ and $a$ are, in the gauge $N\equiv 1$,
\begin{align}
{n(n-1)\over 2} \left[\Big({\dot a\over a}\Big)^2+{1\over a^2}\right]&=\phantom{-}8\pi G\, \rho,\label{f1}\\
(n-1)\,{\ddot a\over a}+{(n-1)(n-2)\over 2} \left[\Big({\dot a\over a}\Big)^2+{1\over a^2}\right]&=-8\pi  G\,p, \label{f2}
\end{align}
where dots stand for derivatives with respect to cosmological time $t$, while $\rho$ and $p$ are respectively the energy density and pressure in the universe. The second equation can be replaced by
\be
(n-1)\,{\ddot a\over a}= -8\pi G\Big[{n-2\over n}\, \rho+p\Big].
\label{accel}
\ee
Alternatively, taking the time derivative of \Eq{f1}, one can show that \Eq{f2} can be replaced by 
\be
\dot \rho+n\, {\dot a\over a}\, (\rho+p)=0.
\label{conser}
\ee
This equation is nothing but the continuity equation for the stress-energy tensor, which in the present case expresses the conservation of energy in a universe undergoing adiabatic evolution. 

Let us now assume that the perfect fluid filling the entire universe satisfies the state equation~(\ref{seq}), where $w\in[-1,1]$ is a constant fluid index. In this case,  \Eq{conser} can be readily integrated, giving 
\be
\rho={C\over a^{n(1+w)}},
\ee
where $C> 0$ is a constant. Moreover, \Eq{accel} reduces to 
\be
(n-1)\, {\ddot a\over a}=8\pi G(w_{\rm c}-w)\rho,~~\quad \mathrm{where}~~\quad w_{\rm c}=-1+{2\over n}\in(-1,0].
\ee
This shows that as a function of cosmological time $t$, the evolution is accelerating when $-1\le w<w_{\rm c}$ and decelerating when $w_{\rm c}<w\le 1$. When the fluid index is critical, $w=w_{\rm c}$, the evolution is linear in $t$. 

In order to find explicitly the evolution of the scale factor, it proves relevant to work in the conformal gauge, $N=a$. The Friedmann equation~(\ref{f1})  becomes 
\be
{n(n-1)\over 2} \left[\Big({a'\over a}\Big)^2+1\right]=8\pi G\, a^2\rho,
\label{frieta}
\ee
where primes denote derivatives with respect to conformal time $\eta$. Substituting for $\rho$, we obtain
\be
\mbox{for $w\neq w_{\rm c}:$} \quad \Big({a'\over a}\Big)^2=\Big({a\over a_0}\Big)^{n(w_{\rm c}-w)}-1,~~\quad \mathrm{where} ~~\quad a_0=\left({n(n-1)\over 16\pi C}\right)^{1\over n(w_{\rm c}-w)}.
\ee
By defining
\be
{a\over a_0}=A^{-{2\over n(w_{\rm c}-w)}},
\ee
the above equation simplifies to
\be
\Big({d A\over d u}\Big)^2+A^2=1, ~~\quad\mathrm{where}~~\quad u={n(w_{\rm c}-w)\over 2}\,\eta,
\ee
leading to the solution
\be
A(u)=|\:\!\!\sin u|, ~~\quad u\in[0,\pi].
\ee
As a result, the scale factor reads
\be
a(\eta)=a_0\!\left(\sin{\eta\over |\gamma|}\right)^\gamma, ~~\quad \eta\in\big[0,|\gamma|\pi\big], 
\ee
where we have defined
\be
\gamma={2\over n(w-w_{\rm c})}.
\ee

For completeness, let us mention that the Friedmann equation~(\ref{frieta}) becomes
\be
\mbox{for $w= w_{\rm c}:$} ~~\quad\Big({a'\over a}\Big)^2={16\pi C\over n(n-1)}-1,
\ee
which admits solutions when the right-hand side is non-negative. When this is the case, one obtains 
\be
a(\eta)=e^{\pm \eta \sqrt{{16\pi C\over n(n-1)}-1}}, ~~\quad \eta\in \R,
\label{acri}
\ee
for either choice of constant sign $\pm$.

\addtocontents{toc}{\protect\end{adjustwidth}}

\providecommand{\href}[2]{#2}\begingroup\raggedright\endgroup

\newpage
\thispagestyle{empty} 
\
\newpage

\includepdf[pages=-]{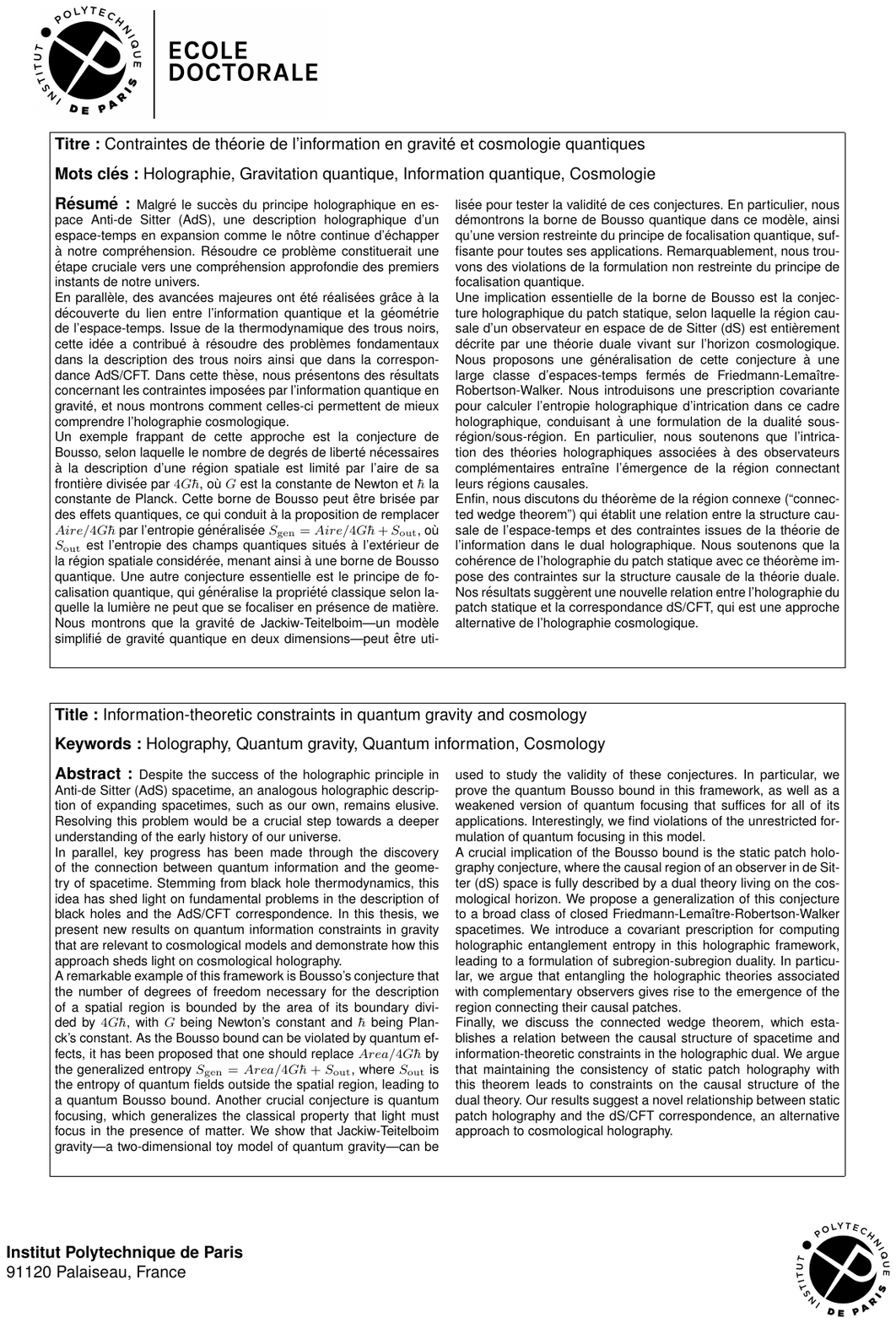}

\end{document}